\documentclass[british,english,aps,prx,twocolumn,floatfix,letterpaper,superscriptaddress,amssymb,eprint]{revtex4-2}
\usepackage[T1]{fontenc}
\usepackage[utf8]{inputenc}
\usepackage{color}
\usepackage{babel}
\usepackage{verbatim}
\usepackage{float}
\usepackage{units}
\usepackage{dsfont}
\usepackage{amsmath}
\usepackage{amssymb}
\usepackage{graphicx}
\usepackage[unicode=true,pdfusetitle,
 bookmarks=true,bookmarksnumbered=false,bookmarksopen=false,
 breaklinks=false,pdfborder={0 0 1},backref=false,colorlinks=false]
 {hyperref}

\makeatletter
\usepackage{braket}
\usepackage{lmodern}
\usepackage{dsfont}
\usepackage{graphicx}
\usepackage{tikz}
\usepackage{mathtools}
\usepackage{tensor}
\definecolor{magenta}{HTML}{FF00FF}
\definecolor{green}{HTML}{00C000}
\definecolor{purple}{HTML}{800080}
\hypersetup{colorlinks,allcolors=blue,breaklinks=true}
\allowdisplaybreaks[1]

\makeatother

\begin{document}
\global\long\def\cev#1{\reflectbox{\ensuremath{\vec{\reflectbox{\ensuremath{#1}}}}}}%

\global\long\def\sidescripts#1#2#3#4#5{\tensor*[_{#2}^{#1}]{#3}{_{{#5}}^{{#4}}}}%

\global\long\def\ar{\vec{\alpha}}%
\global\long\def\al{\cev{\alpha}}%
\global\long\def\arb{\vec{\bar{\alpha}}}%
\global\long\def\alb{\cev{\bar{\alpha}}}%

\global\long\def\br{\vec{\beta}}%
\global\long\def\bl{\cev{\beta}}%
\global\long\def\brb{\vec{\bar{\beta}}}%
\global\long\def\blb{\cev{\bar{\beta}}}%

\global\long\def\gr{\vec{\gamma}}%
\global\long\def\gl{\cev{\gamma}}%
\global\long\def\grb{\vec{\bar{\gamma}}}%
\global\long\def\glb{\cev{\bar{\gamma}}}%

\global\long\def\lambr{\vec{\lambda}}%
\global\long\def\lambl{\cev{\lambda}}%
\global\long\def\lambrb{\vec{\bar{\lambda}}}%
\global\long\def\lamblb{\cev{\bar{\lambda}}}%

\global\long\def\thetar{\vec{\theta}}%
\global\long\def\thetal{\cev{\theta}}%
\global\long\def\thetarb{\vec{\bar{\theta}}}%
\global\long\def\thetalb{\cev{\bar{\theta}}}%

\global\long\def\varthetar{\vec{\vartheta}}%
\global\long\def\varthetal{\cev{\vartheta}}%
\global\long\def\varthetarb{\vec{\bar{\vartheta}}}%
\global\long\def\varthetalb{\cev{\bar{\vartheta}}}%

\global\long\def\p{\prime}%
\global\long\def\pp{\prime\prime}%
\global\long\def\xin{x_{\textrm{in}}}%

\global\long\def\k#1{\ket{#1}}%
\global\long\def\b#1{\bra{#1}}%
\global\long\def\bk#1{\braket{#1}}%
\global\long\def\KK#1{\left\Vert #1\right\rangle }%
\global\long\def\BB#1{\left\langle #1\right\Vert }%

\global\long\def\rmn{\mathfrak{r}}%
\global\long\def\lmn{\mathfrak{l}}%
\global\long\def\mn{\mathfrak{a}}%

\global\long\def\Res#1{\underset{#1}{\mathrm{Res}}}%

\global\long\def\Gin{\Gamma^{\textrm{in}}\left(\omega\right)}%
\global\long\def\GLR{\mathcal{G}_{\mathcal{LR}}}%
\global\long\def\GRL{\mathcal{G}_{\mathcal{RL}}}%
\global\long\def\GLRz{\mathcal{G}_{0}}%
\global\long\def\K{\textrm{K}}%
\global\long\def\bsg{\textrm{bsG}}%

\global\long\def\maxbreath{\left\lceil \frac{1}{z}\right\rceil -2}%

\global\long\def\id{\mathrm{d}}%
\global\long\def\ii{\mathrm{i}}%
\global\long\def\P{\mathrm{P}}%

\global\long\def\eps{\epsilon}%
\global\long\def\beps{\bar{\epsilon}}%

\title{Inelastic decay from integrability}
\author{Amir Burshtein}
\email{burshtein2@mail.tau.ac.il}

\affiliation{Raymond and Beverly Sackler School of Physics and Astronomy, Tel Aviv
University, Tel Aviv 6997801, Israel}
\author{Moshe Goldstein}
\affiliation{Raymond and Beverly Sackler School of Physics and Astronomy, Tel Aviv
University, Tel Aviv 6997801, Israel}
\begin{abstract}
A hallmark of integrable systems is the purely elastic scattering
of their excitations. Such systems possess an extensive number of
locally conserved charges, leading to the conservation of the number
of scattered excitations, as well as their set of individual momenta.
In this work, we show that inelastic decay can nevertheless be observed
in circuit QED realizations of integrable boundary models. We consider
the scattering of microwave photons off impurities in superconducting
circuits implementing the boundary sine-Gordon and Kondo models, which
are both integrable. We show that not only inelastic decay is possible
for the microwave photons, in spite of integrability, and thanks to
a nonlinear relation between them and the elastically-scattered excitations,
but also that integrability in fact provides powerful analytical tools
allowing to obtain exact expressions for response functions describing
the inelastic decay. Using the framework of form factors, we calculate
the total inelastic decay rate and elastic phase shift of the microwave
photons, extracted from a 2-point response function. We then go beyond
linear response and obtain the exact energy-resolved inelastic decay
spectrum, using a novel method to evaluate form factor expansions
of 3-point response functions, which could prove useful in other applications
of integrable quantum field theories. \textcolor{black}{Our results
could be relevant to several recent photon splitting experiments,
and in particular to recent experimental works that provide evidence
for the elusive Schmid-Bulgadaev dissipative quantum phase transition.}
\end{abstract}
\maketitle

\section{Introduction}

Integrability entails exceptional consequences. An extensive number
of local conservation laws, the defining feature of an integrable
system, are at the heart of the celebrated Bethe ansatz \citep{bethe_zur_1931},
which allows for rare exact solutions of 1-dimensional interacting
many-body quantum systems. Its fundamental ingredient is a set of
elementary excitations whose scattering is purely elastic --- in
any scattering process, the number of excitations is conserved, as
well as the set of their individual energies and momenta. This striking
feature has drawn large theoretical interest ever since Bethe's seminal
work, promoting extensions of the method which apply to a large variety
of discrete and continuous 1-dimensional models \citep{lieb_exact_1963,yang_one-dimensional_1966,yang_exact_1967,andrei_solution_1983,caux_bethe_2014,baxter_exactly_1985,korepin_quantum_1993,zamolodchikov_factorized_1979,faddeev_quantum_1995,mussardo_statistical_2020}.

Remarkably, integrability is no longer just a theoretical curiosity.
Recent advances in the fabrication techniques of quantum simulators
have enabled the experimental realization of integrable systems, leading
to an interplay between experiment and theory. The past decade has
seen several theoretical breakthroughs concerning the equilibrium
and out-of-equilibrium dynamics of integrable systems \citep{caux_quench_2016,bertini_transport_2016,castro-alvaredo_emergent_2016,calabrese_introduction_2016}
that go hand-in-hand with surprising experimental observations \citep{kinoshita_quantum_2006,guan_fermi_2013,langen_experimental_2015,bouchoule_generalized_2022}.
Essentially, integrability gives rise to counterintuitive experimental
measurements which push the boundaries of well-established theoretical
frameworks, and improve our understanding of the role of integrability
in an ever-growing list of mechanisms.

Experiments on the quantum simulation of many-body quantum models,
both integrable and non-integrable, have been mostly restricted to
the realm of cold atom systems. Another possible platform for quantum
simulation is that of superconducting circuits. The rapidly evolving
field of circuit quantum electrodynamics (cQED) deals with the simulation
of interacting models by means of Josephson junctions or their flux-tunable
counterparts, the SQUIDs. Yet experimental realizations have up to
now been quite limited, and mostly dedicated the use of superconducting
circuits to the implementation of nonlinear bulk models exhibiting
non-ergodic behavior \citep{xu_MBL_2018,zha_MBL_2020,gong_MBL_2021,guo_MBL_2021,google_MBL_2022,roy_quantum_2019,roy_quantum_2021,roy_soliton_2023}.
The field of cQED reveals its true strength in the simulation of quantum
impurity models. The intrinsically large kinetic inductance of Josephson
junctions allows one to design transmission lines with impedances
on the order of the resistance quantum \citep{kuzmin_quantum_2019,leger_observation_2019,puertas_martinez_tunable_2019},
providing an environment for photons with an effective fine structure
constant of order unity. Furthermore, the nonlinearity of the junctions
provides the means to realize many types of quantum impurities \citep{goldstein_inelastic_2013,le_hur_kondo_2012,kuzmin_inelastic_2021,mehta_down-conversion_2023}
which are strongly-coupled to the photonic environment. Single-photon
spectroscopy then provides highly sensitive tools to investigate the
fine details of the boundary models of interest, across a wide range
of parameters, and probe fundamental phenomena in those many-body
systems.

\begin{figure*}[t]
\begin{centering}
\includegraphics[width=0.8\paperwidth]{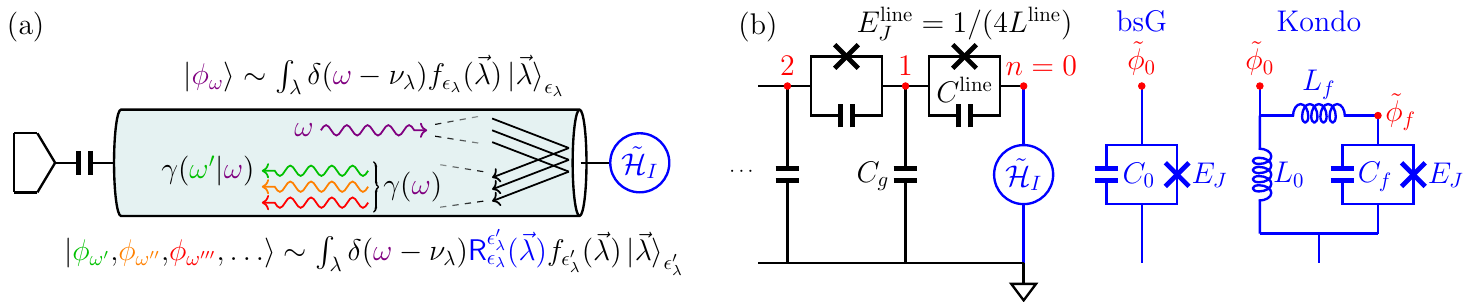}
\par\end{centering}
\caption{\textcolor{black}{\label{fig:cQED}(a) An incoming photon with frequency
$\omega$, injected from the antenna on the left and propagating through
the transmission line in the center, may decay inelastically as it
scatters off the impurity on the right, in spite of the purely elastic
reflection of the fundamental excitations of the integrable models.
Each photon may be represented as a combination of eigenstates composed
of excitations $\protect\k{\protect\lambr}_{\epsilon_{\lambda}}$
with types $\epsilon_{\lambda}$ and rapidities $\protect\lambr$,
where the weights are determined by the form factors $f_{\epsilon_{\lambda}}(\protect\lambr)$,
and the total energy of the excitations in each eigenstate, $\nu_{\lambda}\sim\sum_{i}e^{\lambda_{i}}$,
is equal to the photon frequency $\omega$ (see Section \ref{sec:integrable}
for notations and definitions). The excitations scatter elastically
off the boundary (quantum impurity), picking up phases determined
by the reflection matrix $\mathsf{R}_{\epsilon_{\lambda}}^{\epsilon_{\lambda}^{\protect\p}}(\protect\lambr)$,
such that the outgoing combination no longer represents a single-photon
state, but rather a multi-photon state. The measured observables ---
the total inelastic decay rate $\gamma\left(\omega\right)$ and the
energy-resolved inelastic decay spectrum $\gamma\left(\omega^{\protect\p}\lvert\omega\right)$,
as well as the elastic phase shift $\delta\left(\omega\right)$ (not
depicted here) --- all shed light on the fundamental properties of
the impurity models. (b) Implementation of Eq. (\ref{eq:H_discrete})
with the bsG and Kondo impurities (Eqs. (\ref{eq:HIbSG}) and (\ref{eq:HIK}),
respectively) in a cQED setup. The array of Josephson junctions and
capacitors implements a high-impedance transmission line, thanks to
the kinetic inductance of the Josephson junctions.}}
\end{figure*}

The starting point of this work is an experimentally observed phenomenon
that is seemingly at odds with integrability. Recent experiments have
demonstrated that photons propagating in a high-impedance transmission
line, setting an environment with a large effective light-matter coupling,
scatter inelastically off a quantum impurity with a very high probability
\citep{kuzmin_inelastic_2021,mehta_down-conversion_2023}. These observations
were reproduced in Ref. \citep{leger_revealing_2022}. It appears
that such photon splitting, depicted in Fig. \ref{fig:cQED}, has
nothing to do with integrability; as stated above, scattering processes
in integrable systems are highly restricted by the extensive number
of local conservation laws, forbidding particle production. The flexibility
of the circuit elements provides us with tools to check this assumption,
by tuning the impurity parameters to those of integrable boundary
models. Strikingly, experiments show that inelastic decay persists
even when the impurity parameters are pushed towards those of the
boundary sine-Gordon (bsG) model \citep{kuzmin_observation_2023},
which is known to be integrable \citep{ghoshal_boundary_1994}. This
appears to contradict the defining feature of integrability, and raises
a fundamental question --- how can inelastic decay emerge in a system
governed by purely elastic scattering rules?

Previous theoretical treatments of such scattering experiments used
different forms of weak/strong coupling expansions \citep{le_hur_kondo_2012,goldstein_inelastic_2013,gheeraert_particle_2018,houzet_critical_2020,burshtein_photon-instanton_2021,leger_revealing_2022,mehta_theory_2022,houzet_microwave_2023}
which, despite providing successful quantitative predictions, are
intrinsically limited in two manners. First and foremost, weak or
strong coupling expansions do not address the apparent discrepancy
between integrability and inelastic decay; in order to settle it,
one needs to explicitly use the framework of integrable systems, and
show how the purely elastic excitations underlying the theory can
give rise to the observed photon splitting. We note that inelastic
effects have also been considered both theoretically and experimentally
in integrable strongly correlated electronic quantum impurity systems
\citep{mallet_scaling_2006,borda_theory_2007,freton_out_of_eq_2014},
yet a direct consideration of the elastically scattered excitations
from the integrability picture has been missing. Second, in case the
impurity term in the Hamiltonian is relevant, perturbation theory
and strong coupling expansions are only applicable either above or
below a certain renormalization group (RG) energy scale, regardless
of the impurity's strength. Wilson's numerical RG (NRG) can in principle
cover the entire frequency range, but is not easy to apply accurately
for bosonic baths \citep{bera_dynamics_2016}. This again calls for
the use of the integrability framework, as it provides powerful analytical
tools allowing for an exact solution at all frequencies, linking the
scaling laws above and below the RG scale. An exact low energy solution
is especially desirable for the bsG model in the context of the Schmid-Bulgadaev
transition \citep{schmid_diffusion_1983,bulgadaev_phase_1984} ---
a 40-years-old predicted quantum phase transition, \textcolor{black}{whose
lack of clear experimental proof sparked a recent debate \citep{murani_absence_2020,hakonen_comment_2021,murani_reply_2021,masuki_absence_2022,sepulcre_comment_2022,masuki_reply_2022,subero_bolometric_2022,giacomelli_emergent_2023}.}
A low energy theory of the scattering of photons in the bsG model
could supplement a recent experimental study, which seeks signatures
of the transition in the inelastic and elastic scattering rates of
microwave photons \citep{kuzmin_observation_2023}.

In this work, we show how inelastic decay can be described via the
language of integrability. \textcolor{black}{The principle idea is
summarized in Fig. \ref{fig:cQED}(a), which depicts a generic integrable
quantum field theory realized in a cQED setup.} We rely on the nonlinear
relation between the microwave photons, which are observed to scatter
inelastically, and the elementary excitations of the integrable system,
whose scattering is purely elastic. This nonlinear relation is encoded
in the form factors of the models --- the matrix elements of the
bosonic field operator in the basis of the fundamental excitations
of the integrable theory \citep{smirnov_form_1992}. Building upon
Refs. \citep{fendley_exact_1994,lesage_form_1996,lesage_time_1996},
we use the form factors to obtain exact expressions for the reflection
coefficient of the microwave photons, which encodes the inelastic
and elastic decay rates. We then go beyond linear response and calculate
the exact energy-resolved inelastic decay spectrum, by devising a
method to evaluate a 3-point response function using form factors.
\textcolor{black}{This is a non-perturbative and rapidly-convergent
diagrammatic approach with clear physical intuition, analogous to
Wick's theorem in free theories. }Our technique yields more general
results as compared to existing methods for the calculation of form
factor expansions of multipoint correlation functions \citep{babujian_multipoint_2017,squarcini_multipoint_2021,fava_divergent_2023},
and could prove useful in other contexts of integrable quantum field
theories.

\textcolor{black}{We apply our methods for the bsG and Kondo models,
both of which qualify as special cases of Fig. \ref{fig:cQED}(a),
with implementations depicted in Fig. \ref{fig:cQED}(b). We illuminate
the fundamental physics governing each model through the lens of the
exact scattering rates, and show how they act as probes which discriminate
between the two models, as well as between the regimes of low and
high energies. In particular, we analyze our non-perturbative results
in light of the Schmid-Bulgadaev transition, and show how one may
observe signatures of the transition in the calculated rates, which
have been recently measured in an experiment \citep{kuzmin_observation_2023}.}

\textcolor{black}{The rest of the paper is organized as follows. We
begin by describing the bsG and Kondo models in Section \ref{sec:models},
discussing their realizations in a cQED environment and the observables
commonly measured in experiments. In Section \ref{sec:integrable},
we summarize the quintessential features and tools used in the treatment
of massless integrable quantum field theories with a boundary, and
introduce some notations that are utilized in this work. Armed with
this analytical power, we calculate the exact reflection coefficient
of the microwave photons in Section \ref{sec:total_rate}, from which
we extract the total inelastic decay rate and elastic phase shift.
We show how inelastic decay can emerge from the purely elastic scattering
rules of the fundamental excitations, discuss the key features and
asymptotic behavior of our results, and relate them to recent experiments.
The exact energy-resolved inelastic decay spectrum is obtained in
Section \ref{sec:spec}, where we introduce a method to calculate
a 3-point response function using form factors. The technical details
of our method are discussed in Section \ref{subsec:3pt_details} and
the Appendixes, the physical intuition underlying our diagrammatic
approach is emphasized in Section \ref{subsec:diagrams}, and the
advantage of the decay spectrum over the total inelastic decay rate
as a diagnostic tool is explained in Section \ref{subsec:spec_res}.
We conclude in Section \ref{sec:conclusions}.}

\section{\label{sec:models}Models and observables in a cQED setup}

In this work, we consider two different yet related quantum impurity
models --- the bsG and Kondo models. We begin by showing how they
can be derived from cQED setups. The implementation of these models
using superconducting circuits allows one to probe their fundamental
properties in experiments via spectroscopy, which measures the elastic
and inelastic scattering rates of microwave photons.

Consider a very long array of $N+1\gg1$ superconducting grains with
lattice spacing $a$ (total length $\ell=Na$), linked to each other
by Josephson junctions $E_{J}^{\textrm{line}}$ and capacitors $C^{\textrm{line}}$,
and to the ground by capacitors $C_{g}$. The circuit Hamiltonian
reads (setting $\hbar=e=1$)
\begin{align}
\tilde{\mathcal{H}}= & \sum_{n,m=0}^{N}2\left[\mathrm{C}^{-1}\right]_{n,m}\tilde{Q}_{n}\tilde{Q}_{m}-\sum_{n=1}^{N}E_{J}^{\textrm{line}}\cos\left(\tilde{\phi}_{n}-\tilde{\phi}_{n-1}\right)\nonumber \\
 & +\tilde{\mathcal{H}}_{I},\label{eq:H_discrete}
\end{align}
where $\tilde{\phi}_{n}$ and $\tilde{Q}_{n}$ are the superconducting
phases and charges of the grains, respectively, $\tilde{\mathcal{H}}_{I}$
is the Hamiltonian of the impurity at $n=0$, and $\mathrm{C}^{-1}$
is the inverse of the capacitance matrix $\mathrm{C}$, which is given
by $\left[\mathrm{C}\right]_{n,m}=\left(C_{g}+2C^{\textrm{line}}\right)\delta_{n,m}-C^{\textrm{line}}\left(\delta_{n,m+1}+\delta_{n,m-1}\right)$
in the bulk. The generic circuit implementing Eq. (\ref{eq:H_discrete})
is depicted in Fig. \ref{fig:cQED}. Choosing the line parameters
such that $E_{J}^{\textrm{line}}/E_{C}^{\textrm{line}}\gg1$, where
$E_{C}^{\textrm{line}}=1/\left(2C^{\textrm{line}}\right)$, both phase
slips and anharmonic effects are strongly suppressed in the intergrain
Josephson junctions, leading to an approximately quadratic transmission
line. In the thermodynamic and continuum limits, $a\rightarrow0$
and $N\rightarrow\infty$ with $\ell=Na$ fixed, the Hamiltonian reads
\begin{equation}
\tilde{\mathcal{H}}=\frac{v}{2\pi}\int_{-\ell}^{0}\id x\left[\frac{1}{2z}\left(\partial_{x}\tilde{\phi}\left(x\right)\right)^{2}+2z\left(\pi\tilde{\rho}\left(x\right)\right)^{2}\right]+\tilde{\mathcal{H}}_{I},\label{eq:H_generic}
\end{equation}
where $\tilde{\rho}\left(x\right)=\tilde{Q}_{n=x/a}/a$ is the charge
density field. The array velocity is given by $v=a/\sqrt{L^{\textrm{line}}C_{g}}$,
and $L^{\textrm{line}}=1/\left(4E_{J}^{\textrm{line}}\right)$ is
the effective inductance of the array junctions. The Luttinger parameter
of the system is given by the dimensionless line impedance, $z=Z/R_{Q}$,
where $Z=\sqrt{L^{\textrm{line}}/C_{g}}$ and $R_{Q}=h/\left(2e\right)^{2}=\pi/2$
is the resistance quantum. The kinetic inductance of the Josephson
arrays, $L^{\textrm{line}}$, enables the implementation of large
impedances, $z\sim1$, two orders of magnitude larger than typical
impedances in classical LC resonator lines \citep{kuzmin_quantum_2019,kuzmin_superstrong_2019}.
The impedance plays the role of an effective fine structure constant,
determining the strength of light-matter interaction in such cQED
setups; thus, achieving $z\sim1$ is the key to high-probability inelastic
decay on the single-photon level\textbf{ }\citep{kuzmin_inelastic_2021,mehta_down-conversion_2023}.

\subsection{The boundary sine-Gordon Hamiltonian}

The bsG model can be implemented by connecting the transmission line
to a Cooper pair box,
\begin{equation}
\tilde{\mathcal{H}}_{I}=4E_{C}a^{2}\tilde{\rho}^{2}\left(x=0\right)-E_{J}\cos\left(\tilde{\phi}\left(x=0\right)\right),\label{eq:HIbSG}
\end{equation}
where the charging energy $E_{C}=1/\left(2C_{0}\right)$ is significantly
larger than the Josephson energy, $E_{C}\gg E_{J}$. The capacitance
$C_{0}$ introduces a cutoff frequency, $\Gamma_{0}=1/\left(ZC_{0}\right)=4E_{C}/\left(\pi z\right)$,
which is the elastic linewidth of the impurity, equal to one over
the RC time defined by the line impedance and impurity capacitance.
\textcolor{black}{Another high-energy cutoff is imposed by the plasma
frequency of the Josephson array, $\omega_{p}=1/\sqrt{L^{\textrm{line}}C^{\textrm{line}}}$;
the UV cutoff is then set by the smaller of these two scales, $\Lambda\sim\min\left\{ \Gamma_{0},\omega_{p}\right\} $,
and is assumed to be much larger than any other energy scale in the
problem. We consider the scaling limit, $E_{J},\Lambda\rightarrow\infty$,
such that the ratio $E_{J}^{\star}\equiv\left(E_{J}/\Lambda^{z}\right)^{1/\left(1-z\right)}$
remains finite for $z<1$; at $z>1$, $E_{J}^{\star}=0$. $E_{J}^{\star}$
is an emergent RG scale, marking the characteristic frequency below
which the impurity term cannot treated as a perturbation, and the
phase at the boundary $\tilde{\phi}\left(x=0\right)$ is pinned to
one of the minima of the boundary cosine (note that this implies that
perturbation theory remains valid at all frequencies for $z>1$).}
At zero temperature, $E_{J}^{\star}$ is the only remaining energy
scale in the problem. In this limit, the capacitive term is effectively
eliminated from the impurity Hamiltonian. Redefining the fields as
$\phi=\tilde{\phi}/\sqrt{2z},\rho=\tilde{\rho}\sqrt{2z}$, we find
\begin{align}
\mathcal{H}_{\bsg}= & \frac{v}{2\pi}\int_{-\ell}^{0}\id x\left[\left(\partial_{x}\phi\left(x\right)\right)^{2}+\left(\pi\rho\left(x\right)\right)^{2}\right]\nonumber \\
 & -E_{J}\cos\left(\sqrt{2z}\phi\left(x=0\right)\right).\label{eq:HbSG}
\end{align}
Taking the limit $\ell\rightarrow\infty$, we arrive at the Hamiltonian
of the bsG model.

\subsection{The Kondo Hamiltonian}

There are several ways to arrive at the Kondo model from cQED Hamiltonians.
One is to connect the transmission line to a fluxonium\textbf{ }\citep{manucharyan_fluxonium_2009}
at external half flux quantum, with the impurity Hamiltonian
\begin{equation}
\tilde{\mathcal{H}}_{I}=\frac{2\tilde{Q}_{f}^{2}}{C_{f}}+E_{J}\cos\left(\tilde{\phi}_{f}\right)+\frac{\left(\tilde{\phi}_{f}-\tilde{\phi}_{0}\right)^{2}}{8L_{f}}+\frac{\tilde{\phi}_{0}^{2}}{8L_{0}}.\label{eq:HIK}
\end{equation}
This Hamiltonian realizes a double-well potential for the flux $\tilde{\phi}_{f}$.
In Appendix \ref{app:kondo_hamiltonian}, it is shown that the line
and the fluxonium give rise to the spin-boson Hamiltonian,
\begin{align}
\mathcal{H}_{\textrm{SB}}= & \frac{v}{2\pi}\int_{-\ell}^{0}\id x\left[\left(\partial_{x}\phi\left(x\right)\right)^{2}+\left(\pi\rho\left(x\right)\right)^{2}\right]\nonumber \\
 & -J\mathcal{S}_{x}-\sqrt{2z^{\p}}v\pi\rho\left(x=0,t\right)\mathcal{S}_{z},\label{eq:HSB}
\end{align}
where $\mathcal{S}_{z}$ is a pseudo-spin operator whose two eigenvalues,
$\pm1/2$, correspond to the two potential wells. The $\mathcal{S}_{x}$
term describes the tunneling between the wells, where $J$ is the
tunneling matrix element, equal to the splitting between the two lowest
energy levels of the fluxonium. The $\mathcal{S}_{z}$ term describes
an effective capacitive coupling of the fluxonium to the array. Note
that the coupling coefficient is written in terms of some parameter
$z^{\p}$, which is proportional but not identical to the normalized
impedance $z$ (see Appendix \ref{app:kondo_hamiltonian} for details);
for brevity, from here on we revert to writing $z$ instead of $z^{\p}$.
The unitary transformation $\mathcal{H}\rightarrow\mathcal{U}^{-1}\mathcal{H}\mathcal{U}$
with $\mathcal{U}=e^{\ii\phi\left(x=0,t\right)\mathcal{S}_{z}}$ leads
to the Kondo Hamiltonian \citep{leggett_dynamics_1987},
\begin{align}
\mathcal{H}_{\K}^{\p}= & \frac{v}{2\pi}\int_{-\ell}^{0}\id x\left[\left(\partial_{x}\phi\left(x\right)\right)^{2}+\left(\pi\rho\left(x\right)\right)^{2}\right]\nonumber \\
 & -\frac{J}{2}\left(\mathcal{S}_{+}e^{\ii\phi\left(x=0\right)}+\mathcal{S}_{-}e^{-\ii\phi\left(x=0\right)}\right)\nonumber \\
 & -\left(\sqrt{2z}-1\right)v\pi\rho\left(x=0,t\right)\mathcal{S}_{z}.
\end{align}
A similar implementation, using a flux qubit instead of a fluxonium,
has been realized in Ref. \citep{forn_diaz_ultrastrong_2017}. The
same Hamiltonian can also be derived from the effective low-energy
description of two transmission lines coupled by a Cooper pair box,
with Josephson junctions connecting the capacitors to the lines \citep{goldstein_inelastic_2013},
\textcolor{black}{although this implementation could suffer from some
drawbacks \citep{kaur_spin-boson_2021}. }A slightly different transformation,
$\mathcal{V}=e^{\sqrt{2z}\ii\phi\left(x=0,t\right)\mathcal{S}_{z}}$,
eliminates the capacitive coupling term:
\begin{align}
\mathcal{H}_{\K}= & \frac{v}{2\pi}\int_{-\ell}^{0}\id x\left[\left(\partial_{x}\phi\left(x\right)\right)^{2}+\left(\pi\rho\left(x\right)\right)^{2}\right]\nonumber \\
 & -\frac{J}{2}\left(\mathcal{S}_{+}e^{\sqrt{2z}\ii\phi\left(x=0\right)}+\mathcal{S}_{-}e^{-\sqrt{2z}\ii\phi\left(x=0\right)}\right).\label{eq:HK}
\end{align}
This will be the form used in the following, again in the limit of
a semi-infinite lead, $\ell\rightarrow\infty$. Introducing a cutoff
frequency $\Lambda$ to our model, the scaling limit reads again $J,\Lambda\rightarrow\infty$
with finite $E_{J}^{\star}=\left(J/\Lambda^{z}\right)^{1/\left(1-z\right)}$.

\subsection{Definition of the scattering rates}

Both Hamiltonians may be written in the form
\begin{equation}
\mathcal{H}=\frac{v}{2\pi}\int_{-\ell}^{0}\id x\left[\left(\partial_{x}\phi\left(x\right)\right)^{2}+\left(\pi\rho\left(x\right)\right)^{2}\right]+\mathcal{H}_{I}.\label{eq:H_generic_redef}
\end{equation}
The bulk Hamiltonian is quadratic and diagonalized by plane waves,
with the dispersion relation $\omega_{q}=vq$ and mode spacing $\Delta=\pi v/\ell$,
so that $q=n\Delta$ where $n>0$ is an integer. The mode expansion
of the fields reads
\begin{align}
\phi\left(x\right)= & \ii\sum_{q}\sqrt{\frac{\pi}{q\ell}}\left(b_{q}-b_{q}^{\dagger}\right)\cos\left(qx\right),\nonumber \\
\rho\left(x\right)= & \sum_{q}\sqrt{\frac{q}{\pi\ell}}\left(b_{q}+b_{q}^{\dagger}\right)\cos\left(qx\right),\label{eq:field_modes}
\end{align}
where $b_{q}^{\dagger},b_{q}$ are bosonic creation and annihilation
operators.

The implementation of the Hamiltonians (\ref{eq:HbSG}) and (\ref{eq:HK})
in cQED setups provides direct access to their properties in a controlled
environment. In this work, we focus on scattering experiments at zero
temperature, where a single microwave photon at frequency $\omega$
is injected from the open end of the system (the antenna on the left
in Fig. \ref{fig:cQED}), and propagates towards the impurity. One
may then measure response functions; the simplest one is the reflection
coefficient $r\left(x,x^{\p};\omega\right)$ of a photon with frequency
$\omega$, injected at $x^{\p}$ and measured at $x$. The dependence
on $x,x^{\p}$ should amount to a trivial phase term, $r\left(x,x^{\p};\omega\right)=e^{-\ii\omega\left(x+x^{\p}\right)/v}r\left(\omega\right)$,
and $r\left(\omega\right)$ defines the inelastic decay rate $\gamma\left(\omega\right)$
and elastic phase shift $\delta\left(\omega\right)$:
\begin{equation}
r\left(\omega\right)=e^{-2\ii\delta\left(\omega\right)}e^{-\gamma\left(\omega\right)/2}.\label{eq:r_rates}
\end{equation}
There are several ways to calculate the reflection coefficient. One
is to solve a scattering problem for the microwave photons, and read
off $r\left(\omega\right)$ from the scattering matrix of the theory
\citep{houzet_priv}. Here we adopt a different approach, and calculate
$r\left(\omega\right)$ from the conductance in the half-infinite
line, by means of a Landauer-Buttiker-like formula. The AC conductance
is given by the Kubo formula:
\begin{align}
\mathcal{G}\left(x,x^{\p};\omega\right)= & \lim_{\eta\rightarrow0^{+}}\frac{z}{\pi\omega}\nonumber \\
 & \hspace{-1.5cm}\times\int_{0}^{\infty}\id te^{\ii\left(\omega+\ii\eta\right)t}\left\langle \left[\partial_{t}\phi\left(x,t\right),\partial_{t}\phi\left(x^{\p},0\right)\right]\right\rangle .\label{eq:conductance}
\end{align}
Introducing the right and left current operators,
\begin{equation}
\mathcal{R}=\partial_{t}\phi-v\partial_{x}\phi,\quad\mathcal{L}=\partial_{t}\phi+v\partial_{x}\phi,\label{eq:R_L}
\end{equation}
we find
\begin{align}
\mathcal{G}\left(x,x^{\p};\omega\right)= & \frac{1}{4}\sum_{\mathcal{A},\mathcal{B}=\mathcal{R},\mathcal{L}}\mathcal{G}_{\mathcal{AB}}\left(x,x^{\p};\omega\right),\nonumber \\
\mathcal{G}_{\mathcal{AB}}\left(x,x^{\p};\omega\right)= & \lim_{\eta\rightarrow0^{+}}\frac{z}{\pi\omega}\nonumber \\
 & \hspace{-1.5cm}\times\int_{0}^{\infty}\id te^{\ii\left(\omega+\ii\eta\right)t}\left\langle \left[\mathcal{A}\left(x,t\right),\mathcal{B}\left(x^{\p},0\right)\right]\right\rangle .\label{eq:GAB}
\end{align}
The terms $\mathcal{G}_{\mathcal{RR}}$, $\mathcal{G}_{\mathcal{LL}}$
do not involve the boundary, and are therefore determined by the perfect
conductance in a Luttinger liquid with coupling parameter $z$ \citep{kane_transmission_1992},
\begin{align}
\mathcal{G}_{\mathcal{RR}}\left(x,x^{\p};\omega\right)= & \GLRz e^{-\ii\omega\left(x+x^{\p}\right)/v}\Theta\left(x-x^{\p}\right),\nonumber \\
\mathcal{G}_{\mathcal{LL}}\left(x,x^{\p};\omega\right)= & \GLRz e^{-\ii\omega\left(x+x^{\p}\right)/v}\Theta\left(x^{\p}-x\right),
\end{align}
where $\GLRz=2z$. The heaviside step functions, $\Theta\left(\pm\left(x-x^{\p}\right)\right)$,
ensure that the responses are causal, and $\mathcal{G}_{\mathcal{RR}}\left(x,x^{\p};\omega\right)+\mathcal{G}_{\mathcal{LL}}\left(x,x^{\p};\omega\right)=\GLRz e^{-\ii\omega\left(x+x^{\p}\right)/v}.$
The term $\GRL$ corresponds to a non-causal response function, as
it measures a right moving current which results from a given left
moving current, and therefore vanishes for all $x,x^{\p}$, as we
verify explicitly in Section \ref{sec:total_rate} below. We thus
have
\begin{equation}
\mathcal{G}\left(x,x^{\p};\omega\right)=\frac{1}{4}\left(\GLRz e^{-\ii\omega\left(x+x^{\p}\right)/v}+\GLR\left(x,x^{\p};\omega\right)\right).
\end{equation}
We can now define the reflection coefficient. The conductance $\mathcal{G}$
should be equal to $z$ in the absence of an impurity ($E_{J}^{\star}\rightarrow0$),
and vanish for an impenetrable boundary ($E_{J}^{\star}\rightarrow\infty$),
corresponding to $r\rightarrow1$ and $r\rightarrow-1$, respectively.
We therefore write
\begin{equation}
r\left(x,x^{\p};\omega\right)=\frac{\GLR\left(x,x^{\p};\omega\right)}{\GLRz}.\label{eq:landauer}
\end{equation}

The Kubo formula relates the conductance, and hence the elastic phase
shift and inelastic decay rate, to a 2-point response function. Higher-order
response functions yield more refined rates; in this work, we focus
on the energy-resolved inelastic decay spectrum, $\gamma\left(\omega^{\p}\lvert\omega\right)$,
which is the rate of the decay process $\omega\rightarrow\omega^{\p}+\sum_{i}\omega_{i}^{\pp}$
for any choice of $\omega_{i}^{\pp}$ such that $\omega=\omega^{\p}+\sum_{i}\omega_{i}^{\pp}$.
The spectrum $\gamma\left(\omega^{\p}\lvert\omega\right)$, which
may be evaluated in an experiment by injecting a photon $\omega$
and measuring the reflected power at frequency $\omega^{\prime}$
(see Fig. \ref{fig:cQED}), corresponds to a 3-point response function,
as shown in Ref. \citep{goldstein_inelastic_2013} and discussed in
Section \ref{sec:spec}.

Measurements of the total inelastic decay rate and spectrum shed light
on the fundamental properties of the models. At large enough frequencies,
we expect to find Luttinger liquid scaling laws, such as $\gamma\left(\omega\right)\sim\omega^{2z-2}$;
the exponents, which could be probed directly in such experiments,
indicate whether the boundary operators in (\ref{eq:HbSG}) and (\ref{eq:HK})
are relevant ($z<1$) or irrelevant ($z>1$). In the bsG model, this
is of particular importance in the context of the elusive Schmid-Bulgadaev
quantum phase transition \citep{schmid_diffusion_1983,bulgadaev_phase_1984},
where the relevant and irrelevant cases correspond, respectively,
to the superconducting and insulating phases of the boundary Josephson
junction. Measuring $r\left(\omega\right)$ using single photon spectroscopy
thus offers a non-invasive approach to probe the dynamics of the boundary
junction, avoiding DC transport measurements which were mostly used
in previous works \citep{yagi_phase_1997,penttila_superconductor_1999,penttila_experiments_2001,watanabe_quantum_2003},
and potentially settling a recent debate concerning the presence of
the transition \citep{murani_absence_2020,hakonen_comment_2021,murani_reply_2021,masuki_absence_2022,sepulcre_comment_2022,masuki_reply_2022}.
The measurements must be supplemented by theory, notably in the low
energy sector, where a perturbative solution does not apply since
the boundary cosine in (\ref{eq:HbSG}) is relevant for $z<1$. Signatures
of the transition could also be found in the spectrum $\gamma\left(\omega^{\p}\lvert\omega\right)$
\citep{houzet_critical_2020}. A non-perturbative calculation of the
total inelastic decay rate $\gamma\left(\omega\right)$, the elastic
phase shift $\delta\left(\omega\right)$, and the inelastic decay
spectrum $\gamma\left(\omega^{\p}\lvert\omega\right)$, for both bsG
and Kondo models, is the objective of this work.

An exact solution for the quantities above is possible thanks to the
integrability of the models. As explained in the next Section, the
Hamiltonians (\ref{eq:HbSG}) and (\ref{eq:HK}) are characterized
by an underlying purely elastic dynamics, and may be solved using
variations of the Bethe ansatz which apply to field theories. As discussed
above, this pure elasticity seems to be at odds with the observed
photon splitting. To this end, it should be mentioned that realistic
implementations of both models would inevitably suffer from integrability-breaking
perturbations. To name a few, stray capacitances, array disorder,
and finite lattice spacing, all unavoidable in a cQED environment,
would break the integrability of the systems. In addition, the models
in Eqs. (\ref{eq:HbSG}) and (\ref{eq:HK}) lack a cutoff frequency,
which would be present in any physical realization. Therefore, one
may be tempted to attribute the inelastic decay of photons observed
in cQED experiments to the presence of integrability-breaking terms
in the Hamiltonian. However, it is clear in these experiments that
inelastic decay persists even when the circuit parameters are pushed
towards those of the integrable systems. Particularly, inelastic decay
rates observed for a transmon impurity \citep{kuzmin_inelastic_2021}\textbf{
}are of the same order of magnitude as those measured for a Cooper
pair box \citep{kuzmin_observation_2023}, where the charging energy
of the impurity Josephson junction is pushed to larger and larger
values and the Hamiltonian of the system approaches Eq. (\ref{eq:HbSG}).
\textcolor{black}{Furthermore, we stress that while experimental systems
are not exactly integrable, the use of the integrability formalism
is still justified within reasonable assumptions. This is discussed
in Appendix \ref{app:exp_considerations}.}

\textcolor{black}{Finally, we note that the integrability formalism
can only be applied for $z<1$ (see explanation below). Fortunately,
for $z>1$, the RG scale vanishes, and perturbation theory holds for
$z>1$ at all frequencies (assuming that the boundary Josephson energy
is much smaller than the UV cutoff), so that an exact solution is
unnecessary. From here on, we restrict ourselves to the regime $z<1$.}

\section{\label{sec:integrable}Overview of massless integrable quantum field
theories with a boundary}

In this section, we briefly sketch the main ingredients needed for
the calculation of correlation functions in the bsG and Kondo models.
These ingredients are the S-matrices, reflection matrices, and form
factors. Here we only outline the general structure of the calculation
mechanism and introduce some useful notations, relegating explicit
expressions to the Appendixes. More details may be found in several
papers and reviews \citep{dorey_exact_1997,bombardelli_s-matrices_2016,lesage_form_1996,fendley_exact_1994,ghoshal_boundary_1994,smirnov_form_1992}.

It is useful to think of the quadratic bulk Hamiltonian of Eqs. (\ref{eq:HbSG})
and (\ref{eq:HK}) as the limiting case of a more general Hamiltonian,
\begin{align}
\mathcal{H}^{\textrm{bulk}}= & \frac{1}{2\pi}\int_{-\ell}^{0}\id x\left[\left(\partial_{x}\phi\left(x\right)\right)^{2}+\left(\pi\rho\left(x\right)\right)^{2}\right.\nonumber \\
 & \left.-E_{J}^{\textrm{bulk}}\cos\left(\sqrt{8z}\phi\left(x\right)\right)\right].
\end{align}
For an infinite transmission line (as opposed to the half-infinite
line above), this would be the Hamiltonian of the bulk sine-Gordon
model, with a coupling parameter $\beta=\sqrt{8\pi z}$, which can
be implemented in a cQED environment by connecting the superconducting
grains to the ground via Josephson junctions \citep{roy_quantum_2021}.
As mentioned above, we focus on the regime $z<1$ ($\beta<\sqrt{8\pi}$).

The nonlinear equations of motion of the generalized bulk Hamiltonian
are solved by solitons and antisolitons --- wave packets that propagate
in the nonlinear medium without dispersion. In the attractive regime,
$z<1/2$ ($\beta<\sqrt{4\pi}$), a pair of a soliton and an antisoliton
may form a bound state, called a breather, which also propagates in
the bulk without dispersion. The energy and momentum of a soliton
($\epsilon=+$), an antisoliton ($\epsilon=-$), or a breather ($\epsilon=m$,
where $m$ is an integer satisfying $1\le m\le\maxbreath$), can be
parameterized by its rapidity $\theta$ as $E=M_{\epsilon}\cosh\theta$
and $P=M_{\epsilon}\sinh\theta$, where from here on we set the velocity
to unity, $v=1$. The excitation mass, $M_{\epsilon}$, scales with
the bulk Josephson energy, $M_{\epsilon}\sim\left(E_{J}^{\textrm{bulk}}\right)^{1/\left(2-2z\right)}$
\citep{zamolodchikov_mass_1995}. We therefore refer to (\ref{eq:HbSG})
and (\ref{eq:HK}) as massless Hamiltonians; writing $\theta=\pm\left(A+\lambda\right)$
with $A\rightarrow\infty$ such that $M_{+}e^{A}/2\rightarrow1$,
the energy and momentum become those of chiral wave packets, $E=\mu_{\epsilon}e^{\lambda}$
and $P=\pm\mu_{\epsilon}e^{\lambda}$, where a plus (minus) sign corresponds
to a right (left) mover, and $\mu_{\epsilon}=M_{\epsilon}/M_{+}$
($\mu_{\pm}=1,\mu_{m}=2\sin\left(m\xi/2\right)$) is the bulk mass
ratio.

The presence of an extensive (infinite) number of locally conserved
charges in the integrable field theories leads to purely elastic scatterings.
Any two solutions to the equations of motion with energies $E_{1,2}$
and momenta $P_{1,2}$ will maintain their individual energies and
momenta when they scatter off one another. The same holds for reflections
off the boundary --- the momentum and energy of any reflected solution
are equal to the incoming ones. This property makes the solitons and
breathers a natural basis to diagonalize the Hamiltonians (\ref{eq:HbSG})
and (\ref{eq:HK}). Quantizing these field theories, a classical
field configuration of type $\epsilon$ and rapidity $\theta$ is
a assigned a field excitation, which can be added to a given state
by a creation operator $Z_{\epsilon}^{\dagger}\left(\theta\right)$,
or removed from a state by an annihilation operator $Z^{\epsilon}\left(\theta\right)$.
These elementary excitations are a defining feature of integrable
quantum field theories, and form eigenstates of the Hamiltonian, $\k{\theta_{n},\ldots,\theta_{1}}_{\epsilon_{n}\ldots\epsilon_{1}}=Z_{\epsilon_{n}}^{\dagger}\left(\theta_{n}\right)\ldots Z_{\epsilon_{1}}^{\dagger}\left(\theta_{1}\right)\k 0$,
where $\k 0$ is the vacuum state, $Z_{\epsilon}\left(\theta\right)\k 0=0$.
The eigenstates are normalized such that $^{\epsilon_{2}}\bk{\theta_{2}|\theta_{1}}_{\epsilon_{1}}=2\pi\delta_{\epsilon_{1}}^{\epsilon_{2}}\delta\left(\theta_{1}-\theta_{2}\right)$.
The dynamics of the excitations is strongly constrained by the extensive
number of local conservation laws present in the system, as captured
by the Zamolodchikov-Faddeev algebra \citep{zamolodchikov_factorized_1979,faddeev_quantum_1995}:

\begin{align}
Z_{\epsilon_{1}}^{\dagger}\left(\theta_{1}\right)Z_{\epsilon_{2}}^{\dagger}\left(\theta_{2}\right)= & \sum_{\epsilon_{1}^{\p},\epsilon_{2}^{\p}}S_{\epsilon_{1}\epsilon_{2}}^{\epsilon_{1}^{\p}\epsilon_{2}^{\p}}\left(\theta_{1}-\theta_{2}\right)Z_{\epsilon_{2}^{\p}}^{\dagger}\left(\theta_{2}\right)Z_{\epsilon_{1}^{\p}}^{\dagger}\left(\theta_{1}\right),\nonumber \\
Z^{\epsilon_{1}}\left(\theta_{1}\right)Z^{\epsilon_{2}}\left(\theta_{2}\right)= & \sum_{\epsilon_{1}^{\p},\epsilon_{2}^{\p}}S_{\epsilon_{1}^{\p}\epsilon_{2}^{\p}}^{\epsilon_{1}\epsilon_{2}}\left(\theta_{1}-\theta_{2}\right)Z^{\epsilon_{2}^{\p}}\left(\theta_{2}\right)Z^{\epsilon_{1}^{\p}}\left(\theta_{1}\right),\nonumber \\
Z^{\epsilon_{1}}\left(\theta_{1}\right)Z_{\epsilon_{2}}^{\dagger}\left(\theta_{2}\right)= & \sum_{\epsilon_{1}^{\p},\epsilon_{2}^{\p}}S_{\epsilon_{2}\epsilon_{1}^{\p}}^{\epsilon_{2}^{\p}\epsilon_{1}}\left(\theta_{2}-\theta_{1}\right)Z_{\epsilon_{2}^{\p}}^{\dagger}\left(\theta_{2}\right)Z^{\epsilon_{1}^{\p}}\left(\theta_{1}\right)\nonumber \\
 & +2\pi\delta_{\epsilon_{2}}^{\epsilon_{1}}\delta\left(\theta_{1}-\theta_{2}\right).\label{eq:ZF_alg}
\end{align}
The S-matrix $S_{\epsilon_{1}\epsilon_{2}}^{\epsilon_{1}^{\p}\epsilon_{2}^{\p}}\left(\theta_{1}-\theta_{2}\right)$
satisfies the Yang-Baxter equation and several additional symmetry
relations which are summarized in Appendix \ref{app:Smat_props}.
The relations in Eq. (\ref{eq:ZF_alg}) are a manifestation of the
factorization of $n$-body interactions to a product of 2-body interactions
in the integrable system.

The eigenstates $\k{\theta_{n},\ldots,\theta_{1}}_{\epsilon_{n}\ldots\epsilon_{1}}$
form a complete set of states. In the massless limit and in the presence
of a boundary, the complete set is \citep{lesage_form_1996}
\begin{align}
\mathds{1}= & \k 0\b 0+\sum_{n=1}^{\infty}\sum_{\epsilon_{1}\ldots\epsilon_{n}}\int_{-\infty}^{\infty}\frac{\prod_{k=1}^{n}\id\lambda_{k}}{\left(2\pi\right)^{n}n!}\nonumber \\
 & \KK{\lambda_{n},\ldots,\lambda_{1}}_{\epsilon_{n}\ldots\epsilon_{1}}\sidescripts{\epsilon_{1}\ldots\epsilon_{n}}{}{\BB{\lambda_{1},\ldots,\lambda_{n}}}{}{},\label{eq:complete_set_1}
\end{align}
where the states take into account both right and left movers:
\begin{align}
\KK{\lambda_{n},\ldots,\lambda_{1}}_{\epsilon_{n}\ldots\epsilon_{1}}= & \k{\lambda_{n}^{\rmn},\ldots,\lambda_{1}^{\rmn}}_{\epsilon_{n}\ldots\epsilon_{1}}\nonumber \\
 & \hspace{-3.2cm}+\sum_{\epsilon_{1}^{\prime}}\prod_{k=2}^{n}\tilde{S}_{\epsilon_{k}\epsilon_{1}^{\p}}^{\epsilon_{k}\epsilon_{1}^{\p}}R_{\epsilon_{1}}^{\epsilon_{1}^{\prime}}\left(\lambda_{1}\right)\k{\lambda_{1}^{\lmn},\lambda_{n}^{\rmn},\ldots,\lambda_{2}^{\rmn}}_{\epsilon_{1}^{\prime}\epsilon_{n}\ldots\epsilon_{2}}+\ldots\nonumber \\
 & \hspace{-3.2cm}+\sum_{\epsilon_{1}^{\p}\ldots\epsilon_{n}^{\p}}\prod_{l=1}^{n-1}\prod_{k=l+1}^{n}\tilde{S}_{\epsilon_{k}\epsilon_{l}^{\p}}^{\epsilon_{k}\epsilon_{l}^{\p}}\prod_{k=1}^{n}R_{\epsilon_{k}}^{\epsilon_{k}^{\prime}}\left(\lambda_{k}\right)\k{\lambda_{1}^{\lmn},\ldots,\lambda_{n}^{\lmn}}_{\epsilon_{1}^{\prime}\ldots\epsilon_{n}^{\p}}.\label{eq:complete_set_2}
\end{align}
The superscript $\mn$ of a rapidity indicates the momentum carried
by the excitation --- it is $\varsigma_{\mn}\mu_{\epsilon}e^{\lambda}$,
where $\varsigma_{\mathfrak{r}}=1$ and $\varsigma_{\mathfrak{l}}=-1$.
Here $R_{\epsilon}^{\epsilon^{\prime}}\left(\lambda\right)$ is the
boundary reflection matrix, which is non-zero only if $\mu_{\epsilon}=\mu_{\epsilon^{\prime}}$;
therefore, a soliton can become an antisoliton (or vice-versa) upon
reflection of the boundary, but a reflected breather $m$ remains
a breather of the same type. As shown in \citep{fendley_exact_1994},
$R_{\epsilon}^{\epsilon^{\prime}}\left(\lambda\right)$ depends on
the difference $\lambda-\lambda_{B}$, where $\lambda_{B}$ is the
rapidity associated with the energy scale of the boundary, $T_{B}\equiv e^{\lambda_{B}}$,
which is proportional to the RG scale, $T_{B}\sim E_{J}^{\star}\sim E_{J}^{1/\left(1-z\right)}$
(or $T_{B}\sim J^{1/\left(1-z\right)}$), \textcolor{black}{and given
explicitly by \citep{lesage_perturbation_1999}
\begin{equation}
T_{B}=\frac{\Gamma\left(\frac{z}{2\left(1-z\right)}\right)}{\sqrt{\pi}\Gamma\left(\frac{1}{2\left(1-z\right)}\right)}\left(\frac{\pi\varepsilon}{\Gamma\left(z\right)\Lambda^{z}}\right)^{1/\left(1-z\right)},\label{eq:TB_EJs}
\end{equation}
where $\Gamma\left(x\right)$ is the gamma function \citep{abramowitz_handbook_1965},
and $\varepsilon$ is either $E_{J}$ or $J$ for the bsG or Kondo
Hamiltonians, respectively (recall that $\Lambda$ is the UV cutoff
frequency). Note that the reflection matrices are thus defined only
for $z<1$, where $T_{B}$ is finite; hence, this formalism cannot
be used to treat devices with $z>1$.} The reflection matrices for
the bsG and Kondo models are given in Appendix \ref{app:R}. In Eq.
(\ref{eq:complete_set_2}), $\tilde{S}_{\epsilon_{1}\epsilon_{2}}^{\epsilon_{1}\epsilon_{2}}\equiv\lim_{\theta\rightarrow\infty}S_{\epsilon_{1}\epsilon_{2}}^{\epsilon_{1}^{\p}\epsilon_{2}^{\p}}\left(\theta\right)$
denotes the S-matrix for the exchange of a right mover and a left
mover; note that it is diagonal.

The complete set of states can be inserted between any two operators
in a correlation function. For example, consider a 2-point correlator:
\begin{align}
\left\langle \mathcal{O}_{1}\left(x,t\right)\mathcal{O}_{2}\left(0,0\right)\right\rangle = & \left\langle \mathcal{O}_{1}\right\rangle \left\langle \mathcal{O}_{2}\right\rangle +\nonumber \\
 & \hspace{-3cm}\sum_{n=1}^{\infty}\sum_{\epsilon_{1}\ldots\epsilon_{n}}\int\frac{\prod_{k=1}^{n}\id\lambda_{k}}{\left(2\pi\right)^{n}n!}\b 0\mathcal{O}_{1}\left(x,t\right)\KK{\lambda_{n},\ldots,\lambda_{1}}_{\epsilon_{n}\ldots\epsilon_{1}}\nonumber \\
 & \hspace{-3cm}\times\sidescripts{\epsilon_{1}\ldots\epsilon_{n}}{}{\BB{\lambda_{1},\ldots,\lambda_{n}}\mathcal{O}_{2}\left(0,0\right)\k 0}{}{}.
\end{align}
In order to calculate $\left\langle \mathcal{O}_{1}\left(x,t\right)\mathcal{O}_{2}\left(0,0\right)\right\rangle $,
it is sufficient to know the matrix elements of $\mathcal{O}_{1,2}$
in the basis of the excitations. These matrix elements are called
form factors, and are denoted by $f_{\epsilon_{1}\ldots\epsilon_{n}}^{\mathcal{O}}\left(\lambda_{1}^{\mn},\ldots,\lambda_{n}^{\mn}\right)\equiv\bk{0|\mathcal{O}\left(0,0\right)|\lambda_{n}^{\mn},\ldots,\lambda_{1}^{\mn}}_{\epsilon_{n}\ldots\epsilon_{1}}$.
The form factors satisfy a set of axioms and symmetry properties \citep{smirnov_form_1992}
which are summarized in Appendix \ref{app:FF}. In this work, we only
consider correlation functions of right and left current operators,
defined in Eq. (\ref{eq:R_L}). It is shown in Appendix \ref{app:FF}
that the form factors of $\mathcal{R}$ are non-zero only if all excitations
are right moving, i.e., the first term in Eq. (\ref{eq:complete_set_2}),
and similarly for form factors of $\mathcal{L}$, where only the last
term in Eq. (\ref{eq:complete_set_2}) contributes; all other terms
can be discarded. We therefore denote $\k{\lambda_{n},\ldots,\lambda_{1}}_{\epsilon_{n}\ldots\epsilon_{1}}^{\mn}\equiv\k{\lambda_{n}^{\mn},\ldots,\lambda_{1}^{\mn}}_{\epsilon_{n}\ldots\epsilon_{1}}$,
and omit the superscripts $\mn$ from the form factors, $f_{\epsilon_{1}\ldots\epsilon_{n}}^{\mathcal{A}}\left(\lambda_{1},\ldots,\lambda_{n}\right)\equiv f_{\epsilon_{1}\ldots\epsilon_{n}}^{\mathcal{\mathcal{A}}}\left(\lambda_{1}^{\mn},\ldots,\lambda_{n}^{\mn}\right)$
(where $\mn=\rmn$ and $\mathcal{A}=\mathcal{R}$ or $\mn=\lmn$ and
$\mathcal{A}=\mathcal{L}$), since these are understood to be the
only non-vanishing matrix elements. The crossing relations, given
in Appendix \ref{app:FF}, allow us to evaluate terms of the form
$\sidescripts{\epsilon_{1}\ldots\epsilon_{n}}{\mn}{\bk{\lambda_{1},\ldots,\lambda_{n}|\mathcal{A}\left(0,0\right)|0}}{}{}$,
and more generally $\sidescripts{\epsilon_{1}^{\p}\ldots\epsilon_{l}^{\p}}{\mn}{\bk{\lambda_{1}^{\p},\ldots,\lambda_{l}^{\p}|\mathcal{A}\left(0,0\right)|\lambda_{n},\ldots,\lambda_{1}}}{\mn}{\epsilon_{n}\ldots\epsilon_{1}}$,
which appear in the calculation of multipoint correlation functions.
Note that it is enough to know the matrix elements of the operators
at the origin, since, in the massless limit,
\begin{align}
\sidescripts{\epsilon_{1}^{\p}\ldots\epsilon_{l}^{\p}}{\mn}{\bk{\lambda_{1}^{\p},\ldots,\lambda_{l}^{\p}|\mathcal{A}\left(x,t\right)|\lambda_{n},\ldots,\lambda_{1}}}{\mn}{\epsilon_{n}\ldots\epsilon_{1}}=\nonumber \\
 & \hspace{-6.6cm}\exp\left\{ \ii\left(\sum_{k=1}^{n}\mu_{\epsilon_{k}}e^{\lambda_{k}}-\sum_{k=1}^{l}\mu_{\epsilon_{k}^{\prime}}e^{\lambda_{k}^{\p}}\right)\left(\varsigma_{\mn}x-t\right)\right\} \nonumber \\
 & \hspace{-6.6cm}\times\sidescripts{\epsilon_{1}^{\p}\ldots\epsilon_{l}^{\p}}{\mn}{\bk{\lambda_{1}^{\p},\ldots,\lambda_{l}^{\p}|\mathcal{A}\left(0,0\right)|\lambda_{n},\ldots,\lambda_{1}}}{\mn}{\epsilon_{n}\ldots\epsilon_{1}}.\label{eq:xt_dep}
\end{align}
In the following, if the coordinates of the operator are omitted,
it should be understood that the matrix element is evaluated at the
origin.

Before we proceed, let us introduce a few additional notations that
will simplify the equations to follow. First, ket states formed by
a set $\lambda$ of $n_{\lambda}$ excitations are written as $\k{\lambr}_{\epsilon_{\lambda}}^{\mn}\equiv\k{\lambda_{1},\ldots,\lambda_{n_{\lambda}}}_{\epsilon_{1}\ldots\epsilon_{n_{\lambda}}}^{\mn}$and
$\k{\lambl}_{\epsilon_{\lambda}}^{\mn}\equiv\k{\lambda_{n_{\lambda}},\ldots,\lambda_{1}}_{\epsilon_{n_{\lambda}}\ldots\epsilon_{1}}^{\mn}$,
with their dual bra states written correspondingly. We also introduce
the charge-conjugate state, $\k{\lambr}_{\bar{\epsilon}_{\lambda}}^{\mn}\equiv\k{\lambda_{1},\ldots,\lambda_{n_{\lambda}}}_{\bar{\epsilon}_{1}\ldots\bar{\epsilon}_{n_{\lambda}}}^{\mn}$,
where $\bar{\pm}=\mp$ and $\bar{m}=m$. The complete set of states
in Eqs. (\ref{eq:complete_set_1}), (\ref{eq:complete_set_2}) is
then compactly written as
\begin{align}
\mathds{1}= & \int_{\lambda}\left(\k{\lambl}_{\epsilon_{\lambda}}^{\rmn}+\mathsf{R}_{\epsilon_{\lambda}}^{\epsilon_{\lambda}^{\p}}\left(\lambr\right)\k{\lambr}_{\epsilon_{\lambda}^{\p}}^{\lmn}\right)\nonumber \\
 & \times\left(\sidescripts{\epsilon_{\lambda}}{\rmn}{\b{\lambr}}{}{}+\left(\mathsf{R}_{\epsilon_{\lambda}^{\p}}^{\epsilon_{\lambda}}\left(\lambr\right)\right)^{*}\sidescripts{\epsilon_{\lambda}^{\p}}{\lmn}{\b{\lambl}}{}{}\right),\label{eq:complete_set_3}
\end{align}
where
\begin{equation}
\int_{\lambda}\equiv\sum_{n_{\lambda}=1}^{\infty}\sum_{\epsilon_{1}\ldots\epsilon_{n_{\lambda}}}\int_{-\infty}^{\infty}\frac{\prod_{k=1}^{n_{\lambda}}\id\lambda_{k}}{\left(2\pi\right)^{n_{\lambda}}n_{\lambda}!},\label{eq:sum_abbr}
\end{equation}
and the product of reflection matrices is abbreviated as
\begin{align}
\mathsf{R}_{\epsilon_{\lambda}}^{\epsilon_{\lambda}^{\p}}\left(\lambr\right)\k{\lambr}_{\epsilon_{\lambda}^{\p}}^{\lmn}\equiv & \sum_{\epsilon_{1}^{\p}\ldots\epsilon_{n_{\lambda}}^{\p}}\left(\prod_{l=1}^{n_{\lambda}-1}\prod_{k=l+1}^{n_{\lambda}}\tilde{S}_{\epsilon_{k}\epsilon_{l}^{\p}}^{\epsilon_{k}\epsilon_{l}^{\p}}\right)\left(\prod_{k=1}^{n}R_{\epsilon_{k}}^{\epsilon_{k}^{\prime}}\left(\lambda_{k}\right)\right)\nonumber \\
 & \times\k{\lambda_{1},\ldots,\lambda_{n_{\lambda}}}_{\epsilon_{1}^{\p}\ldots\epsilon_{n_{\lambda}}^{\p}}^{\lmn}.\label{eq:Rprod_def}
\end{align}
The form factors are appropriately abbreviated as $f_{\epsilon_{\lambda}}^{\mathcal{A}}\left(\lambr\right)\equiv f_{\epsilon_{1}\ldots\epsilon_{n_{\lambda}}}^{\mathcal{A}}\left(\lambda_{1},\ldots,\lambda_{n}\right)=\bk{0|\mathcal{A}|\lambl}_{\epsilon_{\lambda}}^{\mn}$.

In many of the following calculations, the rapidities will be shifted
according to their bulk mass ratios $\mu_{\epsilon}=M_{\epsilon}/M_{+}$:
$\lambda_{k}\rightarrow\lambda_{k}-\log\mu_{\epsilon_{k}}$. This
will be denoted by a hat:
\begin{equation}
\hat{f}_{\epsilon_{\lambda}}^{\mathcal{A}}\left(\lambr\right)\equiv f_{\epsilon_{1}\ldots\epsilon_{n_{\lambda}}}^{\mathcal{A}}\left(\lambda_{1}-\log\mu_{\epsilon_{1}},\ldots,\lambda_{n_{\lambda}}-\log\mu_{\epsilon_{n_{\lambda}}}\right),\label{eq:FF_hat_def}
\end{equation}
and similarly for scattering and reflection matrices. We also define
the total energy of the shifted set $\lambr$, $\nu_{\lambda}\equiv\sum_{k=1}^{n_{\lambda}}e^{\lambda_{k}}$.

Finally, it is customary to use a graphical representation for the
matrix elements $\sidescripts{\epsilon_{1}^{\p}\ldots\epsilon_{l}^{\p}}{\mn}{\bk{\lambda_{1}^{\p},\ldots,\lambda_{l}^{\p}|\mathcal{A}|\lambda_{n},\ldots,\lambda_{1}}}{\mn}{\epsilon_{n}\ldots\epsilon_{1}}$:

\begin{equation} 		\sidescripts{\epsilon_{1}^{\p}\ldots\epsilon_{l}^{\p}}{\mn}{\bk{\lambda_{1}^{\p},\ldots,\lambda_{l}^{\p}|\mathcal{A}|\lambda_{n},\ldots,\lambda_{1}}}{\mn}{\epsilon_{n}\ldots\epsilon_{1}} \hspace{-0.1cm} = 	\hspace{-0.1cm}\begin{tikzpicture}[baseline={([yshift=-.5ex]current bounding box.center)},vertex/.style={anchor=base, 			circle,fill=black!25,minimum size=18pt,inner sep=2pt}] 			\pgfmathsetmacro{\R}{0.5} 			\pgfmathsetmacro{\L}{0.8} 			\pgfmathsetmacro{\PHI}{30} 		
			\draw [black, line width=0.8pt] plot [smooth, tension=0.8, very thick] coordinates { ({-cos(\PHI)*\R},{-sin(\PHI)*\R}) ({-cos(\PHI)*\R-\L},{-sin(\PHI)*\R}) }; 			\draw [black, line width=0.8pt] plot [smooth, tension=0.8, very thick] coordinates { ({-cos(\PHI)*\R},{sin(\PHI)*\R}) ({-cos(\PHI)*\R-\L},{sin(\PHI)*\R}) }; 		
			\node[anchor=south] at (-\R-0.4*\L, {sin(\PHI)*\R}) {\textcolor{black}{ $\lambda_1^{\prime\mathfrak{a}},\epsilon^\prime_1$}}; 			\node[anchor=north] at (-\R-0.4*\L, {-sin(\PHI)*\R}) {\textcolor{black}{ $\lambda_l^{\prime\mathfrak{a}},\epsilon^\prime_l$}}; 		
			\filldraw[black] (-\R-0.4*\L,-0.2*\R) circle (0.5pt); 			\filldraw[black] (-\R-0.4*\L,0) circle (0.5pt); 			\filldraw[black] (-\R-0.4*\L,0.2*\R) circle (0.5pt); 		
			\draw [black, line width=0.8pt] plot [smooth, tension=0.8, very thick] coordinates { ({cos(\PHI)*\R},{-sin(\PHI)*\R}) ({cos(\PHI)*\R+\L},{-sin(\PHI)*\R}) }; 			\draw [black, line width=0.8pt] plot [smooth, tension=0.8, very thick] coordinates { ({cos(\PHI)*\R},{sin(\PHI)*\R}) ({cos(\PHI)*\R+\L},{sin(\PHI)*\R}) }; 		
			\node[anchor=south] at (\R+0.4*\L, {sin(\PHI)*\R}) {\textcolor{black}{ $\lambda_1^{\mathfrak{a}},\epsilon_1$}}; 			\node[anchor=north] at (\R+0.4*\L, {-sin(\PHI)*\R}) {\textcolor{black}{ $\lambda_n^{\mathfrak{a}},\epsilon_n$}}; 		
			\filldraw[black] (\R+0.4*\L,-0.2*\R) circle (0.5pt); 			\filldraw[black] (\R+0.4*\L,0) circle (0.5pt); 			\filldraw[black] (\R+0.4*\L,0.2*\R) circle (0.5pt); 		
			\filldraw[color=black, fill=teal!20, very thick](0,0) circle (\R) node{$\mathcal{A}$}; 		\end{tikzpicture}\hspace{-0.1cm}, 	\end{equation}and, using the abbreviated notation,\begin{equation} 	\braket{0|\mathcal{A}|\cev{\lambda}}_{\epsilon}^{\mathfrak{a}} \tensor*[^{\epsilon}_{\mathfrak{a}}]{\braket{\vec{\lambda}|\mathcal{A}|0}}{_{}^{}} = 	\begin{tikzpicture}[baseline={([yshift=-.5ex]current bounding box.center)},vertex/.style={anchor=base, 			circle,fill=black!25,minimum size=18pt,inner sep=2pt}] 		\pgfmathsetmacro{\R}{0.5} 		\pgfmathsetmacro{\L}{1.5} 		\pgfmathsetmacro{\PHI}{30} 	
		\draw [black, line width=0.8pt] plot [smooth, tension=0.8, very thick] coordinates { (\R,0) (\R+\L,0) }; 	
		\node[anchor=south west] at (0.85*\R, -0.1*\R) {\textcolor{black}{$\tensor*[^{}_{}]{\cev{\lambda}}{_{}^{}}$}}; 		\node[anchor=south west] at (0.9*\R, -0.9*\R) {\textcolor{black}{$\epsilon$}}; 		\node[anchor=south east] at (1.05*\L, -0.1*\R) {\textcolor{black}{$\tensor*[^{}_{}]{\vec{\lambda}}{_{}^{}}$}}; 		\node[anchor=south east] at (1.03*\L, -0.9*\R) {\textcolor{black}{$\epsilon$}}; 	
		\filldraw[color=black, fill=teal!20, very thick](0,0) circle (\R) node{$\mathcal{A}$}; 		\filldraw[color=black, fill=teal!20, very thick](\L+\R,0) circle (\R) node{$\mathcal{A}$}; 	\end{tikzpicture}~. \end{equation}The boundary is depicted by\begin{equation} 		\begin{tikzpicture}[baseline={([yshift=-.5ex]current bounding box.center)},vertex/.style={anchor=base, 				circle,fill=black!25,minimum size=18pt,inner sep=2pt}] 			\pgfmathsetmacro{\R}{0.6} 			\pgfmathsetmacro{\L}{1} 			\pgfmathsetmacro{\PHI}{30}			 			\draw [black, line width=0.8pt] plot [smooth, tension=0.8, very thick] coordinates { (0,{-sin(\PHI)*\L}) (\L,0) (2*\L,{sin(\PHI)*\L})}; 			\draw [black, line width=0.8pt] plot [smooth, tension=0.8, very thick] coordinates { (0,{sin(\PHI)*\L}) (\L,0) (2*\L,{-sin(\PHI)*\L})}; 		
			\node[anchor=west] at (-1.4*\L, {sin(\PHI)*\L}) {\textcolor{black}{ $\lambda_{n_\lambda}^{\mathfrak{l}},\epsilon_{n_\lambda}^\prime$}}; 			\node[anchor=west] at (-1.4*\L, {-sin(\PHI)*\L}) {\textcolor{black}{ $\lambda_1^{\mathfrak{l}},\epsilon_1^\prime$}}; 			\node[anchor=south] at (\L, 0.3*\L) {\textcolor{black}{ $\mathsf{R}_{\epsilon_{\lambda}}^{\epsilon_{\lambda}^\prime}(\lambr)$}}; 			\node[anchor=west] at (2*\L, {sin(\PHI)*\L}) {\textcolor{black}{ $\lambda_1^{\mathfrak{r}},\epsilon_1$}}; 			\node[anchor=west] at (2*\L, {-sin(\PHI)*\L}) {\textcolor{black}{ $\lambda_{n_\lambda}^{\mathfrak{r}},\epsilon_{n_\lambda}$}}; 		
			\filldraw[black] (0,-0.1*\L) circle (0.5pt); 			\filldraw[black] (0,0) circle (0.5pt); 			\filldraw[black] (0,0.1*\L) circle (0.5pt); 			\filldraw[black] (2*\L,-0.1*\L) circle (0.5pt); 			\filldraw[black] (2*\L,0) circle (0.5pt); 			\filldraw[black] (2*\L,0.1*\L) circle (0.5pt); 		\end{tikzpicture}. 	\end{equation}Note that the order of the excitations is reversed upon reflection
off the boundary. In the abbreviated notation,\begin{equation} 	\braket{0|\mathcal{L}|\cev{\lambda}}_{\epsilon^\prime}^{\mathfrak{l}} \tensor*[^{\epsilon}_{\mathfrak{r}}]{\braket{\cev{\lambda}|\mathcal{R}|0}}{_{}^{}} = 	\begin{tikzpicture}[baseline={([yshift=-.5ex]current bounding box.center)},vertex/.style={anchor=base, 			circle,fill=black!25,minimum size=18pt,inner sep=2pt}] 		\pgfmathsetmacro{\R}{0.5} 		\pgfmathsetmacro{\L}{1.5} 		\pgfmathsetmacro{\PHI}{30} 	
		\draw [black, line width=0.8pt] plot [smooth, tension=0.8, very thick] coordinates { ({\R},0) ({\R+\L},0) }; 	
		\node[anchor=south west] at (0.85*\R, -0.1*\R) {\textcolor{black}{$\tensor*[^{}_{}]{\cev{\lambda}}{_{}^{}}$}}; 		\node[anchor=south west] at (0.9*\R, -0.9*\R) {\textcolor{black}{$\epsilon^\prime$}}; 		\node[anchor=south east] at (1.05*\L, -0.1*\R) {\textcolor{black}{$\tensor*[^{}_{}]{\cev{\lambda}}{_{}^{}}$}}; 		\node[anchor=south east] at (1.03*\L, -0.9*\R) {\textcolor{black}{$\epsilon$}}; 	
		\filldraw[color=black, fill=teal!20, very thick](0,0) circle (\R) node{$\mathcal{L}$}; 		\filldraw[color=black, fill=teal!20, very thick](\L+\R,0) circle (\R) node{$\mathcal{R}$}; 	\end{tikzpicture}~. \end{equation} 

The form factors in the massless sine-Gordon model are given in Appendix
\ref{app:FF}. The form factors, scattering matrices, and reflection
matrices are all of the ingredients we need to calculate our desired
response functions.

\section{\label{sec:total_rate}The total inelastic decay rate and elastic
phase shift}

We are now equipped with all of the necessary tools to calculate response
functions. In this section, we calculate the reflection coefficient
$r\left(\omega\right)$, which defines the total inelastic decay rate
$\gamma\left(\omega\right)$ and the elastic phase shift $\delta\left(\omega\right)$
(Eq. (\ref{eq:r_rates})), using Eqs. (\ref{eq:landauer}) and (\ref{eq:GAB}).

\subsection{\label{subsec:AC_conductance}The AC conductance}

The conductance $\GLR\left(x,x^{\p};\omega\right)$ has been calculated
in Ref. \citep{lesage_form_1996} for the boundary sinh-Gordon and
bsG models. The conductance of the bsG model with $z=1/3$ is of particular
interest, as it describes the tunneling of fractionally-charged excitations
in a fractional quantum Hall sample at filling $\nu=1/3$. Closely-related
expressions were also obtained for the Kondo model. Here we retrace
the steps of Ref. \citep{lesage_form_1996} and calculate $\GLR\left(x,x^{\p};\omega\right)$
for the bsG and Kondo models with general coupling parameters. We
emphasize the key steps in the calculation of the 2-point response
function, laying the groundwork for the derivation of the 3-point
response function in the next Section. We also show that the non-causal
response function $\GRL\left(x,x^{\p};\omega\right)$, measuring the
response of a right moving current to a left moving current perturbation,
is manifestly zero in all models, for any coupling parameter and impurity
strength, as it should be.

The Kubo formula relates the conductance $\GLR\left(x,x^{\p};\omega\right)$
to a 2-point response function of the current operators. Each of the
two correlators in Eq. (\ref{eq:GAB}) can be evaluated via a form
factors expansion, by inserting the complete set of states in Eq.
(\ref{eq:complete_set_3}). As we will now show, it is necessary to
combine the two correlators to obtain a single unified expansion for
$\GLR\left(x,x^{\p};\omega\right)$. Let us start with $\left\langle \mathcal{L}\left(x,t\right)\mathcal{R}\left(x^{\p},0\right)\right\rangle $;
using Eqs. (\ref{eq:complete_set_3}), (\ref{eq:xt_dep}), (\ref{eq:FF_cross_conj}),
and (\ref{eq:FF_lr}), we find\begin{align} 		\left\langle \mathcal{L}\left(x,t\right)\mathcal{R}\left(x^{\prime},0\right)\right\rangle & \nonumber \\ 		&\hspace{-2.7cm}=\int_{\lambda}\mathsf{R}_{\epsilon_{\lambda}}^{\epsilon_{\lambda}^{\p}}\left(\lambr\right)\braket{0|\mathcal{L}\left(x,t\right)|\lambr}_{\epsilon_{\lambda}^{\p}}^{\lmn}\tensor*[_{\rmn}^{\epsilon_{\lambda}}]{\braket{\lambr|\mathcal{R}\left(x^{\p},0\right)|0}}{_{{}}^{{}}} \nonumber \\ 		&\hspace{-2.7cm} = \int_\lambda \mathsf{R}_{\epsilon_{\lambda}}^{\epsilon^{\prime}_{\lambda}}\left(\lambr\right)e^{-\mathrm{i}\sum_{k=1}^{n_\lambda}\mu_{\epsilon_{k}}e^{\lambda_{k}}\left(x+x^{\prime}+t\right)} 		\times \begin{tikzpicture}[baseline={([yshift=-.5ex]current bounding box.center)},vertex/.style={anchor=base, 			circle,fill=black!25,minimum size=18pt,inner sep=2pt}] 		\pgfmathsetmacro{\R}{0.5} 		\pgfmathsetmacro{\L}{0.8} 		\pgfmathsetmacro{\PHI}{30}			 		\draw [black, line width=0.8pt] plot [smooth, tension=0, very thick] coordinates {(0,0) (\L,0)}; 		\node[anchor=south west] at (-0.15*\R, 0) {\textcolor{black}{ $\lambr$}}; 		\node[anchor=south east] at (\L + 0.1*\R, 0) {\textcolor{black}{ $\lambr$}}; 		\node[anchor=south west] at (-0.15*\R, -0.9*\R) {\textcolor{black}{ $\epsilon^\p$}}; 		\node[anchor=south east] at (\L + 0.1*\R, -0.9*\R) {\textcolor{black}{ $\epsilon$}}; 		\filldraw[color=black, fill=teal!20, very thick](-\R,0) circle (\R) node{$\mathcal{L}$}; 		\filldraw[color=black, fill=teal!20, very thick](\L+\R,0) circle (\R) node{$\mathcal{R}$}; 	\end{tikzpicture} \nonumber \\ 	&\hspace{-2.7cm}=\int_{\lambda}\mathsf{R}_{\epsilon_{\lambda}}^{\epsilon_{\lambda}^{\p}}\left(\lambr\right)e^{-\ii\sum_{k=1}^{n_\lambda}\mu_{\epsilon_{k}}e^{\lambda_{k}}\left(x+x^{\p}+t\right)}f_{\bar{\epsilon}_{\lambda}^{\p}}^{\mathcal{R}}\left(\lambr\right)\left(f_{\epsilon_{\lambda}}^{\mathcal{R}}\left(\lambr\right)\right)^{*}. 	\end{align}Shifting
the rapidities according to the bulk mass ratios of their corresponding
excitations, $\lambda_{k}\rightarrow\lambda_{k}-\log\mu_{\epsilon_{k}}$,
leads to
\begin{align}
\left\langle \mathcal{L}\left(x,t\right)\mathcal{R}\left(x^{\p},0\right)\right\rangle = & \int_{\lambda}\hat{\mathsf{R}}_{\epsilon_{\lambda}}^{\epsilon_{\lambda}^{\p}}\left(\lambr\right)e^{-\ii\nu_{\lambda}\left(x+x^{\p}+t\right)}\nonumber \\
 & \hspace{-1cm}\times\hat{f}_{\bar{\epsilon}_{\lambda}^{\p}}^{\mathcal{R}}\left(\lambr\right)\left(\hat{f}_{\epsilon_{\lambda}}^{\mathcal{R}}\left(\lambr\right)\right)^{*}.
\end{align}
We now introduce an auxiliary rapidity, $\kappa$:
\begin{align}
\left\langle \mathcal{L}\left(x,t\right)\mathcal{R}\left(x^{\p},0\right)\right\rangle = & \int_{-\infty}^{\infty}\id\kappa e^{\kappa}\delta\left(\nu_{\lambda}-e^{\kappa}\right)\int_{\lambda}\hat{\mathsf{R}}_{\epsilon_{\lambda}}^{\epsilon_{\lambda}^{\p}}\left(\lambr\right)\nonumber \\
 & \hspace{-1cm}\times e^{-\ii\nu_{\lambda}\left(x+x^{\p}+t\right)}\hat{f}_{\bar{\epsilon}_{\lambda}^{\p}}^{\mathcal{R}}\left(\lambr\right)\left(\hat{f}_{\epsilon_{\lambda}}^{\mathcal{R}}\left(\lambr\right)\right)^{*}.
\end{align}
Shifting the rapidities by $\kappa$, $\lambda_{k}\rightarrow\lambda_{k}+\kappa$,
we obtain
\begin{align}
\left\langle \mathcal{L}\left(x,t\right)\mathcal{R}\left(x^{\p},0\right)\right\rangle = & \int_{-\infty}^{\infty}\id\kappa e^{-\ii e^{\kappa}\left(x+x^{\p}+t\right)}e^{2\kappa}\nonumber \\
 & \hspace{-3cm}\times\int_{\lambda}\hat{\mathsf{R}}_{\epsilon_{\lambda}}^{\epsilon_{\lambda}^{\p}}\left(\lambr+\kappa\right)\delta\left(\nu_{\lambda}-1\right)\hat{f}_{\bar{\epsilon}_{\lambda}^{\p}}^{\mathcal{R}}\left(\lambr\right)\left(\hat{f}_{\epsilon_{\lambda}}^{\mathcal{R}}\left(\lambr\right)\right)^{*},\label{eq:LR_corr}
\end{align}
where we used Lorentz invariance, Eq. (\ref{eq:FF_lorentz}), and
the notation $\lambr+\kappa=\left\{ \lambda_{1}+\kappa,\ldots,\lambda_{n_{\lambda}}+\kappa\right\} $.
For the second correlator of the commutator, $\left\langle \mathcal{R}\left(x^{\p},0\right)\mathcal{L}\left(x,t\right)\right\rangle $,
following the same steps, we find
\begin{align}
\left\langle \mathcal{R}\left(x^{\p},0\right)\mathcal{L}\left(x,t\right)\right\rangle = & \int_{-\infty}^{\infty}\id\kappa e^{\ii e^{\kappa}\left(x+x^{\p}+t\right)}e^{2\kappa}\nonumber \\
 & \hspace{-2.8cm}\times\int_{\lambda}\left(\hat{\mathsf{R}}_{\epsilon_{\lambda}}^{\epsilon_{\lambda}^{\p}}\left(\lambr+\kappa\right)\right)^{*}\delta\left(\nu_{\lambda}-1\right)\hat{f}_{\bar{\epsilon}_{\lambda}^{\p}}^{\mathcal{R}}\left(\lambr\right)\left(\hat{f}_{\epsilon_{\lambda}}^{\mathcal{R}}\left(\lambr\right)\right)^{*}.\label{eq:RL_corr}
\end{align}
Now, we shift $\kappa\rightarrow\kappa-\ii\pi$ in Eq. (\ref{eq:RL_corr}).
Using $\hat{\mathsf{R}}_{\epsilon_{\lambda}}^{\epsilon_{\lambda}^{\p}}\left(\lambr+\ii\pi\right)=\left(\hat{\mathsf{R}}_{\epsilon_{\lambda}}^{\epsilon_{\lambda}^{\p}}\left(\lambr\right)\right)^{*}$
(see Appendix \ref{app:R}), we find
\begin{align}
\left\langle \mathcal{R}\left(x^{\p},0\right)\mathcal{L}\left(x,t\right)\right\rangle = & \int_{-\infty+\ii\pi}^{\infty+\ii\pi}\id\kappa e^{-\ii e^{\kappa}\left(x+x^{\p}+t\right)}e^{2\kappa}\nonumber \\
 & \hspace{-2.8cm}\times\int_{\lambda}\hat{\mathsf{R}}_{\epsilon_{\lambda}}^{\epsilon_{\lambda}^{\p}}\left(\lambr+\kappa\right)\delta\left(\nu_{\lambda}-1\right)\hat{f}_{\bar{\epsilon}_{\lambda}^{\p}}^{\mathcal{R}}\left(\lambr\right)\left(\hat{f}_{\epsilon_{\lambda}}^{\mathcal{R}}\left(\lambr\right)\right)^{*}.\label{eq:LR_corr_2}
\end{align}
Next, plug Eqs. (\ref{eq:LR_corr}) and (\ref{eq:LR_corr_2}) to Eq.
(\ref{eq:GAB}). Integrating over time yields
\begin{align}
\GLR\left(x,x^{\p};\omega\right)= & \lim_{\eta\rightarrow0^{+}}\frac{1}{8\pi\omega}\left(\int_{-\infty}^{\infty}\id\kappa+\int_{\infty+\ii\pi}^{-\infty+\ii\pi}\id\kappa\right)\nonumber \\
 & \hspace{-1cm}\times e^{2\kappa}\frac{\ii e^{-\ii e^{\kappa}\left(x+x^{\p}\right)}}{\omega-e^{\kappa}+\ii\eta}\int_{\lambda}\hat{\mathsf{R}}_{\epsilon_{\lambda}}^{\epsilon_{\lambda}^{\p}}\left(\lambr+\kappa\right)\nonumber \\
 & \hspace{-1cm}\times\delta\left(\nu_{\lambda}-1\right)\hat{f}_{\bar{\epsilon}_{\lambda}^{\p}}^{\mathcal{R}}\left(\lambr\right)\left(\hat{f}_{\epsilon_{\lambda}}^{\mathcal{R}}\left(\lambr\right)\right)^{*}.\label{eq:GLR_contour}
\end{align}
The integration contour of $\kappa$ can be closed by adding the edges
at $\kappa=\sigma+\ii\chi$, with $\sigma\rightarrow\pm\infty$ and
$0\le\chi\le\pi$, where the integrand decays rapidly (recall that
$x,x^{\p}<0$). In the presence of bound states, $z<1/2$, there could
be poles of the reflection matrices enclosed by the contour; however,
the residue of the integrand at $\kappa=\kappa_{0}$ gives rise to
the exponential $e^{e^{\mathrm{Re}\left\{ \kappa_{0}\right\} }\sin\left(\mathrm{Im}\left\{ \kappa_{0}\right\} \right)\left(x+x^{\p}\right)}$,
where $\sin\left(\mathrm{Im}\left\{ \kappa_{0}\right\} \right)>0$,
hence this term vanishes in the limit of a half-infinite line for
large enough $\left|x+x^{\p}\right|$. Therefore, the only pole contributing
to the integral is at $\kappa=\log\omega+\ii\eta$:
\begin{align}
\GLR\left(x,x^{\p};\omega\right)= & \frac{1}{4}e^{-\ii\omega\left(x+x^{\p}\right)}\int_{\lambda}\hat{\mathsf{R}}_{\epsilon_{\lambda}}^{\epsilon_{\lambda}^{\p}}\left(\lambr+\log\omega\right)\nonumber \\
 & \hspace{-1cm}\times\delta\left(\nu_{\lambda}-1\right)\hat{f}_{\bar{\epsilon}_{\lambda}^{\p}}^{\mathcal{R}}\left(\lambr\right)\left(\hat{f}_{\epsilon_{\lambda}}^{\mathcal{R}}\left(\lambr\right)\right)^{*}.\label{eq:GLR_result}
\end{align}
Before we proceed to the calculation of the reflection coefficient,
it is instructive to consider the non-causal conductance $\GRL\left(x,x^{\p};\omega\right)$,
and show that it vanishes. An identical treatment would lead to
\begin{align}
\GRL\left(x,x^{\p};\omega\right)= & \lim_{\eta\rightarrow0^{+}}\frac{1}{8\pi\omega}\left(\int_{-\infty}^{\infty}\id\kappa+\int_{\infty-\ii\pi}^{-\infty-\ii\pi}\id\kappa\right)\nonumber \\
 & \hspace{-1cm}\times e^{2\kappa}\frac{\ii e^{\ii e^{\kappa}\left(x+x^{\p}\right)}}{\omega-e^{\kappa}+\ii\eta}\int_{\lambda}\left(\hat{\mathsf{R}}_{\epsilon_{\lambda}}^{\epsilon_{\lambda}^{\p}}\left(\lambr+\kappa\right)\right)^{*}\nonumber \\
 & \hspace{-1cm}\times\delta\left(\nu_{\lambda}-1\right)\hat{f}_{\bar{\epsilon}_{\lambda}^{\p}}^{\mathcal{R}}\left(\lambr\right)\left(\hat{f}_{\epsilon_{\lambda}}^{\mathcal{R}}\left(\lambr\right)\right)^{*}.
\end{align}
Note the three crucial differences compared to Eq. (\ref{eq:GLR_contour}):
the complex conjugation of the reflection matrices, the sign in the
space-dependent exponential, and the integral $\int_{\infty-\ii\pi}^{-\infty-\ii\pi}\id\kappa$,
which runs along the line $\mathrm{Im}\kappa=-\ii\pi$. This time,
the pole at $\kappa=\omega+\ii\eta$ is not enclosed by the contour,
leading to $\GRL\left(x,x^{\p};\omega\right)=0$.

\subsection{\label{subsec:origin_of_decay}The reflection coefficient and the
origin of inelastic decay}

We now return to Eq. (\ref{eq:GLR_result}) and extract the reflection
coefficient using Eq. (\ref{eq:landauer}). Reshifting the rapidities
as $\lambda_{k}\rightarrow\lambda_{k}-\log\omega$ yields
\begin{equation}
r\left(\omega\right)=\frac{1}{\omega}\int_{\lambda}\hat{\mathsf{R}}_{\epsilon_{\lambda}}^{\epsilon_{\lambda}^{\p}}\left(\lambr\right)\delta\left(\nu_{\lambda}-\omega\right)\hat{f}_{\bar{\epsilon}_{\lambda}^{\p}}^{\mathcal{R}}\left(\lambr\right)\left(\hat{f}_{\epsilon_{\lambda}}^{\mathcal{R}}\left(\lambr\right)\right)^{*}.\label{eq:r_coeff}
\end{equation}
Recall that $\nu_{\lambda}=\sum_{k=1}^{n_{\lambda}}e^{\lambda_{k}}$
is the sum of energies of the excitations in the set $\lambda$. The
physical interpretation of the expression above is clear --- the
reflection coefficient of a photon with frequency $\omega$ is given
by the sum over all excitations with energies summing up to $\omega$.
Note that this result is general, and applies to any integrable boundary
model --- the choice of model specifies the reflection matrices and
form factors. The only energy scale, $T_{B}$, is encoded within the
reflection matrices. While the sum in (\ref{eq:r_coeff}) runs over
all possible number of excitations $n_{\lambda}$, the terms decay
very rapidly with $n_{\lambda}$, and in practice it is enough to
calculate only a few terms to obtain $r\left(\omega\right)$ with
excellent accuracy. The accuracy of the expansion can be evaluated
using the high frequency behavior of $r\left(\omega\right)$; using
$\hat{\mathsf{R}}_{\epsilon_{\lambda}}^{\epsilon_{\lambda}^{\p}}\left(\lambr\rightarrow\infty\right)\rightarrow\delta_{\bar{\epsilon}_{\lambda}}^{\epsilon_{\lambda}^{\p}}$
(Eq. (\ref{eq:Rprod_high_energy})), we find $r\left(\omega\gg T_{B}\right)\rightarrow\sum_{\epsilon_{\lambda}}r_{\epsilon_{\lambda}}^{0}=1$
in all models, where
\begin{equation}
r_{\epsilon_{\lambda}}^{0}\equiv\int_{-\infty}^{\infty}\frac{\prod_{k=1}^{n_{\lambda}}\id\lambda_{k}}{\left(2\pi\right)^{n_{\lambda}}n_{\lambda}!}\delta\left(\nu_{\lambda}-1\right)\left|\hat{f}_{\epsilon_{\lambda}}^{\mathcal{R}}\left(\lambr\right)\right|^{2}.\label{eq:r0}
\end{equation}
Then, calculating the truncated sum $\sum_{\epsilon_{\lambda}}^{\p}r_{\epsilon_{\lambda}}^{0}$
provides an estimation for the accuracy of the result, and sets an
upper bound on the contribution of the remaining terms (since $r_{\epsilon_{\lambda}}^{0}>0$
for all $\epsilon_{\lambda}$). At low frequencies, we have $r_{\K}\left(\omega\ll T_{B}\right)\rightarrow\sum_{\epsilon_{\lambda}}r_{\epsilon_{\lambda}}^{0}=1$
and $r_{\bsg}\left(\omega\ll T_{B}\right)\rightarrow-\sum_{\epsilon_{\lambda}}r_{\epsilon_{\lambda}}^{0}=-1$
(using Eq. (\ref{eq:R_prod_low_energy_bSG})). It is shown in Fig.
\ref{fig:r0} in Appendix \ref{app:FF} that, indeed, only a few terms
are needed to obtain $1-\sum_{\epsilon_{\lambda}}^{\p}r_{\epsilon_{\lambda}}^{0}<10^{-2}$
for most values of $z$.

The leading terms for the bsG and Kondo models are as follows. First,
a soliton-antisoliton pair contributes
\begin{align}
r_{+-}\left(\omega\right)= & \frac{1}{\omega}\int_{-\infty}^{\infty}\frac{\id\lambda_{1}\id\lambda_{2}}{\left(2\pi\right)^{2}}\delta\left(e^{\lambda_{1}}+e^{\lambda_{2}}-\omega\right)\nonumber \\
 & \hspace{-2cm}\times\left[e^{-\frac{\ii\pi}{2z}}R_{+}^{-}\left(\lambda_{1}\right)R_{-}^{+}\left(\lambda_{2}\right)-e^{\frac{\ii\pi}{2z}}R_{+}^{+}\left(\lambda_{1}\right)R_{+}^{+}\left(\lambda_{2}\right)\right]\nonumber \\
 & \hspace{-2cm}\times\left|f_{+-}^{\mathcal{R}}\left(\lambda_{1},\lambda_{2}\right)\right|^{2}.\label{eq:r_2s}
\end{align}
This is the leading term for $z\ge1/2$. In the attractive regime,
$z<1/2$, the dominant contribution comes from a single breather:
\begin{equation}
r_{m}\left(\omega\right)=\frac{\left|f_{m}^{\mathcal{R}}\left(0\right)\right|^{2}}{2\pi\mu_{m}^{2}}R_{m}^{m}\left(\log\left(\frac{\omega}{\mu_{m}}\right)\right),\label{eq:r_b}
\end{equation}
where $1\le m\le\maxbreath$ is odd; the largest contribution is of
$m=1$. Two other non-negligible contributions are
\begin{align}
r_{+-1}\left(\omega\right)= & \frac{1}{\omega}\int_{-\infty}^{\infty}\frac{\id\lambda_{1}\id\lambda_{2}\id\lambda_{3}}{\left(2\pi\right)^{3}}\delta\left(e^{\lambda_{1}}+e^{\lambda_{2}}+e^{\lambda_{2}}-\omega\right)\nonumber \\
 & \hspace{-1.7cm}\times\left[e^{-\frac{\ii\pi}{2z}}R_{+}^{-}\left(\lambda_{1}\right)R_{-}^{+}\left(\lambda_{2}\right)+e^{\frac{\ii\pi}{2z}}R_{+}^{+}\left(\lambda_{1}\right)R_{+}^{+}\left(\lambda_{2}\right)\right]\nonumber \\
 & \hspace{-1.7cm}\times R_{1}^{1}\left(\lambda_{3}-\log\mu_{1}\right)\left|f_{+-1}^{\mathcal{R}}\left(\lambda_{1},\lambda_{2},\lambda_{3}-\log\mu_{1}\right)\right|^{2},\label{eq:r_2s1b}
\end{align}
and
\begin{align}
r_{12}\left(\omega\right)= & \frac{1}{\omega}\int_{-\infty}^{\infty}\frac{\id\lambda_{1}\id\lambda_{2}}{\left(2\pi\right)^{2}}\delta\left(e^{\lambda_{1}}+e^{\lambda_{2}}-\omega\right)\nonumber \\
 & \hspace{-1cm}\times R_{1}^{1}\left(\lambda_{1}-\log\mu_{1}\right)R_{2}^{2}\left(\lambda_{2}-\log\mu_{2}\right)\nonumber \\
 & \hspace{-1cm}\times\left|f_{12}^{\mathcal{R}}\left(\lambda_{1}-\log\mu_{1},\lambda_{2}-\log\mu_{2}\right)\right|^{2}.\label{eq:r_2b}
\end{align}
In the following, we evaluate the reflection coefficient at integer
$p=1/z$ as
\begin{equation}
r\left(\omega\right)\approx\sum_{m=1}^{p-2}r_{m}\left(\omega\right)+r_{+-}\left(\omega\right)+r_{12}\left(\omega\right)+r_{+-1}\left(\omega\right).\label{eq:r_eval}
\end{equation}
The evaluation of $r_{+-1}$ is significantly more complicated for
non-integer $1/z$. However, Fig. \ref{fig:r0} shows that $r\left(\omega\right)\approx\sum_{m}r_{m}\left(\omega\right)+r_{+-}\left(\omega\right)+r_{12}\left(\omega\right)$
is still a very good approximation ($1-r_{+-}^{0}-\sum_{m}r_{m}^{0}-r_{12}^{0}\ll1$)
for $z\lesssim1/4$ (where the contribution of $r_{12}^{\mathcal{R}}$
becomes larger than that of $r_{+-1}^{\mathcal{R}}$), and also for
$z>1/2$, as long as $z$ is not too large (where $r_{+-+-}$ is non-negligible).
The inelastic decay rate and elastic phase shift are then readily
extracted following Eq. (\ref{eq:r_rates}).

Eq. (\ref{eq:r_coeff}) reveals the origin of the inelastic scattering.
The inelastic decay rate of a photon $\omega$ is
\begin{equation}
\gamma\left(\omega\right)=-\log\left|r\left(\omega\right)\right|^{2}\approx1-\left|r\left(\omega\right)\right|^{2},
\end{equation}
assuming $1-\left|r\left(\omega\right)\right|^{2}\ll1$. We thus identify
the coherent sum in Eq. (\ref{eq:r_coeff}) as the source of the photon
decay. A plane wave mode at frequency $\omega$ impinging on the boundary
can be formally written as a sum of eigenstates, $\k{\lambl}_{\epsilon_{\lambda}}^{\rmn}$,
with appropriate weights and phases. Note that this is a nonlinear
relation --- plane waves cannot be expressed as a sum of individual
solitons and breathers, and must be spanned using all eigenstates
with any number of excitations. While each excitation in each of the
eigenstates is reflected elastically off the boundary, it picks up
a phase, determined by the reflection matrix, that depends on its
type and energy. These relative phases between the eigenstates alter
the specific weights of the decomposition of the incoming photon.
Therefore, the reflected excitations no longer form a single photon
at frequency $\omega$, but rather a set of photons with frequencies
$\omega_{i}$, such that $\sum_{i}\omega_{i}=\omega$.

The nonlinear relation between the photons and elementary excitations
is generally implicit, and is hidden within the form factors of the
theory, which are the matrix elements of the derivatives of the bosonic
field $\phi$ in the basis of solitons and breathers. An explicit
relation can be found at the free-fermion point, $z=1/2$, using refermionization
\citep{gogolin_bosonization_2004}, as discussed in Appendix \ref{app:referm}.
Introducing a fermionic field $\psi\left(x\right)\sim e^{\ii\phi\left(x\right)}$,
we find a one-to-one correspondence between $\psi,\psi^{\dagger}$
and solitons and antisolitons by expanding the fermionic field as
\begin{equation}
\psi\left(x;t\right)=\frac{1}{\sqrt{2\ell}}\sum_{k}\psi_{k}\left(t\right)e^{\ii kx}=\frac{1}{\sqrt{2\ell}}\sum_{k}\psi_{k}e^{\ii k\left(x-t\right)},\label{eq:eigenmdoes}
\end{equation}
with $k=n\Delta$ with $n\in\mathbb{Z}$. The commutation relations
of $\psi_{k},\psi_{k}^{\dagger}$ satisfy the Zamolodchikov-Faddeev
algebra (Eq. (\ref{eq:ZF_alg})) at $z=1/2$, where the S-matrix becomes
trivial, $S_{+-}^{+-}=S_{++}^{++}=-1$. We therefore identify $\psi_{k>0}$
with a soliton with energy $k$, and $\psi_{-k<0}^{\dagger}$ with
an antisoliton with energy $k$, establishing an explicit nonlinear
relation between the solitons and the bosonic field $\phi$. The calculation
of $r\left(\omega\right)$ simplifies considerably at the free-fermion
point, as the sole contribution to the reflection coefficient is Eq.
(\ref{eq:r_2s}), and the particularly compact expressions for the
reflection matrices (Eqs. (\ref{eq:R_bSG_z_half})-(\ref{eq:Rs_K}))
and form factors (Eq. (\ref{eq:f+-_z_half})) allow for closed analytical
expressions:
\begin{align}
r_{\bsg}\left(\omega;z=\frac{1}{2}\right)= & 1-\frac{2\ii T_{B}}{\omega}\log\left(1-\frac{\ii\omega}{T_{B}}\right),\nonumber \\
r_{\K}\left(\omega;z=\frac{1}{2}\right)= & 1-\frac{4\ii T_{B}}{\omega+2\ii T_{B}}\log\left(1-\frac{\ii\omega}{T_{B}}\right).\label{eq:r_half_exact}
\end{align}
The same expressions are derived in Appendix \ref{app:referm} using
refermionization.\vspace{-0.2cm}

\subsection{\label{subsec:results_total}Results}

\begin{figure}[t]
\begin{centering}
\includegraphics[width=1\columnwidth]{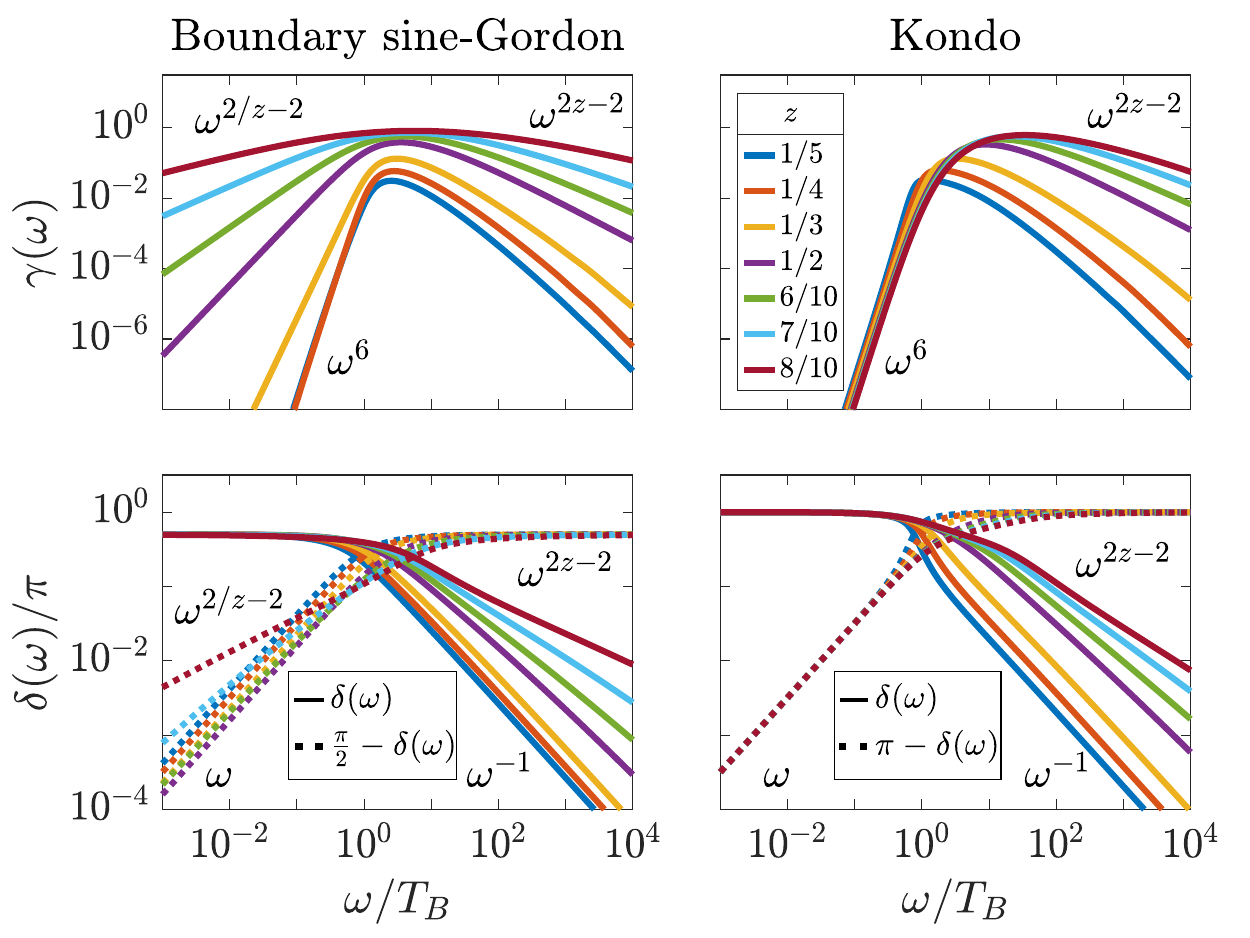}
\par\end{centering}
\caption{\label{fig:total_rates}Total inelastic rate $\gamma\left(\omega\right)$
and phase shift $\delta\left(\omega\right)$ for the bsG and Kondo
models and several values of $z$. The power laws at low and high
frequencies are denoted on the plots. In the phase shift panels, we
plot both $\delta\left(\omega\right)$ and $\pi/2-\delta\left(\omega\right)$
or $\pi-\delta\left(\omega\right)$ (for the bsG and Kondo models,
respectively). We use Eq. (\ref{eq:r_eval}) to evaluate $r\left(\omega\right)$
at integer $p=1/z$, and $r\left(\omega\right)=r_{+-}\left(\omega\right)$
at noninteger $z>1/2$.}
\end{figure}

The rates $\gamma\left(\omega\right)$ and $\delta\left(\omega\right)$
in both models, for several values of $z$, are displayed in Fig.
\ref{fig:total_rates}. The inelastic rate in both models follows
a Luttinger liquid power law at high frequencies, $\gamma\left(\omega\gg T_{B}\right)\sim\omega^{2z-2}$,
which may be obtained from the Hamiltonians (\ref{eq:HIbSG}) and
(\ref{eq:HIK}) by means of perturbation theory \citep{leger_revealing_2022,goldstein_inelastic_2013,kuzmin_observation_2023}.
\textcolor{black}{Note that Eq. (\ref{eq:TB_EJs}) restores the dependence
of the rates on the UV cutoff $\Lambda$ (as long as $\omega\ll\Lambda$),
which is present in a perturbative treatment and should be considered
for quantitative comparisons with experimental measurements, not only
in the high frequency regime, but also in the low frequency regime,
$\omega\ll E_{J}^{\star}$.} Indeed, perturbation theory is invalid
for the Hamiltonians (\ref{eq:HIbSG}) and (\ref{eq:HIK}) for $z<1$
and $\omega\ll E_{J}^{\star}$. The low frequency power laws for the
rates are model dependent, and could be predicted from the expansions
near the strong coupling fixed points. In the Kondo model, the dominant
contribution stems from a quartic density term, $\rho^{4}$, leading
to $\gamma_{\K}\left(\omega\ll T_{B}\right)\sim\omega^{6}$ for all
$z<1$ \citep{goldstein_inelastic_2013}. In bsG, the leading expansion
terms are the quartic phase term $\phi^{4}$, giving rise to the same
$\omega^{6}$ power law as the quartic density term, and the dual
cosine $\cos\left(\pi\rho\right)$, which generates instantons between
the minima of the cosine potential, and leads to a $\omega^{2/z-2}$
behavior \citep{burshtein_photon-instanton_2021}; hence, $\gamma_{\bsg}\left(\omega\ll T_{B};z\ge1/4\right)\sim\omega^{2/z-2}$
and $\gamma_{\bsg}\left(\omega\ll T_{B};z<1/4\right)\sim\omega^{6}$.

The asymptotic behavior of the phase shift $\delta\left(\omega\right)$
may be obtained from Eq. (\ref{eq:r_coeff}), rewritten here as
\begin{equation}
r\left(\omega\right)=\int_{\lambda}\hat{\mathsf{R}}_{\epsilon_{\lambda}}^{\epsilon_{\lambda}^{\p}}\left(\lambr+\log\omega\right)\delta\left(\nu_{\lambda}-1\right)\hat{f}_{\bar{\epsilon}_{\lambda}^{\p}}^{\mathcal{R}}\left(\lambr\right)\left(\hat{f}_{\epsilon_{\lambda}}^{\mathcal{R}}\left(\lambr\right)\right)^{*}.
\end{equation}
As discussed in Ref. \citep{lesage_form_1996}, one may expand the
reflection matrices as a power series in $\omega/T_{B}$ or $T_{B}/\omega$,
for low or high frequencies, respectively. In the bsG model, the soliton
matrices $R_{\pm}^{\pm/\mp}\left(\lambda\right)$ expand as double
power series in $e^{\lambda}$ and $e^{\left(1/z-1\right)\lambda}$,
and the soliton matrices $R_{m}^{m}\left(\lambda\right)$ expand as
power series in $e^{\lambda}$. We find $\delta_{\bsg}\left(\omega\gg T_{B};z\ge1/2\right)\sim\omega^{2z-2}$
and $\delta_{\bsg}\left(\omega\gg T_{B};z<1/2\right)\sim\omega^{-1}$
at high frequencies (in agreement with perturbation theory), and $\pi/2-\delta_{\bsg}\left(\omega\ll T_{B};z\ge2/3\right)\sim\omega^{2/z-2}$
and $\pi/2-\delta_{\bsg}\left(\omega\ll T_{B};z<2/3\right)\sim\omega$
at low frequencies, which are again anticipated by the expansion near
the strong coupling fixed point. The reflection matrices of the Kondo
model are considerably simpler and all expand as power series in $e^{\lambda}$,
leading to $\delta_{\K}\left(\omega\gg T_{B};z\ge1/2\right)\sim\omega^{2z-2}$,
$\delta_{\K}\left(\omega\gg T_{B};z<1/2\right)\sim\omega^{-1}$, and
$\pi-\delta_{\K}\left(\omega\ll T_{B}\right)\sim\omega$. We note
that all of the scaling laws of the scattering rates $\gamma_{\bsg}\left(\omega\right)$
and $\delta_{\bsg}\left(\omega\right)$ in the bsG model, at frequencies
both above and below the RG scale, are in agreement with a recent
theoretical study combining perturbation theory at $\omega\gg T_{B}$
and strong coupling expansions at $\omega\ll T_{B}$ \citep{houzet_microwave_2023};
our exact calculation provides the full crossover between the two
regimes.

To conclude this section, we relate the limiting cases of the inelastic
and elastic rates to the Schmid-Bulgadaev transition. Approaching
the transition, $z\rightarrow1$, we find that both low and high frequency
power laws, $2/z-2$ and $2z-2$, respectively, tend towards 0. In
other words, the rate $\gamma_{\bsg}\left(\omega\right)$ and phase
shift $\delta_{\bsg}\left(\omega\right)$ become frequency-independent
at the transition point, $z=1$. Perturbation theory at $z>1$ shows
that $\gamma_{\bsg}\left(\omega\right)\sim\omega^{2z-2}$ for both
low and high frequencies (since the boundary cosine term of Eq. (\ref{eq:HIbSG})
becomes irrelevant at $z>1$), and $\delta_{\bsg}\left(\omega;1<z\le3/2\right)\sim\omega^{2z-2}$
and $\delta_{\bsg}\left(\omega;z>3/2\right)\sim\omega$, also for
all frequencies \citep{kuzmin_observation_2023}. Crucially, the phase
shift at $\omega=0$ jumps from $\pi/2$ to $0$ across the transition,
and changes sign at finite frequencies. The sign change of $\delta_{\bsg}\left(\omega\right)$
and the different trends of $\gamma_{\bsg}\left(\omega\right)$ on
the two sides of the transition, both at finite frequencies, were
recently observed using single photon spectroscopy, providing evidence
for the long-sought-after quantum phase transition \citep{kuzmin_observation_2023}.

\section{\label{sec:spec}The energy-resolved inelastic decay spectrum}

\subsection{The spectrum as a 3-point response function}

The inelastic decay spectrum, $\gamma\left(\omega^{\p}\lvert\omega\right)$,
measures the production rate of photons at frequency $\omega^{\p}$
due to a splitting of a photon with frequency $\omega>\omega^{\p}$.
It is related to the total inelastic decay rate, $\gamma\left(\omega\right)$,
by means of an energy-conservation sum rule:
\begin{equation}
\int_{0}^{\omega}\omega^{\p}\gamma\left(\omega^{\p}\lvert\omega\right)\id\omega^{\p}=\omega\gamma\left(\omega\right).\label{eq:sumrule}
\end{equation}
It was shown in Ref. \citep{goldstein_inelastic_2013} that, for the
generic impurity setup in Eq. (\ref{eq:H_generic_redef}), $\gamma\left(\omega^{\p}\lvert\omega\right)$
can be found by calculating a 3-point response function:
\begin{equation}
\gamma\left(\omega^{\prime}\lvert\omega\right)=\lim_{\eta\rightarrow0^{+}}\dfrac{2\eta\ell}{\omega}G_{\omega^{\p}}^{\mathrm{cqq}}\left(\omega+\ii\eta,-\omega+\ii\eta\right),\label{eq:inel_spec}
\end{equation}
which is defined in the time domain as
\begin{align}
G_{\omega^{\p}}^{\mathrm{cqq}}\left(t-t^{\p},t-t^{\pp}\right)= & -\Theta\left(t-t^{\p}\right)\Theta\left(t^{\p}-t^{\pp}\right)\nonumber \\
 & \hspace{-3cm}\times\left\langle \left[\left[n_{q^{\p}}\left(t\right),\rho\left(\xin,t^{\p}\right)\right],\rho\left(\xin,t^{\pp}\right)\right]\right\rangle +\left\{ t^{\p}\leftrightarrow t^{\pp}\right\} ,
\end{align}
where $\xin<0$ is an arbitrary point in the half-infinite lead, and
$n_{q^{\p}}=b_{q^{\p}}^{\dagger}b_{q^{\p}}$ is the occupation number
of the mode $q^{\p}$, with frequency $\omega^{\p}=vq^{\p}=q^{\p}$.
The Fourier transform is
\begin{align}
G_{\omega^{\p}}^{\mathrm{cqq}}\left(\omega_{1},\omega_{2}\right)= & \int_{-\infty}^{\infty}\id t^{\p}e^{\ii\omega_{1}\left(t-t^{\p}\right)}\int_{-\infty}^{\infty}\id t^{\pp}e^{\ii\omega_{2}\left(t-t^{\pp}\right)}\nonumber \\
 & \times G_{\omega^{\p}}^{\mathrm{cqq}}\left(t-t^{\p},t-t^{\pp}\right).
\end{align}
Let us sketch the main steps leading to this expression.

The goal is to calculate the time-averaged rate of change of $n_{q^{\p}}$
in response to an incoming photon with frequency $\omega$, injected
at $\xin<0$, impinging at the boundary. Note that we are looking
for a second order response, since the change in the photonic number
$n_{q^{\p}}$ is proportional to the flux of incoming energy. The
photon at $\omega$ is injected by applying an AC voltage, $\mathcal{H}_{\mathrm{AC}}=V\left(t\right)e^{\eta t}\rho\left(\xin\right)$,
where $V\left(t\right)=2V_{0}\cos\left(\omega t\right)$ and $\eta\rightarrow0^{+}$.
The time dependence of $\mathcal{H}_{\mathrm{AC}}$ makes the Keldysh
formalism natural for our purpose \citep{kamenev_field_2011,chou_equilibrium_1985};
the second order Kubo formula for $n_{q^{\p}}$ reads
\begin{equation}
\left\langle n_{q^{\p}}\left(t\right)\right\rangle =\frac{1}{2}\int_{-\infty}^{\infty}\id t^{\p}\id t^{\p\p}G_{\omega^{\p}}^{\mathrm{cqq}}\left(t-t^{\p},t-t^{\pp}\right)V\left(t^{\p}\right)V\left(t^{\pp}\right).\label{eq:nq_avg}
\end{equation}
The notation $\mathrm{c},\mathrm{q}$ corresponds to the ``classical''
and ``quantum'' fields in the Keldysh formalism, that is, the sum
and difference, respectively, of the fields on the forward and backward
time contours. We are concerned with the so-called ``fully-retarded''
multipoint correlator $G_{\omega^{\p}}^{\mathrm{cqq}}$ \citep{kugler_multipoint_2021},
which measures the causal response of $n_{q^{\p}}$ to the perturbing
density $\rho$. To find the time-averaged photon production rate,
we take the derivative of Eq. (\ref{eq:nq_avg}) and discard the oscillating
exponentials:
\begin{equation}
\frac{\id}{\id t}\overline{\left\langle n_{q^{\p}}\left(t\right)\right\rangle }=2\eta\left|V_{0}\right|^{2}G_{\omega^{\p}}^{\mathrm{cqq}}\left(\omega+\ii\eta,-\omega+\ii\eta\right).
\end{equation}
Then, in order to obtain $\gamma\left(\omega^{\p}\lvert\omega\right)$,
we multiply the above by the density of modes, $1/\Delta=\ell/\pi$,
and divide by the average power of the source that propagates towards
the impurity, $\omega\left|V_{0}^{2}\right|/\pi$. We thus arrive
at Eq. (\ref{eq:inel_spec}). Using time-translation invariance and
applying simple algebraic manipulations, we may write the 3-point
response function as
\begin{align}
G_{\omega^{\p}}^{\mathrm{cqq}}\left(\omega+\ii\eta,-\omega+\ii\eta\right)= & -2\int_{0}^{\infty}\id t^{\p}\id t^{\pp}e^{-2\eta t^{\p}}e^{-\eta t^{\pp}}\nonumber \\
 & \hspace{-3.5cm}\times\cos\left(\omega t^{\pp}\right)\left\langle \left[\left[n_{q^{\p}}\left(0\right),\rho\left(\xin,-t^{\p}\right)\right],\rho\left(\xin,-t^{\p}-t^{\pp}\right)\right]\right\rangle .\label{eq:Gcqq}
\end{align}
Note the $\eta\rightarrow0^{+}$ prefactor in Eq. (\ref{eq:inel_spec}),
which implies that we need to look for contributions to $G_{\omega^{\p}}^{\mathrm{cqq}}$
that are singular in $\eta$.

In order to calculate $G_{\omega^{\p}}^{\mathrm{cqq}}$ using form
factors, we need to express $n_{q^{\p}},\rho$ in terms of the current
operators $\mathcal{R},\mathcal{L}$. From Eqs. (\ref{eq:field_modes}),
(\ref{eq:R_L}), and the quantization $q=n\Delta$, we may write $b_{q}=b_{q}^{\mathcal{R}}+b_{q}^{\mathcal{L}}$,
with
\begin{equation}
b_{q}^{\mathcal{A}}\left(t\right)=\frac{1}{2\sqrt{\pi q\ell}}\int_{-\ell}^{0}\id xe^{-\ii\varsigma_{\mathcal{A}}qx}\mathcal{A}\left(x,t\right),\label{eq:bqA}
\end{equation}
where $\omega=vq=q$, $\varsigma_{\mathcal{R}}=1$, and $\varsigma_{\mathcal{L}}=-1$.
The operators $b_{q}^{\mathcal{R}}$ and $b_{q}^{\mathcal{L}}$ annihilate
right and left moving photons with frequency $q$, respectively. The
density $\rho$ and occupation number $n_{q^{\p}}$ may then be decomposed
into their chiral parts. The density can be written as $\rho=\rho^{\mathcal{R}}+\rho^{\mathcal{L}}$,
where
\begin{equation}
\rho^{\mathcal{A}}\left(\xin,t\right)=\frac{1}{2\pi}\mathcal{A}\left(\xin,t\right),\label{eq:rho_chiral}
\end{equation}
whereas the occupation number may be expressed as $n_{q^{\p}}=\sum_{\mathcal{A},\mathcal{B}=\mathcal{R},\mathcal{L}}n_{q^{\p}}^{\mathcal{AB}}$,
with $n_{q^{\p}}^{\mathcal{AB}}=b_{q^{\p}}^{\mathcal{A}\dagger}b_{q^{\p}}^{\mathcal{B}}$:
\begin{align}
n_{q^{\p}}^{\mathcal{AB}}\left(t\right)= & \frac{1}{4\pi\omega^{\p}\ell}\int_{-\ell}^{0}\id x_{1}\id x_{2}e^{\ii\omega^{\p}\left(\varsigma_{\mathcal{A}}x_{1}+\varsigma_{\mathcal{B}}x_{2}\right)}\nonumber \\
 & \hspace{1.5cm}\times\mathcal{\mathcal{A}}\left(x_{1},t\right)\mathcal{B}\left(x_{2},t\right).
\end{align}
The decomposition of $G_{\omega^{\p}}^{\mathrm{cqq}}$ into its chiral
parts is then
\begin{align}
G_{\omega^{\p}}^{\mathrm{cqq}}\left(\omega+\ii\eta,-\omega+\ii\eta\right)\nonumber \\
 & \hspace{-3cm}=\sum_{\mathcal{A},\mathcal{B},\mathcal{C},\mathcal{D}=\mathcal{R},\mathcal{L}}G_{\omega^{\p};\mathcal{ABCD}}^{\mathrm{cqq}}\left(\omega+\ii\eta,-\omega+\ii\eta\right),
\end{align}
with
\begin{align}
G_{\omega^{\p};\mathcal{ABCD}}^{\mathrm{cqq}}\left(t-t^{\p},t-t^{\pp}\right)= & -\Theta\left(t-t^{\p}\right)\Theta\left(t^{\p}-t^{\pp}\right)\nonumber \\
 & \hspace{-4cm}\times\left\langle \left[\left[n_{q^{\p}}^{\mathcal{AB}}\left(t\right),\rho^{\mathcal{C}}\left(\xin,t^{\p}\right)\right],\rho^{\mathcal{D}}\left(\xin,t^{\pp}\right)\right]\right\rangle +\left\{ t^{\p}\leftrightarrow t^{\pp}\right\} .\label{eq:Gcqq_RL_decomp}
\end{align}
Fortunately, there is no need to calculate all 16 terms $G_{\omega^{\p};\mathcal{ABCD}}^{\mathrm{cqq}}$.
First, in an experimental setup, the photon at $\omega$ is injected
such that it propagates towards the boundary; we should therefore
only consider terms with $\rho^{\mathcal{C}},\rho^{\mathcal{D}}=\rho^{\mathcal{R}}$.
This leaves us with the calculation of 4 terms, $G_{\omega^{\p};\mathcal{ABRR}}^{\mathrm{cqq}}$.
Furthermore, since the produced photon propagates to the left, away
from the impurity, we expect only $G_{\omega^{\p};\mathcal{LLRR}}^{\mathrm{cqq}}$
to contribute, measuring the response of the left moving occupation
number to the right moving source. In the following, we calculate
$G_{\omega^{\p};\mathcal{LLRR}}^{\mathrm{cqq}}$, and show that the
other 3 terms indeed vanish.\vspace{-0.5cm}

\subsection{\label{subsec:3pt_details}Calculating the 3-point response function
using form factors}

The calculation of a 3-point response function using form factors
is considerably more involved than that of a 2-point function. The
3-point response function $G_{\omega^{\p};\mathcal{LLRR}}^{\mathrm{cqq}}$
comprises 4 correlators, each involving 4 current operators. We therefore
must insert 3 complete sets of states in each correlator, giving rise
to mixed matrix elements of the form $\sidescripts{\epsilon_{\vartheta}}{\mn}{\bk{\varthetar|\mathcal{A}|\thetal}}{\mn}{\epsilon_{\theta}}$.
These mixed elements, evaluated using the crossing relations in Eq.
(\ref{eq:cross_rels_3}), lead to a series of terms with a different
structure in each of the 4 correlators.

In order to make sense of these complicated expressions, let us recall
the key step in the derivation of the 2-point function $\GLR\left(t\right)$.
It comprises only two correlators, $\left\langle \mathcal{L}\left(t\right)\mathcal{R}\left(0\right)\right\rangle $
and $\left\langle \mathcal{R}\left(0\right)\mathcal{L}\left(t\right)\right\rangle $,
which were both evaluated by inserting a single complete set of states.
The result, however, was not obtained by considering each correlator
on its own, but rather by combining the two form factor expansions;
we had to take the difference of these two correlators, allowing us
to close the integration contour in Eq. (\ref{eq:GLR_contour}), in
order to arrive at the anticipated delta function in the result, Eq.
(\ref{eq:GLR_result}). Similarly, it was necessary to combine the
two correlators of $\GRL$ to show that the non-casual conductance
vanishes. It is therefore crucial to find some convenient way to combine
and unify the 4 correlators of $G_{\omega^{\p};\mathcal{LLRR}}^{\mathrm{cqq}}$;
given a term in the form factor expansion of one of the correlators,
we must find a way to identify its 3 counterparts in the other correlators.
In the following, we accomplish this and show how to identify a quartet
of terms from the expansions of the four correlators that need to
be summed up together, by labeling the excitations according to their
``origins'' and ``destinations'' in each correlator, leading to
a unified general expression for $G_{\omega^{\p};\mathcal{LLRR}}^{\mathrm{cqq}}$.

We begin by denoting the 4 terms of the double commutator as

\begin{widetext}
\begin{align}
\mathrm{I}= & \int_{0}^{\infty}\id t^{\p}\id t^{\pp}e^{-2\eta t^{\p}}e^{-\eta t^{\pp}}\cos\left(\omega t^{\pp}\right)\int_{-\ell}^{0}\id x_{1}\id x_{2}e^{-\ii\omega^{\p}\left(x_{1}-x_{2}\right)}\left\langle \mathcal{L}\left(x_{1},0\right)\mathcal{L}\left(x_{2},0\right)\mathcal{R}\left(\xin,-t^{\p}\right)\mathcal{R}\left(\xin,-t^{\p}-t^{\pp}\right)\right\rangle ,\nonumber \\
\mathrm{II}= & \int_{0}^{\infty}\id t^{\p}\id t^{\pp}e^{-2\eta t^{\p}}e^{-\eta t^{\pp}}\cos\left(\omega t^{\pp}\right)\int_{-\ell}^{0}\id x_{1}\id x_{2}e^{-\ii\omega^{\p}\left(x_{1}-x_{2}\right)}\left\langle \mathcal{R}\left(\xin,-t^{\p}-t^{\pp}\right)\mathcal{R}\left(\xin,-t^{\p}\right)\mathcal{L}\left(x_{1},0\right)\mathcal{L}\left(x_{2},0\right)\right\rangle ,\nonumber \\
\mathrm{III}= & \int_{0}^{\infty}\id t^{\p}\id t^{\pp}e^{-2\eta t^{\p}}e^{-\eta t^{\pp}}\cos\left(\omega t^{\pp}\right)\int_{-\ell}^{0}\id x_{1}\id x_{2}e^{-\ii\omega^{\p}\left(x_{1}-x_{2}\right)}\left\langle \mathcal{R}\left(\xin,-t^{\p}\right)\mathcal{L}\left(x_{1},0\right)\mathcal{L}\left(x_{2},0\right)\mathcal{R}\left(\xin,-t^{\p}-t^{\pp}\right)\right\rangle ,\nonumber \\
\mathrm{IV}= & \int_{0}^{\infty}\id t^{\p}\id t^{\pp}e^{-2\eta t^{\p}}e^{-\eta t^{\pp}}\cos\left(\omega t^{\pp}\right)\int_{-\ell}^{0}\id x_{1}\id x_{2}e^{-\ii\omega^{\p}\left(x_{1}-x_{2}\right)}\left\langle \mathcal{R}\left(\xin,-t^{\p}-t^{\pp}\right)\mathcal{L}\left(x_{1},0\right)\mathcal{L}\left(x_{2},0\right)\mathcal{R}\left(\xin,-t^{\p}\right)\right\rangle ,\label{eq:I_II_III_IV_terms}
\end{align}
so that
\begin{equation}
G_{\omega^{\p};\mathcal{LLRR}}^{\mathrm{cqq}}\left(\omega+\ii\eta,-\omega+\ii\eta\right)=\frac{-1}{8\pi^{3}\omega^{\p}\ell}\left(\mathrm{I}+\mathrm{II}-\mathrm{III}-\mathrm{IV}\right).\label{eq:3p_RL_decomp_I_II_III_IV}
\end{equation}
Consider $\mathrm{III}$ above and insert complete sets of states
between its operators:
\begin{align}
\mathrm{III}= & \int_{0}^{\infty}\id t^{\p}\id t^{\pp}e^{-2\eta t^{\p}}e^{-\eta t^{\pp}}\cos\left(\omega t^{\pp}\right)\int_{-\ell}^{0}\id x_{1}\id x_{2}e^{-\ii\omega^{\p}\left(x_{1}-x_{2}\right)}\int_{\lambda_{i}}\left(\mathsf{R}_{\epsilon_{\lambda_{1}}}^{\epsilon_{\lambda_{1}}^{\p}}\left(\lambr_{1}\right)\right)^{*}\mathsf{R}_{\epsilon_{\lambda_{3}}^{\p}}^{\epsilon_{\lambda_{3}}}\left(\lambr_{3}\right)\nonumber \\
 & \hspace{-0.5cm}\times\bk{0|\mathcal{R}\left(\xin,-t^{\p}\right)|\lambr_{1}}_{\epsilon_{\lambda_{1}}^{\p}}^{\rmn}\sidescripts{\epsilon_{\lambda_{1}}}{\lmn}{\bk{\lambr_{1}|\mathcal{L}\left(x_{1},0\right)|\lambr_{2}}}{\lmn}{\epsilon_{\lambda_{2}}}\sidescripts{\epsilon_{\lambda_{2}}}{\lmn}{\bk{\lambl_{2}|\mathcal{L}\left(x_{2},0\right)|\lambl_{3}}}{\lmn}{\epsilon_{\lambda_{3}}}\sidescripts{\epsilon_{\lambda_{3}}^{\p}}{\rmn}{\bk{\lambl_{3}|\mathcal{R}\left(\xin,-t^{\p}-t^{\pp}\right)|0}}{}{}.\label{eq:III_raw}
\end{align}
Using Eq. (\ref{eq:xt_dep}), we get complex exponentials from the
coordinates of the current operators. We assume a half-infinite line
and extend the lower integration limits of $x_{1,2}$ to $-\ell\rightarrow-\infty$,
introducing an infinitesimal parameter $\eta$ to assure convergence:
\begin{equation}
\int_{-\infty}^{0}\id x_{1}e^{-\ii\left(\omega^{\p}+\nu_{\lambda_{2}}-\nu_{\lambda_{1}}+\ii\eta\right)x_{1}}\int_{-\infty}^{0}\id x_{2}e^{\ii\left(\omega^{\p}+\nu_{\lambda_{2}}-\nu_{\lambda_{3}}-\ii\eta\right)x_{2}}=\frac{1}{\omega^{\p}+\nu_{\lambda_{2}}-\nu_{\lambda_{1}}+\ii\eta}\frac{1}{\omega^{\p}+\nu_{\lambda_{2}}-\nu_{\lambda_{3}}-\ii\eta}.\label{eq:x_denoms}
\end{equation}
Integrating over $x_{1,2}$ and $t^{\p}$ and shifting the rapidities
according to the bulk mass ratios of their excitations, as was done
in the calculation of the reflection coefficient, then yields
\begin{align}
\mathrm{III}= & \int_{0}^{\infty}\id t^{\pp}\cos\left(\omega t^{\pp}\right)\int_{\lambda_{i}}\frac{\ii e^{\ii\left(\nu_{\lambda_{1}}-\nu_{\lambda_{3}}\right)\xin}e^{-\ii\left(\nu_{\lambda_{3}}-\ii\eta\right)t^{\pp}}}{\nu_{\lambda_{1}}-\nu_{\lambda_{3}}+2\ii\eta}\frac{\left(\hat{\mathsf{R}}_{\epsilon_{\lambda_{1}}}^{\epsilon_{\lambda_{1}}^{\p}}\left(\lambr_{1}\right)\right)^{*}\hat{\mathsf{R}}_{\epsilon_{\lambda_{3}}^{\p}}^{\epsilon_{\lambda_{3}}}\left(\lambr_{3}\right)}{\left(\omega^{\p}+\nu_{\lambda_{2}}-\nu_{\lambda_{1}}+\ii\eta\right)\left(\omega^{\p}+\nu_{\lambda_{2}}-\nu_{\lambda_{3}}-\ii\eta\right)}\nonumber \\
 & \hspace{1cm}\times\bk{0|\hat{\mathcal{R}}|\lambr_{1}}_{\epsilon_{\lambda_{1}}^{\p}}^{\rmn}\sidescripts{\epsilon_{\lambda_{1}}}{\lmn}{\bk{\lambr_{1}|\hat{\mathcal{L}}|\lambr_{2}}}{\lmn}{\epsilon_{\lambda_{2}}}\sidescripts{\epsilon_{\lambda_{2}}}{\lmn}{\bk{\lambl_{2}|\hat{\mathcal{L}}|\lambl_{3}}}{\lmn}{\epsilon_{\lambda_{3}}}\sidescripts{\epsilon_{\lambda_{3}}^{\p}}{\rmn}{\bk{\lambl_{3}|\hat{\mathcal{R}}|0}}{}{}.\label{eq:III_raw_2}
\end{align}
where a hat over the operators $\mathcal{R},\mathcal{L}$ indicates
that the rapidities in the matrix elements are $\lambda-\log\mu_{\epsilon}$.
Now, consider the mixed matrix elements, which need to be evaluated
using the crossing relations in Eq. (\ref{eq:cross_rels_3}). The
idea is to partition the sets $\lambda_{1,2,3}$ into smaller subsets,
$\alpha_{ij}$ ($i,j=1,2$), $\beta$, and $\gamma$, that label the
excitations according to the operators they are connected to: the
excitations in the set $\alpha_{ij}$ connect $\rho_{i}^{\mathcal{R}}$
($\rho_{1}^{\mathcal{R}}\equiv\rho^{\mathcal{R}}\left(\xin,-t^{\p}\right)$,
$\rho_{2}^{\mathcal{R}}\equiv\rho^{\mathcal{R}}\left(\xin,-t^{\p}-t^{\pp}\right)$)
to $b_{q^{\p}}^{\mathcal{L}\dagger}$ ($j=1$) or $b_{q^{\p}}^{\mathcal{L}}$
($j=2$), the set $\beta$ connects the $b_{q^{\p}}^{\mathcal{L}},b_{q^{\p}}^{\mathcal{L}\dagger}$
operators to each other, and the set $\gamma$ connects the $\rho^{\mathcal{R}}$
operators. We find
\begin{align}
\mathrm{III}= & \int_{0}^{\infty}\id t^{\pp}\cos\left(\omega t^{\pp}\right)\int_{\alpha_{ij},\beta,\gamma}\frac{\ii e^{\ii\left(\nu_{11}+\nu_{12}-\nu_{21}-\nu_{22}\right)\xin}e^{-\ii\left(\nu_{21}+\nu_{22}+\nu_{\gamma}-\ii\eta\right)t^{\pp}}\left(\hat{\mathsf{R}}_{\epsilon_{11}}^{\epsilon_{11}^{\p}}\left(\ar_{11}\right)\hat{\mathsf{R}}_{\epsilon_{12}}^{\epsilon_{12}^{\p}}\left(\ar_{12}\right)\right)^{*}\hat{\mathsf{R}}_{\epsilon_{21}^{\p}}^{\epsilon_{21}}\left(\ar_{21}\right)\hat{\mathsf{R}}_{\epsilon_{22}^{\p}}^{\epsilon_{22}}\left(\ar_{22}\right)}{\left(\nu_{11}+\nu_{12}-\nu_{21}-\nu_{22}+2\ii\eta\right)\left(\omega^{\p}+\nu_{21}-\nu_{11}+\nu_{\beta}+\ii\eta\right)\left(\omega^{\p}+\nu_{12}-\nu_{22}+\nu_{\beta}-\ii\eta\right)}\nonumber \\
 & \hspace{-0.6cm}\times\bk{0|\hat{\mathcal{R}}|\gr,\ar_{12},\ar_{11}}_{\epsilon_{\gamma}^{\p}\epsilon_{12}^{\p}\epsilon_{11}^{\p}}^{\rmn}\sidescripts{\epsilon_{11}}{\lmn}{\bk{\ar_{11}+\ii\delta|\hat{\mathcal{L}}|\al_{21},\br}}{\lmn}{\epsilon_{21}\epsilon_{\beta}}\sidescripts{\epsilon_{12}\epsilon_{\beta}}{\lmn}{\bk{\ar_{12}+\ii\delta,\bl-\ii\delta|\hat{\mathcal{L}}|\al_{22}}}{\lmn}{\epsilon_{22}}\sidescripts{\epsilon_{21}^{\p}\epsilon_{22}^{\p}\epsilon_{\gamma}^{\p}}{\rmn}{\bk{\al_{21},\al_{22},\gl|\hat{\mathcal{R}}|0}}{}{}.\label{eq:III_1}
\end{align}
The derivation of Eq. (\ref{eq:III_1}) using the crossing relations
is detailed in Appendix \ref{app:3p_details}. Here we shorten the
notations, $\epsilon_{\alpha_{ij}}\rightarrow\epsilon_{ij}$ and $\nu_{\alpha_{ij}}\rightarrow\nu_{ij}$.
Counting the number of possibilities to partition the sets $\lambda_{i}$
to $\alpha_{ij},\beta,\gamma$, we see that the integration measure
becomes
\begin{align}
\int_{\lambda_{i}}= & \prod_{i=1,2,3}\left[\sum_{n_{\lambda_{i}},\epsilon_{\lambda_{i}}}\int_{-\infty}^{\infty}\frac{\prod_{k=1}^{n_{\lambda_{i}}}\id\left(\lambda_{i}\right)_{k}}{\left(2\pi\right)^{n_{\lambda_{i}}}n_{\lambda_{i}}!}\right]\nonumber \\
\rightarrow\int_{\alpha_{ij},\beta,\gamma}= & \prod_{i,j=1,2}\left[\sum_{n_{ij},\epsilon_{ij}}\int_{-\infty}^{\infty}\frac{\prod_{k=1}^{n_{ij}}\id\left(\alpha_{ij}\right)_{k}}{\left(2\pi\right)^{n_{ij}}n_{ij}!}\right]\sum_{n_{\beta},\epsilon_{\beta}}\left[\int_{-\infty}^{\infty}\frac{\prod_{k=1}^{n_{\beta}}\id\beta_{k}}{\left(2\pi\right)^{n_{\beta}}n_{\beta}!}\right]\sum_{n_{\gamma},\epsilon_{\gamma}}\left[\int_{-\infty}^{\infty}\frac{\prod_{k=1}^{n_{\gamma}}\id\gamma_{k}}{\left(2\pi\right)^{n_{\gamma}}n_{\gamma}!}\right],
\end{align}
that is, the factorials and $2\pi$ factors in the denominators translate
naturally from the $\lambda$ sets to the smaller subsets. The other
correlators, $\mathrm{I},\mathrm{II},\mathrm{IV}$, may be treated
similarly, with appropriate labeling of the excitations. The 4 correlators,
written explicitly in Eqs. (\ref{eq:I})-(\ref{eq:IV}), are conveniently
represented by the graphical notation:\begin{align} 		\label{eq:diagrams_1_I}&\mathrm{I}\quad&\begin{tikzpicture}[baseline=-0.5ex] 			\pgfmathsetmacro{\R}{0.6} 			\pgfmathsetmacro{\H}{3.5*\R} 			\pgfmathsetmacro{\D}{5} 			\pgfmathsetmacro{\PHI}{30}			 			\begin{scope} 				\draw [red, very thick] plot [smooth, tension=0.8] coordinates { (2*\D+\R,0) (2.2*\D, {-0.5*sin(\PHI)*\R}) ({3*\D - cos(\PHI)*\R},{-sin(\PHI)*\R})}; 				\draw [purple, very thick] plot [smooth, tension=0.8] coordinates { ({\D + cos(\PHI)*\R},{-sin(\PHI)*\R}) (1.5*\D,0) (2*\D - 0.7*\R,1.2*\R) (3*\D - \R,0)}; 				\draw [orange, very thick] plot [smooth, tension=0.8] coordinates { ({cos(\PHI)*\R},{-sin(\PHI)*\R}) (1.2*\D,-1.5*\R) (1.8*\D, 1.5*\R) ({3*\D - cos(\PHI)*\R}, {sin(\PHI)*\R})}; 				\draw [blue, very thick] plot [smooth, tension=0.8] coordinates { (\D+\R,0) (2*\D-\R,0)}; 				\draw [green, very thick] plot [smooth, tension=0.8] coordinates { ({cos(\PHI)*\R}, {sin(\PHI)*\R}) (\D,1.5*\R) (1.5*\D, 0) ({2*\D - \R*cos(\PHI)},{-sin(\PHI)*\R})}; 				\draw [magenta, very thick] plot [smooth, tension=0.8] coordinates { (\R,0) (\D - \R,0)};				 				\filldraw[color=black, fill=teal!20, very thick](0,0) circle (\R) node{$b_{q^\prime}^{\mathcal{L}\dagger}$}; 				\filldraw[color=black, fill=teal!20, very thick](\D,0) circle (\R) node{$b_{q^\prime}^{\mathcal{L}}$}; 				\filldraw[color=black, fill=teal!20, very thick](2*\D,0) circle (\R) node{$\rho_1^{\mathcal{R}}$}; 				\filldraw[color=black, fill=teal!20, very thick](3*\D,0) circle (\R) node{$\rho_2^{\mathcal{R}}$}; 				\node at (3*\D - 1.7*\R,1.2*\R) {\textcolor{orange}{$\cev{\alpha}_{21}, \eps_{21}^\p$}}; 				\node[anchor=west] at (0.6*\R,-1.1*\R) {\textcolor{orange}{$\cev{\alpha}_{21}, \eps_{21}$}}; 				\node[anchor=east] at (3*\D - 1.9*\R,-0.05*\R) {\textcolor{purple}{$\cev{\alpha}_{22}, \eps_{22}^\p$}}; 				\node[anchor=west] at (\D + 0.8*\R,-0.85*\R) {\textcolor{purple}{$\cev{\alpha}_{22}, \eps_{22}$}}; 				\node[anchor=north] at (3*\D - 1.7*\R,-0.43*\R) {\textcolor{red}{$\cev{\gamma}, \eps_\gamma$}}; 				\node[anchor=north] at (2*\D + 1.7*\R,-0.25*\R) {\textcolor{red}{$\vec{\gamma}, \eps_\gamma$}}; 				\node[anchor=east] at (2*\D - 0.85*\R,-0.9*\R) {\textcolor{green}{$\vec{{\alpha}}_{11}, \beps_{11}^\p$}}; 				\node[anchor=west] at (0.6*\R,1.2*\R) {\textcolor{green}{$\vec{{\alpha}}_{11}, \beps_{11}$}}; 				\node[anchor=east] at (2*\D - 0.85*\R,0.35*\R) {\textcolor{blue}{$\vec{{\alpha}}_{12}, \beps_{12}^\p$}}; 				\node[anchor=west] at (\D + 0.85*\R,0.35*\R) {\textcolor{blue}{$\vec{{\alpha}}_{12}, \beps_{12}$}}; 				\node[anchor=south] at (2.4*\R,-0.15*\R) {\textcolor{magenta}{$\vec{\beta}$}}; 				\node[anchor=north] at (2.4*\R,0.05*\R) {\textcolor{magenta}{$\eps_\beta$}}; 				\node[anchor=south] at (\D - 1.4*\R,-0.15*\R) {\textcolor{magenta}{$\cev{\beta}$}}; 				\node[anchor=north] at (\D - 1.4*\R,0.05*\R) {\textcolor{magenta}{$\eps_\beta$}}; 			\end{scope} 		\end{tikzpicture}~, 		\\ 		\label{eq:diagrams_1_II}&\mathrm{II}\quad&\begin{tikzpicture}[baseline=-0.5ex] 			\pgfmathsetmacro{\R}{0.6} 			\pgfmathsetmacro{\H}{3.5*\R} 			\pgfmathsetmacro{\D}{5} 			\pgfmathsetmacro{\PHI}{30} 			\pgfmathsetmacro{\PHITWO}{30} 			\begin{scope}[shift={(3*\D, 0)}, xscale=-1] 				\draw [red, very thick] plot [smooth, tension=0.8] coordinates { (2*\D+\R,0) (2.2*\D, {-0.5*sin(\PHI)*\R}) ({3*\D - cos(\PHI)*\R},{-sin(\PHI)*\R})}; 				\draw [purple, very thick] plot [smooth, tension=0.9] coordinates { ({\D + cos(\PHI)*\R},{-sin(\PHI)*\R}) (1.5*\D,0) (2*\D - 0.2*\R,1.2*\R) (3*\D - \R,0)}; 				\draw [orange, very thick] plot [smooth, tension=0.8] coordinates { ({cos(\PHI)*\R},{-sin(\PHI)*\R}) (1.2*\D,-1.5*\R) (1.8*\D, 1.5*\R) ({3*\D - cos(\PHI)*\R}, {sin(\PHI)*\R})}; 				\draw [blue, very thick] plot [smooth, tension=0.8] coordinates { (\D+\R,0) (2*\D-\R,0)}; 				\draw [green, very thick] plot [smooth, tension=0.8] coordinates { ({cos(\PHI)*\R}, {sin(\PHI)*\R}) (\D + 0.5*\R,1.5*\R) (1.5*\D, 0) ({2*\D - \R*cos(\PHI)},{-sin(\PHI)*\R})}; 				\draw [magenta, very thick] plot [smooth, tension=0.8] coordinates { (\R,0) (\D - \R,0)};				 				\filldraw[color=black, fill=teal!20, very thick](\D,0) circle (\R) node{$b_{q^\prime}^{\mathcal{L}\dagger}$}; 				\filldraw[color=black, fill=teal!20, very thick](0,0) circle (\R) node{$b_{q^\prime}^{\mathcal{L}}$}; 				\filldraw[color=black, fill=teal!20, very thick](2*\D,0) circle (\R) node{$\rho_1^{\mathcal{R}}$}; 				\filldraw[color=black, fill=teal!20, very thick](3*\D,0) circle (\R) node{$\rho_2^{\mathcal{R}}$}; 			\end{scope} 			\node at (1.6*\R,1.2*\R) {\textcolor{orange}{$\vec{{\alpha}}_{21}, \beps_{21}^\p$}}; 			\node at (3*\D - 1.6*\R,-1.2*\R) {\textcolor{orange}{$\vec{{\alpha}}_{21}, \beps_{21}$}}; 			\node[anchor=west] at (1.7*\R,-0.05*\R) {\textcolor{purple}{$\vec{{\alpha}}_{22}, \beps_{22}^\p$}}; 			\node[anchor=east] at (2*\D - 0.85*\R,-0.75*\R) {\textcolor{purple}{$\vec{{\alpha}}_{22}, \beps_{22}$}}; 			\node[anchor=north] at (1.5*\R,-0.43*\R) {\textcolor{red}{$\vec{{\gamma}}, \beps_{\gamma}$}}; 			\node[anchor=north] at (\D - 1.5*\R,-0.25*\R) {\textcolor{red}{$\cev{{\gamma}}, \beps_{\gamma}$}}; 			\node[anchor=west] at (\D + 0.85*\R,-0.85*\R) {\textcolor{green}{$\cev{\alpha}_{11}, \eps_{11}^\p$}}; 			\node[anchor=east] at (3*\D - 0.7*\R,1.2*\R) {\textcolor{green}{$\cev{\alpha}_{11}, \eps_{11}$}}; 			\node[anchor=west] at (\D + 0.85*\R,0.35*\R) {\textcolor{blue}{$\cev{\alpha}_{12}, \eps_{12}^\p$}}; 			\node[anchor=east] at (2*\D - 0.85*\R,0.35*\R) {\textcolor{blue}{$\cev{\alpha}_{12}, \eps_{12}$}}; 			\node[anchor=south] at (2*\D + 1.7*\R,-0.15*\R) {\textcolor{magenta}{$\vec{{\beta}}, \beps_\beta$}}; 			\node[anchor=south] at (3*\D - 2.4*\R,-0.15*\R) {\textcolor{magenta}{$\cev{{\beta}}, \beps_\beta$}}; 		\end{tikzpicture}~, 		\\ 		\label{eq:diagrams_1_III}&\mathrm{III}\quad&\begin{tikzpicture}[baseline=-0.5ex] 			\pgfmathsetmacro{\R}{0.6} 			\pgfmathsetmacro{\H}{3.5*\R} 			\pgfmathsetmacro{\D}{5} 			\pgfmathsetmacro{\PHI}{30} 			\pgfmathsetmacro{\VARPHI}{75} 			\begin{scope}[shift={(0, 0)}] 				\draw [red, very thick] plot [smooth, tension=0.5] coordinates { ({cos(\VARPHI)*\R},{-sin(\VARPHI)*\R}) (0.25*\D,-0.8*\R) (0.5*\D,0) (\D,1.5*\R) (2*\D,1.5*\R) (2.5*\D,0) (2.75*\D,-0.8*\R) ({3*\D - cos(\VARPHI)*\R},{-sin(\VARPHI)*\R})}; 				\draw [purple, very thick] plot [smooth, tension=0.8] coordinates { (2*\D + \R,0) (3*\D - \R,0)}; 				\draw [orange, very thick] plot [smooth, tension=0.5] coordinates {({\D + cos(\PHI)*\R},{-sin(\PHI)*\R}) (2*\D,-1.5*\R) (2.5*\D,0) (2.75*\D,0.8*\R) ({3*\D - cos(\VARPHI)*\R},{sin(\VARPHI)*\R}) }; 				\draw [blue, very thick] plot [smooth, tension=0.5] coordinates { (\R,0) (0.6*\D, 0.22*\R) (\D,1.2*\R) ({2*\D - cos(\PHI)*\R}, {sin(\PHI)*\R})}; 				\draw [green, very thick] plot [smooth, tension=0.5] coordinates { ({cos(\VARPHI)*\R}, {sin(\VARPHI)*\R}) (0.25*\D,0.8*\R) (0.5*\D,0.2*\R) (\D - \R,0)}; 				\draw [magenta, very thick] plot [smooth, tension=0.8] coordinates { (\D + \R,0) (2*\D - \R,0)}; 				\filldraw[color=black, fill=teal!20, very thick](0,0) circle (\R) node{$\rho_1^{\mathcal{R}}$}; 				\filldraw[color=black, fill=teal!20, very thick](\D,0) circle (\R) node{$b_{q^\prime}^{\mathcal{L}\dagger}$}; 				\filldraw[color=black, fill=teal!20, very thick](2*\D,0) circle (\R) node{$b_{q^\prime}^{\mathcal{L}}$}; 				\filldraw[color=black, fill=teal!20, very thick](3*\D,0) circle (\R) node{$\rho_2^{\mathcal{R}}$}; 			\end{scope} 			\node[anchor=south] at (1.7*\R,0.8*\R) {$\textcolor{green}{\vec{\alpha}_{11}, \eps_{11}^\p}$}; 			\node[anchor=north] at (\D - 1.9*\R,0.1*\R) {$\textcolor{green}{\vec{\alpha}_{11}, \eps_{11}}$}; 			\node[anchor=south] at (1.5*\R,-0.1*\R) {$\textcolor{blue}{\vec{\alpha}_{12}}$}; 			\node[anchor=north] at (1.5*\R,0.1*\R) {$\textcolor{blue}{\eps_{12}^\p}$}; 			\node[anchor=east] at (2*\D - 0.7*\R,1.0*\R) {$\textcolor{blue}{\vec{\alpha}_{12}, \eps_{12}}$}; 			\node[anchor=north] at (1.4*\R,-0.85*\R) {$\textcolor{red}{\vec{\gamma}, \eps_\gamma}$}; 			\node[anchor=north] at (3*\D - 1.6*\R,-0.85*\R) {$\textcolor{red}{\cev{\gamma}, \eps_\gamma}$}; 			\node[anchor=north] at (\D + 1.9*\R,-0.7*\R) {$\textcolor{orange}{\cev{\alpha}_{21}, \eps_{21}}$}; 			\node[anchor=south] at (3*\D - 1.6*\R,0.85*\R) {$\textcolor{orange}{\cev{\alpha}_{21}, \eps_{21}^\p}$}; 			\node[anchor=south] at (2*\D + 1.55*\R,-0.1*\R) {$\textcolor{purple}{\cev{\alpha}_{22}}$}; 			\node[anchor=north] at (2*\D + 1.55*\R,0.1*\R) {$\textcolor{purple}{\eps_{22}}$}; 			\node[anchor=south] at (3*\D - 1.6*\R,-0.1*\R) {$\textcolor{purple}{\cev{\alpha}_{22}}$}; 			\node[anchor=north] at (3*\D - 1.6*\R,0.1*\R) {$\textcolor{purple}{\eps_{22}^\p}$}; 			\node[anchor=south] at (\D + 1.7*\R,-0.15*\R) {$\textcolor{magenta}{\vec{\beta}, \eps_\beta}$}; 			\node[anchor=north] at (2*\D - 1.7*\R,0.1*\R) {$\textcolor{magenta}{\cev{\beta}, \eps_\beta}$}; 		\end{tikzpicture}~, 		\\ 		\label{eq:diagrams_1_IV}&\mathrm{IV}\quad&\begin{tikzpicture}[baseline=-0.5ex] 			\pgfmathsetmacro{\R}{0.6} 			\pgfmathsetmacro{\H}{3.5*\R} 			\pgfmathsetmacro{\D}{5} 			\pgfmathsetmacro{\PHI}{30} 			\pgfmathsetmacro{\VARPHI}{75} 			\begin{scope}[shift={(3*\D, 0)}, xscale=-1] 				\draw [red, very thick] plot [smooth, tension=0.5] coordinates { ({cos(\VARPHI)*\R},{-sin(\VARPHI)*\R}) (0.25*\D,-0.8*\R) (0.5*\D,0) (\D,1.5*\R) (2*\D,1.5*\R) (2.5*\D,0) (2.75*\D,-0.8*\R) ({3*\D - cos(\VARPHI)*\R},{-sin(\VARPHI)*\R})}; 				\draw [purple, very thick] plot [smooth, tension=0.8] coordinates { (2*\D + \R,0) (3*\D - \R,0)}; 				\draw [orange, very thick] plot [smooth, tension=0.5] coordinates {({\D + cos(\PHI)*\R},{-sin(\PHI)*\R}) (2*\D,-1.5*\R) (2.5*\D,0) (2.75*\D,0.8*\R) ({3*\D - cos(\VARPHI)*\R},{sin(\VARPHI)*\R}) }; 				\draw [blue, very thick] plot [smooth, tension=0.5] coordinates { (\R,0) (0.6*\D, 0.22*\R) (\D,1.2*\R) ({2*\D - cos(\PHI)*\R}, {sin(\PHI)*\R})}; 				\draw [green, very thick] plot [smooth, tension=0.5] coordinates { ({cos(\VARPHI)*\R}, {sin(\VARPHI)*\R}) (0.25*\D,0.8*\R) (0.5*\D,0.2*\R) (\D - \R,0)}; 				\draw [magenta, very thick] plot [smooth, tension=0.8] coordinates { (\D + \R,0) (2*\D - \R,0)};				 				\filldraw[color=black, fill=teal!20, very thick](0,0) circle (\R) node{$\rho_1^{\mathcal{R}}$}; 				\filldraw[color=black, fill=teal!20, very thick](2*\D,0) circle (\R) node{$b_{q^\prime}^{\mathcal{L}\dagger}$}; 				\filldraw[color=black, fill=teal!20, very thick](\D,0) circle (\R) node{$b_{q^\prime}^{\mathcal{L}}$}; 				\filldraw[color=black, fill=teal!20, very thick](3*\D,0) circle (\R) node{$\rho_2^{\mathcal{R}}$}; 			\end{scope} 			\node[anchor=south] at (3*\D-1.5*\R,0.8*\R) {$\textcolor{green}{\cev{\alpha}_{11}, \beps_{11}^\p}$}; 			\node[anchor=north] at (2*\D + 2.1*\R,0.1*\R) {$\textcolor{green}{\cev{\alpha}_{11}, \beps_{11}}$}; 			\node[anchor=south] at (3*\D - 1.6*\R,-0.1*\R) {$\textcolor{blue}{\cev{\alpha}_{12}}$}; 			\node[anchor=north] at (3*\D - 1.6*\R,0.1*\R) {$\textcolor{blue}{\beps_{12}^\p}$}; 			\node[anchor=west] at (\D + 0.7*\R,1.0*\R) {$\textcolor{blue}{\cev{\alpha}_{12}, \beps_{12}}$}; 			\node[anchor=north] at (3*\D - \R,-0.85*\R) {$\textcolor{red}{\cev{{\gamma}}, \beps_\gamma}$}; 			\node[anchor=north] at (1.2*\R,-0.85*\R) {$\textcolor{red}{\vec{{\gamma}}, \beps_\gamma}$}; 			\node[anchor=north] at (2*\D - 1.3*\R,-0.6*\R) {$\textcolor{orange}{\vec{{\alpha}}_{21}, \beps_{21}}$}; 			\node[anchor=south] at (1.4*\R,0.8*\R) {$\textcolor{orange}{\vec{{\alpha}}_{21}, \beps_{21}^\p}$}; 			\node[anchor=south] at (\D - 1.5*\R,-0.1*\R) {$\textcolor{purple}{\vec{{\alpha}}_{22}}$}; 			\node[anchor=north] at (\D - 1.5*\R,0.05*\R) {$\textcolor{purple}{\beps_{22}}$}; 			\node[anchor=south] at (1.5*\R,-0.1*\R) {$\textcolor{purple}{\vec{{\alpha}}_{22}}$}; 			\node[anchor=north] at (1.5*\R,0.1*\R) {$\textcolor{purple}{\beps_{22}^\p}$}; 			\node[anchor=north] at (\D + 1.7*\R,0.1*\R) {$\textcolor{magenta}{\vec{{\beta}},\beps_\beta}$}; 			\node[anchor=south] at (2*\D - 1.7*\R,-0.15*\R) {$\textcolor{magenta}{\cev{{\beta}},\beps_\beta}$}; 		\end{tikzpicture}~. 	\end{align}Labeling
the excitations offers a natural way to sum up the correlators, following
a procedure similar to that of the 2-point function. Consider $\mathrm{I},\mathrm{III}$;
we introduce two auxiliary rapidities, $\kappa_{1,2}$, such that
$e^{\kappa_{i}}=\nu_{i1}+\nu_{i2}$, and shift $\ar_{ij}\rightarrow\ar_{ij}+\kappa_{i}$,
leading to
\begin{align}
\mathrm{I}= & \int_{0}^{\infty}\id t^{\pp}\cos\left(\omega t^{\pp}\right)\int_{\alpha_{ij},\beta,\gamma}\mathsf{C}_{\beta}\mathsf{C}_{\gamma}\prod_{i=1}^{2}\delta\left(\nu_{i1}+\nu_{i2}-1\right)\int_{-\infty}^{\infty}\prod_{i=1}^{2}\id\kappa_{i}\nonumber \\
 & \times\frac{\ii e^{-\ii\left(e^{\kappa_{1}}+e^{\kappa_{2}}\right)\xin}e^{-\ii\left(e^{\kappa_{2}}+\nu_{\gamma}-\ii\eta\right)t^{\pp}}}{-e^{\kappa_{1}}-e^{\kappa_{2}}+2\ii\eta}\frac{\hat{\mathsf{R}}_{\epsilon_{11}}^{\epsilon_{11}^{\p}}\left(\ar_{11}+\kappa_{1}\right)\hat{\mathsf{R}}_{\epsilon_{12}}^{\epsilon_{12}^{\p}}\left(\ar_{12}+\kappa_{1}\right)\hat{\mathsf{R}}_{\epsilon_{21}^{\p}}^{\epsilon_{21}}\left(\ar_{21}+\kappa_{2}\right)\hat{\mathsf{R}}_{\epsilon_{22}^{\p}}^{\epsilon_{22}}\left(\ar_{22}+\kappa_{2}\right)}{\left(\omega^{\p}+e^{\kappa_{2}}\nu_{21}+e^{\kappa_{1}}\nu_{11}+\nu_{\beta}+\ii\eta\right)\left(\omega^{\p}-e^{\kappa_{1}}\nu_{12}-e^{\kappa_{2}}\nu_{22}+\nu_{\beta}-\ii\eta\right)}\nonumber \\
 & \times f_{\epsilon_{11}^{\p}\epsilon_{12}^{\p}\epsilon_{\gamma}}^{\mathcal{R}}\left(\al_{11}+\kappa_{1}+\ii\delta,\al_{12}+\kappa_{1}+\ii\delta,\gl+\ii\pi\right)f_{\epsilon_{21}\epsilon_{\beta}\bar{\epsilon}_{11}}^{\mathcal{R}}\left(\al_{21}+\kappa_{2},\br,\ar_{11}+\kappa_{1}\right)\nonumber \\
 & \times f_{\bar{\epsilon}_{\beta}\epsilon_{22}\bar{\epsilon}_{12}}^{\mathcal{R}}\left(\bl-\ii\pi,\al_{22}+\kappa_{2}-\ii\delta,\ar_{12}+\kappa_{1}+\ii\delta\right)f_{\bar{\epsilon}_{\gamma}\bar{\epsilon}_{22}^{\p}\bar{\epsilon}_{21}^{\p}}^{\mathcal{R}}\left(\gr,\ar_{22}+\kappa_{2},\ar_{21}+\kappa_{2}\right),\label{eq:I_kappa}
\end{align}
\begin{align}
\mathrm{III}= & -\int_{0}^{\infty}\id t^{\pp}\cos\left(\omega t^{\pp}\right)\int_{\alpha_{ij},\beta,\gamma}\mathsf{C}_{\beta}\mathsf{C}_{\gamma}\prod_{i=1}^{2}\delta\left(\nu_{i1}+\nu_{i2}-1\right)\int_{-\infty}^{\infty}\prod_{i=1}^{2}\id\kappa_{i}\nonumber \\
 & \times\frac{\ii e^{\ii\left(e^{\kappa_{1}}-e^{\kappa_{2}}\right)\xin}e^{-\ii\left(e^{\kappa_{2}}+\nu_{\gamma}-\ii\eta\right)t^{\pp}}}{e^{\kappa_{1}}-e^{\kappa_{2}}+2\ii\eta}\frac{\left(\hat{\mathsf{R}}_{\epsilon_{11}}^{\epsilon_{11}^{\p}}\left(\ar_{11}+\kappa_{1}\right)\hat{\mathsf{R}}_{\epsilon_{12}}^{\epsilon_{12}^{\p}}\left(\ar_{12}+\kappa_{1}\right)\right)^{*}\hat{\mathsf{R}}_{\epsilon_{21}^{\p}}^{\epsilon_{21}}\left(\ar_{21}+\kappa_{2}\right)\hat{\mathsf{R}}_{\epsilon_{22}^{\p}}^{\epsilon_{22}}\left(\ar_{22}+\kappa_{2}\right)}{\left(\omega^{\p}+e^{\kappa_{2}}\nu_{21}-e^{\kappa_{1}}\nu_{11}+\nu_{\beta}+\ii\eta\right)\left(\omega^{\p}+e^{\kappa_{1}}\nu_{12}-e^{\kappa_{2}}\nu_{22}+\nu_{\beta}-\ii\eta\right)}\nonumber \\
 & \times f_{\epsilon_{11}^{\p}\epsilon_{12}^{\p}\epsilon_{\gamma}}^{\mathcal{R}}\left(\al_{11}+\kappa_{1},\al_{12}+\kappa_{1},\gl\right)f_{\epsilon_{21}\epsilon_{\beta}\bar{\epsilon}_{11}}^{\mathcal{R}}\left(\al_{21}+\kappa_{2},\br,\ar_{11}+\kappa_{1}+\ii\pi-\ii\delta\right)\nonumber \\
 & \times f_{\bar{\epsilon}_{\beta}\epsilon_{22}\bar{\epsilon}_{12}}^{\mathcal{R}}\left(\bl-\ii\pi+\ii\delta,\al_{22}+\kappa_{2},\ar_{12}+\kappa_{1}+\ii\pi-\ii\delta\right)f_{\bar{\epsilon}_{\gamma}\bar{\epsilon}_{22}^{\p}\bar{\epsilon}_{21}^{\p}}^{\mathcal{R}}\left(\gr,\ar_{22}+\kappa_{2},\ar_{21}+\kappa_{2}\right),\label{eq:III_kappa}
\end{align}
where $\mathsf{C}_{\lambda}$ is a sign factor stemming from charge
conjugation, defined in Eq. (\ref{eq:charge_conj_mat}). Shifting
$\kappa_{1}\rightarrow\kappa_{1}+\ii\pi-3\ii\delta$ in $\mathrm{I}$
yields
\begin{align}
\mathrm{I}-\mathrm{III}= & \int_{0}^{\infty}\id t^{\pp}\cos\left(\omega t^{\pp}\right)\int_{\alpha_{ij},\beta,\gamma}\mathsf{C}_{\beta}\mathsf{C}_{\gamma}\prod_{i=1}^{2}\delta\left(\nu_{i1}+\nu_{i2}-1\right)\left(\int_{-\infty}^{\infty}\id\kappa_{1}+\int_{\infty-\ii\pi+3\ii\delta}^{-\infty-\ii\pi+3\ii\delta}\id\kappa_{1}\right)\int_{-\infty}^{\infty}\id\kappa_{2}\nonumber \\
 & \times\frac{\ii e^{\ii\left(e^{\kappa_{1}}-e^{\kappa_{2}}\right)\xin}e^{-\ii\left(e^{\kappa_{2}}+\nu_{\gamma}-\ii\eta\right)t^{\pp}}}{e^{\kappa_{1}}-e^{\kappa_{2}}+2\ii\eta}\frac{\left(\hat{\mathsf{R}}_{\epsilon_{11}}^{\epsilon_{11}^{\p}}\left(\ar_{11}+\kappa_{1}\right)\hat{\mathsf{R}}_{\epsilon_{12}}^{\epsilon_{12}^{\p}}\left(\ar_{12}+\kappa_{1}\right)\right)^{*}\hat{\mathsf{R}}_{\epsilon_{21}^{\p}}^{\epsilon_{21}}\left(\ar_{21}+\kappa_{2}\right)\hat{\mathsf{R}}_{\epsilon_{22}^{\p}}^{\epsilon_{22}}\left(\ar_{22}+\kappa_{2}\right)}{\left(\omega^{\p}+e^{\kappa_{2}}\nu_{21}-e^{\kappa_{1}}\nu_{11}+\nu_{\beta}+\ii\eta\right)\left(\omega^{\p}+e^{\kappa_{2}}\nu_{21}-e^{\kappa_{1}}\nu_{11}+e^{\kappa_{1}}-e^{\kappa_{2}}+\nu_{\beta}-\ii\eta\right)}\nonumber \\
 & \times f_{\epsilon_{11}^{\p}\epsilon_{12}^{\p}\epsilon_{\gamma}}^{\mathcal{R}}\left(\al_{11}+\kappa_{1},\al_{12}+\kappa_{1},\gl\right)f_{\epsilon_{21}\epsilon_{\beta}\bar{\epsilon}_{11}}^{\mathcal{R}}\left(\al_{21}+\kappa_{2},\br,\ar_{11}+\kappa_{1}+\ii\pi-\ii\delta\right)\nonumber \\
 & \times f_{\bar{\epsilon}_{\beta}\epsilon_{22}\bar{\epsilon}_{12}}^{\mathcal{R}}\left(\bl-\ii\pi+\ii\delta,\al_{22}+\kappa_{2},\ar_{12}+\kappa_{1}+\ii\pi-\ii\delta\right)f_{\bar{\epsilon}_{\gamma}\bar{\epsilon}_{22}^{\p}\bar{\epsilon}_{21}^{\p}}^{\mathcal{R}}\left(\gr,\ar_{22}+\kappa_{2},\ar_{21}+\kappa_{2}\right).
\end{align}
Note that we also rewrote the $\omega^{\p}$ denominators using the
delta functions. In order to close the $\kappa_{1}$ contour, we add
the edges at $\kappa_{1}=\sigma+\ii\chi$ with $\sigma\rightarrow\pm\infty$
and $-\pi\le\chi\le0$ --- this is allowed, since the exponential
$e^{\ii\left(e^{\kappa_{1}}-e^{\kappa_{2}}\right)\xin}$ assures that
the integrand decays fast enough at $\sigma\rightarrow\infty$, whereas
the form factors vanish exponentially-fast as $\sigma\rightarrow-\infty$.
Now, the pole at $\kappa_{1}=\kappa_{2}-2\ii\eta$ is enclosed by
the contour; setting $\kappa_{1}=\kappa_{2}$, the $\omega^{\p}$
denominators become
\begin{equation}
\frac{1}{\omega^{\p}+e^{\kappa_{2}}\left(\nu_{21}-\nu_{11}\right)+\nu_{\beta}+\ii\eta}\frac{1}{\omega^{\p}+e^{\kappa_{2}}\left(\nu_{21}-\nu_{11}\right)+\nu_{\beta}-\ii\eta}\rightarrow\frac{\pi}{\eta}\delta\left(\omega^{\p}+e^{\kappa_{2}}\left(\nu_{21}-\nu_{11}\right)+\nu_{\beta}\right),
\end{equation}
that is, the contribution of this pole is singular in $\eta$. In
fact, the pole at $\kappa_{1}=\kappa_{2}-2\ii\eta$ is the sole singular
contribution to the contour integral. The annihilation poles of the
form factors, occuring at $\theta_{i}-\theta_{j}=\ii\pi$ for some
$i>j$ (when the same excitation appears on both bra and ket of a
matrix element, see Eq. (\ref{eq:kin_pole})), are just above or below
the upper and lower boundaries of the contour, and while in the attractive
regime ($z<1/2$) the form factors and reflection matrices have bound
state poles that are enclosed by the contour, their residues are not
singular in $\eta$.

We are now in position to show why $G_{\omega^{\p};\mathcal{LLRR}}^{\mathrm{cqq}}$
is the only non-vanishing contribution to the spectrum. First, $G_{\omega^{\p};\mathcal{RRRR}}^{\mathrm{cqq}}$
is a background term that does not involve the boundary and thus cannot
account for inelastic scattering. Next, consider $G_{\omega^{\p};\mathcal{RLRR}}^{\mathrm{cqq}}$,
that is, the response of $n_{q^{\p}}^{\mathcal{RL}}$ to a right-moving
photon. Replacing $\sidescripts{\epsilon_{\lambda_{1}}}{\lmn}{\bk{\lambr_{1}|\mathcal{L}\left(x_{1},0\right)|\lambr_{2}}}{\lmn}{\epsilon_{\lambda_{2}}}$
with $\sidescripts{\epsilon_{\lambda_{1}}}{\rmn}{\bk{\lambr_{1}|\mathcal{R}\left(x_{1},0\right)|\lambr_{2}}}{\rmn}{\epsilon_{\lambda_{2}}}$
in Eq. (\ref{eq:III_raw}), the sign in the $x_{1}$ exponential in
Eq. (\ref{eq:x_denoms}) is flipped:
\begin{equation}
\int_{-\infty}^{0}\id x_{1}e^{\ii\left(\omega^{\p}+\nu_{\lambda_{2}}-\nu_{\lambda_{1}}-\ii\eta\right)x_{1}}\int_{-\infty}^{0}\id x_{2}e^{\ii\left(\omega^{\p}+\nu_{\lambda_{2}}-\nu_{\lambda_{3}}-\ii\eta\right)x_{2}}=-\frac{1}{\omega^{\p}+\nu_{\lambda_{2}}-\nu_{\lambda_{1}}-\ii\eta}\frac{1}{\omega^{\p}+\nu_{\lambda_{2}}-\nu_{\lambda_{3}}-\ii\eta}.
\end{equation}
The key difference with respect to Eq. (\ref{eq:III_raw}) is the
same relative sign of the $\ii\eta$ terms in the denominators. Therefore,
$G_{\omega^{\p};\mathcal{RLRR}}^{\mathrm{cqq}}$ (and, similarly,
$G_{\omega^{\p};\mathcal{LRRR}}^{\mathrm{cqq}}$) is non-singular
in $\eta$, and its contribution vanishes in the limit $\eta\rightarrow0$.
It is also reassuring to verify that $G_{\omega^{\p};\mathcal{RRLL}}^{\mathrm{cqq}}$,
the non-causal counterpart to $G_{\omega^{\p};\mathcal{LLRR}}^{\mathrm{cqq}}$,
vanishes identically; this is shown in Appendix \ref{app:3p_details}.

The treatment of $\mathrm{II},\mathrm{IV}$ is identical to the above.
Plugging everything to Eq. (\ref{eq:3p_RL_decomp_I_II_III_IV}) and
back to Eq. (\ref{eq:inel_spec}), we find
\begin{align}
\gamma\left(\omega^{\prime}\lvert\omega\right)= & -\frac{2}{\pi\omega^{\p}\omega}\int_{\alpha_{ij},\beta,\gamma}\mathsf{C}_{\beta}\mathsf{C}_{\gamma}\int_{0}^{\infty}\id\Omega\prod_{i=1}^{2}\delta\left(\nu_{i1}+\nu_{i2}+\nu_{\gamma}-\Omega\right)\nonumber \\
 & \times\delta\left(\omega^{\p}+\nu_{21}-\nu_{11}+\nu_{\beta}\right)\int_{0}^{\infty}\id t^{\pp}e^{-\eta t^{\pp}}\cos\left(\omega t^{\pp}\right)\mathrm{Re}\left\{ e^{-\ii\Omega t^{\pp}}\prod\mathsf{F}\prod\mathsf{R}\right\} ,\label{eq:gamma_1}
\end{align}
where the products of form factors and reflection matrices are (denoting
$\pi^{-}\equiv\pi-\delta$)
\begin{align}
\prod\mathsf{F}\equiv & f_{\epsilon_{11}^{\p}\epsilon_{12}^{\p}\epsilon_{\gamma}}^{\mathcal{R}}\left(\al_{11},\al_{12},\gl\right)f_{\epsilon_{21}\epsilon_{\beta}\bar{\epsilon}_{11}}^{\mathcal{R}}\left(\al_{21},\br,\ar_{11}+\ii\pi^{-}\right)f_{\bar{\epsilon}_{\beta}\epsilon_{22}\bar{\epsilon}_{12}}^{\mathcal{R}}\left(\bl-\ii\pi^{-},\al_{22},\ar_{12}+\ii\pi^{-}\right)f_{\bar{\epsilon}_{\gamma}\bar{\epsilon}_{22}^{\p}\bar{\epsilon}_{21}^{\p}}^{\mathcal{R}}\left(\gr,\ar_{22},\ar_{21}\right),\label{eq:Fprod}\\
\prod\mathsf{R}\equiv & \left(\hat{\mathsf{R}}_{\epsilon_{11}}^{\epsilon_{11}^{\p}}\left(\ar_{11}\right)\hat{\mathsf{R}}_{\epsilon_{12}}^{\epsilon_{12}^{\p}}\left(\ar_{12}\right)\right)^{*}\hat{\mathsf{R}}_{\epsilon_{21}^{\p}}^{\epsilon_{21}}\left(\ar_{21}\right)\hat{\mathsf{R}}_{\epsilon_{22}^{\p}}^{\epsilon_{22}}\left(\ar_{22}\right)-\delta_{\bar{\epsilon}_{11}}^{\epsilon_{11}^{\p}}\delta_{\bar{\epsilon}_{12}}^{\epsilon_{12}^{\p}}\delta_{\bar{\epsilon}_{21}}^{\epsilon_{21}^{\p}}\delta_{\bar{\epsilon}_{22}}^{\epsilon_{22}^{\p}}.\label{eq:Rprod}
\end{align}
Note that we subtract the product of Kronecker deltas in $\prod\mathsf{R}$.
This corresponds to subtracting the background term $G_{\omega^{\p};\mathcal{LLLL}}^{\mathrm{cqq}}$,
which does not involve the boundary and therefore does not contribute
to inelastic scattering, and assures that $\gamma\left(\omega^{\p}\lvert\omega\right)$
vanishes as $\omega/T_{B}\rightarrow\infty$ (following Eq. (\ref{eq:Rprod_high_energy}))
or $\omega/T_{B}\rightarrow0$ (following $\mathsf{R}_{\epsilon_{\lambda}}^{\epsilon_{\lambda}^{\p}}\left(\lambr\rightarrow-\infty\right)\rightarrow\delta_{\bar{\epsilon}_{\lambda}}^{\epsilon_{\lambda}^{\p}}$
in the Kondo model, or Eqs. (\ref{eq:R_prod_low_energy_bSG}) and
(\ref{eq:FF_lr}) in the bsG model). The integral over $t^{\pp}$
yields
\begin{equation}
\int_{0}^{\infty}dt^{\pp}e^{-\eta t^{\pp}}\cos\left(\omega t^{\pp}\right)\mathrm{Re}\left\{ e^{-\ii\Omega t^{\pp}}\prod\mathsf{F}\prod\mathsf{R}\right\} =\frac{\pi}{2}\delta\left(\omega-\Omega\right)\mathrm{Re}\left\{ \prod\mathsf{F}\prod\mathsf{R}\right\} -\frac{1}{2}\left[\mathcal{P}\frac{1}{\omega+\Omega}-\mathcal{P}\frac{1}{\omega-\Omega}\right]\mathrm{Im}\left\{ \prod\mathsf{F}\prod\mathsf{R}\right\} ,
\end{equation}
where $\mathcal{P}$ denotes the principal value. To get rid of the
awkward principal value terms, consider another equivalent way to
expand the correlators $\mathrm{I},\ldots,\mathrm{IV}$:

\begin{align} 		&\mathrm{I}\quad&\begin{tikzpicture}[baseline=-0.5ex] 			\pgfmathsetmacro{\R}{0.6} 			\pgfmathsetmacro{\H}{3.5*\R} 			\pgfmathsetmacro{\D}{5} 			\pgfmathsetmacro{\PHI}{30} 			\pgfmathsetmacro{\VARPHI}{60} 			\begin{scope} 				\draw [red, very thick] plot [smooth, tension=0.8] coordinates { (2*\D+\R,0) (2.2*\D, {0.5*sin(\PHI)*\R}) ({3*\D - cos(\VARPHI)*\R},{sin(\VARPHI)*\R})}; 				\draw [blue, very thick] plot [smooth, tension=0.9] coordinates { ({\D + cos(\PHI)*\R},{sin(\PHI)*\R}) (1.5*\D,0) (2*\D - 0.5*\R,-1.2*\R) (3*\D - \R,0)}; 				\draw [green, very thick] plot [smooth, tension=0.8] coordinates { ({cos(\PHI)*\R},{sin(\PHI)*\R}) (1.2*\D,1.5*\R) (1.8*\D, -1.5*\R) ({3*\D - cos(\VARPHI)*\R}, {-sin(\VARPHI)*\R})}; 				\draw [purple, very thick] plot [smooth, tension=0.8] coordinates { (\D+\R,0) (2*\D-\R,0)}; 				\draw [orange, very thick] plot [smooth, tension=0.8] coordinates { ({cos(\PHI)*\R}, {-sin(\PHI)*\R}) (\D + 0.5*\R,-1.5*\R) (1.5*\D, 0) ({2*\D - \R*cos(\PHI)},{sin(\PHI)*\R})}; 				\draw [magenta, very thick] plot [smooth, tension=0.8] coordinates { (\R,0) (\D - \R,0)}; 				\filldraw[color=black, fill=teal!20, very thick](0,0) circle (\R) node{$b_{q^\prime}^{\mathcal{L}\dagger}$}; 				\filldraw[color=black, fill=teal!20, very thick](\D,0) circle (\R) node{$b_{q^\prime}^{\mathcal{L}}$}; 				\filldraw[color=black, fill=teal!20, very thick](2*\D,0) circle (\R) node{$\rho_1^{\mathcal{R}}$}; 				\filldraw[color=black, fill=teal!20, very thick](3*\D,0) circle (\R) node{$\rho_2^{\mathcal{R}}$}; 			\end{scope} 			\node[anchor=south] at (3*\D - 1.75*\R, 0.75*\R) {$\textcolor{red}{\vec{{\gamma}}, \beps_\gamma}$}; 			\node[anchor=south] at (2*\D + 1.7*\R, 0.25*\R) {$\textcolor{red}{\cev{{\gamma}}, \beps_\gamma}$}; 			\node[anchor=south] at (3*\D - 1.7*\R, -0.2*\R) {$\textcolor{blue}{\vec{{\alpha}}_{12}}$}; 			\node[anchor=north] at (3*\D - 1.7*\R, -0.1*\R) {$\textcolor{blue}{\beps_{12}^\p}$}; 			\node[anchor=south] at (\D + 2.0*\R, 0.3*\R) {$\textcolor{blue}{\vec{{\alpha}}_{12}, \beps_{12}}$}; 			\node[anchor=north] at (2*\D - 1.9*\R, 0.1*\R) {$\textcolor{purple}{\cev{\alpha}_{22}, \eps_{22}^\p}$}; 			\node[anchor=north] at (\D + 2.05*\R, 0.1*\R) {$\textcolor{purple}{\cev{\alpha}_{22}, \eps_{22}}$}; 			\node[anchor=south] at (2*\D - 1.9*\R, 0.35*\R) {$\textcolor{orange}{\cev{\alpha}_{21}, \eps_{21}^\p}$}; 			\node[anchor=north] at (1.7*\R, -0.6*\R) {$\textcolor{orange}{\cev{\alpha}_{21}, \eps_{21}}$}; 			\node[anchor=north] at (3*\D - 1.5*\R, -1.1*\R) {$\textcolor{green}{\vec{{\alpha}}_{11}, \beps_{11}^\p}$}; 			\node[anchor=south] at (1.7*\R, 0.7*\R) {$\textcolor{green}{\vec{{\alpha}}_{11}, \beps_{11}}$}; 			\node[anchor=south] at (2.2*\R,-0.15*\R) {$\textcolor{magenta}{\vec{\beta}}$}; 			\node[anchor=north] at (2.2*\R,0.05*\R) {$\textcolor{magenta}{\eps_\beta}$}; 			\node[anchor=south] at (\D - 1.4*\R,-0.15*\R) {$\textcolor{magenta}{\cev{\beta}}$}; 			\node[anchor=north] at (\D - 1.4*\R,0.05*\R) {$\textcolor{magenta}{\eps_\beta}$}; 		\end{tikzpicture}~, \\ 		&\mathrm{II}\quad&\begin{tikzpicture}[baseline=-0.5ex] 			\pgfmathsetmacro{\R}{0.6} 			\pgfmathsetmacro{\H}{3.5*\R} 			\pgfmathsetmacro{\D}{5} 			\pgfmathsetmacro{\PHI}{30} 			\pgfmathsetmacro{\VARPHI}{60} 			\begin{scope}[shift={(3*\D,0)}, xscale=-1] 				\draw [red, very thick] plot [smooth, tension=0.8] coordinates { (2*\D+\R,0) (2.2*\D, {0.5*sin(\PHI)*\R}) ({3*\D - cos(\VARPHI)*\R},{sin(\VARPHI)*\R})}; 				\draw [blue, very thick] plot [smooth, tension=0.9] coordinates { ({\D + cos(\PHI)*\R},{sin(\PHI)*\R}) (1.5*\D,0) (2*\D - 0.7*\R,-1.2*\R) (3*\D - \R,0)}; 				\draw [green, very thick] plot [smooth, tension=0.8] coordinates { ({cos(\PHI)*\R},{sin(\PHI)*\R}) (1.2*\D,1.5*\R) (1.8*\D, -1.5*\R) ({3*\D - cos(\VARPHI)*\R}, {-sin(\VARPHI)*\R})}; 				\draw [purple, very thick] plot [smooth, tension=0.8] coordinates { (\D+\R,0) (2*\D-\R,0)}; 				\draw [orange, very thick] plot [smooth, tension=0.8] coordinates { ({cos(\PHI)*\R}, {-sin(\PHI)*\R}) (\D,-1.5*\R) (1.5*\D, 0) ({2*\D - \R*cos(\PHI)},{sin(\PHI)*\R})}; 				\draw [magenta, very thick] plot [smooth, tension=0.8] coordinates { (\R,0) (\D - \R,0)};	 				\filldraw[color=black, fill=teal!20, very thick](\D,0) circle (\R) node{$b_{q^\prime}^{\mathcal{L}\dagger}$}; 				\filldraw[color=black, fill=teal!20, very thick](0,0) circle (\R) node{$b_{q^\prime}^{\mathcal{L}}$}; 				\filldraw[color=black, fill=teal!20, very thick](2*\D,0) circle (\R) node{$\rho_1^{\mathcal{R}}$}; 				\filldraw[color=black, fill=teal!20, very thick](3*\D,0) circle (\R) node{$\rho_2^{\mathcal{R}}$}; 			\end{scope} 			\node[anchor=south] at (1.6*\R, 0.7*\R) {$\textcolor{red}{\cev{\gamma}, \eps_\gamma}$}; 			\node[anchor=south] at (\D - 1.7*\R, 0.2*\R) {$\textcolor{red}{\vec{\gamma}, \eps_\gamma}$}; 			\node[anchor=south] at (1.6*\R, -0.2*\R) {$\textcolor{blue}{\cev{\alpha}_{12}}$}; 			\node[anchor=north] at (1.6*\R, -0.05*\R) {$\textcolor{blue}{\eps_{12}^\p}$}; 			\node[anchor=south] at (2*\D - 2.0*\R, 0.35*\R) {$\textcolor{blue}{\cev{\alpha}_{12}, \eps_{12}}$}; 			\node[anchor=north] at (\D + 2.0*\R, 0.1*\R) {$\textcolor{purple}{\vec{{\alpha}}_{22}, \beps_{22}^\p}$}; 			\node[anchor=north] at (2*\D - 1.9*\R, 0.1*\R) {$\textcolor{purple}{\vec{{\alpha}}_{22}, \beps_{22}}$}; 			\node[anchor=south] at (\D + 1.9*\R, 0.35*\R) {$\textcolor{orange}{\vec{{\alpha}}_{21}, \beps_{21}^\p}$}; 			\node[anchor=north] at (3*\D - 1.7*\R, -0.8*\R) {$\textcolor{orange}{\vec{{\alpha}}_{21}, \beps_{21}}$}; 			\node[anchor=north] at (2.1*\R, -1.1*\R) {$\textcolor{green}{\cev{\alpha}_{11}, \eps_{11}^\p}$}; 			\node[anchor=south] at (3*\D - 1.7*\R, 0.8*\R) {$\textcolor{green}{\cev{\alpha}_{11}, \eps_{11}}$}; 			\node[anchor=south] at (3*\D - 2.5*\R,-0.1*\R) {$\textcolor{magenta}{\cev{{\beta}}}$}; 			\node[anchor=north] at (3*\D - 2.5*\R,0.05*\R) {$\textcolor{magenta}{\beps_\beta}$}; 			\node[anchor=south] at (2*\D + 1.4*\R,-0.1*\R) {$\textcolor{magenta}{\vec{{\beta}}}$}; 			\node[anchor=north] at (2*\D + 1.4*\R,0.05*\R) {$\textcolor{magenta}{\beps_\beta}$}; 		\end{tikzpicture}~, 		\\ 		&\mathrm{III}\quad&\begin{tikzpicture}[baseline=-0.5ex] 			\pgfmathsetmacro{\R}{0.6} 			\pgfmathsetmacro{\H}{3.5*\R} 			\pgfmathsetmacro{\D}{5} 			\pgfmathsetmacro{\PHI}{30} 			\pgfmathsetmacro{\VARPHI}{75} 			\begin{scope}[shift={(3*\D, 0)}, xscale=-1] 				\draw [red, very thick] plot [smooth, tension=0.5] coordinates { ({cos(\VARPHI)*\R},{-sin(\VARPHI)*\R}) (0.25*\D,-0.8*\R) (0.5*\D,0) (\D,1.5*\R) (2*\D,1.5*\R) (2.5*\D,0) (2.75*\D,-0.8*\R) ({3*\D - cos(\VARPHI)*\R},{-sin(\VARPHI)*\R})}; 				\draw [purple, very thick] plot [smooth, tension=0.8] coordinates { (2*\D + \R,0) (3*\D - \R,0)}; 				\draw [orange, very thick] plot [smooth, tension=0.5] coordinates {({\D + cos(\PHI)*\R},{-sin(\PHI)*\R}) (2*\D,-1.5*\R) (2.5*\D,0) (2.75*\D,0.8*\R) ({3*\D - cos(\VARPHI)*\R},{sin(\VARPHI)*\R}) }; 				\draw [blue, very thick] plot [smooth, tension=0.5] coordinates { (\R,0) (0.6*\D, 0.22*\R) (\D,1.2*\R) ({2*\D - cos(\PHI)*\R}, {sin(\PHI)*\R})}; 				\draw [green, very thick] plot [smooth, tension=0.5] coordinates { ({cos(\VARPHI)*\R}, {sin(\VARPHI)*\R}) (0.25*\D,0.8*\R) (0.5*\D,0.2*\R) (\D - \R,0)}; 				\draw [magenta, very thick] plot [smooth, tension=0.8] coordinates { (\D + \R,0) (2*\D - \R,0)};				 				\filldraw[color=black, fill=teal!20, very thick](0,0) circle (\R) node{$\rho_2^{\mathcal{R}}$}; 				\filldraw[color=black, fill=teal!20, very thick](2*\D,0) circle (\R) node{$b_{q^\prime}^{\mathcal{L}\dagger}$}; 				\filldraw[color=black, fill=teal!20, very thick](\D,0) circle (\R) node{$b_{q^\prime}^{\mathcal{L}}$}; 				\filldraw[color=black, fill=teal!20, very thick](3*\D,0) circle (\R) node{$\rho_1^{\mathcal{R}}$}; 			\end{scope} 			\node[anchor=south] at (3*\D-1.5*\R,0.8*\R) {$\textcolor{green}{\cev{\alpha}_{11}, \beps_{11}^\p}$}; 			\node[anchor=north] at (2*\D + 2.1*\R,0.1*\R) {$\textcolor{green}{\cev{\alpha}_{11}, \beps_{11}}$}; 			\node[anchor=south] at (3*\D - 1.6*\R,-0.1*\R) {$\textcolor{blue}{\cev{\alpha}_{12}}$}; 			\node[anchor=north] at (3*\D - 1.6*\R,0.1*\R) {$\textcolor{blue}{\beps_{12}^\p}$}; 			\node[anchor=west] at (\D + 0.7*\R,1.0*\R) {$\textcolor{blue}{\cev{\alpha}_{12}, \beps_{12}}$}; 			\node[anchor=north] at (3*\D - \R,-0.85*\R) {$\textcolor{red}{\cev{{\gamma}}, \beps_\gamma}$}; 			\node[anchor=north] at (1.2*\R,-0.85*\R) {$\textcolor{red}{\vec{{\gamma}}, \beps_\gamma}$}; 			\node[anchor=north] at (2*\D - 1.3*\R,-0.6*\R) {$\textcolor{orange}{\vec{{\alpha}}_{21}, \beps_{21}}$}; 			\node[anchor=south] at (1.4*\R,0.8*\R) {$\textcolor{orange}{\vec{{\alpha}}_{21}, \beps_{21}^\p}$}; 			\node[anchor=south] at (\D - 1.5*\R,-0.1*\R) {$\textcolor{purple}{\vec{{\alpha}}_{22}}$}; 			\node[anchor=north] at (\D - 1.5*\R,0.05*\R) {$\textcolor{purple}{\beps_{22}}$}; 			\node[anchor=south] at (1.5*\R,-0.1*\R) {$\textcolor{purple}{\vec{{\alpha}}_{22}}$}; 			\node[anchor=north] at (1.5*\R,0.1*\R) {$\textcolor{purple}{\beps_{22}^\p}$}; 			\node[anchor=north] at (\D + 1.7*\R,0.1*\R) {$\textcolor{magenta}{\vec{{\beta}},\beps_\beta}$}; 			\node[anchor=south] at (2*\D - 1.7*\R,-0.15*\R) {$\textcolor{magenta}{\cev{{\beta}},\beps_\beta}$}; 		\end{tikzpicture}~, 		\\ 		&\mathrm{IV}\quad&\begin{tikzpicture}[baseline=-0.5ex] 			\pgfmathsetmacro{\R}{0.6} 			\pgfmathsetmacro{\H}{3.5*\R} 			\pgfmathsetmacro{\D}{5} 			\pgfmathsetmacro{\PHI}{30} 			\pgfmathsetmacro{\VARPHI}{75} 			\begin{scope}[shift={(0, 0)}] 				\draw [red, very thick] plot [smooth, tension=0.5] coordinates { ({cos(\VARPHI)*\R},{-sin(\VARPHI)*\R}) (0.25*\D,-0.8*\R) (0.5*\D,0) (\D,1.5*\R) (2*\D,1.5*\R) (2.5*\D,0) (2.75*\D,-0.8*\R) ({3*\D - cos(\VARPHI)*\R},{-sin(\VARPHI)*\R})}; 				\draw [purple, very thick] plot [smooth, tension=0.8] coordinates { (2*\D + \R,0) (3*\D - \R,0)}; 				\draw [orange, very thick] plot [smooth, tension=0.5] coordinates {({\D + cos(\PHI)*\R},{-sin(\PHI)*\R}) (2*\D,-1.5*\R) (2.5*\D,0) (2.75*\D,0.8*\R) ({3*\D - cos(\VARPHI)*\R},{sin(\VARPHI)*\R}) }; 				\draw [blue, very thick] plot [smooth, tension=0.5] coordinates { (\R,0) (0.6*\D, 0.22*\R) (\D,1.2*\R) ({2*\D - cos(\PHI)*\R}, {sin(\PHI)*\R})}; 				\draw [green, very thick] plot [smooth, tension=0.5] coordinates { ({cos(\VARPHI)*\R}, {sin(\VARPHI)*\R}) (0.25*\D,0.8*\R) (0.5*\D,0.2*\R) (\D - \R,0)}; 				\draw [magenta, very thick] plot [smooth, tension=0.8] coordinates { (\D + \R,0) (2*\D - \R,0)}; 				\filldraw[color=black, fill=teal!20, very thick](0,0) circle (\R) node{$\rho_2^{\mathcal{R}}$}; 				\filldraw[color=black, fill=teal!20, very thick](\D,0) circle (\R) node{$b_{q^\prime}^{\mathcal{L}\dagger}$}; 				\filldraw[color=black, fill=teal!20, very thick](2*\D,0) circle (\R) node{$b_{q^\prime}^{\mathcal{L}}$}; 				\filldraw[color=black, fill=teal!20, very thick](3*\D,0) circle (\R) node{$\rho_1^{\mathcal{R}}$}; 			\end{scope} 			\node[anchor=south] at (1.7*\R,0.8*\R) {$\textcolor{green}{\vec{\alpha}_{11}, \eps_{11}^\p}$}; 			\node[anchor=north] at (\D - 1.9*\R,0.1*\R) {$\textcolor{green}{\vec{\alpha}_{11}, \eps_{11}}$}; 			\node[anchor=south] at (1.5*\R,-0.1*\R) {$\textcolor{blue}{\vec{\alpha}_{12}}$}; 			\node[anchor=north] at (1.5*\R,0.1*\R) {$\textcolor{blue}{\eps_{12}^\p}$}; 			\node[anchor=east] at (2*\D - 0.7*\R,1.0*\R) {$\textcolor{blue}{\vec{\alpha}_{12}, \eps_{12}}$}; 			\node[anchor=north] at (1.4*\R,-0.85*\R) {$\textcolor{red}{\vec{\gamma}, \eps_\gamma}$}; 			\node[anchor=north] at (3*\D - 1.6*\R,-0.85*\R) {$\textcolor{red}{\cev{\gamma}, \eps_\gamma}$}; 			\node[anchor=north] at (\D + 1.9*\R,-0.7*\R) {$\textcolor{orange}{\cev{\alpha}_{21}, \eps_{21}}$}; 			\node[anchor=south] at (3*\D - 1.6*\R,0.85*\R) {$\textcolor{orange}{\cev{\alpha}_{21}, \eps_{21}^\p}$}; 			\node[anchor=south] at (2*\D + 1.55*\R,-0.1*\R) {$\textcolor{purple}{\cev{\alpha}_{22}}$}; 			\node[anchor=north] at (2*\D + 1.55*\R,0.1*\R) {$\textcolor{purple}{\eps_{22}}$}; 			\node[anchor=south] at (3*\D - 1.6*\R,-0.1*\R) {$\textcolor{purple}{\cev{\alpha}_{22}}$}; 			\node[anchor=north] at (3*\D - 1.6*\R,0.1*\R) {$\textcolor{purple}{\eps_{22}^\p}$}; 			\node[anchor=south] at (\D + 1.7*\R,-0.15*\R) {$\textcolor{magenta}{\vec{\beta}, \eps_\beta}$}; 			\node[anchor=north] at (2*\D - 1.7*\R,0.1*\R) {$\textcolor{magenta}{\cev{\beta}, \eps_\beta}$}; 		\end{tikzpicture}~. 	\end{align}Note
that the diagrams $\mathrm{III}$ and $\mathrm{IV}$ above are the
mirror images of the ones in Eqs. (\ref{eq:diagrams_1_III}) and (\ref{eq:diagrams_1_IV}),
where bra states become ket states and vice-versa -- this corresponds
to complex conjugation of both form factors (see Eq. (\ref{eq:FF_conj_bra}))
and reflection matrices. This time, we pair $\mathrm{I}$ with $\mathrm{IV}$
and $\mathrm{II}$ with $\mathrm{III}$. The same steps lead to
\begin{align}
\gamma\left(\omega^{\prime}\lvert\omega\right)= & -\frac{2}{\pi\omega^{\p}\omega}\int_{\alpha_{ij},\beta,\gamma}\mathsf{C}_{\beta}\mathsf{C}_{\gamma}\int_{0}^{\infty}\id\Omega\prod_{i=1}^{2}\delta\left(\nu_{i1}+\nu_{i2}+\nu_{\gamma}-\Omega\right)\nonumber \\
 & \times\delta\left(\omega^{\p}+\nu_{21}-\nu_{11}+\nu_{\beta}\right)\int_{0}^{\infty}\id t^{\pp}e^{-\eta t^{\pp}}\cos\left(\omega t^{\pp}\right)\mathrm{Re}\left\{ e^{\ii\Omega t^{\pp}}\prod\mathsf{F}\prod\mathsf{R}\right\} .\label{eq:gamma_2}
\end{align}
The expressions in Eqs. (\ref{eq:gamma_1}), (\ref{eq:gamma_2}) are
equal; taking their average eliminates the term proportional to $\mathrm{Im}\left\{ \prod\mathsf{F}\prod\mathsf{R}\right\} $,
leaving us with the simple delta function $\delta\left(\omega-\Omega\right)$.
Finally, we arrive at
\begin{equation}
\gamma\left(\omega^{\p}\lvert\omega\right)=-\frac{1}{\omega^{\p}\omega}\int_{\alpha_{ij},\beta,\gamma}\mathsf{C}_{\beta}\mathsf{C}_{\gamma}\prod_{i=1}^{2}\delta\left(\nu_{i1}+\nu_{i2}+\nu_{\gamma}-\omega\right)\delta\left(\omega^{\p}+\nu_{21}-\nu_{11}+\nu_{\beta}\right)\mathrm{Re}\left\{ \prod\mathsf{F}\prod\mathsf{R}\right\} .\label{eq:spectrum_general}
\end{equation}
\end{widetext}

This is a general form factor expansion for the inelastic spectrum.
Similarly to Eq. (\ref{eq:r_coeff}), it involves some intuitive delta
functions: the two functions $\delta\left(\nu_{i1}+\nu_{i2}+\nu_{\gamma}-\omega\right)$
imply that the the energies of the excitations connected to $\rho_{i}^{\mathcal{R}}$
must sum up to the photon energy $\omega$, and $\delta\left(\omega^{\p}+\nu_{21}-\nu_{11}+\nu_{\beta}\right)$
assures that the total energies connected to $b_{q^{\p}}^{\mathcal{L}\dagger}$
and $b_{q^{\p}}^{\mathcal{L}}$ are both equal to $\omega^{\p}$ (note
that $\nu_{21}-\nu_{11}=\nu_{12}-\nu_{22}$).

Before we proceed, let us consider the disconnected case, $\epsilon_{12}=\epsilon_{21}=\epsilon_{\beta}=\epsilon_{\gamma}=\left\{ \right\} $,
for which we find
\begin{equation}
\gamma_{\textrm{disc}}\left(\omega^{\p}\lvert\omega\right)=\delta\left(\omega^{\p}-\omega\right)\left(\left|r\left(\omega\right)\right|^{2}-1\right)=-\gamma\left(\omega\right)\delta\left(\omega^{\p}-\omega\right).\label{eq:gamma_disconnected}
\end{equation}
That is, the disconnected terms do not contribute to the spectrum
at $\omega^{\p}<\omega$, and reproduce the total inelastic decay
rate. That is similar to the calculation in the Keldysh formalism,
where the total rate is also obtained from the disconnected diagrams
(see Appendix \ref{app:referm} and Ref. \citep{goldstein_inelastic_2013}).

\subsection{\textcolor{black}{\label{subsec:diagrams}Leading diagrams}}

\textcolor{black}{Eq. (\ref{eq:spectrum_general}) offers a non-perturbative
and rapidly-convergent diagrammatic approach for calculating the spectrum.
Each term in Eq. (\ref{eq:spectrum_general}) may be represented by
a diagram, from which one may read off the corresponding form factors
and reflection matrices, as illustrated below. The physical intuition
behind this approach is clear; one needs to sum over all processes
with excitations of the integrable theory connecting the bosonic operators,
with energy conservation imposed by the delta functions. Comparing
to refermionization at the free fermion point $z=1/2$, discussed
in Appendix \ref{app:referm}, we draw an analogy between our approach
and Wick's theorem, as explained below.}

All that is left to be done is to draw the leading diagrams and evaluate
their contributions. As in any form factor expansion, such as that
of the reflection coefficient in Eq. (\ref{eq:r_coeff}), we expect
the contributions to decay rapidly with the number of excitations
involved, allowing us to obtain the spectrum with good accuracy using
only a few terms. Our figure of merit is the sum rule in Eq. (\ref{eq:sumrule}),
which should hold for all $\omega$; hence, it is essential that the
asymptotic power laws of $\gamma\left(\omega\right)$ at low and high
frequencies will be recovered by the sum $\int_{0}^{\omega}\frac{\omega^{\p}}{\omega}\gamma\left(\omega^{\p}\lvert\omega\right)\id\omega^{\p}$.

In the following, we focus on $z=1/3$ and $z=1/2$, list the leading
terms in Eq. (\ref{eq:spectrum_general}), and draw their corresponding
diagrams. The details behind the evaluation of these terms, as well
as additional subleading terms, are given in Appendix \ref{app:3p_details}.
In particular, one must be careful to take into account all terms
with comparable contributions, which could be misleading in the presence
of mixed matrix elements, $\sidescripts{\epsilon_{\vartheta}}{\mn}{\bk{\varthetar|\mathcal{A}|\thetal}}{\mn}{\epsilon_{\theta}}$,
due to the annihilation poles of the form factors (Eq. (\ref{eq:kin_pole})).
Appendix \ref{app:3p_details} presents a consistent method to identify
and evaluate all such terms.

First, consider the diagram $\epsilon_{11}=\left\{ \pm\right\} ,\epsilon_{12}=\left\{ \mp\right\} ,\epsilon_{21}=\left\{ \pm\right\} ,\epsilon_{22}=\left\{ \mp\right\} ,\epsilon_{\beta}=\epsilon_{\gamma}=\left\{ \right\} $,
depicted by

\begin{equation} 		\label{eq:gamma_2s_2s_diagram}\sum_{s,s^\p,s^{\p\p}=\pm}\begin{tikzpicture}[baseline={([yshift=-.5ex]current bounding box.center)},vertex/.style={anchor=base, circle,fill=black!25,minimum size=18pt,inner sep=2pt}] 			\pgfmathsetmacro{\R}{0.5} 			\pgfmathsetmacro{\H}{3.5*\R} 			\pgfmathsetmacro{\D}{2} 			\pgfmathsetmacro{\PHI}{30} 		
			\draw [purple, very thick] plot [smooth, tension=0.8] coordinates { (2*\D + \R,0) (3*\D - \R,0)}; 			\draw [orange, very thick] plot [smooth, tension=0.8] coordinates {({\D + cos(\PHI)*\R},{-sin(\PHI)*\R}) (2*\D,-1.3*\R) (2.5*\D,0) ({3*\D - cos(\PHI)*\R},{sin(\PHI)*\R}) }; 			\draw [blue, very thick] plot [smooth, tension=0.8] coordinates {({cos(\PHI)*\R},{-sin(\PHI)*\R}) (0.5*\D,0) (\D,1.3*\R)  ({2*\D - cos(\PHI)*\R},{sin(\PHI)*\R}) }; 			\draw [green, very thick] plot [smooth, tension=0.8] coordinates { ({\R}, {0}) (\D - \R,0)}; 		
			\filldraw[color=black, fill=teal!20, very thick](0,0) circle (\R) node{$\rho_1^{\mathcal{R}}$}; 			\filldraw[color=black, fill=teal!20, very thick](\D,0) circle (\R) node{$b_{q^\prime}^{\mathcal{L}\dagger}$}; 			\filldraw[color=black, fill=teal!20, very thick](2*\D,0) circle (\R) node{$b_{q^\prime}^{\mathcal{L}}$}; 			\filldraw[color=black, fill=teal!20, very thick](3*\D,0) circle (\R) node{$\rho_2^{\mathcal{R}}$}; 		
			\node[anchor=south] at (1.3*\R, 1*\R) {\textcolor{green}{$\omega - \Omega$}}; 			\node[anchor=north] at (2*\R, -0.9*\R) {\textcolor{blue}{$\Omega$}}; 			\node[anchor=north] at (3*\D - 1.2*\R, -0.9*\R) {\textcolor{purple}{$\omega^\prime + \Omega$}}; 			\node[anchor=south] at (3*\D - 1.8*\R, 1*\R) {\textcolor{orange}{$\omega - \omega^\prime - \Omega$}}; 		
			\node[anchor=north] at (\D-1.3*\R, 0) {\textcolor{green}{$s$}}; 			\node[anchor=south] at (2*\D-1.3*\R, 0.7*\R) {\textcolor{blue}{$\bar{s}$}}; 			\node[anchor=south] at (1.4*\R, 0) {\textcolor{green}{$\bar{s}^\p$}}; 			\node[anchor=north] at (1.4*\R, {-0.8*sin(\PHI)*\R}) {\textcolor{blue}{$s^\p$}}; 			\node[anchor=south] at (2*\D+1.3*\R, 0) {\textcolor{purple}{$\bar{s}$}}; 			\node[anchor=north] at (\D+1.3*\R, -0.7*\R) {\textcolor{orange}{$s$}}; 			\node[anchor=north] at (3*\D-1.4*\R, 0) {\textcolor{purple}{$s^{\p\p}$}}; 			\node[anchor=south] at (3*\D-1.4*\R, {0.7*sin(\PHI)*\R}) {\textcolor{orange}{$\bar{s}^{\p\p}$}}; 		\end{tikzpicture}~. 	\end{equation}Its contribution to the spectrum reads
\begin{align}
\gamma_{1}^{\left(1\right)}\left(\omega^{\p}\lvert\omega\right)= & \frac{2}{\omega^{\p}\omega}\int_{0}^{\omega-\omega^{\p}}\frac{\id\Omega}{\left(2\pi\right)^{4}\prod_{i=1}^{4}e^{\lambda_{i}}}\nonumber \\
 & \hspace{-1.5cm}\times\mathrm{Re}\left\{ \left(e^{-\frac{\ii\pi}{2z}}R_{+}^{-}\left(\lambda_{1}\right)R_{+}^{-}\left(\lambda_{2}\right)-e^{\frac{\ii\pi}{2z}}R_{+}^{+}\left(\lambda_{1}\right)R_{+}^{+}\left(\lambda_{2}\right)\right)^{*}\right.\nonumber \\
 & \hspace{-1.5cm}\left.\times\left(e^{-\frac{\ii\pi}{2z}}R_{+}^{-}\left(\lambda_{3}\right)R_{+}^{-}\left(\lambda_{4}\right)-e^{\frac{\ii\pi}{2z}}R_{+}^{+}\left(\lambda_{3}\right)R_{-}^{-}\left(\lambda_{4}\right)\right)-1\right\} \nonumber \\
 & \hspace{-1.5cm}\times\mathrm{Re}\left\{ f_{+-}^{\mathcal{R}}\left(\lambda_{3},\lambda_{1}+\ii\pi^{-}\right)f_{+-}^{\mathcal{R}}\left(\lambda_{4},\lambda_{2}+\ii\pi^{-}\right)\right.\nonumber \\
 & \hspace{-1.5cm}\left.\times f_{+-}^{\mathcal{R}}\left(\lambda_{1},\lambda_{2}\right)f_{+-}^{\mathcal{R}}\left(\lambda_{4},\lambda_{3}\right)\right\} ,\label{eq:gamma_1^(1)}
\end{align}
with $e^{\lambda_{1}}=\omega-\Omega,e^{\lambda_{2}}=\Omega,e^{\lambda_{3}}=\omega-\omega^{\p}-\Omega,e^{\lambda_{4}}=\omega^{\p}+\Omega$,
as specified on the diagram lines. Note that we take the real parts
of $\prod\mathsf{F}$ and $\prod\mathsf{R}$ separately; this follows
from Eq. (\ref{eq:gamma_2s_2s_a_details}) (see Appendix \ref{app:3p_details}
for details). At the free-fermion point, $z=1/2$, all form factors
other than $f_{+-}^{\mathcal{R}}$ vanish (see Appendix \ref{app:FF}),
and we find that this is the only contribution to the spectrum, $\gamma\left(\omega^{\p}\lvert\omega\right)=\gamma_{1}^{\left(1\right)}\left(\omega^{\p}\lvert\omega\right)$.
As shown in Fig. \ref{fig:inel_spec}, the sum rule is indeed perfectly
obeyed in that case, for both bsG and Kondo models. Furthermore, we
show in Appendix \ref{app:referm} that the same results for either
model can be obtained by means of refermionization. \textcolor{black}{There
is a clear connection between Eq. (\ref{eq:gamma_2s_2s_diagram}),
where each $\rho^{\mathcal{R}}$ leg is connected to both $b_{q^{\p}}^{\mathcal{L}\dagger}$
and $b_{q^{\p}}^{\mathcal{L}}$, and the contractions in Eq. (\ref{eq:Gcqq_Keldysh_fermions})
that lead to Eq. (\ref{eq:Gcqq_Fourier}). This agreement not only
highlights the one-to-one correspondence between the solitons and
fermions at $z=1/2$, but also draws an analogy between our diagrammatic
approach and Wick's theorem in free theories, and serves as an essential
sanity check for the general expression in Eq. (\ref{eq:spectrum_general}).}

Varying $z$ in the vicinity of $z=1/2$, we expect $\gamma_{1}^{\left(1\right)}\left(\omega^{\p}\lvert\omega\right)$
to dominate; indeed, for $z\sim1/2$, the sum rule is obeyed with
good precision. The precision deteriorates as we drift away from the
free-fermion point, where the contributions of multi-soliton ($z>1/2$)
or breather ($z<1/2$) form factors become important, and considerably
expand the available phase space beyond the diagram in Eq. (\ref{eq:gamma_2s_2s_diagram}).
\textcolor{black}{However, panel (b) of Fig.}\textbf{\textcolor{black}{{}
}}\textcolor{black}{\ref{fig:inel_spec} shows that the $z$-dependent
power laws of the sums $\int_{0}^{\omega}\frac{\omega^{\p}}{\omega}\gamma\left(\omega^{\p}\lvert\omega\right)\id\omega^{\p}$
are equal to those of $\gamma\left(\omega\right)$; a power law mismatch
would be reflected by a sharp increase or decrease of the ratio at
low or high frequencies.}

Moving on, we concentrate on $z=1/3$, where form factors involving
the $m=1$ breather should play a major role. We find the leading
term to be

\begin{equation} 		\label{eq:gamma_2s_1b_diagram}\sum_{s,s^\p=\pm}\begin{tikzpicture}[baseline={([yshift=-.5ex]current bounding box.center)},vertex/.style={anchor=base, circle,fill=black!25,minimum size=18pt,inner sep=2pt}] 			\pgfmathsetmacro{\R}{0.5} 			\pgfmathsetmacro{\H}{3.5*\R} 			\pgfmathsetmacro{\D}{2} 			\pgfmathsetmacro{\PHI}{30} 		
			\draw [purple, very thick] plot [smooth, tension=0.8] coordinates { (2*\D + \R,0) (3*\D - \R,0)}; 			\draw [orange, very thick] plot [smooth, tension=0.8] coordinates {({\D + cos(\PHI)*\R},{-sin(\PHI)*\R}) (2*\D,-1.3*\R) (2.5*\D,0) ({3*\D - cos(\PHI)*\R},{sin(\PHI)*\R}) }; 			\draw [green, very thick] plot [smooth, tension=0.8] coordinates { ({\R}, 0) (\D - \R,0)}; 			\draw [magenta, very thick] plot [smooth, tension=0.8] coordinates { ({\D + \R*cos(\PHI)},0) ({2*\D - \R*cos(\PHI)},0)}; 		
			\filldraw[color=black, fill=teal!20, very thick](0,0) circle (\R) node{$\rho_1^{\mathcal{R}}$}; 			\filldraw[color=black, fill=teal!20, very thick](\D,0) circle (\R) node{$b_{q^\prime}^{\mathcal{L}\dagger}$}; 			\filldraw[color=black, fill=teal!20, very thick](2*\D,0) circle (\R) node{$b_{q^\prime}^{\mathcal{L}}$}; 			\filldraw[color=black, fill=teal!20, very thick](3*\D,0) circle (\R) node{$\rho_2^{\mathcal{R}}$}; 		
			\node[anchor=south] at (0.5*\D, 0.9*\R) {\textcolor{green}{$\omega$}}; 			\node[anchor=south] at (1.5*\D, 0.9*\R) {\textcolor{magenta}{$\omega - \omega^\prime - \Omega$}}; 			\node[anchor=south] at (2.5*\D, 0.9*\R) {\textcolor{orange}{$\Omega$}}; 			\node[anchor=north] at (2.5*\D, -0.95*\R) {\textcolor{purple}{$\omega - \Omega$}}; 		
			\node[anchor=south] at (0.5*\D, 0) {\textcolor{green}{$1$}}; 			\node[anchor=south] at (1.5*\D, 0) {\textcolor{magenta}{$\bar{s}$}}; 			\node[anchor=south] at (2*\D+1.3*\R, 0) {\textcolor{purple}{$\bar{s}$}}; 			\node[anchor=north] at (\D+1.3*\R, -0.7*\R) {\textcolor{orange}{$s$}}; 			\node[anchor=north] at (3*\D-1.3*\R, 0) {\textcolor{purple}{$s^\p$}}; 			\node[anchor=south] at (3*\D-1.3*\R, {0.6*sin(\PHI)*\R}) {\textcolor{orange}{$\bar{s}^\p$}}; 		\end{tikzpicture}~, 	\end{equation}corresponding to $\alpha_{11}=\left\{ 1\right\} ,\alpha_{12}=\left\{ \right\} ,\alpha_{21}=\left\{ \pm\right\} ,\alpha_{22}=\left\{ \mp\right\} ,\beta=\left\{ \mp\right\} ,\gamma=\left\{ \right\} $.
Its contribution to the spectrum is
\begin{align}
\gamma_{2}^{\left(1\right)}\left(\omega^{\p}\lvert\omega\right)= & -\frac{4}{\omega^{\p}\omega}\int_{0}^{\omega-\omega^{\p}}\frac{\id\Omega}{\left(2\pi\right)^{4}\prod_{i=1}^{4}e^{\lambda_{i}}}\nonumber \\
 & \hspace{-1.8cm}\times\mathrm{Re}\left\{ \left(R_{1}^{1}\left(\lambda_{1}-\log\mu_{1}\right)\right)^{*}\right.\nonumber \\
 & \hspace{-1.8cm}\left.\times\left(e^{-\frac{\ii\pi}{2z}}R_{+}^{-}\left(\lambda_{3}\right)R_{+}^{-}\left(\lambda_{4}\right)-e^{\frac{\ii\pi}{2z}}R_{+}^{+}\left(\lambda_{3}\right)R_{+}^{+}\left(\lambda_{4}\right)\right)-1\right\} \nonumber \\
 & \hspace{-1.8cm}\times\mathrm{Re}\left\{ f_{1}^{\mathcal{R}}\left(\lambda_{1}-\log\mu_{1}\right)f_{+-1}^{\mathcal{R}}\left(\lambda_{3},\lambda_{2},\lambda_{1}+\ii\pi^{-}-\log\mu_{1}\right)\right.\nonumber \\
 & \hspace{-1.8cm}\left.\times f_{+-}^{\mathcal{R}}\left(\lambda_{4},\lambda_{2}+\ii\pi^{-}\right)f_{+-}^{\mathcal{R}}\left(\lambda_{4},\lambda_{3}\right)\right\} ,\label{eq:gamma_2^(1)}
\end{align}
where $e^{\lambda_{1}}=\omega,e^{\lambda_{2}}=\omega-\omega^{\p}-\Omega,e^{\lambda_{3}}=\Omega,e^{\lambda_{4}}=\omega-\Omega$.
Again, we take the real parts of both $\prod\mathsf{F}$ and $\prod\mathsf{R}$,
as detailed in Eq. (\ref{eq:gamma_2s_1b_a_details}). In $\gamma_{2}^{\left(1\right)}\left(\omega^{\p}\lvert\omega\right)$
we take into account all of the necessary combinatorial factors, as
well as the equal contribution of the mirror diagram of (\ref{eq:gamma_2s_1b_diagram}),
which corresponds to $\alpha_{11}=\left\{ \mp\right\} ,\alpha_{12}=\left\{ \pm\right\} ,\alpha_{21}=\left\{ \right\} ,\alpha_{22}=\left\{ 1\right\} ,\beta=\left\{ \mp\right\} ,\gamma=\left\{ \right\} $.

In addition to the two diagrams above, we consider 3 additional diagrams
in the evaluation of the $z=1/3$ spectrum, listed in Appendix \ref{app:3p_details}.
We find that, for $z=1/3$, the resulting spectrum obeys the sum rule
with good accuracy for a wide range of frequencies $\omega$, as shown
in Fig. \ref{fig:inel_spec}. Furthermore, the contributions decay
rapidly, as demonstrated in Fig. \ref{fig:inel_spec_contributions}
in Appendix \ref{app:3p_details}. It is important to note that plugging
each of the terms $\gamma_{i}\left(\omega^{\p}\lvert\omega\right)$
into the sum rule recovers the correct power laws of $\gamma\left(\omega\right)$,
for both $\omega\ll T_{B}$ and $\omega\gg T_{B}$.

\subsection{\label{subsec:spec_res}Results}

\begin{figure}[t]
\begin{centering}
\includegraphics[width=1\columnwidth]{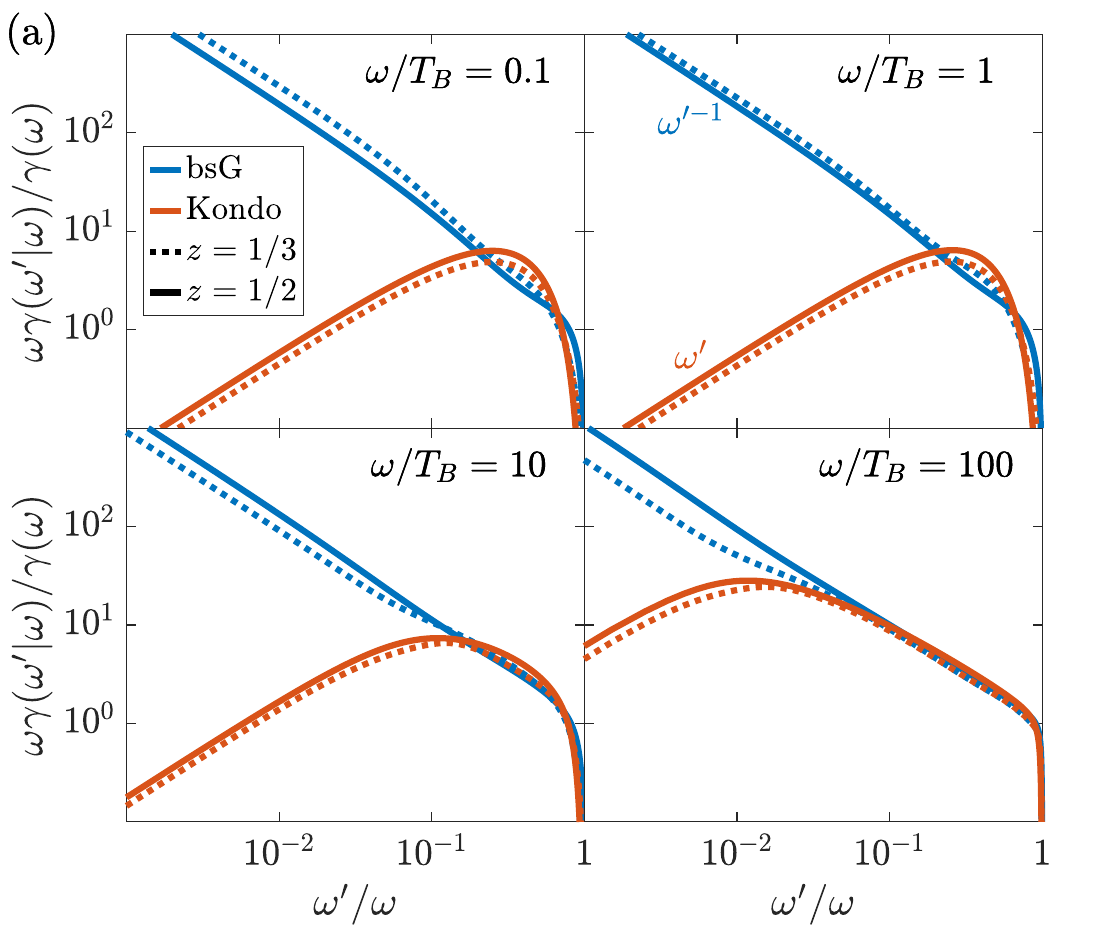}
\par\end{centering}
\begin{centering}
\includegraphics[width=1\columnwidth]{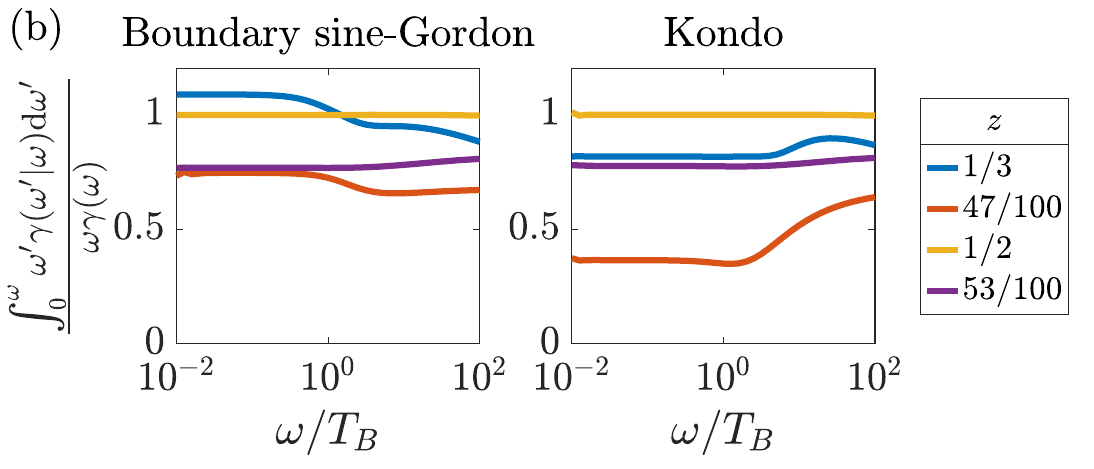}
\par\end{centering}
\caption{\label{fig:inel_spec}(a) The energy-resolved inelastic decay spectrum
as function of $\omega^{\protect\p}$, at several fixed values of
$\omega/T_{B}$, for the bsG and Kondo models and $z=1/3,1/2$. The
diagrams used to evaluate the spectrum are listed in Appendix \ref{app:3p_details}.
(b) The ratio between the LHS and RHS of Eq. (\ref{eq:sumrule}) for
both models and several values of $z$. Note that the power laws of
$\gamma\left(\omega\right)$ are recovered by the sum rule for all
$z$.}
\end{figure}

The spectra for both models at $z=1/3,1/2$, as well as the validation
of Eq. (\ref{eq:sumrule}) for several values of $z$, are displayed
in Fig. \ref{fig:inel_spec}. We have seen in Section \ref{sec:total_rate}
that the bsG and Kondo models exhibit the same Luttinger liquid power
laws at high frequencies, $\omega\gg T_{B}$, but differ below the
RG scale, $\omega\ll T_{B}$. The spectrum, determined from a higher
order response function, gives us more refined information. Indeed,
the difference between the models is emphasized by the spectrum; while
the Kondo spectrum is suppressed at low produced frequencies $\omega^{\prime}\ll T_{B}$,
the bsG spectrum diverges, $\gamma_{\bsg}\left(\omega^{\p}\ll T_{B}\lvert\omega\right)\sim\omega^{\p-1}$,
regardless of the incoming frequency $\omega$ (note that the total
rate, given by the sum rule in Eq. (\ref{eq:sumrule}), is still finite).
This asymptotic behavior results from the term $\gamma_{1}^{\left(1\right)}\left(\omega^{\p}\lvert\omega\right)$
in Eq. (\ref{eq:gamma_1^(1)}) --- one may show that the product
of reflection matrices is different than 1 upon setting $\omega^{\p}=0$,
and the $\omega^{\p-1}$ behavior results from the prefactor of the
integral. In other words, the production of low frequency photons
is favored in a splitting process in the bsG model. The proliferation
of low frequency produced photons results from the inductive coupling
of the half-infinite line to the impurity Josephson junction in Eq.
(\ref{eq:H_discrete}), with $\tilde{\mathcal{H}}_{I}=-E_{J}\cos\left(\tilde{\phi}_{0}\right)$.
The sole source of photon splitting is the boundary cosine term; fluctuations
of the flux $\tilde{\phi}_{0}$ in this nonlinear potential give rise
to the inelastic decay. The inductive coupling to the line translates
fluctuations of $\tilde{\phi}_{0}$ into approximately uniform shifts
of $\tilde{\phi}_{n>0}$, which correspond to low frequency modes.
This behavior is opposed to the Kondo spectrum, which vanishes at
$\omega^{\p}=0$; expanding the product of reflection matrices in
Eq. (\ref{eq:gamma_1^(1)}) at $\omega^{\p}\ll T_{B}$, we find $\gamma_{\K}\left(\omega^{\p}\ll T_{B}\lvert\omega\right)\sim\omega^{\p}$.
This asymptotic behavior turns out to hold for all contributions to
the spectrum and hence to all $z$, in agreement with Ref. \citep{goldstein_inelastic_2013}.

It is also interesting to consider decay processes with $\omega^{\prime}\lesssim\omega$,
where the total energy of the remaining photons is small, $\omega-\omega^{\p}\ll T_{B}$.
One may then apply strong coupling fixed point expansions to extract
the power law dependence of the spectrum as a function of $\omega-\omega^{\p}$.
Analytical expressions for the spectrum, and therefore for the $\omega-\omega^{\prime}$
dependence, may be calculated at the free-fermion point, $z=1/2$
(see Eq. (\ref{eq:spec_referm}), which was obtained using refermionization
and is identical to the form factors result). For the Kondo spectrum,
one finds $\gamma_{\K}\left(\omega^{\p}\lesssim\omega\lvert\omega\right)\sim\left(\omega-\omega^{\prime}\right)^{3}$,
in agreement with the contribution of the strong coupling fixed point
expansion of Ref. \citep{goldstein_inelastic_2013}. In the bsG model,
one finds $\gamma_{\bsg}\left(\omega^{\p}\lesssim\omega\lvert\omega\right)\sim\omega-\omega^{\p}$
at $z=1/2$; interestingly, this power law dependence does not stem
from neither the quartic term of the boundary cosine operator, nor
from the dual cosine term, which both induce a $\left(\omega-\omega^{\prime}\right)^{3}$
power law \citep{burshtein_photon-instanton_2021}. For $z<1/2$ it
appears difficult to numerically evaluate the involved integral to
sufficient accuracy to determine the behavior with enough certainty,
since strong cancellation occurs between different diagrams. We stress
that the spectrum at $z=1/2$ was obtained, for both models, in two
independent methods (form factors and refermionization), which yield
identical results. Furthermore, as evident from Ref. \ref{fig:inel_spec},
the asymptotic power laws of the total decay rate are recovered by
the sum rule in Eq. (\ref{eq:sumrule}) for all values of $z$. It
could be interesting to look further into this issue in a future work.

\section{\label{sec:conclusions}Conclusions}

In this work, we have shown how inelastic decay of microwave photons,
measured in cQED experiments implementing integrable systems, can
emerge from the purely elastic scattering of the excitations from
the integrability picture. Using the framework of form factors, we
identified the origin of the photon splitting as the nonlinear relation
between the microwave photons and the elementary excitations of the
bsG and Kondo models. The form factors, encoding this nonlinear relation,
allowed us to obtain exact results, going beyond previous perturbative
calculations \citep{burshtein_photon-instanton_2021,houzet_critical_2020,leger_revealing_2022,mehta_theory_2022,kuzmin_observation_2023,houzet_microwave_2023}.
Crucially, our results hold at low energies, even if the boundary
impurity terms of the Hamiltonians are relevant, rendering perturbation
theory invalid, as well as when strong coupling expansion fails. The
low energy results for the total inelastic decay rate and the elastic
phase shift distinguish between the bsG and Kondo models, which both
exhibit Luttinger liquid power laws above the RG scale. This distinction
is emphasized by the energy-resolved inelastic decay spectrum, where
$\gamma_{\bsg}\left(\omega^{\p}\ll T_{B}\lvert\omega\right)\sim\omega^{\p-1}$
and $\gamma_{\K}\left(\omega^{\p}\ll T_{B}\lvert\omega\right)\sim\omega^{\p}$
for all $z$; note that this result could not be obtained using perturbation
theory, even for $\omega\gg T_{B}$. As discussed in Section \ref{sec:total_rate},
such exact low energy expressions are particularly useful in the context
of the Schmid-Bulgadaev quantum phase transition, and shed single-photon
light on this intriguing phenomenon.

In the calculation of the energy-resolved inelastic decay spectrum,
we have devised a general method to calculate a 3-point response function
using form factors. While previous works have dealt with the calculation
of multipoint correlation functions in integrable quantum field theories
using form factors \citep{babujian_multipoint_2017,squarcini_multipoint_2021,fava_divergent_2023},
they focused on theories with a single excitation type and only considered
specific contributions. Our method, in contrast, provides a general
expression for all orders, and for any excitation content of the theory.
\textcolor{black}{The physical intuition behind this diagrammatic
approach is clear --- one has to sum over all combinations of excitations
connecting the bosonic operators, imposing appropriate energy conservation.
}As in any form factor expansion, it is sufficient to consider a few
terms to obtain a result with good precision, numerically evaluating
only single or double integrals. The generalization of our method
to 4-point response functions or higher is straight-forward.

Looking forward, there are several possible extensions to this work
which could improve quantitative comparisons with experimental measurements.
The most pressing issue is the incorporation of finite temperatures
into our framework; indeed, realistic temperatures in cQED experiments,
$T\sim50\textrm{ mK}\sim1\textrm{ GHz}$, are usually larger than
the RG scale defined by the impurity. \textcolor{black}{In those cases,
a perturbative approach is valid at any frequency, and diagrammatic
techniques have been shown to provide results that quantitatively
agree with experiments \citep{leger_revealing_2022,kuzmin_observation_2023}.
Yet, the rapid rate of technological advancements in the field of
cQED indicates that $T\lesssim E_{J}^{\star}$ could soon become possible,
thus raising interest in exact finite temperature results at all frequencies.
}We derive such exact results for $z=1/2$ using refermionization
in Appendix \ref{app:referm} (see also Ref. \citep{kuzmin_observation_2023}
for explicit expressions); other values of $z$ have to be treated
within the framework of form factors. In fact, the calculation of
finite temperature correlators using form factors \citep{leclair_finite_1999,saleur_comment_2000,doyon_2007,essler_finite-temperature_2009,pozsgay_form_factor_2010,pozsgay_leclair-mussardo_2018,cubero_thermodynamic_2019}
involves mixed matrix elements of the form $\sidescripts{\epsilon_{\vartheta}}{\mn}{\bk{\varthetar|\mathcal{A}|\thetal}}{\mn}{\epsilon_{\theta}}$,
much like those appearing in the 3-point response function considered
in this work; hence, the methods applied here could also be useful
for evaluating finite temperature response functions. Two other aspects
which should be addressed are the finite volume of the system \citep{pozsgay_form_2008,pozsgay_form_2008-1}
and the presence of integrability-breaking terms \citep{delfino_nonintegrable_1996},
both of which are inevitable in experimental setups.

Furthermore, inelastic decay in cQED setups should not necessarily
emerge from reflection off impurities, and can also occur in nonlinear
bulk models. Our framework could then be extended to the massive bulk
sine-Gordon model, amenable to realization in superconducting circuits
\citep{roy_quantum_2021}, and applied to investigate bulk effects,
such as the superconductor to insulator phase transition in an array
of Josephson junctions \citep{kuzmin_quantum_2019,bard_decay_2018,huan-kuang_theory_2019,houzet_microwave_2019},
through the lens of elastic and inelastic scattering.

Finally, our results should pave a path for tackling other systems,
beyond the scope of cQED experiments. Our framework for calculating
form factor expansions of multipoint response functions could be used
in other contexts of integrable field theories; a particular example
is the calculation of 4-point functions in tunneling experiments between
fractional quantum Hall leads at filling $\nu=1/3$, whose low energy
behavior is captured by the bsG model with $z=1/3$. Another likely
application is in the context of one-dimensional cold atom systems
\citep{bloch_many-body_2008}, which are often integrable.\textcolor{red}{{}
}\textcolor{black}{In particular, we could use our developed formalism
to evaluate multipoint response functions that are measured in post-quench
evolutions and indicate non-gaussian correlations which are ubiquitous
in integrable systems \citep{langen_experimental_2015,schweigler_decay_2021,kukuljan_correlation_2018}.
Multipoint functions could also be used to investigate the onset of
chaos due to weak integrability-breaking terms, which, when treated
perturbatively within the form factors formalism, necessitate the
use of higher-order correlation functions the order of the desired
correlation functions \citep{znidaric_weak_2020,brenes_eigenstate_2020,bulchandani_onset_2022}.}
Our method could also be applied for the calculation of multipoint
correlators in high energy contexts, such as the $\mathcal{N}=4$
supersymmetric Yang-Mills theory in $3+1$ dimensions \citep{dhoker_SYM_2002},
which is suspected to be integrable.\vspace{-0.5cm}

\subsubsection*{Acknowledgements}

We would like to thank Denis M. Basko, Serge Florens, Leonid I. Glazman,
Manuel Houzet, Roman Kuzmin, Vladimir E. Manucharyan, Nicolas Roch,
and Izak Snyman for useful discussions. We also thank Manuel Houzet
and Leonid I. Glazman for sharing with us their unpublished manuscript
\citep{houzet_microwave_2023}, and Manuel Houzet for sharing with
us his unpublished results regarding photon decay at the free-fermion
point, $z=1/2$. Our work has been supported by the Israel Science
Foundation (ISF) and the Directorate for Defense Research and Development
(DDR\&D) Grant No. 3427/21 and by the US-Israel Binational Science
Foundation (BSF) Grant No. 2020072. A.B. is also supported by the
Adams Fellowship Program of the Israel Academy of Sciences and Humanities.

\appendix

\section{\label{app:kondo_hamiltonian}Derivation of the effective Kondo Hamiltonian}

For completeness, we derive the Hamiltonian of the spin-boson model
in Eq. (\ref{eq:HSB}), which is equivalent to the Kondo model via
a unitary transformation and bosonization \citep{leggett_dynamics_1987},
starting from the microscopic Hamiltonian of the discrete transmission
line terminated by a fluxonium qubit.

Consider the discrete version of Eq. (\ref{eq:H_generic}):
\begin{equation}
\tilde{\mathcal{H}}=\sum_{n=1}^{N}\frac{\left(\tilde{\phi}_{n}-\tilde{\phi}_{n-1}\right)^{2}}{8L^{\textrm{line}}}+\sum_{n=0}^{N}\frac{2\tilde{Q}_{n}^{2}}{C_{g}}+\tilde{\mathcal{H}}_{I},
\end{equation}
with the impurity Hamiltonian
\begin{align}
\tilde{\mathcal{H}}_{I}= & \frac{2\tilde{Q}_{f}^{2}}{C_{f}}+\frac{\left(\tilde{\phi}_{f}-\tilde{\phi}_{0}\right)^{2}}{8L_{f}}+\frac{\tilde{\phi}_{0}^{2}}{8L_{0}}\nonumber \\
 & -E_{J}\cos\left(\tilde{\phi}_{f}-2\Phi_{\textrm{ext}}\right),
\end{align}
where $\Phi_{\textrm{ext}}$ is an external magnetic flux, which we
take from here on as half flux quantum, $\Phi_{\textrm{ext}}=\pi\hbar/\left(2e\right)=\pi/2$.
It is useful to rewrite $\tilde{\mathcal{H}}_{I}$ as
\begin{equation}
\tilde{\mathcal{H}}_{I}=\frac{2\tilde{Q}_{f}^{2}}{C_{f}}+E_{J}\cos\left(\tilde{\phi}_{f}\right)+\frac{\tilde{\phi}_{f}^{2}}{8\left(L_{f}+L_{0}\right)}+\frac{\left(\tilde{\phi}_{0}-\alpha\tilde{\phi}_{f}\right)^{2}}{8L_{\parallel}},
\end{equation}
where $\alpha=L_{0}/\left(L_{0}+L_{f}\right)$ and $L_{\parallel}=L_{0}L_{f}/\left(L_{0}+L_{f}\right)$.
This form allows us to decouple the inductive coupling between the
array and the fluxonium by applying a unitary transformation, $\mathcal{U}_{f}=e^{\ii\alpha\tilde{\phi}_{f}\tilde{Q}_{\textrm{tot}}}$,
with $\tilde{Q}_{\textrm{tot}}=\sum_{n=0}^{N}\tilde{Q}_{n}$, which
shifts the array phases, $\tilde{\phi}_{n}\rightarrow\tilde{\phi}_{n}+\alpha\tilde{\phi}_{f}$
for all $0\le n\le N$, as well as the fluxonium charge, $\tilde{Q}_{f}\rightarrow\tilde{Q}_{f}-\alpha\tilde{Q}_{\textrm{tot}}$.
The array-fluxonium coupling becomes capacitive, and the Hamiltonian
reads $\tilde{\mathcal{H}}=\tilde{\mathcal{H}}_{f}+\tilde{\mathcal{H}}_{a}+\tilde{\mathcal{H}}_{c}$,
with
\begin{align}
\tilde{\mathcal{H}}_{f}= & \frac{2\tilde{Q}_{f}^{2}}{C_{f}}+E_{J}\cos\left(\tilde{\phi}_{f}\right)+\frac{\tilde{\phi}_{f}^{2}}{8\left(L_{f}+L_{0}\right)},\nonumber \\
\tilde{\mathcal{H}}_{a}= & \sum_{n=1}^{N}\frac{\left(\tilde{\phi}_{n}-\tilde{\phi}_{n-1}\right)^{2}}{8L^{\textrm{line}}}+\sum_{n=0}^{N}\frac{2\tilde{Q}_{n}^{2}}{C_{g}}+\frac{\tilde{\phi}_{0}^{2}}{8L_{\parallel}}+\frac{2\alpha^{2}\tilde{Q}_{\textrm{tot}}^{2}}{C_{f}},\nonumber \\
\tilde{\mathcal{H}}_{c}= & -\frac{4\alpha\tilde{Q}_{f}\tilde{Q}_{\textrm{tot}}}{C_{f}}.\label{eq:H_fluxonium_terms}
\end{align}

We proceed by diagonalizing the array Hamiltonian $\tilde{\mathcal{H}}_{a}$.
Hamilton's equations read
\begin{align}
\partial_{t}^{2}\tilde{\phi}_{n}= & \frac{\tilde{\phi}_{n+1}+\tilde{\phi}_{n-1}-2\tilde{\phi}_{n}}{C_{g}L^{\textrm{line}}}-\omega_{\parallel}^{2}\tilde{\phi}_{0},\quad n>0,\nonumber \\
\partial_{t}^{2}\tilde{\phi}_{0}= & \frac{\tilde{\phi}_{1}-\tilde{\phi}_{0}}{C_{g}L^{\textrm{line}}}-\left(\frac{1}{C_{g}L_{\parallel}}+\omega_{\parallel}^{2}\right)\tilde{\phi}_{0},
\end{align}
where $\omega_{\parallel}=\alpha/\sqrt{C_{f}L_{\parallel}}$. We look
for an oscillatory solution, $\tilde{\phi}_{n}\sim\sum_{k}\tilde{\phi}_{k}e^{-\ii\omega_{k}t}$.
In order to decouple the $\tilde{\phi}_{0}$ term from the $n>0$
equations, we define new variables, $\tilde{\varphi}_{n}=\tilde{\phi}_{n}-\omega_{\parallel}^{2}\tilde{\phi}_{0}/\omega_{k}^{2}$,
leading to
\begin{align}
\omega_{k}^{2}\tilde{\varphi}_{n}= & \frac{2\tilde{\varphi}_{n}-\tilde{\varphi}_{n+1}-\tilde{\varphi}_{n-1}}{C_{g}L^{\textrm{line}}},\quad n>0,\nonumber \\
\frac{\omega_{k}^{2}\tilde{\varphi}_{0}}{1-\frac{\omega_{\parallel}^{2}}{\omega_{k}^{2}}}= & \frac{\tilde{\varphi}_{0}-\tilde{\varphi}_{1}}{C_{g}L^{\textrm{line}}}+\left(\frac{1}{C_{g}L_{\parallel}}+\omega_{\parallel}^{2}\right)\frac{\tilde{\varphi}_{0}}{1-\frac{\omega_{\parallel}^{2}}{\omega_{k}^{2}}},
\end{align}
In the following, we set the array spacing to unity, $a=1$. The equations
are solved by $\tilde{\varphi}_{n}\sim\sin\left(kn-\delta_{k}\right)$.
The bulk equations, $n>0$, yield the dispersion relation, $\omega_{k}=2v\sin\left(k/2\right)\approx vk$,
where the velocity $v=1/\sqrt{C_{g}L^{\textrm{line}}}$ is assumed
to be much larger than any other energy scale. The $n=0$ equation
yields the phase shift,
\begin{equation}
\tan\delta_{k}=\frac{v\left(\omega_{\parallel}^{2}-\omega_{k}^{2}\right)}{\omega_{k}\left[\frac{1}{C_{g}L_{\parallel}}+\omega_{\parallel}^{2}-\omega_{k}^{2}\right]}\approx\frac{\Gamma_{f}}{\omega_{k}}\left[1-\left(\frac{\omega_{k}}{\omega_{\parallel}}\right)^{2}\right],
\end{equation}
where $\Gamma_{f}=\alpha^{2}/\left(C_{f}Z\right)$ with $Z=\sqrt{L^{\textrm{line}}/C_{g}}$
is the inverse RC time of the fluxonium and the transmission line.
The approximation above holds provided that $\alpha^{2}C_{g}/C_{f}\ll1$
and $L_{\parallel}\gg L^{\textrm{line}}$, so that $1/\sqrt{L_{\parallel}C_{g}}$
is significantly larger than all energy scales other than $v$. We
impose open boundary conditions at $n=N$, leading through the Hamilton
equation for $n=N$ to $\sin\left(kN-\delta_{k}\right)=\sin\left(k\left(N+1\right)-\delta_{k}\right)$,
which yields a quantization condition $kN-\delta_{k}=\pi m+\pi/2$
with $m=0,1,\ldots,N$, and therefore a mode spacing $\Delta\approx\pi v/N$.
In the following, we also need the capacitance matrix $\left[\mathrm{C}\right]_{n,m}$,
obtained by inverting the capacitance energy matrix of $\tilde{\mathcal{H}}_{a}$:%
\begin{equation}
\left[\mathrm{C}\right]_{n,m}=C_{g}\delta_{n,m}+\frac{1}{N}\left[\frac{1}{\frac{1}{C_{g}}+\frac{\alpha^{2}N}{C_{f}}}-C_{g}\right].
\end{equation}
Neglecting $1/N$ corrections, we find that the mode capacitances
(that is, the eigenmode expectation values of the capacitance matrix)
are given by $C_{k}\approx NC_{g}/2$.

The line Hamiltonian may now be quantized by introducing creation
and annihilation operators. The diagonalized phase and charge operators
read
\begin{align}
\phi_{n}\approx & \sum_{k}\frac{-\ii\left(a_{k}-a_{k}^{\dagger}\right)}{\sqrt{NC_{g}\omega_{k}}}\left[\sin\left(kn-\delta_{k}\right)+\frac{\sin\delta_{k}}{1-\left(\frac{\omega_{k}}{\omega_{\parallel}}\right)^{2}}\right],\nonumber \\
Q_{n}= & \sum_{m}\left[\mathrm{C}\right]_{n,m}\partial_{t}\phi_{n}\nonumber \\
\approx & \sum_{k}\sqrt{\frac{C_{g}\omega_{k}}{N}}\sin\left(kn-\delta_{k}\right)\left(a_{k}+a_{k}^{\dagger}\right),
\end{align}
and the diagonalized array Hamiltonian is given by $\mathcal{H}_{a}=\sum_{k}\omega_{k}a_{k}^{\dagger}a_{k}$.
Using the quantization condition, we find
\begin{equation}
\sum_{n=0}^{N}\sin\left(kn-\delta_{k}\right)=\frac{\cos\left(\frac{k}{2}+\delta_{k}\right)}{2\sin\left(\frac{k}{2}\right)}\approx\frac{v}{\omega_{k}}\cos\delta_{k},
\end{equation}
allowing us to express the coupling term $\mathcal{H}_{c}$ in terms
of the eigenmodes:
\begin{equation}
\mathcal{H}_{c}\approx-\frac{2zQ_{f}}{\alpha}\sum_{k}\frac{\sqrt{\frac{2\pi v\omega_{k}}{Nz}}}{\sqrt{\left[1-\left(\frac{\omega_{k}}{\omega_{\parallel}}\right)^{2}\right]^{2}+\left(\frac{\omega_{k}}{\Gamma_{f}}\right)^{2}}}\left(a_{k}+a_{k}^{\dagger}\right).
\end{equation}
The square root in the denominator above imposes a high-frequency
cutoff. To make contact with standard bosonization expressions, we
replace it by an exponential cutoff, $e^{-\omega_{k}/\Lambda}$, where
$\Lambda\sim\min\left\{ \omega_{\parallel},\Gamma_{f}\right\} $.

Finally, we are in position to derive the spin-boson Hamiltonian.
The fluxonium may be approximated as a symmetric double well potential
for the flux $\phi_{f}$ (achieved by tuning the external magnetic
flux to half flux quantum \citep{manucharyan_fluxonium_2009}). We
label the two lowest eigenstates of the fluxonium Hamiltonian as the
eigenstates of the $\mathcal{S}_{x}$ operator, so that $\mathcal{S}_{z}$
eigenstates correspond to wavefunctions with well-defined phase, localized
near either of the minima of the double well. The fluxonium Hamiltonian
then reads $\mathcal{H}_{f}=-J\mathcal{S}_{x}$, where $J$ is the
tunneling matrix element between the two wells, which can be calculated
by WKB or instanton methods \citep{weiss_quantum_1999}. The fluxonium
charge $\tilde{Q}_{f}$ couples the two-level system to the array
via a $\mathcal{S}_{z}$ term \citep{leggett_dynamics_1987},
\begin{equation}
\mathcal{H}_{c}=-\mathcal{S}_{z}\sqrt{2z}\pi v\frac{2q_{f}}{\alpha}\sum_{k}\sqrt{\frac{k}{\pi N}}e^{-\omega_{k}/\Lambda}\left(a_{k}+a_{k}^{\dagger}\right),
\end{equation}
where $q_{f}=\left|\bk{\phi_{f}^{\pm}|Q_{f}|\phi_{f}^{\pm}}\right|$
is the expectation value of the fluxonium charge operator in the eigenfunctions
of either of the symmetric wells. The full Hamiltonian reads
\begin{align}
\mathcal{H}= & \sum_{k}\omega_{k}a_{k}^{\dagger}a_{k}-J\mathcal{S}_{x}\nonumber \\
 & -\mathcal{S}_{z}\sqrt{2z}\pi v\frac{2q_{f}}{\alpha}\sum_{k}\sqrt{\frac{k}{\pi N}}e^{-\omega_{k}/\Lambda}\left(a_{k}+a_{k}^{\dagger}\right).
\end{align}
This $k$-space version of the spin-boson Hamiltonian can now be easily
mapped to the continuum real-space version of Eq. (\ref{eq:HSB}).
Note that the coupling coefficient is given by $\sqrt{2z^{\p}}\pi v$,
with $z^{\p}=z\times\left(2q_{f}/\alpha\right)^{2}$; that is, the
Luttinger parameter of the Kondo model is proportional, but not equivalent,
to the normalized impedance $z$.

\section{\textcolor{black}{\label{app:exp_considerations}Cutoff scales and
deviations from integrability in realistic setups}}

\selectlanguage{british}%
\textcolor{black}{In an experimental environment, IR and UV cutoff
scales are both present. The IR cutoff scale is determined by the
finite length of the array, $\Delta=\pi v/\ell$, and corresponds
to a finite mode spacing that is manifested by well-resolved modes
at frequencies $\omega_{m}=\Delta\left(m-\delta_{0}/\pi\right)$,
where $m=0,1,2,\ldots$ and $\delta_{0}$ is a phase shift set by
the boundary conditions at the far end of the line, away from the
impurity. Our analysis should hold as long as we are concerned with
modes at frequencies $\omega\gg\Delta$; note that in realistic setups
\citep{kuzmin_inelastic_2021,leger_revealing_2022,kuzmin_observation_2023,mehta_down-conversion_2023},
the mode spacing is usually the smallest energy scale, well below
the relevant RG scale. The system remains integrable for a finite
length $\ell$, and one could use finite length form factor techniques
}\foreignlanguage{english}{\textcolor{black}{\citep{pozsgay_form_2008,pozsgay_form_2008-1}}}\textcolor{black}{{}
to investigate the effect of the finite $\ell$ on modes at $\omega\gtrsim\Delta$.}

\textcolor{black}{As mentioned in Section \ref{sec:models}, the UV
cutoff $\Lambda$ is set by the smaller of the inverse RC time of
the impurity and the transmission line, and the plasma frequency of
the line. A finite $\Lambda$ does break integrability; however, usually,
$\Lambda\gg T_{B}$, such that many modes, both below and above the
RG scale, lie well below $\Lambda$, hence our analysis remains valid.
The effect of a finite $\Lambda$ could be explored by treating the
cutoff terms in the Hamiltonian perturbatively, within the form factors
formalism. This would require the calculation of higher order response
functions --- for instance, the total inelastic decay rate $\gamma\left(\omega\right)$
and the phase shift $\delta\left(\omega\right)$ would be given by
a 3- or 4-point response function (as opposed to the 2-point function
considered in Section \ref{sec:total_rate}), which could be calculated
using the formalism developed in this work.}

\textcolor{black}{Finally, note that while the discrete system introduced
in Eq. (\ref{eq:H_discrete}) is not integrable due to the finite
lattice spacing $a$, the associated scale $v/a$ is much larger than
any other scale (including $\Lambda$), and its effect on the results
should be negligible.}
\selectlanguage{english}%

\section{\label{app:Smat_props}S-matrix of the bulk sine-Gordon model}

The S-matrix in an integrable quantum field theory is the key ingredient
in the Zamolodchikov-Faddeev algebra, and reflects the purely elastic
nature of the scattering in the theory. It satisfies several properties;
the first is the Yang-Baxter equation,
\begin{align}
\sum_{\epsilon_{1}^{\prime\prime},\epsilon_{2}^{\prime\prime},\epsilon_{3}^{\prime\prime}}S_{\epsilon_{1}\epsilon_{2}}^{\epsilon_{1}^{\prime\prime}\epsilon_{2}^{\prime\prime}}\left(\theta_{1}-\theta_{2}\right)S_{\epsilon_{1}^{\prime\prime}\epsilon_{3}}^{\epsilon_{1}^{\prime}\epsilon_{3}^{\prime\prime}}\left(\theta_{1}-\theta_{3}\right)S_{\epsilon_{2}^{\prime\prime}\epsilon_{3}^{\prime\prime}}^{\epsilon_{2}^{\prime}\epsilon_{3}^{\prime}}\left(\theta_{2}-\theta_{3}\right)\nonumber \\
 & \hspace{-8cm}=\sum_{\epsilon_{1}^{\prime\prime},\epsilon_{2}^{\prime\prime},\epsilon_{3}^{\prime\prime}}S_{\epsilon_{2}\epsilon_{3}}^{\epsilon_{2}^{\prime\prime}\epsilon_{3}^{\prime\prime}}\left(\theta_{2}-\theta_{3}\right)S_{\epsilon_{1}\epsilon_{3}^{\prime\prime}}^{\epsilon_{1}^{\prime\prime}\epsilon_{3}^{\prime}}\left(\theta_{1}-\theta_{3}\right)S_{\epsilon_{1}^{\prime\prime}\epsilon_{2}^{\prime\prime}}^{\epsilon_{1}^{\prime}\epsilon_{2}^{\prime}}\left(\theta_{1}-\theta_{2}\right),
\end{align}
which ensures the equivalence of the factorization of $n$-body scattering
to a product of 2-body scatterings. Unitarity and crossing symmetry
imply
\begin{align}
\sum_{\epsilon_{1}^{\p\p},\epsilon_{2}^{\p\p}}S_{\epsilon_{1}\epsilon_{2}}^{\epsilon_{1}^{\prime\prime}\epsilon_{2}^{\prime\prime}}\left(\theta\right)S_{\epsilon_{1}^{\prime\prime}\epsilon_{2}^{\prime\prime}}^{\epsilon_{1}^{\prime}\epsilon_{2}^{\prime}}\left(-\theta\right)= & \delta_{\epsilon_{1}}^{\epsilon_{1}^{\p}}\delta_{\epsilon_{2}}^{\epsilon_{2}^{\p}},\\
S_{\epsilon_{1}\epsilon_{2}}^{\epsilon_{1}^{\prime}\epsilon_{2}^{\prime}}\left(\ii\pi-\theta\right)= & S_{\bar{\epsilon}_{2}^{\prime}\epsilon_{1}}^{\bar{\epsilon}_{2}\epsilon_{1}^{\prime}}\left(\theta\right).
\end{align}
Recall that a bar denotes charge conjugation, $\bar{\pm}=\mp$ and
$\bar{m}=m$. The S-matrix of the bulk sine-Gordon model is well-known.
First, for the exchange of two (anti)solitons,
\begin{align}
S_{++}^{++}\left(\theta\right)=S_{--}^{--}\left(\theta\right)= & S_{0}\left(\theta\right),\\
S_{+-}^{+-}\left(\theta\right)=S_{-+}^{-+}\left(\theta\right)= & \dfrac{S_{0}\left(\theta\right)\sin\left(-\dfrac{\pi}{\xi}\ii\theta\right)}{\sin\left(\dfrac{\pi}{\xi}\left(\pi+\ii\theta\right)\right)},\\
S_{+-}^{-+}\left(\theta\right)=S_{-+}^{+-}\left(\theta\right)= & \dfrac{S_{0}\left(\theta\right)\sin\left(\dfrac{\pi^{2}}{\xi}\right)}{\sin\left(\dfrac{\pi}{\xi}\left(\pi+\ii\theta\right)\right)},
\end{align}
where $\xi=\frac{\pi}{1/z-1}$, and
\begin{equation}
S_{0}\left(\theta\right)=-\exp\left\{ -\ii\int_{0}^{\infty}\frac{dx}{x}\frac{\sin\left(x\theta\right)\sinh\left(\frac{\pi-\xi}{2}x\right)}{\sinh\left(\frac{\xi x}{2}\right)\cosh\left(\frac{\pi x}{2}\right)}\right\} .
\end{equation}
For integer $p=1/z$, this sector of the S-matrix becomes diagonal,
since $S_{+-}^{-+}\left(\theta\right)=S_{-+}^{+-}\left(\theta\right)=0$,
and also $S_{+-}^{+-}\left(\theta\right)=S_{-+}^{-+}\left(\theta\right)=\left(-1\right)^{p}S_{0}\left(\theta\right)$.
Note that $S_{0}\left(\theta\right)=-1$ for the free-fermionic case
$z=1/2$. Next, the exchange matrix for a breather and a soliton or
an antisoliton is
\begin{align}
S_{\pm m}^{\pm m}\left(\theta\right)=S_{m\pm}^{m\pm}\left(\theta\right)\nonumber \\
 & \hspace{-3cm}=-\prod_{j=1}^{m}\frac{\ii\cos\left(\frac{\xi}{2}\right)+\sinh\left(\theta-\frac{\ii\xi}{2}\left(m+1-2j\right)\right)}{\ii\cos\left(\frac{\xi}{2}\right)-\sinh\left(\theta-\frac{\ii\xi}{2}\left(m+1-2j\right)\right)}.\label{eq:S_bs}
\end{align}
Finally, the S-matrix for two breathers is
\begin{align}
S_{m_{1}m_{2}}^{m_{1}m_{2}}\left(\theta\right)= & \prod_{j=1}^{\min\left(m_{1},m_{2}\right)-1}\frac{\coth\left(\frac{\theta}{2}-\frac{\ii\xi}{4}\left(\left|m_{1}-m_{2}\right|+2j\right)\right)}{\tanh\left(\frac{\theta}{2}+\frac{\ii\xi}{4}\left(\left|m_{1}-m_{2}\right|+2j\right)\right)}\nonumber \\
 & \hspace{-2cm}\times\frac{\coth\left(\frac{\theta}{2}-\frac{\ii\xi}{4}\left|m_{1}-m_{2}\right|\right)\coth\left(\frac{\theta}{2}-\frac{\ii\xi}{4}\left(m_{1}+m_{2}\right)\right)}{\tanh\left(\frac{\theta}{2}+\frac{\ii\xi}{4}\left|m_{1}-m_{2}\right|\right)\tanh\left(\frac{\theta}{2}+\frac{\ii\xi}{4}\left(m_{1}+m_{2}\right)\right)}.
\end{align}
All other terms of the S-matrix are zero.

It is also useful to obtain expressions for the exchange of a right
mover with a left mover in the massless limit, $\tilde{S}_{\epsilon_{1}\epsilon_{2}}^{\epsilon_{1}^{\p}\epsilon_{2}^{\p}}=\lim_{\theta\rightarrow\infty}S_{\epsilon_{1}\epsilon_{2}}^{\epsilon_{1}^{\p}\epsilon_{2}^{\p}}\left(\theta\right)$,
as such limits of the S-matrix appear in the complete set of states
in Eq. (\ref{eq:complete_set_2}). The S-matrix is diagonal in this
limit, $\tilde{S}_{\epsilon_{1}\epsilon_{2}}^{\epsilon_{1}^{\p}\epsilon_{2}^{\p}}\propto\delta_{\epsilon_{1}}^{\epsilon_{1}^{\p}}\delta_{\epsilon_{2}}^{\epsilon_{2}^{\p}}$,
with
\begin{align}
\tilde{S}_{++}^{++}=\tilde{S}_{--}^{--}=\left(\tilde{S}_{+-}^{+-}\right)^{*}=\left(\tilde{S}_{-+}^{-+}\right)^{*}= & e^{-\frac{\ii\pi}{2z}},\nonumber \\
\tilde{S}_{\pm m}^{\pm m}\left(\theta\right)=\tilde{S}_{m\pm}^{m\pm}\left(\theta\right)=\tilde{S}_{m_{1}m_{2}}^{m_{1}m_{2}}\left(\theta\right)= & 1.\label{eq:SRL}
\end{align}
Also note that, in the massless limit, the S-matrix of two left movers
is the complex conjugate of the S-matrix of two right movers,
\begin{equation}
S\left(\lambda_{1}^{\lmn}-\lambda_{2}^{\lmn}\right)=S^{*}\left(\lambda_{1}^{\rmn}-\lambda_{2}^{\rmn}\right).\label{eq:S_left_right}
\end{equation}

\section{\label{app:R}Boundary reflection matrices in the Kondo and boundary
sine-Gordon models}

The presence of the boundary introduces another component to an integrable
field theory --- the boundary reflection matrix, $R_{\epsilon}^{\epsilon^{\p}}\left(\theta\right)$,
which relates an incoming state $\k{\theta}_{\epsilon}$ with an outgoing
state $\k{-\theta}_{\epsilon^{\p}}$: $\k{\theta}_{\epsilon}=R_{\epsilon}^{\epsilon^{\p}}\left(\theta\right)\k{-\theta}_{\epsilon}$.
First studied for solitons in Ref. \citep{ghoshal_boundary_1994}
and later for bound states in Ref. \citep{ghoshal_bound_1994}, the
reflection matrix can be derived from a set of axioms and properties,
similar to those of the S-matrix, such as boundary unitarity:
\begin{equation}
\sum_{\epsilon^{\p}}R_{\epsilon}^{\epsilon^{\p}}\left(\theta\right)R_{\epsilon^{\p}}^{\epsilon^{\p\p}}\left(-\theta\right)=\delta_{\epsilon}^{\epsilon^{\pp}}.
\end{equation}
The massless limit of the reflection matrices of the bsG model was
later derived in Ref. \citep{fendley_exact_1994}, where it was shown
that $R_{\epsilon}^{\epsilon^{\p}}\left(\lambda\right)$ depends only
on the difference $\lambda-\lambda_{B}$, where $T_{B}=e^{\lambda_{B}}$
is the energy scale associated with the boundary, proportional to
the RG scale $E_{J}^{\star}\sim E_{J}^{1/\left(1-z\right)}$. The
soliton reflection matrices in the bsG model are
\begin{align}
R_{\pm}^{\mp}\left(\lambda\right)= & \ii e^{\left(\frac{1}{z}-1\right)\left(\lambda-\lambda_{B}\right)/2}R_{s}\left(\lambda-\lambda_{B}\right),\nonumber \\
R_{\pm}^{\pm}\left(\lambda\right)= & e^{-\left(\frac{1}{z}-1\right)\left(\lambda-\lambda_{B}\right)/2}R_{s}\left(\lambda-\lambda_{B}\right),\nonumber \\
R_{s}\left(\lambda\right)= & \frac{e^{-\frac{\ii\pi}{4}}}{2\cosh\left(\left(\frac{1}{z}-1\right)\frac{\lambda}{2}-\frac{\ii\pi}{4}\right)}\nonumber \\
 & \hspace{-1cm}\times\exp\left\{ \ii\int_{0}^{\infty}\frac{\id x}{x}\frac{\sin\left(2\lambda x\right)\sinh\left(\left(\pi-\xi\right)x\right)}{\sinh\left(2\xi x\right)\cosh\left(\pi x\right)}\right\} .\label{eq:Rs_bSG}
\end{align}
These matrices simplify considerably at $z=1/2$; since $R_{s}\left(\lambda\right)=e^{\lambda/2}/\left(e^{\lambda}+\ii\right),$
we find
\begin{equation}
R_{\pm}^{\mp}\left(\lambda\right)=\frac{\ii e^{\lambda-\lambda_{B}}}{e^{\lambda-\lambda_{B}}+\ii},\quad R_{\pm}^{\pm}\left(\lambda\right)=\frac{1}{e^{\lambda-\lambda_{B}}+\ii}.\label{eq:R_bSG_z_half}
\end{equation}
For breathers ($z<1/2$), we have
\begin{align}
R_{2m-1}^{2m-1}\left(\lambda\right)= & \tanh\left(\frac{\lambda-\lambda_{B}}{2}-\frac{\ii\pi}{4}\right)\nonumber \\
 & \times\prod_{j=1}^{m-1}\frac{\tanh\left(\frac{\lambda-\lambda_{B}}{2}-\frac{\ii\pi}{4}-\frac{\ii\xi j}{2}\right)}{\tanh\left(\frac{\lambda-\lambda_{B}}{2}-\frac{\ii\pi}{4}+\frac{\ii\xi j}{2}\right)},\\
R_{2m}^{2m}\left(\lambda\right)= & \prod_{j=1}^{m}\frac{\tanh\left(\frac{\lambda-\lambda_{B}}{2}-\frac{\ii\pi}{4}-\frac{\ii\xi}{2}\left(j-\frac{1}{2}\right)\right)}{\tanh\left(\frac{\lambda-\lambda_{B}}{2}-\frac{\ii\pi}{4}+\frac{\ii\xi}{2}\left(j-\frac{1}{2}\right)\right)}.
\end{align}
The reflection matrices of the Kondo model are simpler:
\begin{equation}
R_{\pm}^{\mp}\left(\lambda\right)=e^{\frac{\ii\pi}{4z}}\frac{e^{\lambda-\lambda_{B}}-\ii}{e^{\lambda-\lambda_{B}}+\ii},\quad R_{\pm}^{\pm}\left(\lambda\right)=0,\label{eq:Rs_K}
\end{equation}
and
\begin{equation}
R_{m}^{m}\left(\lambda\right)=\frac{\tanh\left(\frac{\lambda-\lambda_{B}}{2}-\frac{\ii\xi m}{4}\right)}{\tanh\left(\frac{\lambda-\lambda_{B}}{2}+\frac{\ii\xi m}{4}\right)}.
\end{equation}
It is important to note that, in the limit $\lambda\gg\lambda_{B}$,
the reflection matrices become trivial in both models: for a set of
excitations $\lambda$,
\begin{equation}
\mathsf{R}_{\epsilon_{\lambda}}^{\epsilon_{\lambda}^{\p}}\left(\lambr\rightarrow\infty\right)\rightarrow\delta_{\bar{\epsilon}_{\lambda}}^{\epsilon_{\lambda}^{\p}}=\prod_{k=1}^{n_{\lambda}}\delta_{\bar{\epsilon}_{k}}^{\epsilon_{k}^{\p}}.\label{eq:Rprod_high_energy}
\end{equation}
Recall that $\mathsf{R}_{\epsilon_{\lambda}}^{\epsilon_{\lambda}^{\p}}$,
defined in Eq. (\ref{eq:Rprod_def}), involves the product of scattering
matrices of right and left movers given in Eq. (\ref{eq:SRL}), which
ensures that the phase of the product of reflection matrices in that
limit is zero. In the low energy limit, $\lambda\ll\lambda_{B}$,
Eq. (\ref{eq:Rprod_high_energy}) still holds for the Kondo model,
while in the bsG model we have
\begin{equation}
\mathsf{R}_{\epsilon_{\lambda}}^{\epsilon_{\lambda}^{\p}}\left(\lambr\rightarrow-\infty\right)\rightarrow\delta_{\epsilon_{\lambda}}^{\epsilon_{\lambda}^{\p}}\mathsf{C}_{\lambda}=\prod_{k=1}^{n_{\lambda}}\delta_{\epsilon_{k}}^{\epsilon_{k}^{\p}}\prod_{\left\{ k\lvert\epsilon_{k}=m\right\} }\left(-1\right)^{\epsilon_{k}},\label{eq:R_prod_low_energy_bSG}
\end{equation}
where the sign factor $\mathsf{C}_{\lambda}$ is defined in Eq. (\ref{eq:charge_conj_mat}).
Also note that, in both models,
\begin{equation}
\mathsf{R}_{\epsilon_{\lambda}}^{\epsilon_{\lambda}^{\p}}\left(\lambda+\ii\pi\right)=\left(\mathsf{R}_{\epsilon_{\lambda}}^{\epsilon_{\lambda}^{\p}}\left(\lambda\right)\right)^{*}.\label{eq:R_conj}
\end{equation}

\section{\label{app:FF}Form factors in the massless sine-Gordon model}

\subsection{General properties}

The form factors in an integrable quantum field theory can be derived
from a set of axioms and conditions determined by the local conservation
laws of the model, as well as the additional symmetries of the specific
model. First, two excitations can be exchanged via the S-matrix:
\begin{align}
\sum_{\epsilon_{i},\epsilon_{i+1}}S_{\epsilon_{i}^{\p}\epsilon_{i+1}^{\p}}^{\epsilon_{i}\epsilon_{i+1}}\left(\theta_{i}-\theta_{i+1}\right)f_{\ldots\epsilon_{i}\epsilon_{i+1}\ldots}^{\mathcal{O}}\left(\ldots,\theta_{i},\theta_{i+1},\ldots\right)\nonumber \\
 & \hspace{-7cm}=f_{\ldots\epsilon_{i+1}^{\p}\epsilon_{i}^{\p}\ldots}^{\mathcal{O}}\left(\ldots,\theta_{i+1},\theta_{i},\ldots\right).\label{eq:FF_exc}
\end{align}
We also have the following periodicity property:
\begin{equation}
f_{\epsilon_{1}\ldots\epsilon_{n}}^{\mathcal{O}}\left(\theta_{1},\ldots,\theta_{n}+2\pi\ii\right)=f_{\epsilon_{n}\epsilon_{1}\ldots\epsilon_{n-1}}^{\mathcal{O}}\left(\theta_{n},\theta_{1},\ldots,\theta_{n-1}\right).
\end{equation}
Lorentz invariance implies
\begin{equation}
f_{\epsilon_{1}\ldots\epsilon_{n}}^{\mathcal{O}}\left(\theta_{1}+\alpha,\ldots,\theta_{n}+\alpha\right)=e^{s_{\mathcal{O}}\alpha}f_{\epsilon_{1}\ldots\epsilon_{n}}^{\mathcal{O}}\left(\theta_{1},\ldots,\theta_{n}\right),\label{eq:FF_lorentz}
\end{equation}
where $s_{\mathcal{O}}$ is the spin of the operator $\mathcal{O}$.
Specifically, the spin of the current operators is $s_{\mathcal{R}}=s_{\mathcal{L}}=1$.

The form factors are analytical functions of the rapidities in the
strip $0\le\mathrm{Im}\theta\le\pi$, where the only singularities
are two kinds of simple poles. The first are the annihilation poles,
at $\theta_{i}=\theta_{j}+\ii\pi$. The residue of the pole at $\theta_{n}=\theta_{n-1}+\ii\pi$
is given by
\begin{align}
\Res{\theta_{n}=\theta_{n-1}+\ii\pi}f_{\epsilon_{1}\ldots\epsilon_{n}}^{\mathcal{O}}\left(\theta_{1},\ldots,\theta_{n}\right)\nonumber \\
 & \hspace{-4.5cm}=\sum_{\left\{ \epsilon^{\p}\right\} ,\left\{ \tau\right\} }f_{\epsilon_{1}^{\p}\ldots\epsilon_{n-2}^{\p}}^{\mathcal{O}}\left(\theta_{1},\ldots,\theta_{n-2}\right)C_{\epsilon_{n}\epsilon_{n-1}^{\p}}\nonumber \\
 & \hspace{-4cm}\times\left[\prod_{k=1}^{n-1}\delta_{\epsilon_{k}}^{\epsilon_{k}^{\p}}-\delta_{\tau_{n-2}}^{\epsilon_{n-1}}\delta_{\epsilon_{n-1}^{\p}}^{\tau_{0}}\prod_{k=1}^{n-2}S_{\tau_{k}\epsilon_{k}}^{\tau_{k-1}\epsilon_{k}^{\p}}\left(\lambda_{n-1}-\lambda_{k}\right)\right].\label{eq:kin_pole}
\end{align}
Here $C_{\epsilon_{1}\epsilon_{2}}$ is the charge conjugation matrix:
\begin{equation}
\mathfrak{C}Z_{\epsilon}^{\dagger}\left(\theta\right)\mathfrak{C}^{-1}=Z_{\bar{\epsilon}}^{\dagger}\left(\theta\right)C_{\epsilon\bar{\epsilon}},\label{eq:charge_conj}
\end{equation}
where $\mathfrak{C}$ is the charge conjugation operator. In the sine-Gordon
model, $C_{+-}=C_{-+}=1$, $C_{mm}=\left(-1\right)^{m}$, and zero
otherwise. The other residues at $\theta_{i}=\theta_{j}+\ii\pi$ can
be found using Eq. (\ref{eq:FF_exc}). Poles of the second kind indicate
the bound states in the theory. For example, consider a form factor
$f_{\epsilon_{1}\ldots\epsilon_{n-1}\epsilon_{n}}^{\mathcal{O}}\left(\theta_{1},\ldots,\theta_{n-1},\theta_{n}\right)$
in the sine-Gordon model, with $\epsilon_{n-1}=+$ and $\epsilon_{n}=-$;
this form factor has a pole at $\theta_{n}=\theta_{n-1}+\ii\theta^{\left(m\right)}$,
with $\theta^{\left(m\right)}=\pi-\xi m$, corresponding to a breather
of type $m$. Its residue is proportional to the form factor $f_{\epsilon_{1}\ldots\epsilon_{n-2}m}^{\mathcal{O}}\left(\theta_{1},\ldots,\theta_{n-1},\theta_{n}\right)$;
this is known as the bootstrap axiom, which allows one to obtain form
factors of breathers from those of solitons, or higher-order breathers
from lower-order ones.

The crossing relations are needed to evaluate matrix elements of the
form $\sidescripts{\epsilon_{\vartheta_{1}}\ldots\epsilon_{\vartheta_{l}}}{}{\bk{\vartheta_{1},\ldots,\vartheta_{l}|\mathcal{O}|\theta_{n},\ldots,\theta_{1}}}{}{\epsilon_{\theta_{n}}\ldots\epsilon_{\theta_{1}}}$,
which appear in the calculation of multipoint correlation functions.
Following the notations of Ref. \citep{smirnov_form_1992}, we write
$\k{\thetal}_{\epsilon_{\theta}}=\k{\thetal_{\mathrm{a}},\thetal_{\mathrm{b}}}_{\epsilon_{\theta}^{\p}}S\left(\thetal_{\mathrm{a}}\lvert\thetal\right)_{\epsilon_{\theta}}^{\epsilon_{\theta}^{\p}}$,
where $\theta_{\mathrm{a,b}}$ are disjoint sets such that $\theta_{\mathrm{a}}\cup\theta_{\mathrm{b}}=\theta$,
and $S\left(\thetal_{\mathrm{a}}\lvert\thetal\right)_{\epsilon_{\theta}}^{\epsilon_{\theta}^{\p}}$
is the product of S-matrices needed to reorder $\thetal$ as $\thetal_{\mathrm{a}},\thetal_{\mathrm{b}}$;
namely,
\begin{align}
\k{\theta_{n},\ldots,\theta_{1}}_{\epsilon_{\theta_{n}}\ldots\epsilon_{\theta_{1}}}= & S\left(\thetal_{\mathrm{a}}\lvert\thetal\right)_{\epsilon_{\theta}}^{\epsilon_{\theta}^{\p}}\nonumber \\
 & \hspace{-2cm}\times\k{\left(\theta_{\mathrm{a}}\right)_{n_{\mathrm{a}}},\ldots,\left(\theta_{\mathrm{a}}\right)_{1},\left(\theta_{\mathrm{b}}\right)_{n_{\mathrm{b}}},\ldots,\left(\theta_{\mathrm{b}}\right)_{1}}_{\epsilon_{\theta_{n}}^{\p}\ldots\epsilon_{\theta_{1}}^{\p}}.
\end{align}
Similarly, $\sidescripts{\epsilon_{\vartheta}}{}{\b{\varthetar}}{}{}=\sidescripts{\epsilon_{\vartheta}^{\p}}{}{\b{\varthetar_{\mathrm{b}},\varthetar_{\mathrm{a}}}}{}{}S\left(\varthetar\lvert\varthetar_{\mathrm{a}}\right)_{\epsilon_{\vartheta}^{\p}}^{\epsilon_{\vartheta}}$.
We also define $\delta_{\epsilon_{\theta}}^{\epsilon_{\vartheta}}\left(\thetar\lvert\varthetar\right)=\delta_{ln}\prod_{k=1}^{n}2\pi\delta\left(\theta_{k}-\vartheta_{k}\right)\delta_{\epsilon_{\theta_{k}}}^{\epsilon_{\vartheta_{k}}}$.
The crossing relations read
\begin{align}
\sidescripts{\epsilon_{\vartheta}}{}{\bk{\varthetar|\mathcal{O}|\thetal}}{}{\epsilon_{\theta}}= & \sum_{\substack{\theta_{\mathrm{a}}\cup\theta_{\mathrm{b}}=\theta\\
\vartheta_{\mathrm{a}}\cup\vartheta_{\mathrm{b}}=\vartheta
}
}S\left(\varthetar\lvert\varthetar_{\mathrm{a}}\right)_{\epsilon_{\vartheta}^{\p}}^{\epsilon_{\vartheta}}S\left(\thetal_{\mathrm{a}}\lvert\thetal\right)_{\epsilon_{\theta}}^{\epsilon_{\theta}^{\p}}\nonumber \\
 & \hspace{-2cm}\times\delta_{\epsilon_{\theta_{\mathrm{b}}}}^{\epsilon_{\vartheta_{\mathrm{b}}}}\left(\varthetar_{\mathrm{b}}\lvert\thetal_{\mathrm{b}}\right)\times\sidescripts{\epsilon_{\vartheta_{\mathrm{a}}}}{}{\bk{\varthetar_{\mathrm{a}}+\ii\delta|\mathcal{O}|\thetal_{\mathrm{a}}}}{}{\epsilon_{\theta_{\mathrm{a}}}}.\label{eq:cross_rels_1}
\end{align}
The rapidities in the bra state of $\sidescripts{\epsilon_{\vartheta_{\mathrm{a}}}}{}{\bk{\varthetar_{\mathrm{a}}+\ii\delta|\mathcal{O}|\thetal_{\mathrm{a}}}}{}{\epsilon_{\theta_{\mathrm{a}}}}$
are analytically continued so there is no overlap between the bra
and the ket rapidities, and 
\begin{align}
\sidescripts{\epsilon_{\vartheta_{\mathrm{a}}}}{}{\bk{\varthetar_{\mathrm{a}}+\ii\delta|\mathcal{O}|\thetal_{\mathrm{a}}}}{}{\epsilon_{\theta_{\mathrm{a}}}}\nonumber \\
 & \hspace{-2cm}=\bk{0|\mathcal{O}|\varthetar_{\mathrm{a}}+\ii\pi+\ii\delta,\thetal_{\mathrm{a}}}_{\bar{\epsilon}_{\vartheta_{\mathrm{a}}}\epsilon_{\theta_{\mathrm{a}}}}\mathsf{C}_{\epsilon_{\vartheta_{\mathrm{a}}}}\nonumber \\
 & \hspace{-2cm}=\bk{0|\mathcal{O}|\thetal_{\mathrm{a}},\varthetar_{\mathrm{a}}-\ii\pi+\ii\delta}_{\epsilon_{\theta_{\mathrm{a}}}\bar{\epsilon}_{\vartheta_{\mathrm{a}}}}\mathsf{C}_{\epsilon_{\vartheta_{\mathrm{a}}}},\label{eq:cross_rels_no_overlap}
\end{align}
where $\mathsf{C}_{\theta}$ denotes the product of the elements of
the charge conjugation matrix for the excitations in the set $\theta$:
\begin{equation}
\mathsf{C}_{\theta}\equiv\prod_{k=1}^{n}C_{\epsilon_{\theta_{k}}\bar{\epsilon}_{\theta_{k}}}=\prod_{\left\{ k\lvert\epsilon_{\theta_{k}}=m\right\} }\left(-1\right)^{\epsilon_{\theta_{k}}}.\label{eq:charge_conj_mat}
\end{equation}
The infinitesimal imaginary part $\delta\rightarrow0^{+}$ removes
the singularities from $\sidescripts{\epsilon_{\vartheta_{\mathrm{a}}}}{}{\bk{\varthetar_{\mathrm{a}}+\ii\delta|\mathcal{O}|\thetal_{\mathrm{a}}}}{}{\epsilon_{\theta_{\mathrm{a}}}}$.
An equivalent expression for $\sidescripts{\epsilon_{\vartheta}}{}{\bk{\varthetar|\mathcal{O}|\thetal}}{}{\epsilon_{\theta}}$
is
\begin{align}
\sidescripts{\epsilon_{\vartheta}}{}{\bk{\varthetar|\mathcal{O}|\thetal}}{}{\epsilon_{\theta}}= & \sum_{\substack{\theta_{\mathrm{a}}\cup\theta_{\mathrm{b}}=\theta\\
\vartheta_{\mathrm{a}}\cup\vartheta_{\mathrm{b}}=\vartheta
}
}S\left(\varthetar\lvert\varthetar_{\mathrm{b}}\right)_{\epsilon_{\vartheta}^{\p}}^{\epsilon_{\vartheta}}S\left(\thetal_{\mathrm{b}}\lvert\thetal\right)_{\epsilon_{\theta}}^{\epsilon_{\theta}^{\p}}\nonumber \\
 & \hspace{-2cm}\times\delta_{\epsilon_{\theta_{\mathrm{b}}}}^{\epsilon_{\vartheta_{\mathrm{b}}}}\left(\varthetar_{\mathrm{b}}\lvert\thetal_{\mathrm{b}}\right)\times\sidescripts{\epsilon_{\vartheta_{\mathrm{a}}}}{}{\bk{\varthetar_{\mathrm{a}}-\ii\delta|\mathcal{O}|\thetal_{\mathrm{a}}}}{}{\epsilon_{\theta_{\mathrm{a}}}}.\label{eq:cross_rels_2}
\end{align}
The equivalence of Eqs. (\ref{eq:cross_rels_1}) and (\ref{eq:cross_rels_2})
is guaranteed by the axioms (\ref{eq:FF_exc}) and (\ref{eq:kin_pole}).
It is also possible to use a mixed version of the two forms, where
some rapidities are analytically continued with $+\ii\delta$ and
the others with $-\ii\delta$. Choosing a specific partition $\vartheta=\vartheta_{\mathrm{A}}\cup\vartheta_{\mathrm{B}}$,
we may write
\begin{align}
\sidescripts{\epsilon_{\vartheta}}{}{\bk{\varthetar|\mathcal{O}|\thetal}}{}{\epsilon_{\theta}}= & \sum_{\substack{\theta_{\mathrm{a}}\cup\theta_{\mathrm{b}}\cup\theta_{\mathrm{c}}=\theta}
}\sum_{\substack{\vartheta_{\mathrm{Aa}}\cup\vartheta_{\mathrm{Ab}}=\vartheta_{\mathrm{A}}\\
\vartheta_{\mathrm{Ba}}\cup\vartheta_{\mathrm{Bb}}=\vartheta_{\mathrm{B}}
}
}\nonumber \\
 & \hspace{-2cm}S\left(\varthetar\lvert\varthetar_{\mathrm{Ab}},\varthetar_{\mathrm{Aa}},\varthetar_{\mathrm{Ba}},\varthetar_{\mathrm{Bb}}\right)_{\epsilon_{\vartheta}^{\p}}^{\epsilon_{\vartheta}}S\left(\thetal_{\mathrm{c}},\thetal_{\mathrm{b}},\thetal_{\mathrm{a}}\lvert\thetal\right)_{\epsilon_{\theta}}^{\epsilon_{\theta}^{\p}}\nonumber \\
 & \hspace{-2cm}\times\delta_{\epsilon_{\theta_{\mathrm{a}}}}^{\epsilon_{\vartheta_{\mathrm{Ab}}}}\left(\varthetar_{\mathrm{Ab}}\lvert\thetal_{\mathrm{a}}\right)\delta_{\epsilon_{\theta_{\mathrm{c}}}}^{\epsilon_{\vartheta_{\mathrm{Bb}}}}\left(\varthetar_{\mathrm{Bb}}\lvert\thetal_{\mathrm{c}}\right)\nonumber \\
 & \hspace{-2cm}\times\sidescripts{\epsilon_{\vartheta_{\mathrm{Aa}}}\epsilon_{\vartheta_{\mathrm{Ba}}}}{}{\bk{\varthetar_{\mathrm{Aa}}+\ii\delta,\varthetar_{\mathrm{Ba}}-\ii\delta|\mathcal{O}|\thetal_{\mathrm{b}}}}{}{\epsilon_{\theta_{\mathrm{b}}}},\label{eq:cross_rels_3}
\end{align}
with the graphical representation\begin{align}&\begin{tikzpicture}[baseline={([yshift=-.5ex]current bounding box.center)},vertex/.style={anchor=base, 				circle,fill=black!25,minimum size=18pt,inner sep=2pt}] 			\pgfmathsetmacro{\R}{0.5} 			\pgfmathsetmacro{\L}{1.2} 			\pgfmathsetmacro{\X}{40} 			\pgfmathsetmacro{\Y}{5} 			\pgfmathsetmacro{\Z}{30}			 			\draw [black, line width=0.8pt] plot [smooth, tension=0.8, very thick] coordinates { ({-cos(\X)*\R},{-sin(\X)*\R}) ({-\L},{-sin(\X)*\R}) }; 			\draw [black, line width=0.8pt] plot [smooth, tension=0.8, very thick] coordinates { ({-cos(\Y)*\R},{-sin(\Y)*\R}) ({-\L},{-sin(\Y)*\R}) }; 			\draw [black, line width=0.8pt] plot [smooth, tension=0.8, very thick] coordinates { ({-cos(\Y)*\R},{sin(\Y)*\R}) ({-\L},{sin(\Y)*\R}) }; 			\draw [black, line width=0.8pt] plot [smooth, tension=0.8, very thick] coordinates { ({-cos(\X)*\R},{sin(\X)*\R}) ({-\L},{sin(\X)*\R}) };			 			\filldraw[black] (-\R-0.25*\L,-0.45*\R) circle (0.5pt); 			\filldraw[black] (-\R-0.25*\L,-0.35*\R) circle (0.5pt); 			\filldraw[black] (-\R-0.25*\L,-0.25*\R) circle (0.5pt); 			\filldraw[black] (-\R-0.25*\L,0.45*\R) circle (0.5pt); 			\filldraw[black] (-\R-0.25*\L,0.35*\R) circle (0.5pt); 			\filldraw[black] (-\R-0.25*\L,0.25*\R) circle (0.5pt);			 			\begin{scope}[xscale=-1] 			\filldraw[black] (-\R-0.25*\L,-0.15*\R) circle (0.5pt); 			\filldraw[black] (-\R-0.25*\L,0*\R) circle (0.5pt); 			\filldraw[black] (-\R-0.25*\L,0.15*\R) circle (0.5pt); 			\end{scope}			 			\draw [black, line width=0.8pt] plot [smooth, tension=0.8, very thick] coordinates { ({cos(\Z)*\R},{-sin(\Z)*\R}) ({\L},{-sin(\Z)*\R}) }; 			\draw [black, line width=0.8pt] plot [smooth, tension=0.8, very thick] coordinates { ({cos(\Z)*\R},{sin(\Z)*\R}) ({\L},{sin(\Z)*\R}) }; 			\node[anchor=east] at (-\L, 0.375*\R) {$\vartheta_{\mathrm{A}}$}; 			\node[anchor=east] at (-\L, -0.375*\R) {$\vartheta_{\mathrm{B}}$}; 			\node[anchor=west] at (\L, 0) {$\theta$};			 			\filldraw[color=black, fill=teal!20, very thick](0,0) circle (\R) node{$\mathcal{O}$}; 		\end{tikzpicture} \nonumber \\ 	&\hspace{1cm}= \sum \left\{ 	\begin{tikzpicture}[baseline={([yshift=-.5ex]current bounding box.center)},vertex/.style={anchor=base, 			circle,fill=black!25,minimum size=18pt,inner sep=2pt}] 		\pgfmathsetmacro{\R}{0.5} 		\pgfmathsetmacro{\L}{1.2} 		\pgfmathsetmacro{\H}{0.5*\R} 		\pgfmathsetmacro{\X}{40} 		\pgfmathsetmacro{\Y}{5} 		\pgfmathsetmacro{\Z}{20} 		\pgfmathsetmacro{\W}{28}		 		\draw [black, line width=0.8pt] plot [smooth, tension=0.8, very thick] coordinates { ({-\L},{-sin(\X)*\R}) (0, -1.2*\R) ({\L},{-sin(\X)*\R}) }; 		\draw [black, line width=0.8pt] plot [smooth, tension=0.8, very thick] coordinates { ({-\L},{-sin(\X)*\R - \H}) (0, -1.2*\R - \H) ({\L},{-sin(\X)*\R - \H}) };		 		\draw [black, line width=0.8pt] plot [smooth, tension=0.8, very thick] coordinates { ({-\L},{sin(\X)*\R}) (0, 1.2*\R) ({\L},{sin(\X)*\R}) }; 		\draw [black, line width=0.8pt] plot [smooth, tension=0.8, very thick] coordinates { ({-\L},{sin(\X)*\R + \H}) (0, 1.2*\R + \H) ({\L},{sin(\X)*\R + \H}) };		 		\draw [black, line width=0.8pt] plot [smooth, tension=0.8, very thick] coordinates { ({-cos(\W)*\R},{-sin(\W)*\R}) ({-\L},{-sin(\W)*\R}) }; 		\draw [black, line width=0.8pt] plot [smooth, tension=0.8, very thick] coordinates { ({-cos(\Y)*\R},{-sin(\Y)*\R}) ({-\L},{-sin(\Y)*\R}) };		 		\draw [black, line width=0.8pt] plot [smooth, tension=0.8, very thick] coordinates { ({-cos(\Y)*\R},{sin(\Y)*\R}) ({-\L},{sin(\Y)*\R}) }; 		\draw [black, line width=0.8pt] plot [smooth, tension=0.8, very thick] coordinates { ({-cos(\W)*\R},{sin(\W)*\R}) ({-\L},{sin(\W)*\R}) };		 		\filldraw[black] (-\R-0.25*\L,-0.375*\R) circle (0.5pt); 		\filldraw[black] (-\R-0.25*\L,-0.275*\R) circle (0.5pt); 		\filldraw[black] (-\R-0.25*\L,-0.175*\R) circle (0.5pt); 		\filldraw[black] (-\R-0.25*\L,0.375*\R) circle (0.5pt); 		\filldraw[black] (-\R-0.25*\L,0.275*\R) circle (0.5pt); 		\filldraw[black] (-\R-0.25*\L,0.175*\R) circle (0.5pt);		 		\begin{scope}[xscale=-1] 			\filldraw[black] (-\R-0.25*\L,-0.15*\R) circle (0.5pt); 			\filldraw[black] (-\R-0.25*\L,0*\R) circle (0.5pt); 			\filldraw[black] (-\R-0.25*\L,0.15*\R) circle (0.5pt); 		\end{scope}	 		\filldraw[black] (0,1.2*\R + 0.3*\H) circle (0.5pt); 		\filldraw[black] (0,1.2*\R + 0.5*\H) circle (0.5pt); 		\filldraw[black] (0,1.2*\R + 0.7*\H) circle (0.5pt); 		\begin{scope}[yscale=-1] 			\filldraw[black] (0,1.2*\R + 0.3*\H) circle (0.5pt); 			\filldraw[black] (0,1.2*\R + 0.5*\H) circle (0.5pt); 			\filldraw[black] (0,1.2*\R + 0.7*\H) circle (0.5pt); 		\end{scope}		 		\draw [black, line width=0.8pt] plot [smooth, tension=0.8, very thick] coordinates { ({cos(\Z)*\R},{-sin(\Z)*\R}) ({\L},{-sin(\Z)*\R}) }; 		\draw [black, line width=0.8pt] plot [smooth, tension=0.8, very thick] coordinates { ({cos(\Z)*\R},{sin(\Z)*\R}) ({\L},{sin(\Z)*\R}) }; 		\node[anchor=east] at (-\L, 0.9*\R) {$\vartheta_{\mathrm{Ab}}$}; 		\node[anchor=east] at (-\L, 0.275*\R) {$\vartheta_{\mathrm{Aa}} + \mathrm{i}\delta$}; 		\node[anchor=east] at (-\L, -0.275*\R) {$\vartheta_{\mathrm{Ba}} - \mathrm{i}\delta$}; 		\node[anchor=east] at (-\L, -0.85*\R) {$\vartheta_{\mathrm{Bb}}$};		 		\node[anchor=west] at (\L, 0.85*\R) {$\theta_{\mathrm{a}}$}; 		\node[anchor=west] at (\L, 0) {$\theta_{\mathrm{b}}$}; 		\node[anchor=west] at (\L, -0.9*\R) {$\theta_{\mathrm{c}}$};		 		\filldraw[color=black, fill=teal!20, very thick](0,0) circle (\R) node{$\mathcal{O}$}; 	\end{tikzpicture} \right\},\end{align}where
the sum runs over all partitions and permutations of $\vartheta_{\mathrm{A,B}},\theta$.

The crossing relations are particularly simple for matrix elements
of the form $\sidescripts{\epsilon_{\theta}}{}{\bk{\thetar|\mathcal{O}|0}}{}{}$:
\begin{equation}
\sidescripts{\epsilon_{\theta}}{}{\bk{\thetar|\mathcal{O}|0}}{}{}=\bk{0|\mathcal{O}|\thetar+\ii\pi}_{\bar{\epsilon}_{\theta}}\mathsf{C}_{\theta}=e^{i\pi s_{\mathcal{O}}}f_{\bar{\epsilon}_{\theta}}^{\mathcal{O}}\left(\thetal\right)\mathsf{C}_{\theta}.
\end{equation}
If $\mathcal{O}$ is hermitian, then
\begin{equation}
\left(f_{\epsilon_{\theta}}^{\mathcal{O}}\left(\thetar\right)\right)^{*}=\sidescripts{\epsilon_{\theta}}{}{\bk{\thetar|\mathcal{O}^{\dagger}|0}}{}{}=e^{i\pi s_{\mathcal{O}}}f_{\bar{\epsilon}_{\theta}}^{\mathcal{O}}\left(\thetal\right)\mathsf{C}_{\theta}.\label{eq:FF_conj_bra}
\end{equation}
Specifying to the hermitian current operators, $\mathcal{A}=\mathcal{R},\mathcal{L}$
(where $s_{\mathcal{A}}=1$), and inserting the identity operator
$\mathds{1}=\mathfrak{C}^{-1}\mathfrak{C}$ between each pair of creation
operators, we find 
\begin{align}
\left(f_{\epsilon_{\theta}}^{\mathcal{A}}\left(\thetar\right)\right)^{*} & =-f_{\bar{\epsilon}_{\theta}}^{\mathcal{A}}\left(\thetal\right)\mathsf{C}_{\theta}\nonumber \\
 & \hspace{-1.5cm}=-\mathsf{C}_{\theta}\bk{0|\mathfrak{C}^{-1}\mathfrak{C}\mathcal{A}\mathfrak{C}^{-1}\mathfrak{C}Z_{\bar{\epsilon}_{1}}^{\dagger}\left(\theta_{1}\right)\mathfrak{C}^{-1}\ldots\mathfrak{C}Z_{\bar{\epsilon}_{n}}^{\dagger}\left(\theta_{n}\right)\mathfrak{C}^{-1}\mathfrak{C}|0}\nonumber \\
 & \hspace{-1.5cm}=\mathsf{C}_{\theta}^{2}\bk{0|\mathcal{A}Z_{\epsilon_{1}}^{\dagger}\left(\theta_{1}\right)\ldots Z_{\epsilon_{n}}^{\dagger}\left(\theta_{n}\right)|0}\nonumber \\
 & \hspace{-1.5cm}=f_{\epsilon_{\theta}}^{\mathcal{A}}\left(\thetal\right),\label{eq:FF_conj}
\end{align}
where we used Eqs. (\ref{eq:charge_conj}), (\ref{eq:charge_conj_mat}),
$\mathsf{C}_{\theta}^{2}=1$, and $\mathfrak{C}\mathcal{A}\mathfrak{C}^{-1}=-\mathcal{A}$.
We thus have
\begin{equation}
\sidescripts{\epsilon_{\theta}}{}{\bk{\thetar|\mathcal{A}|0}}{}{}=\left(f_{\epsilon_{\theta}}^{\mathcal{A}}\left(\thetar\right)\right)^{*}.\label{eq:FF_cross_conj}
\end{equation}

\subsection{Form factors of the current operators in the massless sine-Gordon
model}

The properties above hold for any integrable quantum field theory.
We now focus on form factors of the current operators in the massless
sine-Gordon model. These are defined by $\mathcal{R}=\mathcal{J}_{0}+\mathcal{J}_{1},\mathcal{L}=-\mathcal{J}_{0}+\mathcal{J}_{1}$,
with $\mathcal{J}_{\mu}=-\epsilon_{\mu\nu}\partial^{\nu}\phi$. The
form factors of $\mathcal{J}_{\mu}$ were derived by Smirnov \citep{smirnov_form_1992},
and have the general structure
\begin{equation}
f_{\epsilon_{\theta}}^{\mu}\left(\thetar\right)=\frac{M_{+}}{2}\sum_{k=1}^{n}\left(e^{\theta_{k}}-\left(-1\right)^{\mu}e^{-\theta_{k}}\right)g_{\epsilon_{\theta}}\left(\thetar\right),
\end{equation}
where $g_{\epsilon_{\theta}}\left(\thetar\right)$ is a function which
depends only on the differences $\theta_{j}-\theta_{k}$. The form
factors of $\mathcal{R},\mathcal{L}$ are then
\begin{align}
f_{\epsilon_{\theta}}^{\mathcal{A}}\left(\thetar\right)= & M_{+}\sum_{k=1}^{n}e^{\varsigma_{\mathcal{A}}\theta_{k}}g_{\epsilon_{\theta}}\left(\thetar\right),
\end{align}
with $\varsigma_{\mathcal{R}}=-\varsigma_{\mathcal{L}}=1$. The function
$g_{\epsilon_{\theta}}\left(\thetar\right)$ vanishes when one of
the rapidity differences approaches infinity, $\lim_{\left|\theta_{j}-\theta_{k}\right|\rightarrow\infty}g_{\epsilon_{\theta}}\left(\thetar\right)=0$.
Thus, setting $\theta_{k}=\pm\left(A+\lambda_{k}\right)$ with $A\rightarrow\infty$
and $M_{+}e^{A}/2\rightarrow1$ in the massless limit, we see that
$f_{\epsilon_{\theta}}^{\mathcal{R}}\left(\thetar\right)$ and $f_{\epsilon_{\theta}}^{\mathcal{L}}\left(\thetar\right)$
are non-zero only if all excitations are right or left movers, respectively.
This justifies keeping only the first and last rows of Eq. (\ref{eq:complete_set_2}),
which correspond to $f_{\epsilon_{\lambda}}^{\mathcal{R}}\left(\lambr\right)=\bk{0|\mathcal{R}|\lambr}_{\epsilon_{\lambda}}^{\rmn}$
and $f_{\epsilon_{\lambda}^{\p}}^{\mathcal{L}}\left(\lambr\right)=\bk{0|\mathcal{R}|\lambr}_{\epsilon_{\lambda}^{\p}}^{\lmn}$.
The form factors of $\mathcal{R}$ and $\mathcal{L}$ are related
by complex conjugation and charge conjugation:
\begin{equation}
f_{\epsilon_{\lambda}}^{\mathcal{L}}\left(\lambr\right)=\left(f_{\bar{\epsilon}_{\lambda}}^{\mathcal{R}}\left(\lambr\right)\right)^{*}=f_{\bar{\epsilon}_{\lambda}}^{\mathcal{R}}\left(\lambl\right)=-f_{\epsilon_{\lambda}}^{\mathcal{R}}\left(\thetal\right)\mathsf{C}_{\lambda}.\label{eq:FF_lr}
\end{equation}
If the imaginary part of a rapidity is not zero, then one has to take
the complex conjugate of the rapidity as well: 
\begin{equation}
f_{\epsilon_{\lambda}}^{\mathcal{L}}\left(\ldots,\lambda+\ii\pi,\ldots\right)\rightarrow f_{\epsilon_{\lambda}}^{\mathcal{R}}\left(\ldots,\lambda-\ii\pi,\ldots\right).
\end{equation}

A complete list of the form factors in the massive sine-Gordon model
may be found in Ref. \citep{smirnov_form_1992}. Here we summarize
the form factors of the right current operator $\mathcal{R}$ that
are used in this work, taking the massless limit of the sine-Gordon
model. First, the form factor of two solitons is given by
\begin{equation}
f_{+-}^{\mathcal{R}}\left(\lambda_{1},\lambda_{2}\right)=\frac{4\pi d\xi e^{\lambda_{1}/2}e^{\lambda_{2}/2}\zeta\left(\lambda_{1}-\lambda_{2}\right)}{\sqrt{2z}\cosh\left(\frac{1-z}{2z}\left(\lambda_{1}-\lambda_{2}+\ii\pi\right)\right)},
\end{equation}
where $d=1/\left(2c\xi\right)$, $\zeta\left(\lambda\right)=c\sinh\left(\frac{\lambda}{2}\right)e^{I\left(\lambda\right)}$,
and\begin{widetext}
\begin{align}
e^{I\left(\lambda\right)}= & \exp\left\{ \int_{0}^{\infty}\frac{\id x}{x}e^{-2N\pi x}\left(1+N-Ne^{-2\pi x}\right)\frac{\sin^{2}\left(\left(\lambda+\ii\pi\right)\frac{x}{2}\right)\sinh\left(\left(\pi-\xi\right)\frac{x}{2}\right)}{\sinh\left(\frac{\xi x}{2}\right)\sinh\left(\pi x\right)\cosh\left(\frac{\pi x}{2}\right)}\right\} \nonumber \\
 & \times\prod_{k=1}^{N}\left[\frac{\Gamma\left(1+\frac{\pi}{\xi}\left(2k+1-\frac{\ii\lambda}{\pi}\right)\right)\Gamma\left(\frac{\pi}{\xi}\left(2k+1-\frac{\ii\lambda}{\pi}\right)\right)\Gamma\left(\frac{\pi}{\xi}\left(2k-1+\frac{\ii\lambda}{\pi}\right)\right)\Gamma\left(1+\frac{\pi}{\xi}\left(2k-1+\frac{\ii\lambda}{\pi}\right)\right)}{\Gamma\left(1+\frac{\pi}{\xi}\left(2k-\frac{\ii\lambda}{\pi}\right)\right)\Gamma\left(\frac{\pi}{\xi}\left(2k+2-\frac{\ii\lambda}{\pi}\right)\right)\Gamma\left(\frac{\pi}{\xi}\left(2k+\frac{\ii\lambda}{\pi}\right)\right)\Gamma\left(1+\frac{\pi}{\xi}\left(2k-2+\frac{\ii\lambda}{\pi}\right)\right)}\right.\nonumber \\
 & \times\left.\left(\frac{\Gamma\left(\frac{\pi}{\xi}\left(2k+1\right)\right)\Gamma\left(1+\frac{\pi}{\xi}\left(2k-1\right)\right)}{\Gamma\left(\frac{2k\pi}{\xi}\right)\Gamma\left(1+\frac{2k\pi}{\xi}\right)}\right)^{2}\right]^{k},\\
c= & \left(4-\frac{4}{z}\right)^{1/4}\exp\left\{ \frac{1}{4}\int_{0}^{\infty}\frac{\id x}{x}\frac{\sinh\left(\frac{\pi x}{2}\right)\sinh\left(\left(\frac{1}{z}-2\right)\frac{\xi x}{2}\right)}{\sinh\left(\frac{\xi x}{2}\right)\cosh^{2}\left(\frac{\pi x}{2}\right)}\right\} .
\end{align}
Note that the expression for $e^{I\left(\lambda\right)}$, which is
independent of the parameter $N$, is valid as long as the integral
converges, which depends on the imaginary part of $\lambda$ (the
integral expression for $e^{I\left(\lambda\right)}$ is derived from
an infinite product of Gamma functions \citep{weisz_exact_1977}).
The rate of convergence may be improved by increasing $N$. We also
have $f_{-+}^{\mathcal{R}}\left(\lambda_{1},\lambda_{2}\right)=-f_{+-}^{\mathcal{R}}\left(\lambda_{1},\lambda_{2}\right)$.
If $z<1/2$, $f_{+-}^{\mathcal{R}}\left(\lambda_{1},\lambda_{2}\right)$
has poles in the regime $0\le\mathrm{Im}\left\{ \lambda_{2}-\lambda_{1}\right\} \le\pi$,
corresponding to the breathers in the theory; using the bootstrap
principle, we obtain the form factors for single breathers,
\begin{equation}
f_{m}^{\mathcal{R}}\left(\lambda\right)=\frac{4\xi\left(-1\right)^{\left(m-1\right)/2}\sin\left(\frac{\theta^{\left(m\right)}}{2}\right)e^{I\left(-\ii\theta^{\left(m\right)}\right)}e^{\lambda}}{\sqrt{2z\frac{\xi}{\pi}\sin\left(\frac{\pi^{2}}{\xi}\right)S_{0}\left(\ii\theta^{\left(m\right)}\right)}},
\end{equation}
where $m$ is odd ($f_{m}^{\mathcal{R}}=0$ for even $m$), and $\theta^{\left(m\right)}=\pi-\xi m$.
If $p=1/z$ is an integer, $S_{0}$ has a pole at $\theta=\ii\theta^{\left(m\right)}$,
and then the argument of the square root has to be evaluated in the
limit $\xi\rightarrow\pi/\left(p-1\right)$,
\begin{equation}
\frac{\xi}{\pi}\sin\left(\frac{\pi^{2}}{\xi}\right)S_{0}\left(\ii\theta^{\left(m\right)}\right)\rightarrow\Res{\theta=\ii\theta^{\left(m\right)}}S_{0}\left(\theta\right)=2\cot\left(\frac{\xi m}{2}\right)\prod_{j=1}^{m-1}\cot^{2}\left(\frac{\xi j}{2}\right).
\end{equation}
In fact, one may obtain $f_{1}^{\mathcal{R}}$, and also $f_{111}^{\mathcal{R}},f_{11111}^{\mathcal{R}},\ldots$
from the correspondence between the sole excitation of the sinh-Gordon
model and the $m=1$ breather of the sine-Gordon model. Here we only
need $f_{111}^{\mathcal{R}}$,
\begin{equation}
f_{111}^{\mathcal{R}}\left(\lambda_{1},\lambda_{2},\lambda_{3}\right)=\frac{8\xi}{\sqrt{2z}}\cos^{2}\left(\frac{\xi}{2}\right)\left[\sqrt{2\sin\left(\frac{\xi}{2}\right)}\exp\left\{ -\int_{0}^{\xi}\frac{x\id x}{2\pi\sin\left(x\right)}\right\} \right]^{3}\left(e^{\lambda_{1}}+e^{\lambda_{2}}+e^{\lambda_{3}}\right)e^{\lambda_{1}}e^{\lambda_{2}}e^{\lambda_{3}}\prod_{i<j}\frac{F\left(\lambda_{i}-\lambda_{j}\right)}{e^{\lambda_{i}}+e^{\lambda_{j}}},
\end{equation}
where
\begin{align}
F\left(\lambda\right)= & \exp\left\{ 4\int_{0}^{\infty}\frac{\id x}{x}\frac{\sinh\left(\pi x\right)\sinh\left(\xi x\right)\sinh\left(\left(\pi+\xi\right)x\right)}{\sinh^{2}\left(2\pi x\right)}\right\} \prod_{k=1}^{N}\left[\frac{\left(1+\frac{\left(\frac{\ii}{2}+\frac{\lambda}{2\pi}\right)^{2}}{\left(k-\frac{1}{2}\right)^{2}}\right)\left(1+\frac{\left(\frac{\ii}{2}+\frac{\lambda}{2\pi}\right)^{2}}{\left(k+\frac{1}{2}+\frac{\xi}{2\pi}\right)^{2}}\right)\left(1+\frac{\left(\frac{\ii}{2}+\frac{\lambda}{2\pi}\right)^{2}}{\left(k-\frac{\xi}{2\pi}\right)^{2}}\right)}{\left(1+\frac{\left(\frac{\ii}{2}+\frac{\lambda}{2\pi}\right)^{2}}{\left(k+\frac{1}{2}\right)^{2}}\right)\left(1+\frac{\left(\frac{\ii}{2}+\frac{\lambda}{2\pi}\right)^{2}}{\left(k-\frac{1}{2}-\frac{\xi}{2\pi}\right)^{2}}\right)\left(1+\frac{\left(\frac{\ii}{2}+\frac{\lambda}{2\pi}\right)^{2}}{\left(k+\frac{\xi}{2\pi}\right)^{2}}\right)}\right]^{k}\nonumber \\
 & \times\exp\left\{ \int_{0}^{\infty}\frac{\id x}{x}\frac{\sinh\left(\pi x\right)\sinh\left(\xi x\right)\sinh\left(\left(\pi+\xi\right)x\right)}{\sinh^{2}\left(2\pi x\right)}e^{-4N\pi x}\sin\left(\left(\lambda+\ii\pi\right)\frac{\pi x}{2}\right)\right\} .
\end{align}
Again, $F\left(\lambda\right)$ does not depend on $N$, which is
a useful parameter to increase the rate of convergence. The pole of
$f_{111}^{\mathcal{R}}$ at $\lambda_{3}-\lambda_{2}=\ii\xi$ yields
the form factor $f_{12}^{\mathcal{R}}$,
\begin{align}
f_{12}^{\mathcal{R}}\left(\lambda_{1},\lambda_{2}\right)= & \frac{2\ii\xi\cos\left(\frac{\xi}{2}\right)\sqrt{\tan\left(\xi\right)}}{\sqrt{z}}\left[\sqrt{2\sin\left(\frac{\xi}{2}\right)}\exp\left\{ -\int_{0}^{\xi}\frac{x\id x}{2\pi\sin\left(x\right)}\right\} \right]^{3}\frac{\left(e^{\lambda_{1}}+2\cos\left(\frac{\xi}{2}\right)e^{\lambda_{2}}\right)e^{\lambda_{1}}e^{\lambda_{2}}}{e^{2\lambda_{1}}+e^{2\lambda_{2}}+2\cos\left(\frac{\xi}{2}\right)e^{\lambda_{1}}e^{\lambda_{2}}}\nonumber \\
 & \times\frac{F\left(\lambda_{2}-\lambda_{1}+\frac{\ii\xi}{2}\right)F\left(\lambda_{2}-\lambda_{1}-\frac{\ii\xi}{2}\right)}{F\left(\ii\left(\pi+\xi\right)\right)}.
\end{align}
The form factors above hold for any value of the coupling constant
$z$. We also need $f_{+-+-}^{\mathcal{R}},f_{+-m}^{\mathcal{R}}$,
which are significantly more complicated for numerical evaluation.
Luckily, their expressions simplify for integer $p=1/z$,
\begin{align}
f_{+-+-}^{\mathcal{R}}\left(\lambda_{1},\lambda_{2},\lambda_{3},\lambda_{4}\right)= & \frac{8\pi^{2}d^{2}\xi}{\sqrt{2z}}\left(-1\right)^{p-1}e^{\left(\lambda_{1}+\lambda_{2}+\lambda_{3}+\lambda_{4}\right)/2}\frac{\prod_{i<j}\zeta\left(\lambda_{i}-\lambda_{j}\right)\times H\left(\lambda_{1},\lambda_{2},\lambda_{3},\lambda_{4}\right)}{\sinh\left(\left(p-1\right)\left(\lambda_{4}-\lambda_{1}\right)\right)\sinh\left(\left(p-1\right)\left(\lambda_{2}-\lambda_{3}\right)\right)}\nonumber \\
 & \hspace{-4cm}\times\left[\frac{1}{\cosh\left(\frac{p-1}{2}\left(\lambda_{2}-\lambda_{1}\right)\right)\sinh\left(\frac{p-1}{2}\left(\lambda_{4}-\lambda_{3}\right)\right)}+\frac{1}{\sinh\left(\frac{p-1}{2}\left(\lambda_{2}-\lambda_{1}\right)\right)\cosh\left(\frac{p-1}{2}\left(\lambda_{4}-\lambda_{3}\right)\right)}\right],
\end{align}
\begin{align}
f_{+-m}^{\mathcal{R}}\left(\lambda_{1},\lambda_{2},\lambda_{3}\right)= & \frac{8\pi^{2}d^{2}\xi}{\sqrt{2z\Res{\theta=\ii\theta^{\left(m\right)}}S_{0}\left(\theta\right)}}\frac{\left(-1\right)^{p}e^{\left(\lambda_{1}+\lambda_{2}\right)/2}e^{\lambda_{3}}}{\sinh\left(\left(p-1\right)\left(\lambda_{2}-\lambda_{1}\right)\right)}\nonumber \\
 & \hspace{-3cm}\times\frac{\zeta\left(\lambda_{1}-\lambda_{2}\right)\prod_{j=1}^{2}\zeta\left(\lambda_{j}-\lambda_{3}-\frac{\ii\theta^{\left(m\right)}}{2}\right)\zeta\left(\lambda_{j}-\lambda_{3}+\frac{\ii\theta^{\left(m\right)}}{2}\right)}{\sinh\left(\left(p-1\right)\left(\lambda_{3}-\lambda_{1}+\frac{\ii\theta^{\left(m\right)}}{2}\right)\right)}\frac{\zeta\left(-\ii\theta^{\left(m\right)}\right)H\left(\lambda_{1},\lambda_{2},\lambda_{3}-\frac{\ii\theta^{\left(m\right)}}{2},\lambda_{3}+\frac{\ii\theta^{\left(m\right)}}{2}\right)}{\sinh\left(\left(p-1\right)\left(\lambda_{2}-\lambda_{3}+\frac{\ii\theta^{\left(m\right)}}{2}\right)\right)},
\end{align}
where
\begin{equation}
H\left(\lambda_{1},\lambda_{2},\lambda_{3},\lambda_{4}\right)=\dfrac{1}{2\pi\ii}\int_{-2\pi\ii}^{0}\mathrm{d}\alpha e^{-\alpha}\prod_{k=1}^{4}\prod_{j=1}^{p-2}\left[2\sinh\left(\dfrac{1}{2}\left(\alpha-\lambda_{k}-\dfrac{\ii\pi j}{p-1}+\dfrac{\ii\pi}{4}\right)\right)\right].
\end{equation}
\end{widetext}The form factors become particularly simple at the
free-fermion point $z=1/2$,
\begin{equation}
f_{+-}^{\mathcal{R}}\left(\lambda_{1},\lambda_{2}\right)=2\pi\ii e^{\lambda_{1}/2}e^{\lambda_{2}/2},\label{eq:f+-_z_half}
\end{equation}
and all other form factors are zero.

\begin{figure}[b]
\begin{centering}
\includegraphics[width=1\columnwidth]{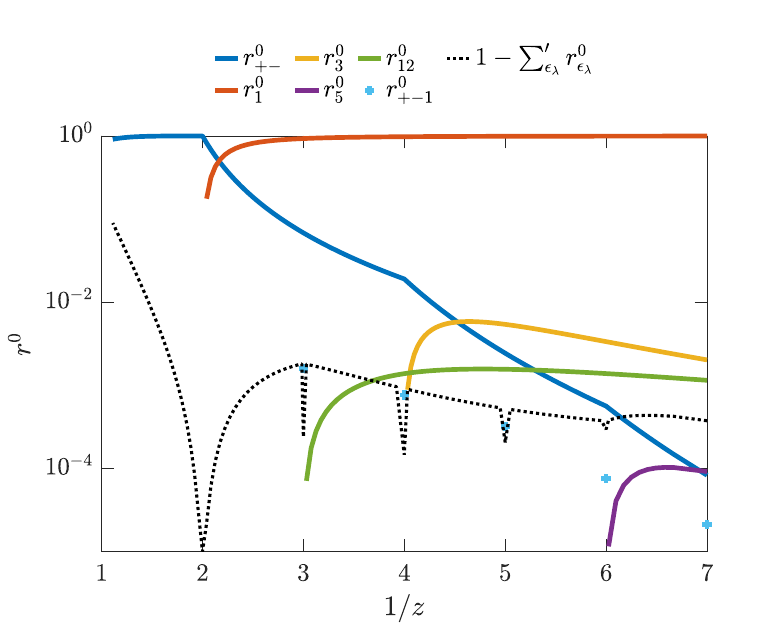}
\par\end{centering}
\caption{\label{fig:r0}The contributions of $f_{m}^{\mathcal{R}},f_{+-}^{\mathcal{R}},f_{12}^{\mathcal{R}},f_{+-1}^{\mathcal{R}}$
(the latter evaluated only for integer $p=1/z$) to the reflection
coefficient of the free theory, $r^{0}=1$, given in Eqs. (\ref{eq:r_2s})-(\ref{eq:r_2b})
with $\omega\rightarrow\infty$. Here $\sum_{\epsilon_{\lambda}}^{\protect\p}r_{\epsilon_{\lambda}}^{0}=\sum_{m=1}^{\protect\maxbreath}r_{m}^{0}+r_{+-}^{0}+r_{12}^{0}+r_{+-1}^{0}$;
the dips of $1-\sum_{\epsilon_{\lambda}}^{\protect\p}r_{\epsilon_{\lambda}}^{0}$
are at integer $p=1/z$, where $r_{+-1}^{0}$ is evaluated.}
\end{figure}

The contributions of the form factors decay rapidly with the number
of excitations. This is demonstrated in Fig. \ref{fig:r0}, which
shows the contributions of $f_{m}^{\mathcal{R}}$, $f_{+-}^{\mathcal{R}}$,
$f_{12}^{\mathcal{R}}$, $f_{+-1}^{\mathcal{R}}$ to the reflection
coefficient in the free theory, $r_{\epsilon_{\lambda}}^{0}$, defined
in Eq. (\ref{eq:r0}), as function of $1/z$. Fig. \ref{fig:r0} shows
that the full sum, $\sum_{\epsilon_{\lambda}}r_{\epsilon_{\lambda}}^{0}=1$,
is well approximated by the terms above, particularly in the attractive
regime, $z<1/2$, where the dominant contributions are those of the
breathers. Also note the singularities of $r_{+-}^{0}$ at $z=1/3,1/5,\ldots$,
where additional poles of $f_{+-}^{\mathcal{R}}\left(\lambda_{1},\lambda_{2}\right)$
enter the ``physical strip'' $0\le\mathrm{Im}\left\{ \lambda_{2}-\lambda_{1}\right\} \le\pi$
and odd breathers join the spectrum ($f_{+-}^{\mathcal{R}}\left(\lambda_{1},\lambda_{2}\right)$
does not have a pole at $\lambda_{2}-\lambda_{1}=\ii\theta^{\left(2m\right)}$,
so $f_{2m}^{\mathcal{R}}=0$).

\section{\label{app:3p_details}Details of the calculation of the inelastic
spectrum}

\subsection{\label{subsec:3p_details_cross_rels_details}Derivation of Eq. (\ref{eq:III_1})}

Consider Eq. (\ref{eq:III_raw_2}). Our goal is to write the product
of matrix elements, using the crossing relations in Eq. (\ref{eq:cross_rels_3}),
in a way that will mark the excitations according to the operators
they are connected to. First, we have
\begin{align}
\bk{0|\hat{\mathcal{R}}|\lambr_{1}}_{\epsilon_{\lambda_{1}}^{\p}}^{\rmn}\sidescripts{\epsilon_{\lambda_{1}}}{\lmn}{\bk{\lambr_{1}|\hat{\mathcal{L}}|\lambr_{2}}}{\lmn}{\epsilon_{\lambda_{2}}}\nonumber \\
 & \hspace{-4.5cm}=\sum_{\substack{\alpha_{11}\cup\lambda_{1\mathrm{b}}=\lambda_{1}\\
\lambda_{2\mathrm{a}}\cup\lambda_{2\mathrm{b}}=\lambda_{2}
}
}S\left(\lambr_{2\mathrm{a}}^{\lmn}\lvert\lambr_{2}^{\lmn}\right)_{\epsilon_{\lambda_{2}}}^{\epsilon_{\lambda_{2}}^{\p}}\bk{0|\hat{\mathcal{R}}|\lambr_{1\mathrm{b}},\ar_{11}}_{\epsilon_{\lambda_{1\mathrm{b}}}^{\p}\epsilon_{11}^{\p}}^{\rmn}\nonumber \\
 & \hspace{-4.5cm}\times\sidescripts{\epsilon_{11}}{\lmn}{\bk{\ar_{11}+\ii\delta|\hat{\mathcal{L}}|\lambr_{2\mathrm{a}}}}{\lmn}{\epsilon_{\lambda_{2\mathrm{a}}}}\delta_{\epsilon_{\lambda_{1\mathrm{b}}}}^{\epsilon_{\lambda_{2\mathrm{b}}}}\left(\lambr_{1\mathrm{b}}\lvert\lambr_{2\mathrm{b}}\right).
\end{align}
Here $\alpha_{11}$ is the group of rapidities which is not passed
on to the other operators, and connects the first operator ($\rho_{1}^{\mathcal{R}}=\rho^{\mathcal{R}}\left(\xin,-t^{\p}\right)$
in $\mathrm{III}$) to the second operator ($b_{q^{\p}}^{\mathcal{L}\dagger}$).
Note that $S\left(\lambr_{1}\lvert\ar_{11}\right)_{\epsilon_{\lambda_{1}}}^{\epsilon_{\lambda_{1}}^{\p}}$
does not appear in the expression above, since we simultaneously order
$\sidescripts{\epsilon_{\lambda_{1}}}{\lmn}{\b{\lambr_{1}}}{}{}$
and $\k{\lambr_{1}}_{\epsilon_{\lambda_{1}}^{\p}}^{\rmn}$ (recall
Eq. (\ref{eq:S_left_right})). We now multiply the above by $\sidescripts{\epsilon_{\lambda_{2}}}{\lmn}{\bk{\lambl_{2}|\hat{\mathcal{L}}|\lambl_{3}}}{\lmn}{\epsilon_{\lambda_{3}}}$,
and use $S\left(\lambr_{2\mathrm{a}}^{\lmn}\lvert\lambr_{2}^{\lmn}\right)_{\epsilon_{\lambda_{2}}}^{\epsilon_{\lambda_{2}}^{\p}}$
(which is the scattering matrix for left movers) and $\delta_{\epsilon_{\lambda_{1\mathrm{b}}}}^{\epsilon_{\lambda_{2\mathrm{b}}}}\left(\lambr_{1\mathrm{b}}\lvert\lambr_{2\mathrm{b}}\right)$
to write
\begin{align}
\bk{0|\hat{\mathcal{R}}|\lambr_{1}}_{\epsilon_{\lambda_{1}}^{\p}}^{\rmn}\sidescripts{\epsilon_{\lambda_{1}}}{\lmn}{\bk{\lambr_{1}|\hat{\mathcal{L}}|\lambr_{2}}}{\lmn}{\epsilon_{\lambda_{2}}}\sidescripts{\epsilon_{\lambda_{2}}}{\lmn}{\bk{\lambl_{2}|\hat{\mathcal{L}}|\lambl_{3}}}{\lmn}{\epsilon_{\lambda_{3}}}\nonumber \\
 & \hspace{-6cm}=\sum_{\substack{\alpha_{11}\cup\lambda_{1\mathrm{b}}=\lambda_{1}\\
\lambda_{2\mathrm{a}}\cup\lambda_{1\mathrm{b}}=\lambda_{2}
}
}\bk{0|\hat{\mathcal{R}}|\lambr_{1\mathrm{b}},\ar_{11}}_{\epsilon_{\lambda_{1\mathrm{b}}}^{\p}\epsilon_{11}^{\p}}^{\rmn}\nonumber \\
 & \hspace{-6cm}\times\sidescripts{\epsilon_{11}}{\lmn}{\bk{\ar_{11}+\ii\delta|\hat{\mathcal{L}}|\lambr_{2\mathrm{a}}}}{\lmn}{\epsilon_{\lambda_{2\mathrm{a}}}}\sidescripts{\epsilon_{\lambda_{2}}}{\lmn}{\bk{\lambl_{2\mathrm{a}},\lambl_{1\mathrm{b}}|\hat{\mathcal{L}}|\lambl_{3}}}{\lmn}{\epsilon_{\lambda_{3}}}.
\end{align}
Finally, we evaluate $\sidescripts{\epsilon_{\lambda_{2}}}{\lmn}{\bk{\lambl_{2\mathrm{a}},\lambl_{1\mathrm{b}}|\hat{\mathcal{L}}|\lambl_{3}}}{\lmn}{\epsilon_{\lambda_{3}}}$
using Eq. (\ref{eq:cross_rels_3}) and the partition $\vartheta_{\mathrm{A}}=\lambda_{2\mathrm{a}}$,
$\vartheta_{\mathrm{B}}=\lambda_{1\mathrm{b}}$:
\begin{align}
\sidescripts{\epsilon_{\lambda_{2}}}{\lmn}{\bk{\lambl_{2\mathrm{a}},\lambl_{1\mathrm{b}}|\hat{\mathcal{L}}|\lambl_{3}}}{\lmn}{\epsilon_{\lambda_{3}}}= & \sum_{\alpha_{ij},\beta,\gamma}S\left(\lambl_{3\mathrm{a}}^{\lmn},\al_{22}^{\lmn},\lambl_{3\mathrm{b}}^{\lmn}\lvert\lambl_{3}^{\lmn}\right)_{\epsilon_{\lambda_{3}}}^{\epsilon_{\lambda_{3}}^{\p}}\nonumber \\
 & \hspace{-2.8cm}\times S\left(\lambl_{2\mathrm{a}}^{\lmn},\lambl_{1\mathrm{b}}^{\lmn}\lvert\gr^{\lmn},\ar_{12}^{\lmn},\bl^{\lmn},\ar_{21}^{\lmn}\right)_{\epsilon_{\lambda_{2\mathrm{a}}}^{\p}\epsilon_{\lambda_{1\mathrm{b}}}^{\p}}^{\epsilon_{\lambda_{2\mathrm{a}}}\epsilon_{\lambda_{1\mathrm{b}}}}\nonumber \\
 & \hspace{-2.8cm}\times\delta_{\epsilon_{\lambda_{3\mathrm{a}}}}^{\epsilon_{21}}\left(\ar_{21}\lvert\lambl_{3\mathrm{a}}\right)\delta_{\epsilon_{\lambda_{3\mathrm{b}}}}^{\epsilon_{\gamma}}\left(\gr\lvert\lambl_{3\mathrm{b}}\right)\nonumber \\
 & \hspace{-2.8cm}\times\sidescripts{\epsilon_{12}\epsilon_{\beta}}{\lmn}{\bk{\ar_{12}+\ii\delta,\bl-\ii\delta|\hat{\mathcal{L}}|\al_{22}}}{\lmn}{\epsilon_{22}}.
\end{align}
Plugging everything back into Eq. (\ref{eq:III_raw_2}) and using
the delta functions and scattering matrices to reorder the other matrix
elements, we arrive at Eq. (\ref{eq:III_1}).

\subsection{Explicit expressions for the diagrams (\ref{eq:diagrams_1_I})-(\ref{eq:diagrams_1_IV})}

The diagrams (\ref{eq:diagrams_1_I})-(\ref{eq:diagrams_1_IV}) correspond
to the following terms:\begin{widetext}

\begin{align}
\mathrm{I}= & \int_{0}^{\infty}\id t^{\pp}\int_{\alpha_{ij},\beta,\gamma}\frac{\ii e^{-\ii\left(\nu_{11}+\nu_{12}+\nu_{21}+\nu_{22}\right)\xin}e^{-\ii\left(\nu_{21}+\nu_{22}+\nu_{\gamma}-\ii\eta\right)t^{\pp}}\cos\left(\omega t^{\pp}\right)\prod_{i,j=1}^{2}\hat{\mathsf{R}}_{\epsilon_{ij}}^{\epsilon_{ij}^{\p}}\left(\ar_{ij}\right)\mathsf{C}_{\beta}\mathsf{C}_{\gamma}}{\left(-\nu_{11}-\nu_{12}-\nu_{21}-\nu_{22}+2\ii\eta\right)\left(\omega^{\p}+\nu_{21}+\nu_{11}+\nu_{\beta}+\ii\eta\right)\left(\omega^{\p}-\nu_{12}-\nu_{22}+\nu_{\beta}-\ii\eta\right)}\nonumber \\
 & \hspace{-0.5cm}\times f_{\epsilon_{11}^{\p}\epsilon_{12}^{\p}\epsilon_{\gamma}}^{\mathcal{R}}\left(\al_{11}+\ii\delta,\al_{12}+\ii\delta,\gl+\ii\pi\right)f_{\epsilon_{21}\epsilon_{\beta}\bar{\epsilon}_{11}}^{\mathcal{R}}\left(\al_{21},\br,\ar_{11}\right)f_{\bar{\epsilon}_{\beta}\epsilon_{22}\bar{\epsilon}_{12}}^{\mathcal{R}}\left(\bl-\ii\pi,\al_{22}-\ii\delta,\ar_{12}+\ii\delta\right)f_{\bar{\epsilon}_{\gamma}\bar{\epsilon}_{22}^{\p}\bar{\epsilon}_{21}^{\p}}^{\mathcal{R}}\left(\gr,\ar_{22},\ar_{21}\right),\label{eq:I}\\
\mathrm{II}= & \int_{0}^{\infty}\id t^{\pp}\int_{\alpha_{ij},\beta,\gamma}\frac{\ii e^{\ii\left(\nu_{11}+\nu_{12}+\nu_{21}+\nu_{22}\right)\xin}e^{\ii\left(\nu_{21}+\nu_{22}+\nu_{\gamma}+\ii\eta\right)t^{\pp}}\cos\left(\omega t^{\pp}\right)\left(\prod_{i,j=1}^{2}\hat{\mathsf{R}}_{\epsilon_{ij}}^{\epsilon_{ij}^{\p}}\left(\ar_{ij}\right)\right)^{*}\mathsf{C}_{\beta}\mathsf{C}_{\gamma}}{\left(\nu_{11}+\nu_{12}+\nu_{21}+\nu_{22}+2\ii\eta\right)\left(\omega^{\p}-\nu_{12}-\nu_{22}+\nu_{\beta}+\ii\eta\right)\left(\omega^{\p}+\nu_{21}+\nu_{11}+\nu_{\beta}-\ii\eta\right)}\nonumber \\
 & \hspace{-0.5cm}\times f_{\epsilon_{\gamma}\epsilon_{12}^{\p}\epsilon_{11}^{\p}}^{\mathcal{R}}\left(\gr-\ii\pi+\ii\delta,\ar_{12},\ar_{11}\right)f_{\bar{\epsilon}_{11}\epsilon_{\beta}\epsilon_{21}}^{\mathcal{R}}\left(\al_{11},\bl,\ar_{21}\right)f_{\bar{\epsilon}_{\beta}\bar{\epsilon}_{12}\epsilon_{22}}^{\mathcal{R}}\left(\br-\ii\pi,\al_{12}-\ii\delta,\ar_{22}+\ii\delta\right)f_{\bar{\epsilon}_{21}^{\p}\bar{\epsilon}_{22}^{\p}\bar{\epsilon}_{\gamma}}^{\mathcal{R}}\left(\al_{21},\al_{22},\gl\right),\\
\mathrm{III}= & -\int_{0}^{\infty}\id t^{\pp}\int_{\alpha_{ij},\beta,\gamma}\frac{\ii e^{\ii\left(\nu_{11}+\nu_{12}-\nu_{21}-\nu_{22}\right)\xin}e^{-\ii\left(\nu_{21}+\nu_{22}+\nu_{\gamma}-\ii\eta\right)t^{\pp}}\cos\left(\omega t^{\pp}\right)\left(\prod_{j=1}^{2}\hat{\mathsf{R}}_{\epsilon_{1j}}^{\epsilon_{1j}^{\p}}\left(\ar_{1j}\right)\right)^{*}\prod_{j=1}^{2}\hat{\mathsf{R}}_{\epsilon_{2j}}^{\epsilon_{2j}^{\p}}\left(\ar_{2j}\right)\mathsf{C}_{\beta}\mathsf{C}_{\gamma}}{\left(\nu_{11}+\nu_{12}-\nu_{21}-\nu_{22}+2\ii\eta\right)\left(\omega^{\p}+\nu_{21}-\nu_{11}+\nu_{\beta}+\ii\eta\right)\left(\omega^{\p}+\nu_{12}-\nu_{22}+\nu_{\beta}-\ii\eta\right)}\nonumber \\
 & \hspace{-0.5cm}\times f_{\epsilon_{11}^{\p}\epsilon_{12}^{\p}\epsilon_{\gamma}}^{\mathcal{R}}\left(\al_{11},\al_{12},\gl\right)f_{\epsilon_{21}\epsilon_{\beta}\bar{\epsilon}_{11}}^{\mathcal{R}}\left(\al_{21},\br,\ar_{11}+\ii\pi-\ii\delta\right)f_{\bar{\epsilon}_{\beta}\epsilon_{22}\bar{\epsilon}_{12}}^{\mathcal{R}}\left(\bl-\ii\pi+\ii\delta,\al_{22},\ar_{12}+\ii\pi-\ii\delta\right)f_{\bar{\epsilon}_{\gamma}\bar{\epsilon}_{22}^{\p}\bar{\epsilon}_{21}^{\p}}^{\mathcal{R}}\left(\gr,\ar_{22},\ar_{21}\right),\\
\mathrm{IV}= & -\int_{0}^{\infty}\id t^{\pp}\int_{\alpha_{ij},\beta,\gamma}\frac{\ii e^{\ii\left(\nu_{21}+\nu_{22}-\nu_{11}-\nu_{12}\right)\xin}e^{\ii\left(\nu_{21}+\nu_{22}+\nu_{\gamma}+\ii\eta\right)t^{\pp}}\cos\left(\omega t^{\pp}\right)\prod_{j=1}^{2}\hat{\mathsf{R}}_{\epsilon_{1j}}^{\epsilon_{1j}^{\p}}\left(\ar_{1j}\right)\left(\prod_{j=1}^{2}\hat{\mathsf{R}}_{\epsilon_{2j}}^{\epsilon_{2j}^{\p}}\left(\ar_{2j}\right)\right)^{*}\mathsf{C}_{\beta}\mathsf{C}_{\gamma}}{\left(\nu_{21}+\nu_{22}-\nu_{11}-\nu_{12}+2\ii\eta\right)\left(\omega^{\p}+\nu_{12}-\nu_{22}+\nu_{\beta}+\ii\eta\right)\left(\omega^{\p}+\nu_{21}-\nu_{11}+\nu_{\beta}-\ii\eta\right)}\nonumber \\
 & \hspace{-0.5cm}\times f_{\epsilon_{\gamma}\epsilon_{12}^{\p}\epsilon_{11}^{\p}}^{\mathcal{R}}\left(\gr,\ar_{12},\ar_{11}\right)f_{\bar{\epsilon}_{11}\epsilon_{\beta}\epsilon_{21}}^{\mathcal{R}}\left(\al_{11}+\ii\delta,\bl+\ii\pi,\ar_{21}+\ii\pi\right)f_{\bar{\epsilon}_{\beta}\bar{\epsilon}_{12}\epsilon_{22}}^{\mathcal{R}}\left(\br-\ii\delta,\al_{12}+\ii\delta,\ar_{22}+\ii\pi\right)f_{\bar{\epsilon}_{21}^{\p}\bar{\epsilon}_{22}^{\p}\bar{\epsilon}_{\gamma}}^{\mathcal{R}}\left(\al_{21},\al_{22},\gl\right).\label{eq:IV}
\end{align}
\end{widetext}These terms follow from the crossing relations, as
shown explicitly for $\mathrm{III}$ in the Subsection above. It is
crucial to keep track of the infinitesimals $\pm\ii\delta$, which
determine the position of the annihilation poles of the form factors
with respect to the integration contour.

\subsection{The non-causal response function $G_{\omega^{\protect\p};\mathcal{RRLL}}^{\mathrm{cqq}}$}

The non-causal response function $G_{\omega^{\p};\mathcal{RRLL}}^{\mathrm{cqq}}$
gives the response of a right-mode occupation $n_{q^{\p}}^{\mathcal{RR}}$
to an injected photon moving away from the boundary. It is a crucial
sanity check to verify that it vanishes. The analogues of Eqs. (\ref{eq:I_kappa}),
(\ref{eq:III_kappa}) are, in this case,\begin{widetext}
\begin{align}
\mathrm{I}= & \int_{0}^{\infty}\id t^{\pp}\cos\left(\omega t^{\pp}\right)\int_{\alpha_{ij},\beta,\gamma}\mathsf{C}_{\beta}\mathsf{C}_{\gamma}\prod_{i=1}^{2}\delta\left(\nu_{i1}+\nu_{i2}-1\right)\int_{-\infty}^{\infty}\prod_{i=1}^{2}\id\kappa_{i}\nonumber \\
 & \times\frac{\ii e^{\ii\left(e^{\kappa_{1}}+e^{\kappa_{2}}\right)\xin}e^{-\ii\left(e^{\kappa_{2}}+\nu_{\gamma}-\ii\eta\right)t^{\pp}}}{-e^{\kappa_{1}}-e^{\kappa_{2}}+2\ii\eta}\frac{\left(\hat{\mathsf{R}}_{\epsilon_{11}}^{\epsilon_{11}^{\p}}\left(\ar_{11}+\kappa_{1}\right)\hat{\mathsf{R}}_{\epsilon_{12}}^{\epsilon_{12}^{\p}}\left(\ar_{12}+\kappa_{1}\right)\hat{\mathsf{R}}_{\epsilon_{21}^{\p}}^{\epsilon_{21}}\left(\ar_{21}+\kappa_{2}\right)\hat{\mathsf{R}}_{\epsilon_{22}^{\p}}^{\epsilon_{22}}\left(\ar_{22}+\kappa_{2}\right)\right)^{*}}{\left(\omega^{\p}+e^{\kappa_{2}}\nu_{21}+e^{\kappa_{1}}\nu_{11}+\nu_{\beta}-\ii\eta\right)\left(\omega^{\p}-e^{\kappa_{1}}\nu_{12}-e^{\kappa_{2}}\nu_{22}+\nu_{\beta}+\ii\eta\right)}\nonumber \\
 & \times f_{\bar{\epsilon}_{\gamma}\bar{\epsilon}_{12}^{\p}\bar{\epsilon}_{11}^{\p}}^{\mathcal{R}}\left(\gr-\ii\pi,\ar_{12}+\kappa_{1}-\ii\delta,\ar_{11}+\kappa_{1}-\ii\delta\right)f_{\epsilon_{11}\bar{\epsilon}_{\beta}\bar{\epsilon}_{21}}^{\mathcal{R}}\left(\al_{11}+\kappa_{1},\bl,\ar_{21}+\kappa_{2}\right)\nonumber \\
 & \times f_{\epsilon_{12}\bar{\epsilon}_{22}\epsilon_{\beta}}^{\mathcal{R}}\left(\al_{12}+\kappa_{1}-\ii\delta,\ar_{22}+\kappa_{2}+\ii\delta,\br+\ii\pi\right)f_{\epsilon_{21}^{\p}\epsilon_{22}^{\p}\epsilon_{\gamma}}^{\mathcal{R}}\left(\al_{21}+\kappa_{2},\al_{22}+\kappa_{2},\gl\right),
\end{align}
\begin{align}
\mathrm{III}= & -\int_{0}^{\infty}\id t^{\pp}\cos\left(\omega t^{\pp}\right)\int_{\alpha_{ij},\beta,\gamma}\mathsf{C}_{\beta}\mathsf{C}_{\gamma}\prod_{i=1}^{2}\delta\left(\nu_{i1}+\nu_{i2}-1\right)\int_{-\infty}^{\infty}\prod_{i=1}^{2}\id\kappa_{i}\nonumber \\
 & \times\frac{\ii e^{\ii\left(e^{\kappa_{1}}-e^{\kappa_{2}}\right)\xin}e^{-\ii\left(e^{\kappa_{2}}+\nu_{\gamma}-\ii\eta\right)t^{\pp}}}{e^{\kappa_{1}}-e^{\kappa_{2}}+2\ii\eta}\frac{\hat{\mathsf{R}}_{\epsilon_{11}}^{\epsilon_{11}^{\p}}\left(\ar_{11}+\kappa_{1}\right)\hat{\mathsf{R}}_{\epsilon_{12}}^{\epsilon_{12}^{\p}}\left(\ar_{12}+\kappa_{1}\right)\left(\hat{\mathsf{R}}_{\epsilon_{21}^{\p}}^{\epsilon_{21}}\left(\ar_{21}+\kappa_{2}\right)\hat{\mathsf{R}}_{\epsilon_{22}^{\p}}^{\epsilon_{22}}\left(\ar_{22}+\kappa_{2}\right)\right)^{*}}{\left(\omega^{\p}+e^{\kappa_{2}}\nu_{21}-e^{\kappa_{1}}\nu_{11}+\nu_{\beta}-\ii\eta\right)\left(\omega^{\p}+e^{\kappa_{1}}\nu_{12}-e^{\kappa_{2}}\nu_{22}+\nu_{\beta}+\ii\eta\right)}\nonumber \\
 & \times f_{\bar{\epsilon}_{\gamma}\bar{\epsilon}_{12}^{\p}\bar{\epsilon}_{11}^{\p}}^{\mathcal{R}}\left(\gr,\ar_{12}+\kappa_{1},\ar_{11}+\kappa_{1}\right)f_{\epsilon_{11}\bar{\epsilon}_{\beta}\bar{\epsilon}_{21}}^{\mathcal{R}}\left(\al_{11}+\kappa_{1}-\ii\pi+\ii\delta,\bl,\ar_{21}+\kappa_{2}\right)\nonumber \\
 & \times f_{\epsilon_{12}\bar{\epsilon}_{22}\epsilon_{\beta}}^{\mathcal{R}}\left(\al_{12}+\kappa_{1}-\ii\pi+\ii\delta,\ar_{22}+\kappa_{2},\br+\ii\pi-\ii\delta\right)f_{\epsilon_{21}^{\p}\epsilon_{22}^{\p}\epsilon_{\gamma}}^{\mathcal{R}}\left(\al_{21}+\kappa_{2},\al_{22}+\kappa_{2},\gl\right).
\end{align}
\end{widetext}Shifting $\kappa_{1}\rightarrow\kappa_{1}-\ii\pi+3\ii\delta$
in $\mathrm{I}$ allows to close the contour. However, now the pole
at $\kappa_{1}=\kappa_{2}-2\ii\eta$ is not enclosed by the contour;
hence, there are no singular contributions to the integral. This holds
for $\mathrm{II},\mathrm{IV}$ as well, which means that $G_{\omega^{\p};\mathcal{RRLL}}^{\mathrm{cqq}}$
does not contribute to the spectrum, as expected. Note that the correlators
comprising the response function do not necessarily vanish on their
own, and we must consider their combined contribution to show that
the response is zero.

\subsection{Identifying and evaluating the leading contributions}

Eq. (\ref{eq:spectrum_general}) provides a general expression for
the inelastic spectrum by means of a form factor expansion. The presence
of breathers in the attractive regime, $z<1/2$, leads to many possible
terms, whose contributions are expected to decay rapidly with the
number of excitations. The diagrammatic representation serves as a
convenient tool to identify the leading terms. It is important to
recognize that many of the form factors vanish due to the $\mathrm{U}\left(1\right)$
symmetry of the bulk sine-Gordon model; namely, for a mixed matrix
element of the form $\sidescripts{\epsilon_{\vartheta}}{\mn}{\bk{\varthetar|\mathcal{A}|\thetal}}{\mn}{\epsilon_{\theta}}$,
the total topological charge must vanish, $\sum_{\left\{ k\lvert\theta_{k}=\pm\right\} }\epsilon_{\theta_{k}}-\sum_{\left\{ k\lvert\vartheta_{k}=\pm\right\} }\epsilon_{\vartheta_{k}}=0$.
Furthermore, $f_{2m}^{\mathcal{R}}=0$ for all $m\ge1$, and $f_{m_{1},m_{2}}^{\mathcal{R}}=0$
for even $m_{1}+m_{2}$, allowing us to exclude many of the terms
in the expansion. The delta functions in Eq. (\ref{eq:spectrum_general})
are also useful in the exclusion of several terms --- for example,
a term with $\epsilon_{11}=\epsilon_{\gamma}=\left\{ \right\} $ and
$\epsilon_{12}=\left\{ m\right\} $ is forbidden, since the delta
function $\delta\left(\omega-\nu_{11}-\nu_{12}-\nu_{\gamma}\right)$
implies $\nu_{12}=\omega$, hence the argument of the delta function
$\delta\left(\omega^{\p}+\nu_{12}-\nu_{22}+\nu_{\beta}\right)=\delta\left(\omega^{\p}+\omega-\nu_{22}+\nu_{\beta}\right)$
can never be equal to 0.

While one looks for terms with as little number of excitations as
possible, one must be careful when mixed matrix elements of the form
$\sidescripts{\epsilon_{\vartheta}}{\mn}{\bk{\varthetar+\ii\delta|\mathcal{A}|\thetal}}{\mn}{\epsilon_{\theta}}$
are involved, due to the presence of the annihilation poles; if $\vartheta_{i}$
gets close to $\theta_{j}$ for some $i,j$, the effective order of
the form factor is reduced by 2 (see Eq. (\ref{eq:kin_pole})). That
means that, in the evaluation of mixed matrix elements using Eq. (\ref{eq:cross_rels_3}),
it is not enough to consider only the terms which eliminate the maximal
number of excitations on both sides of the matrix element (i.e. terms
in Eq. (\ref{eq:cross_rels_3}) with the largest number of delta functions),
since the other terms in the sum are just as important. To be concrete,
consider the following example:
\begin{align}
\sidescripts +{\rmn}{\bk{\vartheta_{1}|\mathcal{R}|\theta_{2},\theta_{1}}}{\rmn}{1+}\nonumber \\
 & \hspace{-2cm}=2\pi\delta\left(\vartheta_{1}-\theta_{1}\right)f_{1}^{\mathcal{R}}\left(\theta_{2}\right)+f_{-+1}^{\mathcal{R}}\left(\vartheta_{1}+\ii\pi^{-},\theta_{1},\theta_{2}\right)\nonumber \\
 & \hspace{-2cm}=2\pi\delta\left(\vartheta_{1}-\theta_{1}\right)S_{1+}^{1+}\left(\theta_{2}-\theta_{1}\right)f_{1}^{\mathcal{R}}\left(\theta_{2}\right)\nonumber \\
 & \hspace{-1.6cm}+f_{-+1}^{\mathcal{R}}\left(\vartheta_{1}+\ii\pi^{+},\theta_{1},\theta_{2}\right),\label{eq:cross_rels_example}
\end{align}
where $\pi^{\pm}=\pi\pm\delta$, and $S_{1+}^{1+}$ is the S-matrix
for a breather and a soliton, given in Eq. (\ref{eq:S_bs}). The contribution
of both terms in each of the rows to some correlation function is
of the same order, since
\begin{align}
f_{-+1}^{\mathcal{R}}\left(\vartheta_{1}+\ii\pi^{\pm},\theta_{1},\theta_{2}\right)\nonumber \\
 & \hspace{-3.5cm}=\frac{\Res{\theta_{1}=\vartheta_{1}}f_{-+1}^{\mathcal{R}}\left(\vartheta_{1}+\ii\pi,\theta_{1},\theta_{2}\right)}{\theta_{1}-\vartheta_{1}\mp\ii\delta}+\textrm{subleading term}\nonumber \\
 & \hspace{-3.5cm}=\Res{\theta_{1}=\vartheta_{1}}f_{-+1}^{\mathcal{R}}\left(\vartheta_{1}+\ii\pi,\theta_{1},\theta_{2}\right)\left(\mathcal{P}\frac{1}{\theta_{1}-\vartheta_{1}}\pm\ii\pi\delta\left(\theta_{1}-\vartheta_{1}\right)\right)\nonumber \\
 & \hspace{-3.1cm}+\textrm{subleading term},\label{eq:delta_residue}
\end{align}
and the residue of $f_{-+1}^{\mathcal{R}}$ is proportional to $f_{1}^{\mathcal{R}}$
(see Eq. (\ref{eq:kin_pole})). The delta function $\delta\left(\theta_{1}-\vartheta_{1}\right)$
in Eq. (\ref{eq:delta_residue}) can be eliminated by taking the average
of the two equivalent forms in Eq. (\ref{eq:cross_rels_example}):
\begin{align}
\sidescripts +{\rmn}{\bk{\vartheta_{1}|\mathcal{R}|\theta_{2},\theta_{1}}}{\rmn}{1+}\nonumber \\
 & \hspace{-2.5cm}=\pi\delta\left(\vartheta_{1}-\theta_{1}\right)\left(1+S_{1+}^{1+}\left(\theta_{2}-\theta_{1}\right)\right)f_{1}^{\mathcal{R}}\left(\theta_{2}\right)\nonumber \\
 & \hspace{-2.1cm}+\mathcal{P}\frac{\Res{\theta_{1}=\vartheta_{1}}f_{-+1}^{\mathcal{R}}\left(\vartheta_{1}+\ii\pi,\theta_{1},\theta_{2}\right)}{\theta_{1}-\vartheta_{1}}+\textrm{subleading term}.\label{eq:FF_avg}
\end{align}
Now we can expect the principal value term to be subleading with respect
to the first term.

The same averaging should be applied for the diagrams drawn in Subsection
\ref{subsec:spec_res}. To illustrate this, consider the leading contribution
to the $z=1/3$ spectrum, depicted in Eq. (\ref{eq:gamma_2s_1b_diagram}).
From the above, we understand that we need to consider its ``prior''
diagram, from which it originates,

\begin{align} 	\label{eq:cross_rels_diagram}	\sum_{s_1,s_2,s_2^\p}&\begin{tikzpicture}[baseline={([yshift=-0.5ex]current bounding box.center)},vertex/.style={anchor=base, circle,fill=black!25,minimum size=18pt,inner sep=2pt}] 			\pgfmathsetmacro{\R}{0.5} 			\pgfmathsetmacro{\H}{2.9*\R} 			\pgfmathsetmacro{\D}{2} 			\pgfmathsetmacro{\PHI}{30}	 			\draw [green, very thick] plot [smooth, tension=0.8] coordinates { ({\R}, 0) (\D - \R,0)}; 			\draw [black, very thick] plot [smooth, tension=0.8] coordinates { ({\D + cos(\PHI)*\R},{-sin(\PHI)*\R}) ({2*\D - cos(\PHI)*\R},{-sin(\PHI)*\R})}; 			\draw [black, very thick] plot [smooth, tension=0.8] coordinates { ({\D + cos(\PHI)*\R},{sin(\PHI)*\R}) ({2*\D - cos(\PHI)*\R},{sin(\PHI)*\R})}; 			\draw [black, very thick] plot [smooth, tension=0.8] coordinates { ({2*\D + cos(\PHI)*\R},{-sin(\PHI)*\R}) ({3*\D - cos(\PHI)*\R},{sin(\PHI)*\R})}; 			\draw [black, very thick] plot [smooth, tension=0.8] coordinates { ({2*\D + cos(\PHI)*\R},{sin(\PHI)*\R}) ({3*\D - cos(\PHI)*\R},{-sin(\PHI)*\R})};			 			\filldraw[color=black, fill=teal!20, very thick](0,0) circle (\R) node{$\rho_1^{\mathcal{R}}$}; 			\filldraw[color=black, fill=teal!20, very thick](\D,0) circle (\R) node{$b_{q^\prime}^{\mathcal{L}\dagger}$}; 			\filldraw[color=black, fill=teal!20, very thick](2*\D,0) circle (\R) node{$b_{q^\prime}^{\mathcal{L}}$}; 			\filldraw[color=black, fill=teal!20, very thick](3*\D,0) circle (\R) node{$\rho_2^{\mathcal{R}}$};			 			\node[anchor=south] at (0.5*\D, 0) {\textcolor{green}{$1$}}; 			\node[anchor=north] at (1.5*\D, {1.3*\R}) {\textcolor{black}{$s_1$}}; 			\node[anchor=north] at (1.5*\D, {-sin(\PHI)*\R}) {\textcolor{black}{$\bar{s}_1$}}; 			\node[anchor=north] at (2*\D+1.3*\R, {1.3*\R}) {\textcolor{black}{$s_2$}}; 			\node[anchor=north] at (2*\D+1.3*\R, {-sin(\PHI)*\R}) {\textcolor{black}{$\bar{s}_2$}}; 			\node[anchor=north] at (3*\D-1.3*\R, {1.5*\R}) {\textcolor{black}{$s_2^\p$}}; 			\node[anchor=north] at (3*\D-1.3*\R, {-0.8*sin(\PHI)*\R}) {\textcolor{black}{$\bar{s}_2^\p$}}; 		\end{tikzpicture} 		\nonumber \\ 		=4\sum_{s,s^\p} 		&\begin{tikzpicture}[baseline={([yshift=-0.5ex]current bounding box.center)},vertex/.style={anchor=base, circle,fill=black!25,minimum size=18pt,inner sep=2pt}] 			\pgfmathsetmacro{\R}{0.5} 			\pgfmathsetmacro{\H}{2.9*\R} 			\pgfmathsetmacro{\D}{2} 			\pgfmathsetmacro{\PHI}{30}			 			\begin{scope}[shift={(0,0)}] 				\draw [purple, very thick] plot [smooth, tension=0.8] coordinates { ({2*\D + cos(\PHI)*\R},{sin(\PHI)*\R}) ({3*\D - cos(\PHI)*\R},{-sin(\PHI)*\R})}; 				\draw [orange, very thick] plot [smooth, tension=0.8] coordinates {({\D + cos(\PHI)*\R},{-sin(\PHI)*\R}) (2*\D,-1.3*\R) (2.5*\D,0) ({3*\D - cos(\PHI)*\R},{sin(\PHI)*\R}) }; 				\draw [green, very thick] plot [smooth, tension=0.8] coordinates { ({\R}, 0) (\D - \R,0)}; 				\draw [magenta, very thick] plot [smooth, tension=0.8] coordinates { ({\D + \R*cos(\PHI)},{sin(\PHI)*\R}) ({2*\D - \R*cos(\PHI)},{sin(\PHI)*\R})};				 				\filldraw[color=black, fill=teal!20, very thick](0,0) circle (\R) node{$\rho_1^{\mathcal{R}}$}; 				\filldraw[color=black, fill=teal!20, very thick](\D,0) circle (\R) node{$b_{q^\prime}^{\mathcal{L}\dagger}$}; 				\filldraw[color=black, fill=teal!20, very thick](2*\D,0) circle (\R) node{$b_{q^\prime}^{\mathcal{L}}$}; 				\filldraw[color=black, fill=teal!20, very thick](3*\D,0) circle (\R) node{$\rho_2^{\mathcal{R}}$};				 				\node[anchor=south] at (0.5*\D, 0) {\textcolor{green}{$1$}}; 				\node[anchor=south] at (1.5*\D, {sin(\PHI)*\R}) {\textcolor{magenta}{$s$, $-\mathrm{i}\delta$}}; 				\node[anchor=north] at (2*\D+1.3*\R, {1.2*\R}) {\textcolor{purple}{$s$}}; 				\node[anchor=north] at (\D+1.3*\R, -0.7*\R) {\textcolor{orange}{$\bar{s}$}}; 				\node[anchor=north] at (3*\D-1.3*\R, {-0.6*\R}) {\textcolor{purple}{$\bar{s}^\p$}}; 				\node[anchor=north] at (3*\D-1.3*\R, {1.4*\R}) {\textcolor{orange}{$s^\p$}}; 			\end{scope} 		\end{tikzpicture} 		\nonumber \\ 		+\sum_{s_1,s_2,s_2^\p} 		&\begin{tikzpicture}[baseline={([yshift=-0.5ex]current bounding box.center)},vertex/.style={anchor=base, circle,fill=black!25,minimum size=18pt,inner sep=2pt}] 			\pgfmathsetmacro{\R}{0.5} 			\pgfmathsetmacro{\H}{2.9*\R} 			\pgfmathsetmacro{\D}{2} 			\pgfmathsetmacro{\PHI}{30}			 			\begin{scope}[shift={(0,0)}] 				\draw [green, very thick] plot [smooth, tension=0.8] coordinates { ({\R}, 0) (\D - \R,0)}; 				\draw [magenta, very thick] plot [smooth, tension=0.8] coordinates { ({\D + cos(\PHI)*\R},{-sin(\PHI)*\R}) ({2*\D - cos(\PHI)*\R},{-sin(\PHI)*\R})}; 				\draw [magenta, very thick] plot [smooth, tension=0.8] coordinates { ({\D + cos(\PHI)*\R},{sin(\PHI)*\R}) ({2*\D - cos(\PHI)*\R},{sin(\PHI)*\R})}; 				\draw [purple, very thick] plot [smooth, tension=0.8] coordinates { ({2*\D + cos(\PHI)*\R},{-sin(\PHI)*\R}) ({3*\D - cos(\PHI)*\R},{sin(\PHI)*\R})}; 				\draw [purple, very thick] plot [smooth, tension=0.8] coordinates { ({2*\D + cos(\PHI)*\R},{sin(\PHI)*\R}) ({3*\D - cos(\PHI)*\R},{-sin(\PHI)*\R})};				 				\filldraw[color=black, fill=teal!20, very thick](0,0) circle (\R) node{$\rho_1^{\mathcal{R}}$}; 				\filldraw[color=black, fill=teal!20, very thick](\D,0) circle (\R) node{$b_{q^\prime}^{\mathcal{L}\dagger}$}; 				\filldraw[color=black, fill=teal!20, very thick](2*\D,0) circle (\R) node{$b_{q^\prime}^{\mathcal{L}}$}; 				\filldraw[color=black, fill=teal!20, very thick](3*\D,0) circle (\R) node{$\rho_2^{\mathcal{R}}$};				 				\node[anchor=south] at (0.5*\D, 0) {\textcolor{green}{$1$}}; 				\node[anchor=north] at (1.5*\D, {1.5*\R}) {\textcolor{magenta}{$s_1$, $-\mathrm{i}\delta$}}; 				\node[anchor=north] at (1.5*\D, {-sin(\PHI)*\R}) {\textcolor{magenta}{$\bar{s}_1$, $-\mathrm{i}\delta$}}; 				\node[anchor=north] at (2*\D+1.3*\R, {1.3*\R}) {\textcolor{purple}{$s_2$}}; 				\node[anchor=north] at (2*\D+1.3*\R, {-sin(\PHI)*\R}) {\textcolor{purple}{$\bar{s}_2$}}; 				\node[anchor=north] at (3*\D-1.3*\R, {1.5*\R}) {\textcolor{purple}{$s_2^\p$}}; 				\node[anchor=north] at (3*\D-1.3*\R, {-0.8*sin(\PHI)*\R}) {\textcolor{purple}{$\bar{s}_2^\p$}}; 			\end{scope} 		\end{tikzpicture} 		\nonumber \\ 		=4\sum_{s,s^\p} 		&\begin{tikzpicture}[baseline={([yshift=-0.5ex]current bounding box.center)},vertex/.style={anchor=base, circle,fill=black!25,minimum size=18pt,inner sep=2pt}] 			\pgfmathsetmacro{\R}{0.5} 			\pgfmathsetmacro{\H}{2.9*\R} 			\pgfmathsetmacro{\D}{2} 			\pgfmathsetmacro{\PHI}{30}			 			\begin{scope}[shift={(0,0)}] 				\draw [purple, very thick] plot [smooth, tension=0.8] coordinates { ({2*\D + cos(\PHI)*\R},{-sin(\PHI)*\R}) ({3*\D - cos(\PHI)*\R},{sin(\PHI)*\R})}; 				\draw [orange, very thick] plot [smooth, tension=0.8] coordinates {({\D + cos(\PHI)*\R},{sin(\PHI)*\R}) (2*\D,1.3*\R) (2.5*\D,0) ({3*\D - cos(\PHI)*\R},{-sin(\PHI)*\R}) }; 				\draw [green, very thick] plot [smooth, tension=0.8] coordinates { ({\R}, 0) (\D - \R,0)}; 				\draw [magenta, very thick] plot [smooth, tension=0.8] coordinates { ({\D + \R*cos(\PHI)},{-sin(\PHI)*\R}) ({2*\D - \R*cos(\PHI)},{-sin(\PHI)*\R})};				 				\filldraw[color=black, fill=teal!20, very thick](0,0) circle (\R) node{$\rho_1^{\mathcal{R}}$}; 				\filldraw[color=black, fill=teal!20, very thick](\D,0) circle (\R) node{$b_{q^\prime}^{\mathcal{L}\dagger}$}; 				\filldraw[color=black, fill=teal!20, very thick](2*\D,0) circle (\R) node{$b_{q^\prime}^{\mathcal{L}}$}; 				\filldraw[color=black, fill=teal!20, very thick](3*\D,0) circle (\R) node{$\rho_2^{\mathcal{R}}$};				 				\node[anchor=south] at (0.5*\D, 0) {\textcolor{green}{$1$}}; 				\node[anchor=north] at (1.5*\D, {-sin(\PHI)*\R}) {\textcolor{magenta}{$s$, $+\mathrm{i}\delta$}}; 				\node[anchor=north] at (2*\D+1.3*\R, {-0.5*\R}) {\textcolor{purple}{$s$}}; 				\node[anchor=north] at (\D+1.3*\R, 1.6*\R) {\textcolor{orange}{$\bar{s}$}}; 				\node[anchor=north] at (3*\D-1.3*\R, 1.6*\R) {\textcolor{purple}{$\bar{s}^\p$}}; 				\node[anchor=north] at (3*\D-1.3*\R, {-0.3*\R}) {\textcolor{orange}{$s^\p$}}; 			\end{scope} 		\end{tikzpicture} 		\nonumber \\ 		+\sum_{s_1,s_2,s_2^\p} 		&\begin{tikzpicture}[baseline={([yshift=-0.5ex]current bounding box.center)},vertex/.style={anchor=base, circle,fill=black!25,minimum size=18pt,inner sep=2pt}] 			\pgfmathsetmacro{\R}{0.5} 			\pgfmathsetmacro{\H}{2.9*\R} 			\pgfmathsetmacro{\D}{2} 			\pgfmathsetmacro{\PHI}{30}	 			\begin{scope}[shift={(0,0)}] 				\draw [green, very thick] plot [smooth, tension=0.8] coordinates { ({\R}, 0) (\D - \R,0)}; 				\draw [magenta, very thick] plot [smooth, tension=0.8] coordinates { ({\D + cos(\PHI)*\R},{-sin(\PHI)*\R}) ({2*\D - cos(\PHI)*\R},{-sin(\PHI)*\R})}; 				\draw [magenta, very thick] plot [smooth, tension=0.8] coordinates { ({\D + cos(\PHI)*\R},{sin(\PHI)*\R}) ({2*\D - cos(\PHI)*\R},{sin(\PHI)*\R})}; 				\draw [purple, very thick] plot [smooth, tension=0.8] coordinates { ({2*\D + cos(\PHI)*\R},{-sin(\PHI)*\R}) ({3*\D - cos(\PHI)*\R},{sin(\PHI)*\R})}; 				\draw [purple, very thick] plot [smooth, tension=0.8] coordinates { ({2*\D + cos(\PHI)*\R},{sin(\PHI)*\R}) ({3*\D - cos(\PHI)*\R},{-sin(\PHI)*\R})};				 				\filldraw[color=black, fill=teal!20, very thick](0,0) circle (\R) node{$\rho_1^{\mathcal{R}}$}; 				\filldraw[color=black, fill=teal!20, very thick](\D,0) circle (\R) node{$b_{q^\prime}^{\mathcal{L}\dagger}$}; 				\filldraw[color=black, fill=teal!20, very thick](2*\D,0) circle (\R) node{$b_{q^\prime}^{\mathcal{L}}$}; 				\filldraw[color=black, fill=teal!20, very thick](3*\D,0) circle (\R) node{$\rho_2^{\mathcal{R}}$};				 				\node[anchor=south] at (0.5*\D, 0) {\textcolor{green}{$1$}}; 				\node[anchor=north] at (1.5*\D, {1.5*\R}) {\textcolor{magenta}{$\bar{s}_1$, $+\mathrm{i}\delta$}}; 				\node[anchor=north] at (1.5*\D, {-sin(\PHI)*\R}) {\textcolor{magenta}{$s_1$, $+\mathrm{i}\delta$}}; 				\node[anchor=north] at (2*\D+1.3*\R, {1.4*\R}) {\textcolor{purple}{$\bar{s}_2$}}; 				\node[anchor=north] at (2*\D+1.3*\R, {-sin(\PHI)*\R}) {\textcolor{purple}{$s_2$}}; 				\node[anchor=north] at (3*\D-1.3*\R, {1.5*\R}) {\textcolor{purple}{$\bar{s}_2^\p$}}; 				\node[anchor=north] at (3*\D-1.3*\R, {-0.8*sin(\PHI)*\R}) {\textcolor{purple}{$s_2^\p$}}; 			\end{scope}			 		\end{tikzpicture}~. 	\end{align}The
equivalence of the two forms is again a result of the consistency
of Eq. (\ref{eq:cross_rels_3}) for any choice of $\vartheta_{\mathrm{A}},\vartheta_{\mathrm{B}}$.
Now, there is a delta function term in the second diagram of each
of the two forms, ``hidden'' within the form factor associated with
$b_{q^{\p}}^{\mathcal{L}}$, whose contribution cannot be neglected.
The solution is the same as in Eq. (\ref{eq:FF_avg}) --- average
the two forms to get rid of the delta function terms. The two forms
are related by flipping the order of excitations in each bra and ket
states, which corresponds to taking the complex conjugate of the form
factors (but not of the reflection matrices). Hence, we may get rid
of the delta function by taking the real part of $\prod\mathsf{F}$
in Eq. (\ref{eq:Fprod}):\begin{align} \label{eq:gamma_2s_1b_a_details} 	\frac{1}{2} 	&\begin{tikzpicture}[baseline={([yshift=-0.5ex]current bounding box.center)},vertex/.style={anchor=base, circle,fill=black!25,minimum size=18pt,inner sep=2pt}] 		\pgfmathsetmacro{\R}{0.5} 		\pgfmathsetmacro{\H}{2.9*\R} 		\pgfmathsetmacro{\D}{2} 		\pgfmathsetmacro{\PHI}{30}			 		\begin{scope}[shift={(0,0)}] 			\draw [purple, very thick] plot [smooth, tension=0.8] coordinates { ({2*\D + cos(\PHI)*\R},{sin(\PHI)*\R}) ({3*\D - cos(\PHI)*\R},{-sin(\PHI)*\R})}; 			\draw [orange, very thick] plot [smooth, tension=0.8] coordinates {({\D + cos(\PHI)*\R},{-sin(\PHI)*\R}) (2*\D,-1.3*\R) (2.5*\D,0) ({3*\D - cos(\PHI)*\R},{sin(\PHI)*\R}) }; 			\draw [green, very thick] plot [smooth, tension=0.8] coordinates { ({\R}, 0) (\D - \R,0)}; 			\draw [magenta, very thick] plot [smooth, tension=0.8] coordinates { ({\D + \R*cos(\PHI)},{sin(\PHI)*\R}) ({2*\D - \R*cos(\PHI)},{sin(\PHI)*\R})};				 			\filldraw[color=black, fill=teal!20, very thick](0,0) circle (\R) node{$\rho_1^{\mathcal{R}}$}; 			\filldraw[color=black, fill=teal!20, very thick](\D,0) circle (\R) node{$b_{q^\prime}^{\mathcal{L}\dagger}$}; 			\filldraw[color=black, fill=teal!20, very thick](2*\D,0) circle (\R) node{$b_{q^\prime}^{\mathcal{L}}$}; 			\filldraw[color=black, fill=teal!20, very thick](3*\D,0) circle (\R) node{$\rho_2^{\mathcal{R}}$};				 			\node[anchor=south] at (0.5*\D, 0) {\textcolor{green}{$1$}}; 			\node[anchor=south] at (1.5*\D, {0.8*sin(\PHI)*\R}) {\textcolor{magenta}{$s$, $-\mathrm{i}\delta$}}; 			\node[anchor=north] at (2*\D+1.3*\R, {1.2*\R}) {\textcolor{purple}{$s$}}; 			\node[anchor=north] at (\D+1.3*\R, -0.7*\R) {\textcolor{orange}{$\bar{s}$}}; 			\node[anchor=north] at (3*\D-1.3*\R, {-0.6*\R}) {\textcolor{purple}{$\bar{s}^\p$}}; 			\node[anchor=north] at (3*\D-1.3*\R, {1.4*\R}) {\textcolor{orange}{$s^\p$}}; 		\end{scope} 	\end{tikzpicture} 	\nonumber \\ 	+\frac{1}{2}&\begin{tikzpicture}[baseline={([yshift=-0.5ex]current bounding box.center)},vertex/.style={anchor=base, circle,fill=black!25,minimum size=18pt,inner sep=2pt}] 		\pgfmathsetmacro{\R}{0.5} 		\pgfmathsetmacro{\H}{2.9*\R} 		\pgfmathsetmacro{\D}{2} 		\pgfmathsetmacro{\PHI}{30}			 		\begin{scope}[shift={(0,0)}] 			\draw [purple, very thick] plot [smooth, tension=0.8] coordinates { ({2*\D + cos(\PHI)*\R},{-sin(\PHI)*\R}) ({3*\D - cos(\PHI)*\R},{sin(\PHI)*\R})}; 			\draw [orange, very thick] plot [smooth, tension=0.8] coordinates {({\D + cos(\PHI)*\R},{sin(\PHI)*\R}) (2*\D,1.3*\R) (2.5*\D,0) ({3*\D - cos(\PHI)*\R},{-sin(\PHI)*\R}) }; 			\draw [green, very thick] plot [smooth, tension=0.8] coordinates { ({\R}, 0) (\D - \R,0)}; 			\draw [magenta, very thick] plot [smooth, tension=0.8] coordinates { ({\D + \R*cos(\PHI)},{-sin(\PHI)*\R}) ({2*\D - \R*cos(\PHI)},{-sin(\PHI)*\R})};				 			\filldraw[color=black, fill=teal!20, very thick](0,0) circle (\R) node{$\rho_1^{\mathcal{R}}$}; 			\filldraw[color=black, fill=teal!20, very thick](\D,0) circle (\R) node{$b_{q^\prime}^{\mathcal{L}\dagger}$}; 			\filldraw[color=black, fill=teal!20, very thick](2*\D,0) circle (\R) node{$b_{q^\prime}^{\mathcal{L}}$}; 			\filldraw[color=black, fill=teal!20, very thick](3*\D,0) circle (\R) node{$\rho_2^{\mathcal{R}}$};				 			\node[anchor=south] at (0.5*\D, 0) {\textcolor{green}{$1$}}; 			\node[anchor=north] at (1.5*\D, {-0.9*sin(\PHI)*\R}) {\textcolor{magenta}{$s$, $+\mathrm{i}\delta$}}; 			\node[anchor=north] at (2*\D+1.3*\R, {-0.5*\R}) {\textcolor{purple}{$s$}}; 			\node[anchor=north] at (\D+1.3*\R, 1.6*\R) {\textcolor{orange}{$\bar{s}$}}; 			\node[anchor=north] at (3*\D-1.3*\R, 1.6*\R) {\textcolor{purple}{$\bar{s}^\p$}}; 			\node[anchor=north] at (3*\D-1.3*\R, {-0.3*\R}) {\textcolor{orange}{$s^\p$}}; 		\end{scope} 	\end{tikzpicture} \nonumber \\ 	=-&\mathrm{Re}\big\{f_{1}^{\mathcal{R}}\left(\alpha_{11}\right)f_{s\bar{s}1}^{\mathcal{R}}\left(\alpha_{21},\beta,\alpha_{11}+\ii\pi\right) \nonumber \\ 	&\hspace{0.5cm}\times f_{s\bar{s}}^{\mathcal{R}}\left(\beta-\ii\pi^{-},\alpha_{22}\right)f_{s^\p\bar{s}^\p}^{\mathcal{R}}\left(\alpha_{22},\alpha_{21}\right)\big\}, \end{align}leading
to Eq. (\ref{eq:gamma_2^(1)}), and\begin{align} \label{eq:gamma_2s_1b_b_details} 	\frac{1}{2} 	&\begin{tikzpicture}[baseline={([yshift=-0.5ex]current bounding box.center)},vertex/.style={anchor=base, circle,fill=black!25,minimum size=18pt,inner sep=2pt}] 		\pgfmathsetmacro{\R}{0.5} 		\pgfmathsetmacro{\H}{2.9*\R} 		\pgfmathsetmacro{\D}{2} 		\pgfmathsetmacro{\PHI}{30}			 		\begin{scope}[shift={(0,0)}] 			\draw [green, very thick] plot [smooth, tension=0.8] coordinates { ({\R}, 0) (\D - \R,0)}; 			\draw [magenta, very thick] plot [smooth, tension=0.8] coordinates { ({\D + cos(\PHI)*\R},{-sin(\PHI)*\R}) ({2*\D - cos(\PHI)*\R},{-sin(\PHI)*\R})}; 			\draw [magenta, very thick] plot [smooth, tension=0.8] coordinates { ({\D + cos(\PHI)*\R},{sin(\PHI)*\R}) ({2*\D - cos(\PHI)*\R},{sin(\PHI)*\R})}; 			\draw [purple, very thick] plot [smooth, tension=0.8] coordinates { ({2*\D + cos(\PHI)*\R},{-sin(\PHI)*\R}) ({3*\D - cos(\PHI)*\R},{sin(\PHI)*\R})}; 			\draw [purple, very thick] plot [smooth, tension=0.8] coordinates { ({2*\D + cos(\PHI)*\R},{sin(\PHI)*\R}) ({3*\D - cos(\PHI)*\R},{-sin(\PHI)*\R})};				 			\filldraw[color=black, fill=teal!20, very thick](0,0) circle (\R) node{$\rho_1^{\mathcal{R}}$}; 			\filldraw[color=black, fill=teal!20, very thick](\D,0) circle (\R) node{$b_{q^\prime}^{\mathcal{L}\dagger}$}; 			\filldraw[color=black, fill=teal!20, very thick](2*\D,0) circle (\R) node{$b_{q^\prime}^{\mathcal{L}}$}; 			\filldraw[color=black, fill=teal!20, very thick](3*\D,0) circle (\R) node{$\rho_2^{\mathcal{R}}$};				 			\node[anchor=south] at (0.5*\D, 0) {\textcolor{green}{$1$}}; 			\node[anchor=north] at (1.5*\D, {1.5*\R}) {\textcolor{magenta}{$s_1$, $-\mathrm{i}\delta$}}; 			\node[anchor=north] at (1.5*\D, {-sin(\PHI)*\R}) {\textcolor{magenta}{$\bar{s}_1$, $-\mathrm{i}\delta$}}; 			\node[anchor=north] at (2*\D+1.3*\R, {1.3*\R}) {\textcolor{purple}{$s_2$}}; 			\node[anchor=north] at (2*\D+1.3*\R, {-sin(\PHI)*\R}) {\textcolor{purple}{$\bar{s}_2$}}; 			\node[anchor=north] at (3*\D-1.3*\R, {1.5*\R}) {\textcolor{purple}{$s_2^\p$}}; 			\node[anchor=north] at (3*\D-1.3*\R, {-0.9*sin(\PHI)*\R}) {\textcolor{purple}{$\bar{s}_2^\p$}}; 		\end{scope} 	\end{tikzpicture} 	\nonumber \\ 	+\frac{1}{2}&\begin{tikzpicture}[baseline={([yshift=-0.5ex]current bounding box.center)},vertex/.style={anchor=base, circle,fill=black!25,minimum size=18pt,inner sep=2pt}] 		\pgfmathsetmacro{\R}{0.5} 		\pgfmathsetmacro{\H}{2.9*\R} 		\pgfmathsetmacro{\D}{2} 		\pgfmathsetmacro{\PHI}{30}	 		\begin{scope}[shift={(0,0)}] 			\draw [green, very thick] plot [smooth, tension=0.8] coordinates { ({\R}, 0) (\D - \R,0)}; 			\draw [magenta, very thick] plot [smooth, tension=0.8] coordinates { ({\D + cos(\PHI)*\R},{-sin(\PHI)*\R}) ({2*\D - cos(\PHI)*\R},{-sin(\PHI)*\R})}; 			\draw [magenta, very thick] plot [smooth, tension=0.8] coordinates { ({\D + cos(\PHI)*\R},{sin(\PHI)*\R}) ({2*\D - cos(\PHI)*\R},{sin(\PHI)*\R})}; 			\draw [purple, very thick] plot [smooth, tension=0.8] coordinates { ({2*\D + cos(\PHI)*\R},{-sin(\PHI)*\R}) ({3*\D - cos(\PHI)*\R},{sin(\PHI)*\R})}; 			\draw [purple, very thick] plot [smooth, tension=0.8] coordinates { ({2*\D + cos(\PHI)*\R},{sin(\PHI)*\R}) ({3*\D - cos(\PHI)*\R},{-sin(\PHI)*\R})};				 			\filldraw[color=black, fill=teal!20, very thick](0,0) circle (\R) node{$\rho_1^{\mathcal{R}}$}; 			\filldraw[color=black, fill=teal!20, very thick](\D,0) circle (\R) node{$b_{q^\prime}^{\mathcal{L}\dagger}$}; 			\filldraw[color=black, fill=teal!20, very thick](2*\D,0) circle (\R) node{$b_{q^\prime}^{\mathcal{L}}$}; 			\filldraw[color=black, fill=teal!20, very thick](3*\D,0) circle (\R) node{$\rho_2^{\mathcal{R}}$};				 			\node[anchor=south] at (0.5*\D, 0) {\textcolor{green}{$1$}}; 			\node[anchor=north] at (1.5*\D, {1.5*\R}) {\textcolor{magenta}{$\bar{s}_1$, $+\mathrm{i}\delta$}}; 			\node[anchor=north] at (1.5*\D, {-0.8*sin(\PHI)*\R}) {\textcolor{magenta}{$s_1$, $+\mathrm{i}\delta$}}; 			\node[anchor=north] at (2*\D+1.3*\R, {1.4*\R}) {\textcolor{purple}{$\bar{s}_2$}}; 			\node[anchor=north] at (2*\D+1.3*\R, {-sin(\PHI)*\R}) {\textcolor{purple}{$s_2$}}; 			\node[anchor=north] at (3*\D-1.3*\R, {1.5*\R}) {\textcolor{purple}{$\bar{s}_2^\p$}}; 			\node[anchor=north] at (3*\D-1.3*\R, {-0.6*sin(\PHI)*\R}) {\textcolor{purple}{$s_2^\p$}}; 		\end{scope}			 	\end{tikzpicture} \nonumber \\ 	=&\mathrm{Re}\big\{f_{1}^{\mathcal{R}}\left(\alpha_{11}\right)f_{s_1\bar{s}_1 1}^{\mathcal{R}}\left(\beta_{1},\beta_{2},\alpha_{11}+\ii\pi^{-}\right) \nonumber \\ 	&\hspace{0.5cm}\times f_{s_1\bar{s}_1\bar{s}_2 s_2}^{\mathcal{R}}\left(\beta_{2}-\ii\pi^{-},\beta_{1}-i\pi^{-},\left(\alpha_{22}\right)_{2},\left(\alpha_{22}\right)_{1}\right) \nonumber \\ 	&\hspace{0.5cm}\times f_{\bar{s}_{2}^\p s_{2}^\p}^{\mathcal{R}}\left(\left(\alpha_{22}\right)_{1},\left(\alpha_{22}\right)_{2}\right)\big\}, \end{align}whose
contribution to the spectrum is
\begin{align}
\gamma_{2}^{\left(2\right)}\left(\omega^{\p}\lvert\omega\right)= & -\frac{2}{\omega^{\p}\omega}\int_{0}^{\omega}\id\Omega_{1}\int_{0}^{\omega-\omega^{\p}}\id\Omega_{2}\frac{1}{\left(2\pi\right)^{5}\prod_{i=1}^{5}e^{\lambda_{i}}}\nonumber \\
 & \hspace{-1.8cm}\times\mathrm{Re}\left\{ \left(R_{1}^{1}\left(\lambda_{1}-\log\mu_{1}\right)\right)^{*}\right.\nonumber \\
 & \hspace{-1.8cm}\left.\times\left(e^{-\frac{\ii\pi}{2z}}R_{+}^{-}\left(\lambda_{2}\right)R_{+}^{-}\left(\lambda_{3}\right)-e^{\frac{\ii\pi}{2z}}R_{+}^{+}\left(\lambda_{2}\right)R_{+}^{+}\left(\lambda_{3}\right)\right)-1\right\} \nonumber \\
 & \hspace{-1.8cm}\times\mathrm{Re}\left\{ f_{1}^{\mathcal{R}}\left(\lambda_{1}-\log\mu_{1}\right)f_{+-1}^{\mathcal{R}}\left(\lambda_{4},\lambda_{5},\lambda_{1}+\ii\pi^{-}-\log\mu_{1}\right)\right.\nonumber \\
 & \hspace{-1.8cm}\left.\times f_{+-+-}^{\mathcal{R}}\left(\lambda_{5},\lambda_{4},\lambda_{2}+\ii\pi^{-},\lambda_{3}+i\pi^{-}\right)f_{+-}^{\mathcal{R}}\left(\lambda_{3},\lambda_{2}\right)\right\} ,\label{eq:gamma_2^(2)}
\end{align}
with $e^{\lambda_{1}}=\omega$, $e^{\lambda_{2}}=\Omega_{1}$, $e^{\lambda_{3}}=\omega-\Omega_{1}$,
$e^{\lambda_{4}}=\Omega_{2}$, $e^{\lambda_{5}}=\omega-\omega^{\p}-\Omega_{2}$.
While the integration contour passes close to (but not through) the
poles of $f_{+-+-}^{\mathcal{R}}$, taking the real part of the product
of form factors means that this term may be evaluated numerically
as a principle value integral. Numerical inspection shows that this
term is negligible compared to $\gamma_{2}^{\left(1\right)}$ in Eq.
(\ref{eq:gamma_2^(1)}).

It is not always possible to get rid of all of the hidden delta function
terms. For example, consider the ``prior'' diagrams of Eq. (\ref{eq:gamma_2s_2s_diagram}):

\begin{align} \label{eq:cross_rels_diagrams_2s_2s}		\sum_{\{s\},\{s^\p\}}&\begin{tikzpicture}[baseline={([yshift=-0.5ex]current bounding box.center)},vertex/.style={anchor=base, circle,fill=black!25,minimum size=18pt,inner sep=2pt}] 			\pgfmathsetmacro{\R}{0.5} 			\pgfmathsetmacro{\H}{2.9*\R} 			\pgfmathsetmacro{\D}{2} 			\pgfmathsetmacro{\PHI}{30} 			\draw [black, very thick] plot [smooth, tension=0.8] coordinates { ({\D + cos(\PHI)*\R},{-sin(\PHI)*\R}) ({2*\D - cos(\PHI)*\R},{-sin(\PHI)*\R})}; 			\draw [black, very thick] plot [smooth, tension=0.8] coordinates { ({\D + cos(\PHI)*\R},{sin(\PHI)*\R}) ({2*\D - cos(\PHI)*\R},{sin(\PHI)*\R})}; 			\draw [black, very thick] plot [smooth, tension=0.8] coordinates { ({cos(\PHI)*\R},{-sin(\PHI)*\R}) ({\D - cos(\PHI)*\R},{sin(\PHI)*\R})}; 			\draw [black, very thick] plot [smooth, tension=0.8] coordinates { ({cos(\PHI)*\R},{sin(\PHI)*\R}) ({\D - cos(\PHI)*\R},{-sin(\PHI)*\R})}; 			\draw [black, very thick] plot [smooth, tension=0.8] coordinates { ({2*\D + cos(\PHI)*\R},{-sin(\PHI)*\R}) ({3*\D - cos(\PHI)*\R},{sin(\PHI)*\R})}; 			\draw [black, very thick] plot [smooth, tension=0.8] coordinates { ({2*\D + cos(\PHI)*\R},{sin(\PHI)*\R}) ({3*\D - cos(\PHI)*\R},{-sin(\PHI)*\R})};			 			\filldraw[color=black, fill=teal!20, very thick](0,0) circle (\R) node{$\rho_1^{\mathcal{R}}$}; 			\filldraw[color=black, fill=teal!20, very thick](\D,0) circle (\R) node{$b_{q^\prime}^{\mathcal{L}\dagger}$}; 			\filldraw[color=black, fill=teal!20, very thick](2*\D,0) circle (\R) node{$b_{q^\prime}^{\mathcal{L}}$}; 			\filldraw[color=black, fill=teal!20, very thick](3*\D,0) circle (\R) node{$\rho_2^{\mathcal{R}}$};			 			\node[anchor=north] at (1.5*\D, {1.3*\R}) {\textcolor{black}{$s_2$}}; 			\node[anchor=north] at (1.5*\D, {-sin(\PHI)*\R}) {\textcolor{black}{$\bar{s}_2$}}; 			\node[anchor=north] at (\D-1.3*\R, {1.3*\R}) {\textcolor{black}{$s_1$}}; 			\node[anchor=north] at (\D-1.3*\R, {-sin(\PHI)*\R}) {\textcolor{black}{$\bar{s}_1$}}; 			\node[anchor=north] at (2*\D+1.3*\R, {1.3*\R}) {\textcolor{black}{$s_3$}}; 			\node[anchor=north] at (2*\D+1.3*\R, {-sin(\PHI)*\R}) {\textcolor{black}{$\bar{s}_3$}}; 			\node[anchor=north] at (3*\D-1.3*\R, {1.52*\R}) {\textcolor{black}{$s_3^\p$}}; 			\node[anchor=north] at (3*\D-1.3*\R, {-0.8*sin(\PHI)*\R}) {\textcolor{black}{$\bar{s}_3^\p$}}; 			\node[anchor=north] at (1.3*\R, {1.5*\R}) {\textcolor{black}{$s_1^\p$}}; 			\node[anchor=north] at (1.3*\R, {-0.8*sin(\PHI)*\R}) {\textcolor{black}{$\bar{s}_1^\p$}}; 		\end{tikzpicture} 	\nonumber \\ 	=8\sum_{s,s_1^\p,s_2^\p}&\begin{tikzpicture}[baseline={([yshift=-0.5ex]current bounding box.center)},vertex/.style={anchor=base, circle,fill=black!25,minimum size=18pt,inner sep=2pt}] 		\pgfmathsetmacro{\R}{0.5} 		\pgfmathsetmacro{\H}{2.9*\R} 		\pgfmathsetmacro{\D}{2} 		\pgfmathsetmacro{\PHI}{30} 			\begin{scope}[shift={(0,0)}] 				\draw [purple, very thick] plot [smooth, tension=0.8] coordinates { (2*\D + \R,0) (3*\D - \R,0)}; 				\draw [orange, very thick] plot [smooth, tension=0.8] coordinates {({\D + cos(\PHI)*\R},{-sin(\PHI)*\R}) (2*\D,-1.3*\R) (2.5*\D,0) ({3*\D - cos(\PHI)*\R},{sin(\PHI)*\R}) }; 				\draw [blue, very thick] plot [smooth, tension=0.8] coordinates {({cos(\PHI)*\R},{-sin(\PHI)*\R}) (0.5*\D,0) (\D,1.3*\R)  ({2*\D - cos(\PHI)*\R},{sin(\PHI)*\R}) }; 				\draw [green, very thick] plot [smooth, tension=0.8] coordinates { ({\R}, {0}) (\D - \R,0)};				 				\filldraw[color=black, fill=teal!20, very thick](0,0) circle (\R) node{$\rho_1^{\mathcal{R}}$}; 				\filldraw[color=black, fill=teal!20, very thick](\D,0) circle (\R) node{$b_{q^\prime}^{\mathcal{L}\dagger}$}; 				\filldraw[color=black, fill=teal!20, very thick](2*\D,0) circle (\R) node{$b_{q^\prime}^{\mathcal{L}}$}; 				\filldraw[color=black, fill=teal!20, very thick](3*\D,0) circle (\R) node{$\rho_2^{\mathcal{R}}$};				 				\node[anchor=north] at (\D-1.3*\R, 0) {\textcolor{green}{$\bar{s}$}}; 				\node[anchor=south] at (2*\D-1.3*\R, 0.7*\R) {\textcolor{blue}{$s$}}; 				\node[anchor=south] at (1.3*\R, 0) {\textcolor{green}{$s_1^\p$}}; 				\node[anchor=north] at (1.3*\R, {-0.8*sin(\PHI)*\R}) {\textcolor{blue}{$\bar{s}_1^\p$}}; 				\node[anchor=south] at (2*\D+1.3*\R, 0) {\textcolor{purple}{$s$}}; 				\node[anchor=north] at (\D+1.3*\R, -0.7*\R) {\textcolor{orange}{$\bar{s}$}}; 				\node[anchor=north] at (3*\D-1.3*\R, 0) {\textcolor{purple}{$\bar{s}_2^\p$}}; 				\node[anchor=south] at (3*\D-1.3*\R, {0.7*sin(\PHI)*\R}) {\textcolor{orange}{$s_2^\p$}}; 			\end{scope} 		\end{tikzpicture} 	\nonumber \\ +4\sum_{s_1,s_2,s_1^\p,s_2^\p}&\begin{tikzpicture}[baseline={([yshift=-0.5ex]current bounding box.center)},vertex/.style={anchor=base, circle,fill=black!25,minimum size=18pt,inner sep=2pt}] \pgfmathsetmacro{\R}{0.5} \pgfmathsetmacro{\H}{2.9*\R} \pgfmathsetmacro{\D}{2} \pgfmathsetmacro{\PHI}{30} 			\begin{scope}[shift={(0,0)}] 				\draw [purple, very thick] plot [smooth, tension=0.8] coordinates { ({2*\D + cos(\PHI)*\R},{-sin(\PHI)*\R}) ({3*\D - cos(\PHI)*\R},{sin(\PHI)*\R})}; 				\draw [purple, very thick] plot [smooth, tension=0.8] coordinates { ({2*\D + cos(\PHI)*\R},{sin(\PHI)*\R}) ({3*\D - cos(\PHI)*\R},{-sin(\PHI)*\R})}; 				\draw [magenta, very thick] plot [smooth, tension=0.8] coordinates { ({\D + cos(\PHI)*\R},{-sin(\PHI)*\R}) ({2*\D - cos(\PHI)*\R},{-sin(\PHI)*\R})}; 				\draw [blue, very thick] plot [smooth, tension=0.8] coordinates {({cos(\PHI)*\R},{-sin(\PHI)*\R}) (0.5*\D,0) (\D,1.3*\R)  ({2*\D - cos(\PHI)*\R},{sin(\PHI)*\R}) }; 				\draw [green, very thick] plot [smooth, tension=0.8] coordinates { ({\R}, {0}) (\D - \R,0)};				 				\filldraw[color=black, fill=teal!20, very thick](0,0) circle (\R) node{$\rho_1^{\mathcal{R}}$}; 				\filldraw[color=black, fill=teal!20, very thick](\D,0) circle (\R) node{$b_{q^\prime}^{\mathcal{L}\dagger}$}; 				\filldraw[color=black, fill=teal!20, very thick](2*\D,0) circle (\R) node{$b_{q^\prime}^{\mathcal{L}}$}; 				\filldraw[color=black, fill=teal!20, very thick](3*\D,0) circle (\R) node{$\rho_2^{\mathcal{R}}$};				 				\node[anchor=north] at (\D-1.3*\R, 0) {\textcolor{green}{$\bar{s}_1$}}; 				\node[anchor=south] at (2*\D-1.3*\R, 0.7*\R) {\textcolor{blue}{$s_1$}}; 				\node[anchor=south] at (1.3*\R, 0) {\textcolor{green}{$s_1^\p$}}; 				\node[anchor=north] at (1.3*\R, {-0.8*sin(\PHI)*\R}) {\textcolor{blue}{$\bar{s}_1^\p$}}; 				\node[anchor=north] at (1.5*\D, {-sin(\PHI)*\R}) {\textcolor{magenta}{$\bar{s}_1$, $-\mathrm{i}\delta$}}; 				\node[anchor=north] at (2*\D+1.3*\R, {1.3*\R}) {\textcolor{purple}{$s_2$}}; 				\node[anchor=north] at (2*\D+1.3*\R, {-sin(\PHI)*\R}) {\textcolor{purple}{$\bar{s}_2$}}; 				\node[anchor=north] at (3*\D-1.3*\R, {1.52*\R}) {\textcolor{purple}{$s_2^\p$}}; 				\node[anchor=north] at (3*\D-1.3*\R, {-0.8*sin(\PHI)*\R}) {\textcolor{purple}{$\bar{s}_2^\p$}}; 			\end{scope} 		\end{tikzpicture} 	\nonumber \\ +4\sum_{s_1,s_2,s_1^\p,s_2^\p}&\begin{tikzpicture}[baseline={([yshift=-0.5ex]current bounding box.center)},vertex/.style={anchor=base, circle,fill=black!25,minimum size=18pt,inner sep=2pt}] \pgfmathsetmacro{\R}{0.5} \pgfmathsetmacro{\H}{2.9*\R} \pgfmathsetmacro{\D}{2} \pgfmathsetmacro{\PHI}{30} 			\begin{scope}[shift={(0, 0)}] 				\draw [magenta, very thick] plot [smooth, tension=0.8] coordinates { ({\D + cos(\PHI)*\R},{sin(\PHI)*\R}) ({2*\D - cos(\PHI)*\R},{sin(\PHI)*\R})}; 				\draw [green, very thick] plot [smooth, tension=0.8] coordinates { ({cos(\PHI)*\R},{-sin(\PHI)*\R}) ({\D - cos(\PHI)*\R},{sin(\PHI)*\R})}; 				\draw [green, very thick] plot [smooth, tension=0.8] coordinates { ({cos(\PHI)*\R},{sin(\PHI)*\R}) ({\D - cos(\PHI)*\R},{-sin(\PHI)*\R})}; 				\draw [purple, very thick] plot [smooth, tension=0.8] coordinates { (2*\D + \R,0) (3*\D - \R,0)}; 				\draw [orange, very thick] plot [smooth, tension=0.8] coordinates {({\D + cos(\PHI)*\R},{-sin(\PHI)*\R}) (2*\D,-1.3*\R) (2.5*\D,0) ({3*\D - cos(\PHI)*\R},{sin(\PHI)*\R}) };				 				\filldraw[color=black, fill=teal!20, very thick](0,0) circle (\R) node{$\rho_1^{\mathcal{R}}$}; 				\filldraw[color=black, fill=teal!20, very thick](\D,0) circle (\R) node{$b_{q^\prime}^{\mathcal{L}\dagger}$}; 				\filldraw[color=black, fill=teal!20, very thick](2*\D,0) circle (\R) node{$b_{q^\prime}^{\mathcal{L}}$}; 				\filldraw[color=black, fill=teal!20, very thick](3*\D,0) circle (\R) node{$\rho_2^{\mathcal{R}}$};				 				\node[anchor=north] at (1.5*\D, {1.5*\R}) {\textcolor{magenta}{$s_2$, $-\mathrm{i}\delta$}}; 				\node[anchor=north] at (\D-1.3*\R, {1.3*\R}) {\textcolor{green}{$s_1$}}; 				\node[anchor=north] at (\D-1.3*\R, {-sin(\PHI)*\R}) {\textcolor{green}{$\bar{s}_1$}}; 				\node[anchor=south] at (2*\D+1.3*\R, 0) {\textcolor{purple}{$s_2$}}; 				\node[anchor=north] at (\D+1.3*\R, -0.8*\R) {\textcolor{orange}{$\bar{s}_2$}}; 				\node[anchor=north] at (3*\D-1.3*\R, 0) {\textcolor{purple}{$\bar{s}_2^\p$}}; 				\node[anchor=south] at (3*\D-1.3*\R, {0.7*sin(\PHI)*\R}) {\textcolor{orange}{$s_2^\p$}}; 				\node[anchor=north] at (1.3*\R, {1.52*\R}) {\textcolor{green}{$s_1^\p$}}; 				\node[anchor=north] at (1.3*\R, {-0.8*sin(\PHI)*\R}) {\textcolor{green}{$\bar{s}_1^\p$}}; 			\end{scope} 		\end{tikzpicture} 	\nonumber \\ +\sum_{\{s\},\{s^\p\}}&\begin{tikzpicture}[baseline={([yshift=-0.5ex]current bounding box.center)},vertex/.style={anchor=base, circle,fill=black!25,minimum size=18pt,inner sep=2pt}] 	\pgfmathsetmacro{\R}{0.5} 	\pgfmathsetmacro{\H}{2.9*\R} 	\pgfmathsetmacro{\D}{2} 	\pgfmathsetmacro{\PHI}{30} 	\draw [magenta, very thick] plot [smooth, tension=0.8] coordinates { ({\D + cos(\PHI)*\R},{-sin(\PHI)*\R}) ({2*\D - cos(\PHI)*\R},{-sin(\PHI)*\R})}; 	\draw [magenta, very thick] plot [smooth, tension=0.8] coordinates { ({\D + cos(\PHI)*\R},{sin(\PHI)*\R}) ({2*\D - cos(\PHI)*\R},{sin(\PHI)*\R})}; 	\draw [green, very thick] plot [smooth, tension=0.8] coordinates { ({cos(\PHI)*\R},{-sin(\PHI)*\R}) ({\D - cos(\PHI)*\R},{sin(\PHI)*\R})}; 	\draw [green, very thick] plot [smooth, tension=0.8] coordinates { ({cos(\PHI)*\R},{sin(\PHI)*\R}) ({\D - cos(\PHI)*\R},{-sin(\PHI)*\R})}; 	\draw [purple, very thick] plot [smooth, tension=0.8] coordinates { ({2*\D + cos(\PHI)*\R},{-sin(\PHI)*\R}) ({3*\D - cos(\PHI)*\R},{sin(\PHI)*\R})}; 	\draw [purple, very thick] plot [smooth, tension=0.8] coordinates { ({2*\D + cos(\PHI)*\R},{sin(\PHI)*\R}) ({3*\D - cos(\PHI)*\R},{-sin(\PHI)*\R})};			 	\filldraw[color=black, fill=teal!20, very thick](0,0) circle (\R) node{$\rho_1^{\mathcal{R}}$}; 	\filldraw[color=black, fill=teal!20, very thick](\D,0) circle (\R) node{$b_{q^\prime}^{\mathcal{L}\dagger}$}; 	\filldraw[color=black, fill=teal!20, very thick](2*\D,0) circle (\R) node{$b_{q^\prime}^{\mathcal{L}}$}; 	\filldraw[color=black, fill=teal!20, very thick](3*\D,0) circle (\R) node{$\rho_2^{\mathcal{R}}$};			 	\node[anchor=north] at (1.5*\D, {1.5*\R}) {\textcolor{magenta}{$s_2$, $-\mathrm{i}\delta$}}; 	\node[anchor=north] at (1.5*\D, {-sin(\PHI)*\R}) {\textcolor{magenta}{$\bar{s}_2$, $-\mathrm{i}\delta$}}; 	\node[anchor=north] at (\D-1.3*\R, {1.3*\R}) {\textcolor{green}{$s_1$}}; 	\node[anchor=north] at (\D-1.3*\R, {-sin(\PHI)*\R}) {\textcolor{green}{$\bar{s}_1$}}; 	\node[anchor=north] at (2*\D+1.3*\R, {1.3*\R}) {\textcolor{purple}{$s_3$}}; 	\node[anchor=north] at (2*\D+1.3*\R, {-sin(\PHI)*\R}) {\textcolor{purple}{$\bar{s}_3$}}; 	\node[anchor=north] at (3*\D-1.3*\R, {1.52*\R}) {\textcolor{purple}{$s_3^\p$}}; 	\node[anchor=north] at (3*\D-1.3*\R, {-0.8*sin(\PHI)*\R}) {\textcolor{purple}{$\bar{s}_3^\p$}}; 	\node[anchor=north] at (1.3*\R, {1.5*\R}) {\textcolor{green}{$s_1^\p$}}; 	\node[anchor=north] at (1.3*\R, {-0.8*sin(\PHI)*\R}) {\textcolor{green}{$\bar{s}_1^\p$}}; \end{tikzpicture}~, 	\end{align}and
its equivalent forms can be drawn as before. We then find\begin{align} \label{eq:gamma_2s_2s_a_details}	\frac{1}{2} 	&\begin{tikzpicture}[baseline={([yshift=-0.5ex]current bounding box.center)},vertex/.style={anchor=base, circle,fill=black!25,minimum size=18pt,inner sep=2pt}] 		\pgfmathsetmacro{\R}{0.5} 		\pgfmathsetmacro{\H}{2.9*\R} 		\pgfmathsetmacro{\D}{2} 		\pgfmathsetmacro{\PHI}{30} 		\begin{scope}[shift={(0,0)}] 			\draw [purple, very thick] plot [smooth, tension=0.8] coordinates { (2*\D + \R,0) (3*\D - \R,0)}; 			\draw [orange, very thick] plot [smooth, tension=0.8] coordinates {({\D + cos(\PHI)*\R},{-sin(\PHI)*\R}) (2*\D,-1.3*\R) (2.5*\D,0) ({3*\D - cos(\PHI)*\R},{sin(\PHI)*\R}) }; 			\draw [blue, very thick] plot [smooth, tension=0.8] coordinates {({cos(\PHI)*\R},{-sin(\PHI)*\R}) (0.5*\D,0) (\D,1.3*\R)  ({2*\D - cos(\PHI)*\R},{sin(\PHI)*\R}) }; 			\draw [green, very thick] plot [smooth, tension=0.8] coordinates { ({\R}, {0}) (\D - \R,0)};				 			\filldraw[color=black, fill=teal!20, very thick](0,0) circle (\R) node{$\rho_1^{\mathcal{R}}$}; 			\filldraw[color=black, fill=teal!20, very thick](\D,0) circle (\R) node{$b_{q^\prime}^{\mathcal{L}\dagger}$}; 			\filldraw[color=black, fill=teal!20, very thick](2*\D,0) circle (\R) node{$b_{q^\prime}^{\mathcal{L}}$}; 			\filldraw[color=black, fill=teal!20, very thick](3*\D,0) circle (\R) node{$\rho_2^{\mathcal{R}}$};				 			\node[anchor=north] at (\D-1.3*\R, 0) {\textcolor{green}{$\bar{s}$}}; 			\node[anchor=south] at (2*\D-1.3*\R, 0.7*\R) {\textcolor{blue}{$s$}}; 			\node[anchor=south] at (1.3*\R, 0) {\textcolor{green}{$s_1^\p$}}; 			\node[anchor=north] at (1.3*\R, {-0.8*sin(\PHI)*\R}) {\textcolor{blue}{$\bar{s}_1^\p$}}; 			\node[anchor=south] at (2*\D+1.3*\R, 0) {\textcolor{purple}{$s$}}; 			\node[anchor=north] at (\D+1.3*\R, -0.7*\R) {\textcolor{orange}{$\bar{s}$}}; 			\node[anchor=north] at (3*\D-1.3*\R, 0) {\textcolor{purple}{$\bar{s}_2^\p$}}; 			\node[anchor=south] at (3*\D-1.3*\R, {0.7*sin(\PHI)*\R}) {\textcolor{orange}{$s_2^\p$}}; 		\end{scope} 	\end{tikzpicture} 	\nonumber \\ 	+\frac{1}{2}&\begin{tikzpicture}[baseline={([yshift=-0.5ex]current bounding box.center)},vertex/.style={anchor=base, circle,fill=black!25,minimum size=18pt,inner sep=2pt}] 		\pgfmathsetmacro{\R}{0.5} 		\pgfmathsetmacro{\H}{2.9*\R} 		\pgfmathsetmacro{\D}{2} 		\pgfmathsetmacro{\PHI}{30} 		\begin{scope}[shift={(0,0)}, yscale=-1] 			\draw [purple, very thick] plot [smooth, tension=0.8] coordinates { (2*\D + \R,0) (3*\D - \R,0)}; 			\draw [orange, very thick] plot [smooth, tension=0.8] coordinates {({\D + cos(\PHI)*\R},{-sin(\PHI)*\R}) (2*\D,-1.3*\R) (2.5*\D,0) ({3*\D - cos(\PHI)*\R},{sin(\PHI)*\R}) }; 			\draw [blue, very thick] plot [smooth, tension=0.8] coordinates {({cos(\PHI)*\R},{-sin(\PHI)*\R}) (0.5*\D,0) (\D,1.3*\R)  ({2*\D - cos(\PHI)*\R},{sin(\PHI)*\R}) }; 			\draw [green, very thick] plot [smooth, tension=0.8] coordinates { ({\R}, {0}) (\D - \R,0)};	 		\end{scope}			 			\filldraw[color=black, fill=teal!20, very thick](0,0) circle (\R) node{$\rho_1^{\mathcal{R}}$}; 			\filldraw[color=black, fill=teal!20, very thick](\D,0) circle (\R) node{$b_{q^\prime}^{\mathcal{L}\dagger}$}; 			\filldraw[color=black, fill=teal!20, very thick](2*\D,0) circle (\R) node{$b_{q^\prime}^{\mathcal{L}}$}; 			\filldraw[color=black, fill=teal!20, very thick](3*\D,0) circle (\R) node{$\rho_2^{\mathcal{R}}$};				 			\node[anchor=south] at (\D-1.3*\R, 0) {\textcolor{green}{$\bar{s}$}}; 			\node[anchor=north] at (2*\D-1.3*\R, -0.7*\R) {\textcolor{blue}{$s$}}; 			\node[anchor=north] at (1.3*\R, 0) {\textcolor{green}{$s_1^\p$}}; 			\node[anchor=south] at (1.3*\R, {0.8*sin(\PHI)*\R}) {\textcolor{blue}{$\bar{s}_1^\p$}}; 			\node[anchor=north] at (2*\D+1.3*\R, 0) {\textcolor{purple}{$s$}}; 			\node[anchor=south] at (\D+1.3*\R, 0.7*\R) {\textcolor{orange}{$\bar{s}$}}; 			\node[anchor=south] at (3*\D-1.3*\R, 0) {\textcolor{purple}{$\bar{s}_2^\p$}}; 			\node[anchor=north] at (3*\D-1.3*\R, {-0.7*sin(\PHI)*\R}) {\textcolor{orange}{$s_2^\p$}};		 	\end{tikzpicture} \nonumber \\ 	=-&\mathrm{Re}\big\{f_{s_1^\p \bar{s}_1^\p}^{\mathcal{R}}\left(\alpha_{11},\alpha_{12}\right)f_{s\bar{s}}^{\mathcal{R}}\left(\alpha_{21},\alpha_{11}+\ii\pi^{-}\right) \nonumber \\ 	&\hspace{0.5cm}\times f_{s\bar{s}}^{\mathcal{R}}\left(\alpha_{12}-\ii\pi^{-},\alpha_{22}\right)f_{s_2^\p \bar{s}_2^\p}^{\mathcal{R}}\left(\alpha_{22},\alpha_{21}\right)\big\}, \end{align}leading
to Eq. (\ref{eq:gamma_1^(1)}). The second and third diagrams in Eq.
(\ref{eq:cross_rels_diagrams_2s_2s}) are mirror images of each other,
and their combined contribution is
\begin{align}
\gamma_{1}^{\left(2\right)}\left(\omega^{\p}\lvert\omega\right)= & -\frac{4}{\omega\omega^{\p}}\int_{0}^{\omega-\omega^{\p}}\id\Omega_{1}\int_{0}^{\omega}\id\Omega_{2}\frac{1}{\left(2\pi\right)^{5}\prod_{i=1}^{5}e^{\lambda_{i}}}\nonumber \\
 & \hspace{-1.8cm}\times\mathrm{Re}\left\{ \left(e^{-\frac{\ii\pi}{2z}}R_{+}^{-}\left(\lambda_{1}\right)R_{+}^{-}\left(\lambda_{2}\right)-e^{\frac{\ii\pi}{2z}}R_{+}^{+}\left(\lambda_{1}\right)R_{+}^{+}\left(\lambda_{2}\right)\right)\right.\nonumber \\
 & \hspace{-1.8cm}\left.\times\left(e^{-\frac{\ii\pi}{2z}}R_{+}^{-}\left(\lambda_{3}\right)R_{+}^{-}\left(\lambda_{4}\right)-e^{\frac{\ii\pi}{2z}}R_{+}^{+}\left(\lambda_{3}\right)R_{+}^{+}\left(\lambda_{4}\right)\right)-1\right\} \nonumber \\
 & \hspace{-1.8cm}\times\mathrm{Re}\left\{ f_{+-}^{\mathcal{R}}\left(\lambda_{2},\lambda_{1}\right)f_{+-}^{\mathcal{R}}\left(\lambda_{5},\lambda_{2}+\ii\pi^{-}\right)\right.\nonumber \\
 & \hspace{-1.8cm}\left.\times f_{+-+-}^{\mathcal{R}}\left(\lambda_{4},\lambda_{3},\lambda_{1}+\ii\pi^{-},\lambda_{5}+i\pi^{+}\right)f_{+-}^{\mathcal{R}}\left(\lambda_{3},\lambda_{4}\right)\right\} ,\label{eq:gamma_1^(2)}
\end{align}
where $e^{\lambda_{1}}=\Omega_{1}$, $e^{\lambda_{2}}=\omega-\Omega_{1}$,
$e^{\lambda_{3}}=\Omega_{2}$, $e^{\lambda_{4}}=\omega-\Omega_{2}$,
$e^{\lambda_{5}}=\omega-\omega^{\p}-\Omega_{1}$. Again, taking the
real part eliminates the delta function terms from $f_{+-+-}^{\mathcal{R}}$,
and the integral can be evaluated as a principle value integral; this
term turns out to be negligble compared to $\gamma_{1}^{\left(1\right)}$
in Eq. (\ref{eq:gamma_1^(1)}). However, the delta function terms
cannot be eliminated from the fourth term in (\ref{eq:cross_rels_diagrams_2s_2s})
by taking its real part, since there are annihilation poles in both
form factors related to $b_{q^{\p}}^{\mathcal{L}\dagger}$ and $b_{q^{\p}}^{\mathcal{L}}$.
While this term involves 6 excitations and thus requires the calculation
of a triple integral, it can be well approximated by considering only
the delta functions, which reduce the number of excitations by 2 and
therefore reduce the triple integral to a single integral:
\begin{align}
\gamma_{1}^{\left(3\right)}\left(\omega^{\p}\lvert\omega\right)\approx & \frac{1}{2\omega^{\p}\omega}\int_{0}^{\omega-\omega^{\p}}\frac{\id\Omega}{\left(2\pi\right)^{4}\prod_{i=1}^{4}e^{\lambda_{i}}}\nonumber \\
 & \hspace{-1.5cm}\times\mathrm{Re}\left\{ \left(e^{-\frac{\ii\pi}{2z}}R_{+}^{-}\left(\lambda_{1}\right)R_{+}^{-}\left(\lambda_{2}\right)-e^{\frac{\ii\pi}{2z}}R_{+}^{+}\left(\lambda_{1}\right)R_{+}^{+}\left(\lambda_{2}\right)\right)^{*}\right.\nonumber \\
 & \hspace{-1.5cm}\left.\times\left(e^{-\frac{\ii\pi}{2z}}R_{+}^{-}\left(\lambda_{3}\right)R_{+}^{-}\left(\lambda_{4}\right)-e^{\frac{\ii\pi}{2z}}R_{+}^{+}\left(\lambda_{3}\right)R_{-}^{-}\left(\lambda_{4}\right)\right)-1\right\} \nonumber \\
 & \hspace{-1.5cm}\times\mathrm{Re}\left\{ f_{+-}^{\mathcal{R}}\left(\lambda_{1},\lambda_{3}+\ii\pi^{-}\right)f_{+-}^{\mathcal{R}}\left(\lambda_{4},\lambda_{2}+\ii\pi^{-}\right)\right.\nonumber \\
 & \hspace{-1.5cm}\times f_{+-}^{\mathcal{R}}\left(\lambda_{2},\lambda_{1}\right)f_{+-}^{\mathcal{R}}\left(\lambda_{3},\lambda_{4}\right)\nonumber \\
 & \hspace{-1.5cm}\times\left(S_{0}\left(\lambda_{4}-\lambda_{1}\right)-S_{0}\left(\lambda_{2}-\lambda_{1}\right)\right)\nonumber \\
 & \hspace{-1.5cm}\left.\times\left(S_{0}\left(\lambda_{1}-\lambda_{4}\right)-S_{0}\left(\lambda_{3}-\lambda_{4}\right)\right)\right\} ,\label{eq:gamma_1^(3)}
\end{align}
with $e^{\lambda_{1}}=\Omega$, $e^{\lambda_{2}}=\omega-\Omega$,
$e^{\lambda_{3}}=\omega^{\p}+\Omega$, $e^{\lambda_{4}}=\omega-\omega^{\p}-\Omega$.
The expression above for $\gamma_{1}^{\left(3\right)}\left(\omega^{\p}\lvert\omega\right)$
holds only for integer $p=1/z$.

\subsection{Terms used to evaluate the spectrum}

\begin{figure}[t]
\begin{centering}
\includegraphics[width=1\columnwidth]{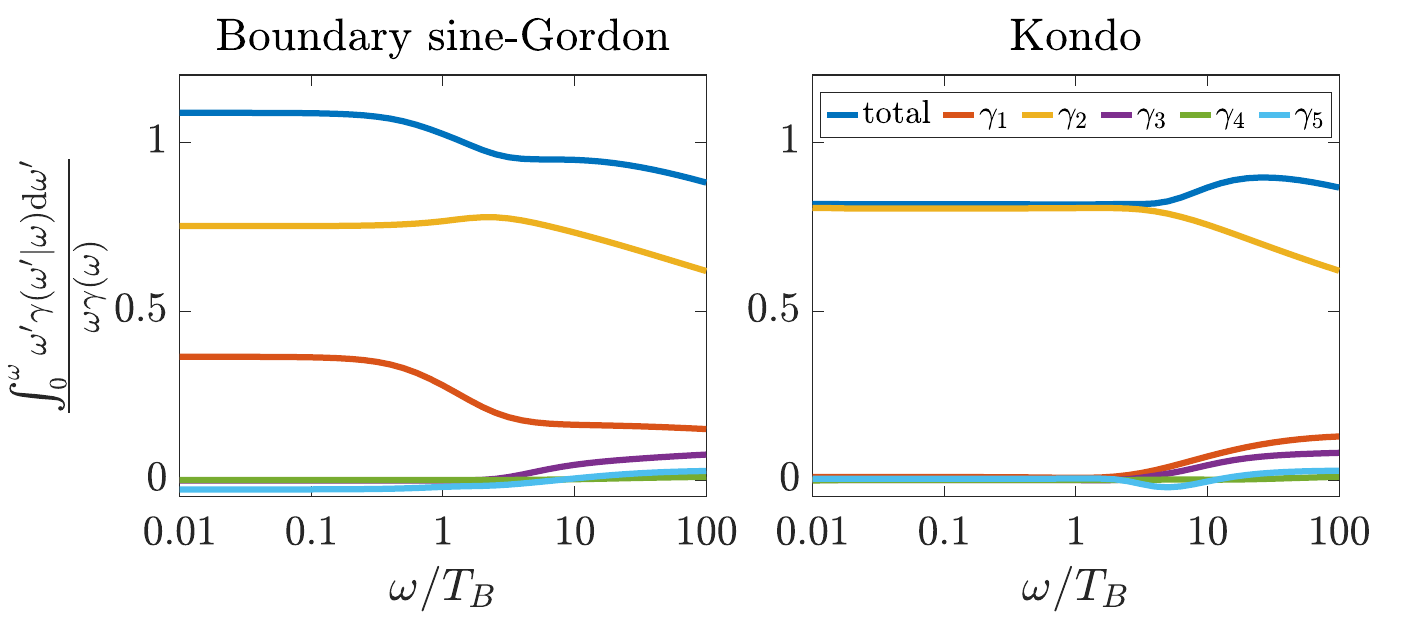}
\par\end{centering}
\caption{\label{fig:inel_spec_contributions}The weight of the contributions
to the $z=1/3$ spectrum, evaluated using the sum rule.}
\end{figure}

We list here all of the diagrams used in Eq. (\ref{eq:spectrum_general})
to evaluate the spectrum. As detailed above, we take the real part
of each of the diagrams, and discard all diagrams which correspond
to principle value integrals, since those are subleading. The contributions
indeed decay rapidly, as illustrated in Fig. \ref{fig:inel_spec_contributions}
for the $z=1/3$ spectrum.

\subsubsection{$\gamma_{1}$}

\begin{align} 	\sum_{\{s\},\{s^\p\}=\pm}&\begin{tikzpicture}[baseline={([yshift=-0.5ex]current bounding box.center)},vertex/.style={anchor=base, circle,fill=black!25,minimum size=18pt,inner sep=2pt}] 		\pgfmathsetmacro{\R}{0.5} 		\pgfmathsetmacro{\H}{2.9*\R} 		\pgfmathsetmacro{\D}{2} 		\pgfmathsetmacro{\PHI}{30} 		\begin{scope}[shift={(0,0)}] 			\draw [purple, very thick] plot [smooth, tension=0.8] coordinates { (2*\D + \R,0) (3*\D - \R,0)}; 			\draw [orange, very thick] plot [smooth, tension=0.8] coordinates {({\D + cos(\PHI)*\R},{-sin(\PHI)*\R}) (2*\D,-1.3*\R) (2.5*\D,0) ({3*\D - cos(\PHI)*\R},{sin(\PHI)*\R}) }; 			\draw [blue, very thick] plot [smooth, tension=0.8] coordinates {({cos(\PHI)*\R},{-sin(\PHI)*\R}) (0.5*\D,0) (\D,1.3*\R)  ({2*\D - cos(\PHI)*\R},{sin(\PHI)*\R}) }; 			\draw [green, very thick] plot [smooth, tension=0.8] coordinates { ({\R}, {0}) (\D - \R,0)};				 			\filldraw[color=black, fill=teal!20, very thick](0,0) circle (\R) node{$\rho_1^{\mathcal{R}}$}; 			\filldraw[color=black, fill=teal!20, very thick](\D,0) circle (\R) node{$b_{q^\prime}^{\mathcal{L}\dagger}$}; 			\filldraw[color=black, fill=teal!20, very thick](2*\D,0) circle (\R) node{$b_{q^\prime}^{\mathcal{L}}$}; 			\filldraw[color=black, fill=teal!20, very thick](3*\D,0) circle (\R) node{$\rho_2^{\mathcal{R}}$};				 			\node[anchor=north] at (\D-1.3*\R, 0) {\textcolor{green}{$\bar{s}$}}; 			\node[anchor=south] at (2*\D-1.1*\R, 0.7*\R) {\textcolor{blue}{$s$}}; 			\node[anchor=south] at (1.3*\R, 0) {\textcolor{green}{$s_1^\p$}}; 			\node[anchor=north] at (1.3*\R, {-0.8*sin(\PHI)*\R}) {\textcolor{blue}{$\bar{s}_1^\p$}}; 			\node[anchor=south] at (2*\D+1.3*\R, 0) {\textcolor{purple}{$s$}}; 			\node[anchor=north] at (\D+1.1*\R, -0.7*\R) {\textcolor{orange}{$\bar{s}$}}; 			\node[anchor=north] at (3*\D-1.3*\R, 0) {\textcolor{purple}{$\bar{s}_2^\p$}}; 			\node[anchor=south] at (3*\D-1.3*\R, {0.7*sin(\PHI)*\R}) {\textcolor{orange}{$s_2^\p$}}; 		\end{scope} 	\end{tikzpicture}~, 	\\ 	\sum_{\{s\},\{s^\p\}=\pm}&\begin{tikzpicture}[baseline={([yshift=-0.5ex]current bounding box.center)},vertex/.style={anchor=base, circle,fill=black!25,minimum size=18pt,inner sep=2pt}] 		\pgfmathsetmacro{\R}{0.5} 		\pgfmathsetmacro{\H}{2.9*\R} 		\pgfmathsetmacro{\D}{2} 		\pgfmathsetmacro{\PHI}{30} 		\draw [magenta, very thick] plot [smooth, tension=0.8] coordinates { ({\D + cos(\PHI)*\R},{-sin(\PHI)*\R}) ({2*\D - cos(\PHI)*\R},{-sin(\PHI)*\R})}; 		\draw [magenta, very thick] plot [smooth, tension=0.8] coordinates { ({\D + cos(\PHI)*\R},{sin(\PHI)*\R}) ({2*\D - cos(\PHI)*\R},{sin(\PHI)*\R})}; 		\draw [green, very thick] plot [smooth, tension=0.8] coordinates { ({cos(\PHI)*\R},{-sin(\PHI)*\R}) ({\D - cos(\PHI)*\R},{sin(\PHI)*\R})}; 		\draw [green, very thick] plot [smooth, tension=0.8] coordinates { ({cos(\PHI)*\R},{sin(\PHI)*\R}) ({\D - cos(\PHI)*\R},{-sin(\PHI)*\R})}; 		\draw [purple, very thick] plot [smooth, tension=0.8] coordinates { ({2*\D + cos(\PHI)*\R},{-sin(\PHI)*\R}) ({3*\D - cos(\PHI)*\R},{sin(\PHI)*\R})}; 		\draw [purple, very thick] plot [smooth, tension=0.8] coordinates { ({2*\D + cos(\PHI)*\R},{sin(\PHI)*\R}) ({3*\D - cos(\PHI)*\R},{-sin(\PHI)*\R})};			 		\filldraw[color=black, fill=teal!20, very thick](0,0) circle (\R) node{$\rho_1^{\mathcal{R}}$}; 		\filldraw[color=black, fill=teal!20, very thick](\D,0) circle (\R) node{$b_{q^\prime}^{\mathcal{L}\dagger}$}; 		\filldraw[color=black, fill=teal!20, very thick](2*\D,0) circle (\R) node{$b_{q^\prime}^{\mathcal{L}}$}; 		\filldraw[color=black, fill=teal!20, very thick](3*\D,0) circle (\R) node{$\rho_2^{\mathcal{R}}$};			 		\node[anchor=north] at (1.5*\D, {1.5*\R}) {\textcolor{magenta}{$s_2$}}; 		\node[anchor=north] at (1.5*\D, {-sin(\PHI)*\R}) {\textcolor{magenta}{$\bar{s}_2$}}; 		\node[anchor=north] at (\D-1.3*\R, {1.3*\R}) {\textcolor{green}{$s_1$}}; 		\node[anchor=north] at (\D-1.3*\R, {-sin(\PHI)*\R}) {\textcolor{green}{$\bar{s}_1$}}; 		\node[anchor=north] at (2*\D+1.3*\R, {1.3*\R}) {\textcolor{purple}{$s_3$}}; 		\node[anchor=north] at (2*\D+1.3*\R, {-sin(\PHI)*\R}) {\textcolor{purple}{$\bar{s}_3$}}; 		\node[anchor=north] at (3*\D-1.3*\R, {1.52*\R}) {\textcolor{purple}{$s_3^\p$}}; 		\node[anchor=north] at (3*\D-1.3*\R, {-0.8*sin(\PHI)*\R}) {\textcolor{purple}{$\bar{s}_3^\p$}}; 		\node[anchor=north] at (1.3*\R, {1.5*\R}) {\textcolor{green}{$s_1^\p$}}; 		\node[anchor=north] at (1.3*\R, {-0.8*sin(\PHI)*\R}) {\textcolor{green}{$\bar{s}_1^\p$}}; 	\end{tikzpicture}~. \end{align}The
two diagrams correspond to Eqs. (\ref{eq:gamma_1^(1)}) and (\ref{eq:gamma_1^(3)}),
respectively. The first diagram is the only nonvanishing term for
$z=1/2$, and is the only diagram used to evaluate the spectrum for
$z\ge1/2$ and $z=0.47$ in Fig. \ref{fig:inel_spec}.

\subsubsection{$\gamma_{2}$}

\begin{equation} 	\sum_{s,s^\p=\pm} 	\begin{tikzpicture}[baseline={([yshift=-0.5ex]current bounding box.center)},vertex/.style={anchor=base, circle,fill=black!25,minimum size=18pt,inner sep=2pt}] 		\pgfmathsetmacro{\R}{0.5} 		\pgfmathsetmacro{\H}{2.9*\R} 		\pgfmathsetmacro{\D}{2} 		\pgfmathsetmacro{\PHI}{30}			 		\begin{scope}[shift={(0,0)}] 			\draw [purple, very thick] plot [smooth, tension=0.8] coordinates { ({2*\D + cos(\PHI)*\R},{sin(\PHI)*\R}) ({3*\D - cos(\PHI)*\R},{-sin(\PHI)*\R})}; 			\draw [orange, very thick] plot [smooth, tension=0.8] coordinates {({\D + cos(\PHI)*\R},{-sin(\PHI)*\R}) (2*\D,-1.3*\R) (2.5*\D,0) ({3*\D - cos(\PHI)*\R},{sin(\PHI)*\R}) }; 			\draw [green, very thick] plot [smooth, tension=0.8] coordinates { ({\R}, 0) (\D - \R,0)}; 			\draw [magenta, very thick] plot [smooth, tension=0.8] coordinates { ({\D + \R*cos(\PHI)},{sin(\PHI)*\R}) ({2*\D - \R*cos(\PHI)},{sin(\PHI)*\R})};				 			\filldraw[color=black, fill=teal!20, very thick](0,0) circle (\R) node{$\rho_1^{\mathcal{R}}$}; 			\filldraw[color=black, fill=teal!20, very thick](\D,0) circle (\R) node{$b_{q^\prime}^{\mathcal{L}\dagger}$}; 			\filldraw[color=black, fill=teal!20, very thick](2*\D,0) circle (\R) node{$b_{q^\prime}^{\mathcal{L}}$}; 			\filldraw[color=black, fill=teal!20, very thick](3*\D,0) circle (\R) node{$\rho_2^{\mathcal{R}}$};				 			\node[anchor=south] at (0.5*\D, 0) {\textcolor{green}{$1$}}; 			\node[anchor=south] at (1.5*\D, {sin(\PHI)*\R}) {\textcolor{magenta}{$s$}}; 			\node[anchor=north] at (2*\D+1.3*\R, {1.2*\R}) {\textcolor{purple}{$s$}}; 			\node[anchor=north] at (2*\D+1.3*\R, -0.7*\R) {\textcolor{orange}{$\bar{s}$}}; 			\node[anchor=north] at (3*\D-1.3*\R, {-0.6*\R}) {\textcolor{purple}{$\bar{s}^\p$}}; 			\node[anchor=north] at (3*\D-1.3*\R, {1.4*\R}) {\textcolor{orange}{$s^\p$}}; 		\end{scope} 	\end{tikzpicture}~. \end{equation}This
diagram corresponds to Eq. (\ref{eq:gamma_2^(1)}).

\subsubsection{$\gamma_{3}$}

\begin{align} 		\label{eq:gamma_2s1b_1b_diagram}\sum_{s,s^\p=\pm}&\begin{tikzpicture}[baseline={([yshift=-1.0ex]current bounding box.center)},vertex/.style={anchor=base, circle,fill=black!25,minimum size=18pt,inner sep=2pt}] 		\pgfmathsetmacro{\R}{0.5} 		\pgfmathsetmacro{\H}{3.5*\R} 		\pgfmathsetmacro{\D}{2} 		\pgfmathsetmacro{\PHI}{15}		 		\draw [purple, very thick] plot [smooth, tension=0.8] coordinates { (2*\D + \R,0) (3*\D - \R,0)}; 		\draw [blue, very thick] plot [smooth, tension=0.8] coordinates {({cos(\PHI)*\R},{-sin(\PHI)*\R}) (0.5*\D,0) (\D,1.2*\R)  ({2*\D - cos(\PHI)*\R},0) }; 		\draw [blue, very thick] plot [smooth, tension=0.8] coordinates {({cos(3*\PHI)*\R},{-sin(3*\PHI)*\R}) (0.5*\D,0) (\D,1.5*\R)  ({2*\D - cos(\PHI)*\R},{sin(\PHI)*\R}) }; 		\draw [green, very thick] plot [smooth, tension=0.8] coordinates { ({\R}, {0}) (\D - \R,0)};		 		\filldraw[color=black, fill=teal!20, very thick](0,0) circle (\R) node{$\rho_1^{\mathcal{R}}$}; 		\filldraw[color=black, fill=teal!20, very thick](\D,0) circle (\R) node{$b_{q^\prime}^{\mathcal{L}\dagger}$}; 		\filldraw[color=black, fill=teal!20, very thick](2*\D,0) circle (\R) node{$b_{q^\prime}^{\mathcal{L}}$}; 		\filldraw[color=black, fill=teal!20, very thick](3*\D,0) circle (\R) node{$\rho_2^{\mathcal{R}}$};		 		\node[anchor=east] at (2*\D-0.8*\R, 1.2*\R) {\textcolor{blue}{$s, \bar{s}$}}; 		\node[anchor=south] at (1.3*\R, 0*\R) {\textcolor{green}{$1$}}; 		\node[anchor=north] at (1.8*\R, {-1.5*sin(\PHI)*\R}) {\textcolor{blue}{$\bar{s}^\p, s^\p$}}; 		\node[anchor=south] at (2.5*\D, 0) {\textcolor{purple}{$1$}}; 	\end{tikzpicture}~, 	\\ 	\sum_{s_1,s_1^\p,s_2=\pm}&\begin{tikzpicture}[baseline={([yshift=-1.2ex]current bounding box.center)},vertex/.style={anchor=base, circle,fill=black!25,minimum size=18pt,inner sep=2pt}] 		\pgfmathsetmacro{\R}{0.5} 		\pgfmathsetmacro{\H}{3.5*\R} 		\pgfmathsetmacro{\D}{2} 		\pgfmathsetmacro{\PHI}{30}		 		\pgfmathsetmacro{\VARPHI}{70} 		\draw [purple, very thick] plot [smooth, tension=0.8] coordinates { (2*\D + \R,0) (3*\D - \R,0)}; 		\draw [green, very thick] plot [smooth, tension=0.8] coordinates { ({\R}, {0}) (\D - \R,0)};		 		\draw [green, very thick] plot [smooth, tension=0.8] coordinates { ({\R*cos(\VARPHI)}, {\R*sin(\VARPHI)}) (0.5*\D - 0.1*\R, {0.8*\R*sin(\VARPHI)}) (0.5*\D + 0.3*\R, {-0.8*\R*sin(\VARPHI)}) ({\D - \R*cos(\VARPHI)},{-\R*sin(\VARPHI)})}; 		\draw [green, very thick] plot [smooth, tension=1] coordinates { ({\R*cos(\VARPHI)}, {-\R*sin(\VARPHI)}) (0.5*\D - 0.1*\R, {-0.8*\R*sin(\VARPHI)}) (0.5*\D + 0.3*\R, {0.8*\R*sin(\VARPHI)}) ({\D - \R*cos(\VARPHI)},{\R*sin(\VARPHI)})}; 		\draw [magenta, very thick] plot [smooth, tension=0.8] coordinates {({\D + \R*cos(\PHI)}, {\R*sin(\PHI)}) ({2*\D - \R*cos(\PHI)}, {\R*sin(\PHI)})}; 		\draw [magenta, very thick] plot [smooth, tension=0.8] coordinates {({\D + \R*cos(\PHI)}, {-\R*sin(\PHI)}) ({2*\D - \R*cos(\PHI)}, {-\R*sin(\PHI)})}; 		\filldraw[color=black, fill=teal!20, very thick](0,0) circle (\R) node{$\rho_1^{\mathcal{R}}$}; 		\filldraw[color=black, fill=teal!20, very thick](\D,0) circle (\R) node{$b_{q^\prime}^{\mathcal{L}\dagger}$}; 		\filldraw[color=black, fill=teal!20, very thick](2*\D,0) circle (\R) node{$b_{q^\prime}^{\mathcal{L}}$}; 		\filldraw[color=black, fill=teal!20, very thick](3*\D,0) circle (\R) node{$\rho_2^{\mathcal{R}}$};		 		\node[anchor=north] at (1.5*\D, -0.45*\R) {\textcolor{magenta}{$\bar{s}_2$}}; 		\node[anchor=north] at (1.5*\D, 0.58*\R) {\textcolor{magenta}{$s_2$}}; 		\node[anchor=south] at (1.3*\R, 0.9*\R) {\textcolor{green}{$1$}}; 		\node[anchor=south] at (1.3*\R, -0.15*\R) {\textcolor{green}{$s_1^\p$}}; 		\node[anchor=south] at (\D - 1.35*\R, -0.15*\R) {\textcolor{green}{$\bar{s}_1$}}; 		\node[anchor=south] at (\D - 1.35*\R, 0.85*\R) {\textcolor{green}{$s_1$}}; 		\node[anchor=north] at (1.3*\R, {0.1*\R}) {\textcolor{green}{$\bar{s}_1^\p$}}; 		\node[anchor=south] at (2.5*\D, 0) {\textcolor{purple}{$1$}}; 	\end{tikzpicture}~. \end{align}

\subsubsection{$\gamma_{4}$}

\begin{align}\label{eq:gamma_3b_1b_diagram}&\begin{tikzpicture}[baseline={([yshift=-1.0ex]current bounding box.center)},vertex/.style={anchor=base, circle,fill=black!25,minimum size=18pt,inner sep=2pt}] 		\pgfmathsetmacro{\R}{0.5} 		\pgfmathsetmacro{\H}{3.5*\R} 		\pgfmathsetmacro{\D}{2} 		\pgfmathsetmacro{\PHI}{15}		 		\draw [purple, very thick] plot [smooth, tension=0.8] coordinates { (2*\D + \R,0) (3*\D - \R,0)}; 		\draw [blue, very thick] plot [smooth, tension=0.8] coordinates {({cos(\PHI)*\R},{-sin(\PHI)*\R}) (0.5*\D,0) (\D,1.2*\R)  ({2*\D - cos(\PHI)*\R},0) }; 		\draw [blue, very thick] plot [smooth, tension=0.8] coordinates {({cos(3*\PHI)*\R},{-sin(3*\PHI)*\R}) (0.5*\D,0) (\D,1.5*\R)  ({2*\D - cos(\PHI)*\R},{sin(\PHI)*\R}) }; 		\draw [green, very thick] plot [smooth, tension=0.8] coordinates { ({\R}, {0}) (\D - \R,0)};		 		\filldraw[color=black, fill=teal!20, very thick](0,0) circle (\R) node{$\rho_1^{\mathcal{R}}$}; 		\filldraw[color=black, fill=teal!20, very thick](\D,0) circle (\R) node{$b_{q^\prime}^{\mathcal{L}\dagger}$}; 		\filldraw[color=black, fill=teal!20, very thick](2*\D,0) circle (\R) node{$b_{q^\prime}^{\mathcal{L}}$}; 		\filldraw[color=black, fill=teal!20, very thick](3*\D,0) circle (\R) node{$\rho_2^{\mathcal{R}}$}; 		\node[anchor=south] at (1.3*\R, 0*\R) {\textcolor{green}{$1$}}; 		\node[anchor=north] at (1.58*\R, {-1.75*sin(\PHI)*\R}) {\textcolor{blue}{$1,1$}}; 		\node[anchor=south] at (2.5*\D, 0) {\textcolor{purple}{$1$}}; 	\end{tikzpicture}~, 	\\ 	&\begin{tikzpicture}[baseline={([yshift=-1.2ex]current bounding box.center)},vertex/.style={anchor=base, circle,fill=black!25,minimum size=18pt,inner sep=2pt}] 		\pgfmathsetmacro{\R}{0.5} 		\pgfmathsetmacro{\H}{3.5*\R} 		\pgfmathsetmacro{\D}{2} 		\pgfmathsetmacro{\PHI}{30}		 		\pgfmathsetmacro{\VARPHI}{70} 		\draw [purple, very thick] plot [smooth, tension=0.8] coordinates { (2*\D + \R,0) (3*\D - \R,0)}; 		\draw [green, very thick] plot [smooth, tension=0.8] coordinates { ({\R}, {0}) (\D - \R,0)};		 		\draw [green, very thick] plot [smooth, tension=0.8] coordinates { ({\R*cos(\VARPHI)}, {\R*sin(\VARPHI)}) ({\D - \R*cos(\VARPHI)},{-\R*sin(\VARPHI)})}; 		\draw [green, very thick] plot [smooth, tension=0.8] coordinates { ({\R*cos(\VARPHI)}, {-\R*sin(\VARPHI)}) ({\D - \R*cos(\VARPHI)},{\R*sin(\VARPHI)})}; 		\draw [magenta, very thick] plot [smooth, tension=0.8] coordinates {({\D + \R*cos(\PHI)}, {\R*sin(\PHI)}) ({2*\D - \R*cos(\PHI)}, {\R*sin(\PHI)})}; 		\draw [magenta, very thick] plot [smooth, tension=0.8] coordinates {({\D + \R*cos(\PHI)}, {-\R*sin(\PHI)}) ({2*\D - \R*cos(\PHI)}, {-\R*sin(\PHI)})}; 		\filldraw[color=black, fill=teal!20, very thick](0,0) circle (\R) node{$\rho_1^{\mathcal{R}}$}; 		\filldraw[color=black, fill=teal!20, very thick](\D,0) circle (\R) node{$b_{q^\prime}^{\mathcal{L}\dagger}$}; 		\filldraw[color=black, fill=teal!20, very thick](2*\D,0) circle (\R) node{$b_{q^\prime}^{\mathcal{L}}$}; 		\filldraw[color=black, fill=teal!20, very thick](3*\D,0) circle (\R) node{$\rho_2^{\mathcal{R}}$};		 		\node at (1.5*\D, -0.1*\R) {\textcolor{magenta}{$1, 1$}}; 		\node[anchor=south] at (1.96*\R, 0.4*\R) {\textcolor{green}{$1, 1, 1$}}; 		\node[anchor=south] at (2.5*\D, 0) {\textcolor{purple}{$1$}}; 	\end{tikzpicture}~.\end{align}

\subsubsection{$\gamma_{5}$}

\begin{align} \label{eq:gamma_2s1b_2s_diagram}\sum_{s,s^\p=\pm}&\begin{tikzpicture}[baseline={([yshift=-0.9ex]current bounding box.center)},vertex/.style={anchor=base, circle,fill=black!25,minimum size=18pt,inner sep=2pt}] 		\pgfmathsetmacro{\R}{0.5} 		\pgfmathsetmacro{\H}{3.5*\R} 		\pgfmathsetmacro{\D}{2} 		\pgfmathsetmacro{\PHI}{60}		 		\draw [red, very thick] plot [smooth, tension=0.8] coordinates { ({cos(\PHI)*\R},{-sin(\PHI)*\R}) (0.5*\D - 0.2*\R,{-0.6*\R}) (0.5*\D + 1.5*\R,1.4*\R) (2.5*\D - 1.5*\R,1.4*\R) (2.5*\D + 0.2*\R,{-0.6*\R}) ({3*\D - cos(\PHI)*\R},{-sin(\PHI)*\R})}; 		\draw [purple, very thick] plot [smooth, tension=0.8] coordinates { (2*\D + \R,0) (3*\D - \R,0)}; 		\draw [blue, very thick] plot [smooth, tension=0.8] coordinates { (\R,0) (0.6*\D, 0.22*\R) (\D,1.2*\R) ({2*\D-\R}, 0)}; 		\draw [green, very thick] plot [smooth, tension=0.8] coordinates { ({cos(\PHI)*\R}, {sin(\PHI)*\R}) (0.5*\D,0.2*\R) (\D - \R,0)};		 		\filldraw[color=black, fill=teal!20, very thick](0,0) circle (\R) node{$\rho_1^{\mathcal{R}}$}; 		\filldraw[color=black, fill=teal!20, very thick](\D,0) circle (\R) node{$b_{q^\prime}^{\mathcal{L}\dagger}$}; 		\filldraw[color=black, fill=teal!20, very thick](2*\D,0) circle (\R) node{$b_{q^\prime}^{\mathcal{L}}$}; 		\filldraw[color=black, fill=teal!20, very thick](3*\D,0) circle (\R) node{$\rho_2^{\mathcal{R}}$};		 		\node[anchor=south] at (1.2*\R, {0.5*\R}) {\textcolor{green}{$1$}}; 		\node[anchor=north] at (1.3*\R, {0.1*\R}) {\textcolor{blue}{$\bar{s}^\p$}}; 		\node[anchor=north] at (1.3*\R, {-0.8*\R}) {\textcolor{red}{$s^\p$}}; 		\node[anchor=north] at (3*\D-1.3*\R, 0.1*\R) {\textcolor{red}{$s^\p$}}; 		\node[anchor=north east] at (2*\D - 1.1*\R, 0.3*\R) {\textcolor{blue}{$s$}}; 		\node[anchor=south] at (2*\D+1.3*\R, {-0.1*\R}) {\textcolor{purple}{$s$}}; 		\node[anchor=south] at (3*\D-1.3*\R, {-0.1*\R}) {\textcolor{purple}{$\bar{s}^\p$}}; 	\end{tikzpicture}~, 	\\ 	\sum_{\{s\},\{s^\p\}=\pm}&\begin{tikzpicture}[baseline={([yshift=-0.2ex]current bounding box.center)},vertex/.style={anchor=base, circle,fill=black!25,minimum size=18pt,inner sep=2pt}] 		\pgfmathsetmacro{\R}{0.5} 		\pgfmathsetmacro{\H}{3.5*\R} 		\pgfmathsetmacro{\D}{2} 		\pgfmathsetmacro{\PHI}{60} 		\pgfmathsetmacro{\VARPHI}{30} 		\draw [purple, very thick] plot [smooth, tension=0.8] coordinates { ({2*\D + \R*cos(\VARPHI)},{\R*sin(\VARPHI)}) ({3*\D - \R*cos(\VARPHI)},{-\R*sin(\VARPHI)})}; 		\draw [purple, very thick] plot [smooth, tension=0.8] coordinates { ({2*\D + \R*cos(\VARPHI)},{-\R*sin(\VARPHI)}) ({3*\D - \R*cos(\VARPHI)},{\R*sin(\VARPHI)})}; 		\draw [green, very thick] plot [smooth, tension=0.8] coordinates { ({cos(\PHI)*\R}, {sin(\PHI)*\R}) ({\D - \R*cos(\PHI)},{-sin(\PHI)*\R})};		 		\draw [green, very thick] plot [smooth, tension=0.8] coordinates { (\R, 0) (\D - \R, 0)}; 		\draw [blue, very thick] plot [smooth, tension=0.8] coordinates { ({cos(\PHI)*\R},{-sin(\PHI)*\R}) (0.5*\D - 0.3*\R,{-0.8*\R}) (0.5*\D + 1*\R,1.4*\R) ({2*\D - cos(\VARPHI)*\R},{\R*sin(\VARPHI)})}; 		\draw [magenta, very thick] plot [smooth, tension=0.8] coordinates { (\D + \R, 0) (2*\D - \R, 0)}; 		\filldraw[color=black, fill=teal!20, very thick](0,0) circle (\R) node{$\rho_1^{\mathcal{R}}$}; 		\filldraw[color=black, fill=teal!20, very thick](\D,0) circle (\R) node{$b_{q^\prime}^{\mathcal{L}\dagger}$}; 		\filldraw[color=black, fill=teal!20, very thick](2*\D,0) circle (\R) node{$b_{q^\prime}^{\mathcal{L}}$}; 		\filldraw[color=black, fill=teal!20, very thick](3*\D,0) circle (\R) node{$\rho_2^{\mathcal{R}}$};		 		\node[anchor=south] at (1.2*\R, {0.5*\R}) {\textcolor{green}{$1$}}; 		\node[anchor=north] at (1.3*\R, {0.1*\R}) {\textcolor{green}{$\bar{s}^\p_1$}}; 		\node[anchor=south] at (\D - 1.35*\R, {-0.1*\R}) {\textcolor{green}{$s_1$}}; 		\node[anchor=north] at (1.3*\R, {-0.9*\R}) {\textcolor{blue}{$s_1^\p$}}; 		\node[anchor=south] at (2*\D - 1.3*\R, {0.6*\R}) {\textcolor{blue}{$\bar{s}_1$}}; 		\node[anchor=south] at (1.5*\D, -0.1*\R) {\textcolor{magenta}{$s_1$}}; 		\node[anchor=south] at (2*\D+1.3*\R, {0.4*\R}) {\textcolor{purple}{$s_2$}}; 		\node[anchor=south] at (3*\D-1.3*\R, {0.27*\R}) {\textcolor{purple}{$s_2^\p$}}; 		\node[anchor=north] at (2*\D+1.3*\R, {-0.4*\R}) {\textcolor{purple}{$\bar{s}_2$}}; 		\node[anchor=north] at (3*\D-1.3*\R, -0.27*\R) {\textcolor{purple}{$\bar{s}_2^\p$}}; 	\end{tikzpicture}~. 	\end{align}

\section{\label{app:referm}Refermionization at $z=\nicefrac{1}{2}$}

The bosonic Hamiltonians (\ref{eq:HbSG}) and (\ref{eq:HK}) may be
mapped into Hamiltonians of free fermions at the special point $z=1/2$
(known as the Toulouse point in the context of the Kondo model \citep{hewson_kondo_1993}).
Solving both by means of refermionization is an important consistency
check for the results of Sections \ref{sec:total_rate} and \ref{sec:spec}.
We introduce a fermionic field $\psi$:
\begin{equation}
\psi\left(x\right)=\frac{1}{\sqrt{2\pi a_{0}}}e^{\ii\pi f^{\dagger}f}e^{\ii\phi\left(x\right)},\label{eq:referm}
\end{equation}
where $a_{0}$ is a short distance cutoff scale, and $f^{\dagger}$
is a fermionic creation operator that anticommutes with $\psi$, necessary
to ensure proper anticommutation relations of $\psi$. It is convenient
to unfold the lead of length $\ell$ to a lead of length $2\ell$,
with the impurity placed at $x=0$. This allows us to expand $\psi$
to its eigenmodes, $\psi\left(x\right)=\frac{1}{\sqrt{2\ell}}\sum_{k}\psi_{k}e^{\ii kx}$,
as written in Eq. (\ref{eq:eigenmdoes}).

The Hamiltonians (\ref{eq:HbSG}) and (\ref{eq:HK}) become quadratic
under this mapping. The Kondo Hamiltonian is simpler; identifying
the pseudo-spin operator in Eq. (\ref{eq:HK}) with $f$, $\mathcal{S}_{-}=f$,
we find
\begin{equation}
\mathcal{H}_{\K}=\sum_{k}k\psi_{k}^{\dagger}\psi_{k}-\alpha\sum_{k}\left(f^{\dagger}\psi_{k}+\psi_{k}^{\dagger}f\right),
\end{equation}
where $\alpha=\frac{J}{2}\sqrt{\frac{\pi a_{0}}{\ell}}$. The bsG
Hamiltonian is slightly more complicated, as straightforward refermionization
of the cosine term in Eq. (\ref{eq:HbSG}) would lead to terms that
are linear in $\psi\left(x=0\right)$ and $\psi^{\dagger}\left(x=0\right)$.
To overcome this problem, one may introduce a spin operator in front
of the cosine term, $E_{J}\cos\left(\phi\left(x=0\right)\right)\rightarrow\mathcal{S}_{x}E_{J}\cos\left(\phi\left(x=0\right)\right)$,
with $\mathcal{S}_{x}=f^{\dagger}+f$, so that the new Hamiltonian
commutes with $\mathcal{S}_{x}$ and is therefore equivalent to the
original Hamiltonian for either $\mathcal{S}_{x}=1$ or $\mathcal{S}_{x}=-1$
(in the latter case, up to a shift of $\phi$) \citep{gogolin_bosonization_2004}.
We then find
\begin{equation}
\mathcal{H}_{\bsg}=\sum_{k}k\psi_{k}^{\dagger}\psi_{k}-\alpha\left(f^{\dagger}-f\right)\sum_{k}\left(\psi_{k}+\psi_{k}^{\dagger}\right),
\end{equation}
where now $\alpha=\frac{E_{J}}{2}\sqrt{\frac{\pi a_{0}}{\ell}}$ (we
relate to the prefactors in both models as $\alpha$ for brevity).
The mapping (\ref{eq:referm}) allows us to calculate exact bosonic
correlations functions using the fermionic propagators. Working in
the Keldysh formalism, we define the matrices $D^{\mathrm{ab}}\left(k,k^{\p};\omega\right)$,
which are the temporal Fourier transform of 
\begin{equation}
D^{\mathrm{ab}}\left(k,k^{\p};t\right)=-\ii\left\langle \mathcal{T}_{K}\begin{pmatrix}\psi_{k}^{\mathrm{a}}\left(t\right)\psi_{k^{\p}}^{\mathrm{b}\dagger}\left(0\right) & \psi_{k}^{\mathrm{a}}\left(t\right)\psi_{k^{\p}}^{\mathrm{b}}\left(0\right)\\
\psi_{k}^{\mathrm{a}\dagger}\left(t\right)\psi_{k^{\p}}^{\mathrm{b}\dagger}\left(0\right) & \psi_{k}^{\mathrm{a}\dagger}\left(t\right)\psi_{k^{\p}}^{\mathrm{b}}\left(0\right)
\end{pmatrix}\right\rangle ,
\end{equation}
where $\mathrm{a,b=c,q}$ denote the classical and quantum components
of the fields, and $\mathcal{T}_{K}$ stands for time-ordering along
the Keldysh contour. The retarded, advanced and Keldysh propagators
correspond to $\mathrm{ab=cq,qc,cc}$, respectively, and $D^{\mathrm{qq}}=0$.
The quadratic fermionic Hamiltonians admit the following exact propagators:
\begin{align}
D_{\bsg}^{\mathrm{cq/qc}}\left(k,k^{\p};\omega\right)= & \delta_{kk^{\p}}\begin{pmatrix}\dfrac{1}{\omega-k\pm\ii\eta} & 0\\
0 & \dfrac{1}{\omega+k\pm\ii\eta}
\end{pmatrix}\nonumber \\
 & \hspace{-3cm}+\begin{pmatrix}\dfrac{T_{\bsg}^{\mathrm{cq/qc}}\left(\omega\right)}{\left(\omega-k\pm\ii\eta\right)\left(\omega-k^{\p}\pm\ii\eta\right)} & \dfrac{T_{\bsg}^{\mathrm{cq/qc}}\left(\omega\right)}{\left(\omega-k\pm\ii\eta\right)\left(\omega+k^{\p}\pm\ii\eta\right)}\\
\dfrac{T_{\bsg}^{\mathrm{cq/qc}}\left(\omega\right)}{\left(\omega+k\pm\ii\eta\right)\left(\omega-k^{\p}\pm\ii\eta\right)} & \dfrac{T_{\bsg}^{\mathrm{cq/qc}}\left(\omega\right)}{\left(\omega+k\pm\ii\eta\right)\left(\omega+k^{\p}\pm\ii\eta\right)}
\end{pmatrix}\label{eq:DbSG}
\end{align}
and
\begin{align}
D_{\K}^{\mathrm{cq/qc}}\left(k,k^{\p};\omega\right)= & \delta_{kk^{\p}}\begin{pmatrix}\dfrac{1}{\omega-k\pm\ii\eta} & 0\\
0 & \dfrac{1}{\omega+k\pm\ii\eta}
\end{pmatrix}\nonumber \\
 & \hspace{-3cm}+\begin{pmatrix}\dfrac{T_{\K}^{\mathrm{cq/qc}}\left(\omega\right)}{\left(\omega-k\pm\ii\eta\right)\left(\omega-k^{\p}\pm\ii\eta\right)} & 0\\
0 & \dfrac{T_{\K}^{\mathrm{cq/qc}}\left(\omega\right)}{\left(\omega+k\pm\ii\eta\right)\left(\omega+k^{\p}\pm\ii\eta\right)}
\end{pmatrix},\label{eq:DK}
\end{align}
where $T_{\bsg}^{\mathrm{cq/qc}}\left(\omega\right)=\frac{2\alpha^{2}}{\omega\pm2\ii\Lambda}$
and $T_{\K}^{\mathrm{cq/qc}}\left(\omega\right)=\frac{\alpha^{2}}{\omega\pm\ii\Lambda/2}$,
the plus (minus) signs correspond to $\mathrm{cq}$ ($\mathrm{qc}$),
and $\Lambda=\frac{2\pi\alpha^{2}}{\Delta}$. The fluctuation-dissipation
theorem states that
\begin{equation}
D^{\mathrm{cc}}\left(k,k^{\p};\omega\right)=\tanh\left(\dfrac{\omega}{2T}\right)\left(D^{\mathrm{cq}}\left(k,k^{\p};\omega\right)-D^{\mathrm{qc}}\left(k,k^{\p};\omega\right)\right),\label{eq:Dcc}
\end{equation}
where $T$ is the temperature.

\subsection{The total inelastic decay rate}

As before, we extract the reflection coefficient from the conductance.
Working in the chiral version, there is only a right moving current,
and the conductance has to be calculated between $x^{\p}<0$ and $x>0$.
In the Keldysh formalism, the retarded correlator is given by
\begin{equation}
\mathcal{G}\left(x>0,x^{\p}<0;t\right)=-\frac{1}{8\pi\omega}\left\langle \mathcal{T}_{K}\mathcal{R}^{\mathrm{c}}\left(x,t\right)\mathcal{R}^{\mathrm{q}}\left(x^{\p},0\right)\right\rangle .
\end{equation}
Note the difference of a minus sign compared to Eq. (\ref{eq:GAB}),
due to the unfolding of the half-infinite line. The current operator
may be written in terms of the fermionic modes $\psi_{k}$ by inverting
Eq. (\ref{eq:bqA}) and using the refermionization relation $b_{q}=\sqrt{\frac{\pi}{\ell q}}\sum_{k}\psi_{k}^{\dagger}\psi_{k+q}$:
\begin{equation}
\mathcal{R}\left(x\right)=\frac{\pi}{\ell}\sum_{k_{1},k_{2}}e^{\ii\left(k_{2}-k_{1}\right)x}\psi_{k_{1}}^{\dagger}\psi_{k_{2}}.
\end{equation}
Using Wick's theorem, we find
\begin{align}
\mathcal{G}\left(x>0,x^{\p}<0;\omega\right)= & -\frac{\pi}{16\omega\ell^{2}}\sum_{k_{1},\ldots,k_{4}}e^{\ii\left(k_{2}-k_{1}\right)x}e^{\ii\left(k_{4}-k_{3}\right)x^{\p}}\nonumber \\
 & \hspace{-2.5cm}\times\sum_{\mathrm{a}=\mathrm{c,q}}\int_{-\infty}^{\infty}\frac{\id\Omega}{2\pi}\left[D_{22}^{\mathrm{ca}}\left(k_{1},k_{4};\Omega\right)D_{11}^{\mathrm{c\bar{a}}}\left(k_{2},k_{3};\omega-\Omega\right)\right.\nonumber \\
 & \hspace{-2.5cm}\left.-D_{21}^{\mathrm{ca}}\left(k_{1},k_{3};\Omega\right)D_{12}^{\mathrm{c\bar{a}}}\left(k_{2},k_{4};\omega-\Omega\right)\right],
\end{align}
where $\bar{\mathrm{c}}=\mathrm{q}$ and vice-versa. The sums over
$k$ may be replaced by integrals, $\sum_{k}\rightarrow\frac{\ell}{\pi}\int_{-\infty}^{\infty}\id k$,
and evaluated by closing the contours in the lower ($k_{1},k_{4}$)
or upper ($k_{2},k_{3}$) half planes, according to the signs of the
corresponding exponentials. Using Eqs. (\ref{eq:DbSG})-(\ref{eq:Dcc}),
we readily find
\begin{align}
\mathcal{G}_{\bsg}\left(x>0,x^{\p}<0;\omega\right)= & \frac{e^{\ii\omega\left(x-x^{\p}\right)}}{8\omega}\nonumber \\
 & \hspace{-2.5cm}\times\int_{-\infty}^{\infty}\id\Omega\left[\tanh\left(\frac{\Omega}{2T}\right)+\tanh\left(\frac{\omega-\Omega}{2T}\right)\right]\nonumber \\
 & \hspace{-2.5cm}\times\left(1-\frac{2\ii\Lambda}{\Omega+2\ii\Lambda}-\frac{2\ii\Lambda}{\omega-\Omega+2\ii\Lambda}\right),\nonumber \\
\mathcal{G}_{\K}\left(x>0,x^{\p}<0;\omega\right)= & \frac{e^{\ii\omega\left(x-x^{\p}\right)}}{8\omega}\nonumber \\
 & \hspace{-2.5cm}\times\int_{-\infty}^{\infty}\id\Omega\left[\tanh\left(\frac{\Omega}{2T}\right)+\tanh\left(\frac{\omega-\Omega}{2T}\right)\right]\nonumber \\
 & \hspace{-2.5cm}\times\left(1-\frac{\ii\Lambda}{\Omega+\ii\Lambda/2}\right)\left(1-\frac{\ii\Lambda}{\omega-\Omega+\ii\Lambda/2}\right).
\end{align}
This result generalizes the conductance (and hence the inelastic and
elastic scattering rates) at $z=1/2$ for finite temperature. At $T=0$,
the $\tanh$ factors become step functions, leading to
\begin{align}
r_{\bsg}\left(\omega;z=\frac{1}{2}\right)= & 1-\frac{4\ii\Lambda}{\omega}\log\left(1-\frac{\ii\omega}{2\Lambda}\right),\nonumber \\
r_{\K}\left(\omega;z=\frac{1}{2}\right)= & 1-\frac{2\ii\Lambda}{\omega+\ii\Lambda}\log\left(1-\frac{\ii\omega}{\Lambda/2}\right).
\end{align}
This result recovers Eq. (\ref{eq:r_half_exact}), as we identify
$\Lambda=T_{B}/2$ and $\Lambda=2T_{B}$ for the bsG and Kondo models,
respectively. The inelastic rate is again given by $\gamma\left(\omega\right)=1-\left|r\left(\omega\right)\right|^{2}$,
and its origin can be understood as before, this time using the language
of bosonization: at $z=1/2$, a photon is comprised of a fermionic
particle-hole pair, both members of which scatter elastically off
the boundary, but pick up different phases while doing so, leading
to the splitting of the incoming photon.

\subsection{The energy-resolved inelastic decay spectrum}

Consider the 3-point fully retarded correlator,
\begin{equation}
G_{\omega^{\p}}^{\mathrm{cqq}}\left(t-t^{\p},t-t^{\pp}\right)=-\left\langle \mathcal{T}_{K}n_{q^{\p}}^{\mathrm{c}}\left(t\right)\rho^{\mathrm{q}}\left(\xin,t^{\p}\right)\rho^{\mathrm{q}}\left(\xin,t^{\pp}\right)\right\rangle ,\label{eq:Gcqq_Keldysh_bosons}
\end{equation}
where $\xin<0$. Note that in the chiral representation of the fermionic
field, the density is defined to be right-moving:
\begin{equation}
\rho^{\mathrm{q}}\left(\xin\right)=\frac{1}{\ell}\sum_{\mathrm{a}=\mathrm{c,q}}\sum_{k_{1},k_{2}}e^{\ii\left(k_{2}-k_{1}\right)\xin}\psi_{k_{1}}^{\mathrm{a}\dagger}\psi_{k_{2}}^{\bar{\mathrm{a}}}.
\end{equation}
The bosonic occupation number $n_{q^{\p}}$ is written in terms of
the fermionic operators as
\begin{align}
n_{q^{\p}}^{\mathrm{c}}= & b_{q^{\p}}^{\mathrm{q}\dagger}b_{q^{\p}}^{\mathrm{q}}+b_{q^{\p}}^{\mathrm{c}\dagger}b_{q^{\p}}^{\mathrm{c}}=b_{q^{\p}}^{\mathrm{q}\dagger}b_{q^{\p}}^{\mathrm{q}}\nonumber \\
 & +\dfrac{\pi}{\omega^{\p}\ell}\sum_{k_{1},k_{2}}\sum_{\mathrm{a_{1},a_{2}=c,q}}\psi_{k_{1}+q^{\p}}^{\mathrm{a_{1}}\dagger}\psi_{k_{1}}^{\mathrm{a_{1}}}\psi_{k_{2}}^{\mathrm{a_{2}}\dagger}\psi_{k_{2}+q^{\p}}^{\mathrm{a_{2}}}.
\end{align}
Note that, plugging the above to Eq. (\ref{eq:Gcqq_Keldysh_bosons}),
the $b_{q^{\p}}^{\mathrm{q}\dagger}b_{q^{\p}}^{\mathrm{q}}$ term
leads to an all-quantum Keldysh correlator, which vanishes identically.
We thus have
\begin{align}
G_{\omega^{\p}}^{\mathrm{cqq}}\left(t-t^{\p},t-t^{\pp}\right)= & -\dfrac{\pi}{\omega^{\p}\ell^{3}}\sum_{k_{1},\ldots,k_{6}}\sum_{\mathrm{a_{1},a_{2},a_{3},a_{4}=c,q}}\nonumber \\
 & \hspace{-2cm}\times e^{\ii\left(k_{4}+k_{6}-k_{3}-k_{5}\right)\xin}\nonumber \\
 & \hspace{-2cm}\times\left\langle \mathcal{T}_{K}\psi_{k_{1}+q^{\p}}^{\mathrm{a_{1}}\dagger}\left(t\right)\psi_{k_{1}}^{\mathrm{a_{1}}}\left(t\right)\psi_{k_{2}}^{\mathrm{a_{2}}\dagger}\left(t\right)\psi_{k_{2}+q^{\p}}^{\mathrm{a_{2}}}\left(t\right)\right.\nonumber \\
 & \hspace{-2cm}\times\left.\psi_{k_{3}}^{\mathrm{a_{3}}\dagger}\left(t^{\p}\right)\psi_{k_{4}}^{\bar{\mathrm{a}}_{3}}\left(t^{\p}\right)\psi_{k_{5}}^{\mathrm{a_{4}}\dagger}\left(t^{\pp}\right)\psi_{k_{6}}^{\bar{\mathrm{a}}_{4}}\left(t^{\pp}\right)\right\rangle .\label{eq:Gcqq_Keldysh_fermions}
\end{align}
Again, we use Wick's theorem to calculate the correlator. The sums
over the $k_{i}$'s belonging to the $\rho$ legs are easy to compute,
since they appear once in the correlator. We define the following
matrices,
\begin{equation}
D^{\mathrm{ab}}\left(k;\omega\right)=\sum_{k^{\p}}D^{\mathrm{ab}}\left(k,k^{\p};\omega\right)\begin{pmatrix}e^{-\ii k^{\p}\xin} & 0\\
0 & e^{\ii k^{\p}\xin}
\end{pmatrix},
\end{equation}
which evaluate as
\begin{align}
D_{\bsg}^{\mathrm{cq}}\left(k;\omega\right)= & D_{0}^{\mathrm{cq}}\left(k;\omega\right)\nonumber \\
 & \hspace{-1.5cm}-e^{-\ii\omega\xin}\begin{pmatrix}\dfrac{\tilde{T}_{\bsg}^{\mathrm{cq}}\left(\omega\right)}{\omega-k+\ii\eta} & \dfrac{\tilde{T}_{\bsg}^{\mathrm{cq}}\left(\omega\right)}{\omega-k+\ii\eta}\\
\dfrac{\tilde{T}_{\bsg}^{\mathrm{cq}}\left(\omega\right)}{\omega+k+\ii\eta} & \dfrac{\tilde{T}_{\bsg}^{\mathrm{cq}}\left(\omega\right)}{\omega+k+\ii\eta}
\end{pmatrix},\\
D_{\K}^{\mathrm{cq}}\left(k;\omega\right)= & D_{0}^{\mathrm{cq}}\left(k;\omega\right)\nonumber \\
 & \hspace{-1.5cm}-e^{-\ii\omega\xin}\begin{pmatrix}\dfrac{\tilde{T}_{\K}^{\mathrm{cq}}\left(\omega\right)}{\omega-k+\ii\eta} & 0\\
0 & \dfrac{\tilde{T}_{\K}^{\mathrm{cq}}\left(\omega\right)}{\omega+k+\ii\eta}
\end{pmatrix},\\
D_{0}^{\mathrm{cq}}\left(k;\omega\right)= & \begin{pmatrix}\dfrac{e^{-\ii k\xin}}{\omega-k+\ii\eta} & 0\\
0 & \dfrac{e^{\ii k\xin}}{\omega+k+\ii\eta}
\end{pmatrix},
\end{align}
where $\tilde{T}^{\mathrm{cq}}\left(\omega\right)=\frac{2\pi\ii}{\Delta}T^{\mathrm{cq}}\left(\omega\right)$,
and $D_{\bsg}^{\mathrm{qc}}\left(k;\omega\right)=D_{\K}^{\mathrm{qc}}\left(k;\omega\right)=0$.
The contractions in Eq. (\ref{eq:Gcqq_Keldysh_fermions}) can be expressed
in terms of $D^{\mathrm{ab}}\left(k;\omega\right)$. Introducing the
following definitions,
\begin{align}
\left(\mathrm{r}_{1},\mathrm{r}_{2},\mathrm{r}_{3},\mathrm{r}_{4}\right)= & \left(\mathrm{a}_{3},\bar{\mathrm{a}}_{3},\mathrm{a}_{4},\bar{\mathrm{a}}_{4}\right),\nonumber \\
\left(t_{1},t_{2},t_{3},t_{4}\right)= & \left(t^{\p},t^{\p},t^{\pp},t^{\pp}\right),\nonumber \\
\left(d_{1},d_{2},d_{3},d_{4}\right)= & \left(1,2,1,2\right),\nonumber \\
\left(\nu_{1},\nu_{2},\nu_{3},\nu_{4}\right)= & \left(\Omega_{1},\omega-\Omega_{1},\Omega_{2},-\omega-\Omega_{2}\right),
\end{align}
we find
\begin{align}
G_{\omega^{\p}}^{\mathrm{cqq}}\left(\omega+\ii\eta,-\omega+\ii\eta\right)= & -\dfrac{\pi}{\omega^{\p}\ell^{3}}\sum_{k_{1},k_{2}}\sum_{\mathrm{a_{3},a_{4}=c,q}}\sum_{\P}\zeta_{\mathrm{P}}\nonumber \\
 & \hspace{-3.5cm}\times\int_{-\infty}^{\infty}\dfrac{\id\Omega_{1}}{2\pi}\int_{-\infty}^{\infty}\dfrac{\id\Omega_{2}}{2\pi}\prod_{i=1}^{4}\left[\tanh\left(\dfrac{\nu_{\P i}}{2T}\right)\right]^{\delta_{\mathrm{r}_{\P i}\mathrm{c}}}\nonumber \\
 & \hspace{-3.5cm}\times\left[D_{2d_{\P1}}^{\mathrm{cq}}\left(k_{1}+q^{\p};\nu_{\P1}\right)D_{1d_{\P2}}^{\mathrm{cq}}\left(k_{1};\nu_{\P2}\right)\right.\nonumber \\
 & \hspace{-3.5cm}\times D_{2d_{\P3}}^{\mathrm{cq}}\left(k_{2};\nu_{\P3}\right)D_{1d_{\P4}}^{\mathrm{cq}}\left(k_{2}+q^{\p};\nu_{\P4}\right)\nonumber \\
 & \hspace{-3.5cm}-\delta_{d_{\P1}2}\delta_{d_{\P2}1}\delta_{d_{\P3}2}\delta_{d_{\P4}1}D_{0;22}^{\mathrm{\mathrm{cq}}}\left(k_{1}+q^{\p};\nu_{\P1}\right)\nonumber \\
 & \hspace{-3.5cm}\times\left.D_{0;11}^{\mathrm{\mathrm{cq}}}\left(k_{1};\nu_{\P2}\right)D_{0;22}^{\mathrm{\mathrm{cq}}}\left(k_{2};\nu_{\P3}\right)D_{0;11}^{\mathrm{cq}}\left(k_{2}+q^{\p};\nu_{\P4}\right)\right],\label{eq:Gcqq_Fourier}
\end{align}
where we sum over all permutations $\P$ of $\left\{ 1,2,3,4\right\} $
($\P i$ denotes the permutation value for the index $i$), and $\zeta_{\P}=1\left(-1\right)$
for an even (odd) permutation. Here we only consider connected terms,
where each $\rho$ leg is connected to both $b_{q^{\p}}^{\dagger}$
and $b_{q^{\p}}$ legs --- namely, $\left|\P1-\P2\right|>1$ or $\left|\P3-\P4\right|>1$
(one may show that the disconnected terms are proportional to $\delta\left(\omega-\omega^{\p}\right)$
and therefore correspond to elastic scattering, similarly to Eq. (\ref{eq:gamma_disconnected})).
Note the subtraction of the background term in Eq. (\ref{eq:Gcqq_Fourier}),
similarly to the subtraction of the Kronecker deltas in Eq. (\ref{eq:Rprod}).
Evaluating the sums over $k_{1,2}$, taking care to close the integration
contours in the half planes allowed by the $\xin$ exponentials, and
keeping only contributions that are singular in $\eta$, we find
\begin{align}
G_{\omega^{\p}}^{\mathrm{cqq}}\left(\omega+\ii\eta,-\omega+\ii\eta\right)=\nonumber \\
 & \hspace{-3.5cm}-\dfrac{1}{\omega^{\p}\eta\ell}\sum_{\mathrm{a_{3},a_{4}=c,q}}\sum_{\P}\zeta_{\mathrm{P}}\int_{-\infty}^{\infty}\id\Omega_{1}\int_{-\infty}^{\infty}\id\Omega_{2}\nonumber \\
 & \hspace{-3.5cm}\times\delta\left(\Omega_{1}+\Omega_{2}+\omega^{\p}\right)\prod_{i=1}^{4}\left[\tanh\left(\dfrac{\nu_{\P i}}{2T}\right)\right]^{\delta_{\mathrm{r}_{\P i}\mathrm{c}}}\nonumber \\
 & \hspace{-3.5cm}\times\left[E^{\mathrm{cq}}\left(\nu_{\P1},\nu_{\P2};d_{\P1},d_{\P2}\right)E^{\mathrm{cq}}\left(\nu_{\P3},\nu_{\P4};d_{\P3},d_{\P4}\right)\right.\nonumber \\
 & \hspace{-3.5cm}\left.-\delta_{d_{\P1}2}\delta_{d_{\P2}1}\delta_{d_{\P3}2}\delta_{d_{\P4}1}\right],
\end{align}
where
\begin{align}
E_{\bsg}^{\mathrm{cq}}\left(\nu_{1},\nu_{2};d_{1},d_{2}\right)= & \left(\delta_{d_{1}2}-\tilde{T}_{\bsg}^{\mathrm{cq}}\left(\nu_{1}\right)\right)\left(\delta_{d_{2}1}-\tilde{T}_{\bsg}^{\mathrm{cq}}\left(\nu_{2}\right)\right),\nonumber \\
E_{\K}^{\mathrm{cq}}\left(\nu_{1},\nu_{2};d_{1},d_{2}\right)= & \delta_{d_{1}2}\delta_{d_{2}1}\left(1-\tilde{T}_{\K}^{\mathrm{cq}}\left(\nu_{1}\right)\right)\left(1-\tilde{T}_{\K}^{\mathrm{cq}}\left(\nu_{2}\right)\right).
\end{align}
We can now plug $G_{\omega^{\p}}^{\mathrm{cqq}}$ into Eq. (\ref{eq:inel_spec}),
thus obtaining the spectrum at $z=1/2$ at finite temperatures. At
$T=0$, the $\tanh$ functions become step functions; direct inspection
of the sums over the permutations shows that
\begin{align}
\gamma_{\bsg}\left(\omega^{\prime}\lvert\omega\right)= & \frac{-2}{\omega\omega^{\prime}}\int_{0}^{\omega-\omega^{\prime}}\id\Omega\left[\left(1-\tilde{T}_{\bsg}^{\mathrm{cq}}\left(\Omega\right)-\tilde{T}_{\bsg}^{\mathrm{cq}}\left(\omega-\Omega\right)\right)\right.\nonumber \\
 & \hspace{-1.5cm}\left.\times\left(1-\tilde{T}_{\bsg}^{\mathrm{cq}}\left(\Omega+\omega^{\prime}-\omega\right)-\tilde{T}_{\bsg}^{\mathrm{cq}}\left(-\Omega-\omega^{\prime}\right)\right)-1\right],\nonumber \\
\gamma_{\K}\left(\omega^{\prime}\lvert\omega\right)= & \frac{-2}{\omega\omega^{\prime}}\int_{0}^{\omega-\omega^{\prime}}\id\Omega\left[\left(1-\tilde{T}_{\K}^{\mathrm{cq}}\left(\Omega\right)\right)\left(1-\tilde{T}_{\K}^{\mathrm{cq}}\left(\omega-\Omega\right)\right)\right.\nonumber \\
 & \hspace{-1.5cm}\left.\times\left(1-\tilde{T}_{\K}^{\mathrm{cq}}\left(\Omega+\omega^{\prime}-\omega\right)\right)\left(1-\tilde{T}_{\K}^{\mathrm{cq}}\left(-\Omega-\omega^{\prime}\right)\right)-1\right],\label{eq:spec_referm}
\end{align}
which are both identical to the form factors results, and are simple
enough to lead to closed analytical expressions. Indeed, one may consider
Eq. (\ref{eq:gamma_1^(1)}), and use the form factor in Eq. (\ref{eq:f+-_z_half})
and the reflection matrices in Eqs. (\ref{eq:R_bSG_z_half}) or (\ref{eq:Rs_K})
to recover the expressions above.

\bibliography{Inelastic_decay_from_integrability_bibliography}

\begin{thebibliography}{107}%
\makeatletter
\providecommand \@ifxundefined [1]{%
 \@ifx{#1\undefined}
}%
\providecommand \@ifnum [1]{%
 \ifnum #1\expandafter \@firstoftwo
 \else \expandafter \@secondoftwo
 \fi
}%
\providecommand \@ifx [1]{%
 \ifx #1\expandafter \@firstoftwo
 \else \expandafter \@secondoftwo
 \fi
}%
\providecommand \natexlab [1]{#1}%
\providecommand \enquote  [1]{``#1''}%
\providecommand \bibnamefont  [1]{#1}%
\providecommand \bibfnamefont [1]{#1}%
\providecommand \citenamefont [1]{#1}%
\providecommand \href@noop [0]{\@secondoftwo}%
\providecommand \href [0]{\begingroup \@sanitize@url \@href}%
\providecommand \@href[1]{\@@startlink{#1}\@@href}%
\providecommand \@@href[1]{\endgroup#1\@@endlink}%
\providecommand \@sanitize@url [0]{\catcode `\\12\catcode `\$12\catcode
  `\&12\catcode `\#12\catcode `\^12\catcode `\_12\catcode `\%12\relax}%
\providecommand \@@startlink[1]{}%
\providecommand \@@endlink[0]{}%
\providecommand \url  [0]{\begingroup\@sanitize@url \@url }%
\providecommand \@url [1]{\endgroup\@href {#1}{\urlprefix }}%
\providecommand \urlprefix  [0]{URL }%
\providecommand \Eprint [0]{\href }%
\providecommand \doibase [0]{https://doi.org/}%
\providecommand \selectlanguage [0]{\@gobble}%
\providecommand \bibinfo  [0]{\@secondoftwo}%
\providecommand \bibfield  [0]{\@secondoftwo}%
\providecommand \translation [1]{[#1]}%
\providecommand \BibitemOpen [0]{}%
\providecommand \bibitemStop [0]{}%
\providecommand \bibitemNoStop [0]{.\EOS\space}%
\providecommand \EOS [0]{\spacefactor3000\relax}%
\providecommand \BibitemShut  [1]{\csname bibitem#1\endcsname}%
\let\auto@bib@innerbib\@empty
\bibitem [{\citenamefont {Bethe}(1931)}]{bethe_zur_1931}%
  \BibitemOpen
  \bibfield  {author} {\bibinfo {author} {\bibfnamefont {H.}~\bibnamefont
  {Bethe}},\ }\bibfield  {title} {\bibinfo {title} {Zur {Theorie} der
  {Metalle}},\ }\href {https://doi.org/10.1007/BF01341708} {\bibfield
  {journal} {\bibinfo  {journal} {Zeitschrift für Physik}\ }\textbf {\bibinfo
  {volume} {71}},\ \bibinfo {pages} {205} (\bibinfo {year} {1931})}\BibitemShut
  {NoStop}%
\bibitem [{\citenamefont {Lieb}\ and\ \citenamefont
  {Liniger}(1963)}]{lieb_exact_1963}%
  \BibitemOpen
  \bibfield  {author} {\bibinfo {author} {\bibfnamefont {E.~H.}\ \bibnamefont
  {Lieb}}\ and\ \bibinfo {author} {\bibfnamefont {W.}~\bibnamefont {Liniger}},\
  }\bibfield  {title} {\bibinfo {title} {Exact {Analysis} of an {Interacting}
  {Bose} {Gas}. {I}. {The} {General} {Solution} and the {Ground} {State}},\
  }\href {https://doi.org/10.1103/PhysRev.130.1605} {\bibfield  {journal}
  {\bibinfo  {journal} {Physical Review}\ }\textbf {\bibinfo {volume} {130}},\
  \bibinfo {pages} {1605} (\bibinfo {year} {1963})}\BibitemShut {NoStop}%
\bibitem [{\citenamefont {Yang}\ and\ \citenamefont
  {Yang}(1966)}]{yang_one-dimensional_1966}%
  \BibitemOpen
  \bibfield  {author} {\bibinfo {author} {\bibfnamefont {C.~N.}\ \bibnamefont
  {Yang}}\ and\ \bibinfo {author} {\bibfnamefont {C.~P.}\ \bibnamefont
  {Yang}},\ }\bibfield  {title} {\bibinfo {title} {One-{Dimensional} {Chain} of
  {Anisotropic} {Spin}-{Spin} {Interactions}. {II}. {Properties} of the
  {Ground}-{State} {Energy} {Per} {Lattice} {Site} for an {Infinite}
  {System}},\ }\href {https://doi.org/10.1103/PhysRev.150.327} {\bibfield
  {journal} {\bibinfo  {journal} {Physical Review}\ }\textbf {\bibinfo {volume}
  {150}},\ \bibinfo {pages} {327} (\bibinfo {year} {1966})}\BibitemShut
  {NoStop}%
\bibitem [{\citenamefont {Yang}(1967)}]{yang_exact_1967}%
  \BibitemOpen
  \bibfield  {author} {\bibinfo {author} {\bibfnamefont {C.~N.}\ \bibnamefont
  {Yang}},\ }\bibfield  {title} {\bibinfo {title} {Some {Exact} {Results} for
  the {Many}-{Body} {Problem} in one {Dimension} with {Repulsive}
  {Delta}-{Function} {Interaction}},\ }\href
  {https://doi.org/10.1103/PhysRevLett.19.1312} {\bibfield  {journal} {\bibinfo
   {journal} {Phys. Rev. Lett.}\ }\textbf {\bibinfo {volume} {19}},\ \bibinfo
  {pages} {1312} (\bibinfo {year} {1967})}\BibitemShut {NoStop}%
\bibitem [{\citenamefont {Andrei}\ \emph {et~al.}(1983)\citenamefont {Andrei},
  \citenamefont {Furuya},\ and\ \citenamefont
  {Lowenstein}}]{andrei_solution_1983}%
  \BibitemOpen
  \bibfield  {author} {\bibinfo {author} {\bibfnamefont {N.}~\bibnamefont
  {Andrei}}, \bibinfo {author} {\bibfnamefont {K.}~\bibnamefont {Furuya}},\
  and\ \bibinfo {author} {\bibfnamefont {J.~H.}\ \bibnamefont {Lowenstein}},\
  }\bibfield  {title} {\bibinfo {title} {Solution of the {Kondo} problem},\
  }\href {https://doi.org/10.1103/RevModPhys.55.331} {\bibfield  {journal}
  {\bibinfo  {journal} {Reviews of Modern Physics}\ }\textbf {\bibinfo {volume}
  {55}},\ \bibinfo {pages} {331} (\bibinfo {year} {1983})}\BibitemShut
  {NoStop}%
\bibitem [{\citenamefont {Gaudin}(2014)}]{caux_bethe_2014}%
  \BibitemOpen
  \bibfield  {author} {\bibinfo {author} {\bibfnamefont {M.}~\bibnamefont
  {Gaudin}},\ }\href {https://doi.org/10.1017/CBO9781107053885} {\emph
  {\bibinfo {title} {The {Bethe} {Wavefunction}}}}\ (\bibinfo  {publisher}
  {Cambridge University Press},\ \bibinfo {address} {Cambridge},\ \bibinfo
  {year} {2014})\BibitemShut {NoStop}%
\bibitem [{\citenamefont {Baxter}(1985)}]{baxter_exactly_1985}%
  \BibitemOpen
  \bibfield  {author} {\bibinfo {author} {\bibfnamefont {R.~J.}\ \bibnamefont
  {Baxter}},\ }\bibfield  {title} {\bibinfo {title} {Exactly {Solved} {Models}
  in {Statistical} {Mechanics}},\ }in\ \href
  {https://doi.org/10.1142/9789814415255_0002} {\emph {\bibinfo {booktitle}
  {Integrable {Systems} in {Statistical} {Mechanics}}}},\ \bibinfo {series}
  {Series on {Advances} in {Statistical} {Mechanics}}, Vol.\ \bibinfo {volume}
  {Volume 1}\ (\bibinfo  {publisher} {World {Scientific}},\ \bibinfo {year}
  {1985})\ pp.\ \bibinfo {pages} {5--63}\BibitemShut {NoStop}%
\bibitem [{\citenamefont {Korepin}\ \emph {et~al.}(1993)\citenamefont
  {Korepin}, \citenamefont {Bogoliubov},\ and\ \citenamefont
  {Izergin}}]{korepin_quantum_1993}%
  \BibitemOpen
  \bibfield  {author} {\bibinfo {author} {\bibfnamefont {V.~E.}\ \bibnamefont
  {Korepin}}, \bibinfo {author} {\bibfnamefont {N.~M.}\ \bibnamefont
  {Bogoliubov}},\ and\ \bibinfo {author} {\bibfnamefont {A.~G.}\ \bibnamefont
  {Izergin}},\ }\href {https://doi.org/10.1017/CBO9780511628832} {\emph
  {\bibinfo {title} {Quantum {Inverse} {Scattering} {Method} and {Correlation}
  {Functions}}}},\ Cambridge {Monographs} on {Mathematical} {Physics}\
  (\bibinfo  {publisher} {Cambridge University Press},\ \bibinfo {address}
  {Cambridge},\ \bibinfo {year} {1993})\BibitemShut {NoStop}%
\bibitem [{\citenamefont {Zamolodchikov}\ and\ \citenamefont
  {Zamolodchikov}(1979)}]{zamolodchikov_factorized_1979}%
  \BibitemOpen
  \bibfield  {author} {\bibinfo {author} {\bibfnamefont {A.~B.}\ \bibnamefont
  {Zamolodchikov}}\ and\ \bibinfo {author} {\bibfnamefont {A.~B.}\ \bibnamefont
  {Zamolodchikov}},\ }\bibfield  {title} {{\selectlanguage {English}\bibinfo
  {title} {Factorized {S}-matrices in two dimensions as the exact solutions of
  certain relativistic quantum field theory models}},\ }\href
  {https://doi.org/10.1016/0003-4916(79)90391-9} {\bibfield  {journal}
  {\bibinfo  {journal} {Annals of Physics}\ }\textbf {\bibinfo {volume}
  {120}},\ \bibinfo {pages} {253} (\bibinfo {year} {1979})}\BibitemShut
  {NoStop}%
\bibitem [{\citenamefont {Faddeev}(1995)}]{faddeev_quantum_1995}%
  \BibitemOpen
  \bibfield  {author} {\bibinfo {author} {\bibfnamefont {L.~D.}\ \bibnamefont
  {Faddeev}},\ }\bibfield  {title} {\bibinfo {title} {Quantum {Completely}
  {Integrable} {Models} in {Field} {Theory}},\ }in\ \href
  {https://doi.org/10.1142/9789812815453_0007} {\emph {\bibinfo {booktitle} {40
  {Years} in {Mathematical} {Physics}}}},\ \bibinfo {series} {World
  {Scientific} {Series} in 20th {Century} {Mathematics}}, Vol.\ \bibinfo
  {volume} {Volume 2}\ (\bibinfo  {publisher} {World {Scientific}},\ \bibinfo
  {year} {1995})\ pp.\ \bibinfo {pages} {187--235}\BibitemShut {NoStop}%
\bibitem [{\citenamefont {Mussardo}(2020)}]{mussardo_statistical_2020}%
  \BibitemOpen
  \bibfield  {author} {\bibinfo {author} {\bibfnamefont {G.}~\bibnamefont
  {Mussardo}},\ }\href {https://doi.org/10.1093/oso/9780198788102.001.0001}
  {\emph {\bibinfo {title} {Statistical {Field} {Theory}: an {Introduction} to
  {Exactly} {Solved} {Models} in {Statistical} {Physics}}}},\ Cambridge
  {Monographs} on {Mathematical} {Physics}\ (\bibinfo  {publisher} {Oxford
  University Press},\ \bibinfo {address} {Oxford},\ \bibinfo {year}
  {2020})\BibitemShut {NoStop}%
\bibitem [{\citenamefont {Caux}(2016)}]{caux_quench_2016}%
  \BibitemOpen
  \bibfield  {author} {\bibinfo {author} {\bibfnamefont {J.-S.}\ \bibnamefont
  {Caux}},\ }\bibfield  {title} {{\selectlanguage {English}\bibinfo {title}
  {The {Quench} {Action}}},\ }\href
  {https://doi.org/10.1088/1742-5468/2016/06/064006} {\bibfield  {journal}
  {\bibinfo  {journal} {Journal of Statistical Mechanics: Theory and
  Experiment}\ }\textbf {\bibinfo {volume} {2016}},\ \bibinfo {pages} {064006}
  (\bibinfo {year} {2016})}\BibitemShut {NoStop}%
\bibitem [{\citenamefont {Bertini}\ \emph {et~al.}(2016)\citenamefont
  {Bertini}, \citenamefont {Collura}, \citenamefont {De~Nardis},\ and\
  \citenamefont {Fagotti}}]{bertini_transport_2016}%
  \BibitemOpen
  \bibfield  {author} {\bibinfo {author} {\bibfnamefont {B.}~\bibnamefont
  {Bertini}}, \bibinfo {author} {\bibfnamefont {M.}~\bibnamefont {Collura}},
  \bibinfo {author} {\bibfnamefont {J.}~\bibnamefont {De~Nardis}},\ and\
  \bibinfo {author} {\bibfnamefont {M.}~\bibnamefont {Fagotti}},\ }\bibfield
  {title} {\bibinfo {title} {Transport in {Out}-of-{Equilibrium} {$XXZ$}
  {Chains}: {Exact} {Profiles} of {Charges} and {Currents}},\ }\href
  {https://doi.org/10.1103/PhysRevLett.117.207201} {\bibfield  {journal}
  {\bibinfo  {journal} {Phys. Rev. Lett.}\ }\textbf {\bibinfo {volume} {117}},\
  \bibinfo {pages} {207201} (\bibinfo {year} {2016})}\BibitemShut {NoStop}%
\bibitem [{\citenamefont {Castro-Alvaredo}\ \emph {et~al.}(2016)\citenamefont
  {Castro-Alvaredo}, \citenamefont {Doyon},\ and\ \citenamefont
  {Yoshimura}}]{castro-alvaredo_emergent_2016}%
  \BibitemOpen
  \bibfield  {author} {\bibinfo {author} {\bibfnamefont {O.~A.}\ \bibnamefont
  {Castro-Alvaredo}}, \bibinfo {author} {\bibfnamefont {B.}~\bibnamefont
  {Doyon}},\ and\ \bibinfo {author} {\bibfnamefont {T.}~\bibnamefont
  {Yoshimura}},\ }\bibfield  {title} {\bibinfo {title} {Emergent
  {Hydrodynamics} in {Integrable} {Quantum} {Systems} {Out} of {Equilibrium}},\
  }\href {https://doi.org/10.1103/PhysRevX.6.041065} {\bibfield  {journal}
  {\bibinfo  {journal} {Phys. Rev. X}\ }\textbf {\bibinfo {volume} {6}},\
  \bibinfo {pages} {041065} (\bibinfo {year} {2016})}\BibitemShut {NoStop}%
\bibitem [{\citenamefont {Calabrese}\ \emph {et~al.}(2016)\citenamefont
  {Calabrese}, \citenamefont {Essler},\ and\ \citenamefont
  {Mussardo}}]{calabrese_introduction_2016}%
  \BibitemOpen
  \bibfield  {author} {\bibinfo {author} {\bibfnamefont {P.}~\bibnamefont
  {Calabrese}}, \bibinfo {author} {\bibfnamefont {F.~H.~L.}\ \bibnamefont
  {Essler}},\ and\ \bibinfo {author} {\bibfnamefont {G.}~\bibnamefont
  {Mussardo}},\ }\bibfield  {title} {{\selectlanguage {English}\bibinfo {title}
  {Introduction to {Quantum} {Integrability} in {Out} of {Equilibrium}
  {Systems}}},\ }\href {https://doi.org/10.1088/1742-5468/2016/06/064001}
  {\bibfield  {journal} {\bibinfo  {journal} {Journal of Statistical Mechanics:
  Theory and Experiment}\ }\textbf {\bibinfo {volume} {2016}},\ \bibinfo
  {pages} {064001} (\bibinfo {year} {2016})}\BibitemShut {NoStop}%
\bibitem [{\citenamefont {Kinoshita}\ \emph {et~al.}(2006)\citenamefont
  {Kinoshita}, \citenamefont {Wenger},\ and\ \citenamefont
  {Weiss}}]{kinoshita_quantum_2006}%
  \BibitemOpen
  \bibfield  {author} {\bibinfo {author} {\bibfnamefont {T.}~\bibnamefont
  {Kinoshita}}, \bibinfo {author} {\bibfnamefont {T.}~\bibnamefont {Wenger}},\
  and\ \bibinfo {author} {\bibfnamefont {D.~S.}\ \bibnamefont {Weiss}},\
  }\bibfield  {title} {{\selectlanguage {English}\bibinfo {title} {A quantum
  {Newton}'s cradle}},\ }\href {https://doi.org/10.1038/nature04693} {\bibfield
   {journal} {\bibinfo  {journal} {Nature}\ }\textbf {\bibinfo {volume}
  {440}},\ \bibinfo {pages} {900} (\bibinfo {year} {2006})}\BibitemShut
  {NoStop}%
\bibitem [{\citenamefont {Guan}\ \emph {et~al.}(2013)\citenamefont {Guan},
  \citenamefont {Batchelor},\ and\ \citenamefont {Lee}}]{guan_fermi_2013}%
  \BibitemOpen
  \bibfield  {author} {\bibinfo {author} {\bibfnamefont {X.-W.}\ \bibnamefont
  {Guan}}, \bibinfo {author} {\bibfnamefont {M.~T.}\ \bibnamefont
  {Batchelor}},\ and\ \bibinfo {author} {\bibfnamefont {C.}~\bibnamefont
  {Lee}},\ }\bibfield  {title} {{\selectlanguage {English}\bibinfo {title}
  {Fermi gases in one dimension: {From} {Bethe} ansatz to experiments}},\
  }\href {https://doi.org/10.1103/RevModPhys.85.1633} {\bibfield  {journal}
  {\bibinfo  {journal} {Reviews of Modern Physics}\ }\textbf {\bibinfo {volume}
  {85}},\ \bibinfo {pages} {1633} (\bibinfo {year} {2013})}\BibitemShut
  {NoStop}%
\bibitem [{\citenamefont {Langen}\ \emph {et~al.}(2015)\citenamefont {Langen},
  \citenamefont {Erne}, \citenamefont {Geiger}, \citenamefont {Rauer},
  \citenamefont {Schweigler}, \citenamefont {Kuhnert}, \citenamefont
  {Rohringer}, \citenamefont {Mazets}, \citenamefont {Gasenzer},\ and\
  \citenamefont {Schmiedmayer}}]{langen_experimental_2015}%
  \BibitemOpen
  \bibfield  {author} {\bibinfo {author} {\bibfnamefont {T.}~\bibnamefont
  {Langen}}, \bibinfo {author} {\bibfnamefont {S.}~\bibnamefont {Erne}},
  \bibinfo {author} {\bibfnamefont {R.}~\bibnamefont {Geiger}}, \bibinfo
  {author} {\bibfnamefont {B.}~\bibnamefont {Rauer}}, \bibinfo {author}
  {\bibfnamefont {T.}~\bibnamefont {Schweigler}}, \bibinfo {author}
  {\bibfnamefont {M.}~\bibnamefont {Kuhnert}}, \bibinfo {author} {\bibfnamefont
  {W.}~\bibnamefont {Rohringer}}, \bibinfo {author} {\bibfnamefont {I.~E.}\
  \bibnamefont {Mazets}}, \bibinfo {author} {\bibfnamefont {T.}~\bibnamefont
  {Gasenzer}},\ and\ \bibinfo {author} {\bibfnamefont {J.}~\bibnamefont
  {Schmiedmayer}},\ }\bibfield  {title} {\bibinfo {title} {Experimental
  observation of a generalized {Gibbs} ensemble},\ }\href
  {https://doi.org/10.1126/science.1257026} {\bibfield  {journal} {\bibinfo
  {journal} {Science}\ }\textbf {\bibinfo {volume} {348}},\ \bibinfo {pages}
  {207} (\bibinfo {year} {2015})}\BibitemShut {NoStop}%
\bibitem [{\citenamefont {Bouchoule}\ and\ \citenamefont
  {Dubail}(2022)}]{bouchoule_generalized_2022}%
  \BibitemOpen
  \bibfield  {author} {\bibinfo {author} {\bibfnamefont {I.}~\bibnamefont
  {Bouchoule}}\ and\ \bibinfo {author} {\bibfnamefont {J.}~\bibnamefont
  {Dubail}},\ }\bibfield  {title} {{\selectlanguage {English}\bibinfo {title}
  {Generalized hydrodynamics in the one-dimensional {Bose} gas: theory and
  experiments}},\ }\href {https://doi.org/10.1088/1742-5468/ac3659} {\bibfield
  {journal} {\bibinfo  {journal} {Journal of Statistical Mechanics: Theory and
  Experiment}\ }\textbf {\bibinfo {volume} {2022}},\ \bibinfo {pages} {014003}
  (\bibinfo {year} {2022})}\BibitemShut {NoStop}%
\bibitem [{\citenamefont {Xu}\ \emph {et~al.}(2018)\citenamefont {Xu},
  \citenamefont {Chen}, \citenamefont {Zeng}, \citenamefont {Zhang},
  \citenamefont {Song}, \citenamefont {Liu}, \citenamefont {Guo}, \citenamefont
  {Zhang}, \citenamefont {Xu}, \citenamefont {Deng}, \citenamefont {Huang},
  \citenamefont {Wang}, \citenamefont {Zhu}, \citenamefont {Zheng},\ and\
  \citenamefont {Fan}}]{xu_MBL_2018}%
  \BibitemOpen
  \bibfield  {author} {\bibinfo {author} {\bibfnamefont {K.}~\bibnamefont
  {Xu}}, \bibinfo {author} {\bibfnamefont {J.-J.}\ \bibnamefont {Chen}},
  \bibinfo {author} {\bibfnamefont {Y.}~\bibnamefont {Zeng}}, \bibinfo {author}
  {\bibfnamefont {Y.-R.}\ \bibnamefont {Zhang}}, \bibinfo {author}
  {\bibfnamefont {C.}~\bibnamefont {Song}}, \bibinfo {author} {\bibfnamefont
  {W.}~\bibnamefont {Liu}}, \bibinfo {author} {\bibfnamefont {Q.}~\bibnamefont
  {Guo}}, \bibinfo {author} {\bibfnamefont {P.}~\bibnamefont {Zhang}}, \bibinfo
  {author} {\bibfnamefont {D.}~\bibnamefont {Xu}}, \bibinfo {author}
  {\bibfnamefont {H.}~\bibnamefont {Deng}}, \bibinfo {author} {\bibfnamefont
  {K.}~\bibnamefont {Huang}}, \bibinfo {author} {\bibfnamefont
  {H.}~\bibnamefont {Wang}}, \bibinfo {author} {\bibfnamefont {X.}~\bibnamefont
  {Zhu}}, \bibinfo {author} {\bibfnamefont {D.}~\bibnamefont {Zheng}},\ and\
  \bibinfo {author} {\bibfnamefont {H.}~\bibnamefont {Fan}},\ }\bibfield
  {title} {\bibinfo {title} {Emulating many-body localization with a
  superconducting quantum processor},\ }\href
  {https://doi.org/10.1103/PhysRevLett.120.050507} {\bibfield  {journal}
  {\bibinfo  {journal} {Phys. Rev. Lett.}\ }\textbf {\bibinfo {volume} {120}},\
  \bibinfo {pages} {050507} (\bibinfo {year} {2018})}\BibitemShut {NoStop}%
\bibitem [{\citenamefont {Zha}\ \emph {et~al.}(2020)\citenamefont {Zha},
  \citenamefont {Bastidas}, \citenamefont {Gong}, \citenamefont {Wu},
  \citenamefont {Rong}, \citenamefont {Yang}, \citenamefont {Ye}, \citenamefont
  {Li}, \citenamefont {Zhu}, \citenamefont {Wang}, \citenamefont {Zhao},
  \citenamefont {Liang}, \citenamefont {Lin}, \citenamefont {Xu}, \citenamefont
  {Peng}, \citenamefont {Schmiedmayer}, \citenamefont {Nemoto}, \citenamefont
  {Deng}, \citenamefont {Munro}, \citenamefont {Zhu},\ and\ \citenamefont
  {Pan}}]{zha_MBL_2020}%
  \BibitemOpen
  \bibfield  {author} {\bibinfo {author} {\bibfnamefont {C.}~\bibnamefont
  {Zha}}, \bibinfo {author} {\bibfnamefont {V.~M.}\ \bibnamefont {Bastidas}},
  \bibinfo {author} {\bibfnamefont {M.}~\bibnamefont {Gong}}, \bibinfo {author}
  {\bibfnamefont {Y.}~\bibnamefont {Wu}}, \bibinfo {author} {\bibfnamefont
  {H.}~\bibnamefont {Rong}}, \bibinfo {author} {\bibfnamefont {R.}~\bibnamefont
  {Yang}}, \bibinfo {author} {\bibfnamefont {Y.}~\bibnamefont {Ye}}, \bibinfo
  {author} {\bibfnamefont {S.}~\bibnamefont {Li}}, \bibinfo {author}
  {\bibfnamefont {Q.}~\bibnamefont {Zhu}}, \bibinfo {author} {\bibfnamefont
  {S.}~\bibnamefont {Wang}}, \bibinfo {author} {\bibfnamefont {Y.}~\bibnamefont
  {Zhao}}, \bibinfo {author} {\bibfnamefont {F.}~\bibnamefont {Liang}},
  \bibinfo {author} {\bibfnamefont {J.}~\bibnamefont {Lin}}, \bibinfo {author}
  {\bibfnamefont {Y.}~\bibnamefont {Xu}}, \bibinfo {author} {\bibfnamefont
  {C.-Z.}\ \bibnamefont {Peng}}, \bibinfo {author} {\bibfnamefont
  {J.}~\bibnamefont {Schmiedmayer}}, \bibinfo {author} {\bibfnamefont
  {K.}~\bibnamefont {Nemoto}}, \bibinfo {author} {\bibfnamefont
  {H.}~\bibnamefont {Deng}}, \bibinfo {author} {\bibfnamefont {W.~J.}\
  \bibnamefont {Munro}}, \bibinfo {author} {\bibfnamefont {X.}~\bibnamefont
  {Zhu}},\ and\ \bibinfo {author} {\bibfnamefont {J.-W.}\ \bibnamefont {Pan}},\
  }\bibfield  {title} {\bibinfo {title} {Ergodic-localized junctions in a
  periodically driven spin chain},\ }\href
  {https://doi.org/10.1103/PhysRevLett.125.170503} {\bibfield  {journal}
  {\bibinfo  {journal} {Phys. Rev. Lett.}\ }\textbf {\bibinfo {volume} {125}},\
  \bibinfo {pages} {170503} (\bibinfo {year} {2020})}\BibitemShut {NoStop}%
\bibitem [{\citenamefont {Gong}\ \emph {et~al.}(2021)\citenamefont {Gong},
  \citenamefont {de~Moraes~Neto}, \citenamefont {Zha}, \citenamefont {Wu},
  \citenamefont {Rong}, \citenamefont {Ye}, \citenamefont {Li}, \citenamefont
  {Zhu}, \citenamefont {Wang}, \citenamefont {Zhao}, \citenamefont {Liang},
  \citenamefont {Lin}, \citenamefont {Xu}, \citenamefont {Peng}, \citenamefont
  {Deng}, \citenamefont {Bayat}, \citenamefont {Zhu},\ and\ \citenamefont
  {Pan}}]{gong_MBL_2021}%
  \BibitemOpen
  \bibfield  {author} {\bibinfo {author} {\bibfnamefont {M.}~\bibnamefont
  {Gong}}, \bibinfo {author} {\bibfnamefont {G.~D.}\ \bibnamefont
  {de~Moraes~Neto}}, \bibinfo {author} {\bibfnamefont {C.}~\bibnamefont {Zha}},
  \bibinfo {author} {\bibfnamefont {Y.}~\bibnamefont {Wu}}, \bibinfo {author}
  {\bibfnamefont {H.}~\bibnamefont {Rong}}, \bibinfo {author} {\bibfnamefont
  {Y.}~\bibnamefont {Ye}}, \bibinfo {author} {\bibfnamefont {S.}~\bibnamefont
  {Li}}, \bibinfo {author} {\bibfnamefont {Q.}~\bibnamefont {Zhu}}, \bibinfo
  {author} {\bibfnamefont {S.}~\bibnamefont {Wang}}, \bibinfo {author}
  {\bibfnamefont {Y.}~\bibnamefont {Zhao}}, \bibinfo {author} {\bibfnamefont
  {F.}~\bibnamefont {Liang}}, \bibinfo {author} {\bibfnamefont
  {J.}~\bibnamefont {Lin}}, \bibinfo {author} {\bibfnamefont {Y.}~\bibnamefont
  {Xu}}, \bibinfo {author} {\bibfnamefont {C.-Z.}\ \bibnamefont {Peng}},
  \bibinfo {author} {\bibfnamefont {H.}~\bibnamefont {Deng}}, \bibinfo {author}
  {\bibfnamefont {A.}~\bibnamefont {Bayat}}, \bibinfo {author} {\bibfnamefont
  {X.}~\bibnamefont {Zhu}},\ and\ \bibinfo {author} {\bibfnamefont {J.-W.}\
  \bibnamefont {Pan}},\ }\bibfield  {title} {\bibinfo {title} {Experimental
  characterization of the quantum many-body localization transition},\ }\href
  {https://doi.org/10.1103/PhysRevResearch.3.033043} {\bibfield  {journal}
  {\bibinfo  {journal} {Phys. Rev. Res.}\ }\textbf {\bibinfo {volume} {3}},\
  \bibinfo {pages} {033043} (\bibinfo {year} {2021})}\BibitemShut {NoStop}%
\bibitem [{\citenamefont {Guo}\ \emph {et~al.}(2021)\citenamefont {Guo},
  \citenamefont {Cheng}, \citenamefont {Sun}, \citenamefont {Song},
  \citenamefont {Li}, \citenamefont {Wang}, \citenamefont {Ren}, \citenamefont
  {Dong}, \citenamefont {Zheng}, \citenamefont {Zhang}, \citenamefont
  {Mondaini}, \citenamefont {Fan},\ and\ \citenamefont {Wang}}]{guo_MBL_2021}%
  \BibitemOpen
  \bibfield  {author} {\bibinfo {author} {\bibfnamefont {Q.}~\bibnamefont
  {Guo}}, \bibinfo {author} {\bibfnamefont {C.}~\bibnamefont {Cheng}}, \bibinfo
  {author} {\bibfnamefont {Z.-H.}\ \bibnamefont {Sun}}, \bibinfo {author}
  {\bibfnamefont {Z.}~\bibnamefont {Song}}, \bibinfo {author} {\bibfnamefont
  {H.}~\bibnamefont {Li}}, \bibinfo {author} {\bibfnamefont {Z.}~\bibnamefont
  {Wang}}, \bibinfo {author} {\bibfnamefont {W.}~\bibnamefont {Ren}}, \bibinfo
  {author} {\bibfnamefont {H.}~\bibnamefont {Dong}}, \bibinfo {author}
  {\bibfnamefont {D.}~\bibnamefont {Zheng}}, \bibinfo {author} {\bibfnamefont
  {Y.-R.}\ \bibnamefont {Zhang}}, \bibinfo {author} {\bibfnamefont
  {R.}~\bibnamefont {Mondaini}}, \bibinfo {author} {\bibfnamefont
  {H.}~\bibnamefont {Fan}},\ and\ \bibinfo {author} {\bibfnamefont
  {H.}~\bibnamefont {Wang}},\ }\bibfield  {title} {\bibinfo {title}
  {Observation of energy-resolved many-body localization},\ }\href
  {https://doi.org/10.1038/s41567-020-1035-1} {\bibfield  {journal} {\bibinfo
  {journal} {Nature Physics}\ }\textbf {\bibinfo {volume} {17}},\ \bibinfo
  {pages} {234} (\bibinfo {year} {2021})}\BibitemShut {NoStop}%
\bibitem [{\citenamefont {Chiaro}\ \emph {et~al.}(2022)\citenamefont {Chiaro},
  \citenamefont {Neill}, \citenamefont {Bohrdt}, \citenamefont {Filippone},
  \citenamefont {Arute}, \citenamefont {Arya}, \citenamefont {Babbush},
  \citenamefont {Bacon}, \citenamefont {Bardin}, \citenamefont {Barends},
  \citenamefont {Boixo}, \citenamefont {Buell}, \citenamefont {Burkett},
  \citenamefont {Chen}, \citenamefont {Chen}, \citenamefont {Collins},
  \citenamefont {Dunsworth}, \citenamefont {Farhi}, \citenamefont {Fowler},
  \citenamefont {Foxen}, \citenamefont {Gidney}, \citenamefont {Giustina},
  \citenamefont {Harrigan}, \citenamefont {Huang}, \citenamefont {Isakov},
  \citenamefont {Jeffrey}, \citenamefont {Jiang}, \citenamefont {Kafri},
  \citenamefont {Kechedzhi}, \citenamefont {Kelly}, \citenamefont {Klimov},
  \citenamefont {Korotkov}, \citenamefont {Kostritsa}, \citenamefont
  {Landhuis}, \citenamefont {Lucero}, \citenamefont {McClean}, \citenamefont
  {Mi}, \citenamefont {Megrant}, \citenamefont {Mohseni}, \citenamefont
  {Mutus}, \citenamefont {McEwen}, \citenamefont {Naaman}, \citenamefont
  {Neeley}, \citenamefont {Niu}, \citenamefont {Petukhov}, \citenamefont
  {Quintana}, \citenamefont {Rubin}, \citenamefont {Sank}, \citenamefont
  {Satzinger}, \citenamefont {White}, \citenamefont {Yao}, \citenamefont {Yeh},
  \citenamefont {Zalcman}, \citenamefont {Smelyanskiy}, \citenamefont {Neven},
  \citenamefont {Gopalakrishnan}, \citenamefont {Abanin}, \citenamefont {Knap},
  \citenamefont {Martinis},\ and\ \citenamefont {Roushan}}]{google_MBL_2022}%
  \BibitemOpen
  \bibfield  {author} {\bibinfo {author} {\bibfnamefont {B.}~\bibnamefont
  {Chiaro}}, \bibinfo {author} {\bibfnamefont {C.}~\bibnamefont {Neill}},
  \bibinfo {author} {\bibfnamefont {A.}~\bibnamefont {Bohrdt}}, \bibinfo
  {author} {\bibfnamefont {M.}~\bibnamefont {Filippone}}, \bibinfo {author}
  {\bibfnamefont {F.}~\bibnamefont {Arute}}, \bibinfo {author} {\bibfnamefont
  {K.}~\bibnamefont {Arya}}, \bibinfo {author} {\bibfnamefont {R.}~\bibnamefont
  {Babbush}}, \bibinfo {author} {\bibfnamefont {D.}~\bibnamefont {Bacon}},
  \bibinfo {author} {\bibfnamefont {J.}~\bibnamefont {Bardin}}, \bibinfo
  {author} {\bibfnamefont {R.}~\bibnamefont {Barends}}, \bibinfo {author}
  {\bibfnamefont {S.}~\bibnamefont {Boixo}}, \bibinfo {author} {\bibfnamefont
  {D.}~\bibnamefont {Buell}}, \bibinfo {author} {\bibfnamefont
  {B.}~\bibnamefont {Burkett}}, \bibinfo {author} {\bibfnamefont
  {Y.}~\bibnamefont {Chen}}, \bibinfo {author} {\bibfnamefont {Z.}~\bibnamefont
  {Chen}}, \bibinfo {author} {\bibfnamefont {R.}~\bibnamefont {Collins}},
  \bibinfo {author} {\bibfnamefont {A.}~\bibnamefont {Dunsworth}}, \bibinfo
  {author} {\bibfnamefont {E.}~\bibnamefont {Farhi}}, \bibinfo {author}
  {\bibfnamefont {A.}~\bibnamefont {Fowler}}, \bibinfo {author} {\bibfnamefont
  {B.}~\bibnamefont {Foxen}}, \bibinfo {author} {\bibfnamefont
  {C.}~\bibnamefont {Gidney}}, \bibinfo {author} {\bibfnamefont
  {M.}~\bibnamefont {Giustina}}, \bibinfo {author} {\bibfnamefont
  {M.}~\bibnamefont {Harrigan}}, \bibinfo {author} {\bibfnamefont
  {T.}~\bibnamefont {Huang}}, \bibinfo {author} {\bibfnamefont
  {S.}~\bibnamefont {Isakov}}, \bibinfo {author} {\bibfnamefont
  {E.}~\bibnamefont {Jeffrey}}, \bibinfo {author} {\bibfnamefont
  {Z.}~\bibnamefont {Jiang}}, \bibinfo {author} {\bibfnamefont
  {D.}~\bibnamefont {Kafri}}, \bibinfo {author} {\bibfnamefont
  {K.}~\bibnamefont {Kechedzhi}}, \bibinfo {author} {\bibfnamefont
  {J.}~\bibnamefont {Kelly}}, \bibinfo {author} {\bibfnamefont
  {P.}~\bibnamefont {Klimov}}, \bibinfo {author} {\bibfnamefont
  {A.}~\bibnamefont {Korotkov}}, \bibinfo {author} {\bibfnamefont
  {F.}~\bibnamefont {Kostritsa}}, \bibinfo {author} {\bibfnamefont
  {D.}~\bibnamefont {Landhuis}}, \bibinfo {author} {\bibfnamefont
  {E.}~\bibnamefont {Lucero}}, \bibinfo {author} {\bibfnamefont
  {J.}~\bibnamefont {McClean}}, \bibinfo {author} {\bibfnamefont
  {X.}~\bibnamefont {Mi}}, \bibinfo {author} {\bibfnamefont {A.}~\bibnamefont
  {Megrant}}, \bibinfo {author} {\bibfnamefont {M.}~\bibnamefont {Mohseni}},
  \bibinfo {author} {\bibfnamefont {J.}~\bibnamefont {Mutus}}, \bibinfo
  {author} {\bibfnamefont {M.}~\bibnamefont {McEwen}}, \bibinfo {author}
  {\bibfnamefont {O.}~\bibnamefont {Naaman}}, \bibinfo {author} {\bibfnamefont
  {M.}~\bibnamefont {Neeley}}, \bibinfo {author} {\bibfnamefont
  {M.}~\bibnamefont {Niu}}, \bibinfo {author} {\bibfnamefont {A.}~\bibnamefont
  {Petukhov}}, \bibinfo {author} {\bibfnamefont {C.}~\bibnamefont {Quintana}},
  \bibinfo {author} {\bibfnamefont {N.}~\bibnamefont {Rubin}}, \bibinfo
  {author} {\bibfnamefont {D.}~\bibnamefont {Sank}}, \bibinfo {author}
  {\bibfnamefont {K.}~\bibnamefont {Satzinger}}, \bibinfo {author}
  {\bibfnamefont {T.}~\bibnamefont {White}}, \bibinfo {author} {\bibfnamefont
  {Z.}~\bibnamefont {Yao}}, \bibinfo {author} {\bibfnamefont {P.}~\bibnamefont
  {Yeh}}, \bibinfo {author} {\bibfnamefont {A.}~\bibnamefont {Zalcman}},
  \bibinfo {author} {\bibfnamefont {V.}~\bibnamefont {Smelyanskiy}}, \bibinfo
  {author} {\bibfnamefont {H.}~\bibnamefont {Neven}}, \bibinfo {author}
  {\bibfnamefont {S.}~\bibnamefont {Gopalakrishnan}}, \bibinfo {author}
  {\bibfnamefont {D.}~\bibnamefont {Abanin}}, \bibinfo {author} {\bibfnamefont
  {M.}~\bibnamefont {Knap}}, \bibinfo {author} {\bibfnamefont {J.}~\bibnamefont
  {Martinis}},\ and\ \bibinfo {author} {\bibfnamefont {P.}~\bibnamefont
  {Roushan}},\ }\bibfield  {title} {\bibinfo {title} {Direct measurement of
  nonlocal interactions in the many-body localized phase},\ }\href
  {https://doi.org/10.1103/PhysRevResearch.4.013148} {\bibfield  {journal}
  {\bibinfo  {journal} {Phys. Rev. Res.}\ }\textbf {\bibinfo {volume} {4}},\
  \bibinfo {pages} {013148} (\bibinfo {year} {2022})}\BibitemShut {NoStop}%
\bibitem [{\citenamefont {Roy}\ and\ \citenamefont
  {Saleur}(2019)}]{roy_quantum_2019}%
  \BibitemOpen
  \bibfield  {author} {\bibinfo {author} {\bibfnamefont {A.}~\bibnamefont
  {Roy}}\ and\ \bibinfo {author} {\bibfnamefont {H.}~\bibnamefont {Saleur}},\
  }\bibfield  {title} {{\selectlanguage {English}\bibinfo {title} {Quantum
  electronic circuit simulation of generalized sine-{Gordon} models}},\ }\href
  {https://doi.org/10.1103/PhysRevB.100.155425} {\bibfield  {journal} {\bibinfo
   {journal} {Phys. Rev. B}\ }\textbf {\bibinfo {volume} {100}},\ \bibinfo
  {pages} {155425} (\bibinfo {year} {2019})}\BibitemShut {NoStop}%
\bibitem [{\citenamefont {Roy}\ \emph {et~al.}(2021)\citenamefont {Roy},
  \citenamefont {Schuricht}, \citenamefont {Hauschild}, \citenamefont
  {Pollmann},\ and\ \citenamefont {Saleur}}]{roy_quantum_2021}%
  \BibitemOpen
  \bibfield  {author} {\bibinfo {author} {\bibfnamefont {A.}~\bibnamefont
  {Roy}}, \bibinfo {author} {\bibfnamefont {D.}~\bibnamefont {Schuricht}},
  \bibinfo {author} {\bibfnamefont {J.}~\bibnamefont {Hauschild}}, \bibinfo
  {author} {\bibfnamefont {F.}~\bibnamefont {Pollmann}},\ and\ \bibinfo
  {author} {\bibfnamefont {H.}~\bibnamefont {Saleur}},\ }\bibfield  {title}
  {{\selectlanguage {English}\bibinfo {title} {The quantum sine-{Gordon} model
  with quantum circuits}},\ }\href
  {https://doi.org/10.1016/j.nuclphysb.2021.115445} {\bibfield  {journal}
  {\bibinfo  {journal} {Nuclear Physics B}\ }\textbf {\bibinfo {volume}
  {968}},\ \bibinfo {pages} {115445} (\bibinfo {year} {2021})}\BibitemShut
  {NoStop}%
\bibitem [{\citenamefont {Roy}\ and\ \citenamefont
  {Lukyanov}(2023)}]{roy_soliton_2023}%
  \BibitemOpen
  \bibfield  {author} {\bibinfo {author} {\bibfnamefont {A.}~\bibnamefont
  {Roy}}\ and\ \bibinfo {author} {\bibfnamefont {S.~L.}\ \bibnamefont
  {Lukyanov}},\ }\bibfield  {title} {\bibinfo {title} {Soliton confinement in a
  quantum circuit},\ }\href {https://doi.org/10.1038/s41467-023-43107-3}
  {\bibfield  {journal} {\bibinfo  {journal} {Nature Communications}\ }\textbf
  {\bibinfo {volume} {14}},\ \bibinfo {pages} {7433} (\bibinfo {year}
  {2023})}\BibitemShut {NoStop}%
\bibitem [{\citenamefont {Kuzmin}\ \emph
  {et~al.}(2019{\natexlab{a}})\citenamefont {Kuzmin}, \citenamefont {Mencia},
  \citenamefont {Grabon}, \citenamefont {Mehta}, \citenamefont {Lin},\ and\
  \citenamefont {Manucharyan}}]{kuzmin_quantum_2019}%
  \BibitemOpen
  \bibfield  {author} {\bibinfo {author} {\bibfnamefont {R.}~\bibnamefont
  {Kuzmin}}, \bibinfo {author} {\bibfnamefont {R.}~\bibnamefont {Mencia}},
  \bibinfo {author} {\bibfnamefont {N.}~\bibnamefont {Grabon}}, \bibinfo
  {author} {\bibfnamefont {N.}~\bibnamefont {Mehta}}, \bibinfo {author}
  {\bibfnamefont {Y.-H.}\ \bibnamefont {Lin}},\ and\ \bibinfo {author}
  {\bibfnamefont {V.~E.}\ \bibnamefont {Manucharyan}},\ }\bibfield  {title}
  {{\selectlanguage {English}\bibinfo {title} {Quantum electrodynamics of a
  superconductor–insulator phase transition}},\ }\href
  {https://doi.org/10.1038/s41567-019-0553-1} {\bibfield  {journal} {\bibinfo
  {journal} {Nature Physics}\ }\textbf {\bibinfo {volume} {15}},\ \bibinfo
  {pages} {930} (\bibinfo {year} {2019}{\natexlab{a}})}\BibitemShut {NoStop}%
\bibitem [{\citenamefont {Léger}\ \emph {et~al.}(2019)\citenamefont {Léger},
  \citenamefont {Puertas-Martínez}, \citenamefont {Bharadwaj}, \citenamefont
  {Dassonneville}, \citenamefont {Delaforce}, \citenamefont {Foroughi},
  \citenamefont {Milchakov}, \citenamefont {Planat}, \citenamefont {Buisson},
  \citenamefont {Naud}, \citenamefont {Hasch-Guichard}, \citenamefont
  {Florens}, \citenamefont {Snyman},\ and\ \citenamefont
  {Roch}}]{leger_observation_2019}%
  \BibitemOpen
  \bibfield  {author} {\bibinfo {author} {\bibfnamefont {S.}~\bibnamefont
  {Léger}}, \bibinfo {author} {\bibfnamefont {J.}~\bibnamefont
  {Puertas-Martínez}}, \bibinfo {author} {\bibfnamefont {K.}~\bibnamefont
  {Bharadwaj}}, \bibinfo {author} {\bibfnamefont {R.}~\bibnamefont
  {Dassonneville}}, \bibinfo {author} {\bibfnamefont {J.}~\bibnamefont
  {Delaforce}}, \bibinfo {author} {\bibfnamefont {F.}~\bibnamefont {Foroughi}},
  \bibinfo {author} {\bibfnamefont {V.}~\bibnamefont {Milchakov}}, \bibinfo
  {author} {\bibfnamefont {L.}~\bibnamefont {Planat}}, \bibinfo {author}
  {\bibfnamefont {O.}~\bibnamefont {Buisson}}, \bibinfo {author} {\bibfnamefont
  {C.}~\bibnamefont {Naud}}, \bibinfo {author} {\bibfnamefont {W.}~\bibnamefont
  {Hasch-Guichard}}, \bibinfo {author} {\bibfnamefont {S.}~\bibnamefont
  {Florens}}, \bibinfo {author} {\bibfnamefont {I.}~\bibnamefont {Snyman}},\
  and\ \bibinfo {author} {\bibfnamefont {N.}~\bibnamefont {Roch}},\ }\bibfield
  {title} {{\selectlanguage {English}\bibinfo {title} {Observation of quantum
  many-body effects due to zero point fluctuations in superconducting
  circuits}},\ }\href {https://doi.org/10.1038/s41467-019-13199-x} {\bibfield
  {journal} {\bibinfo  {journal} {Nature Communications}\ }\textbf {\bibinfo
  {volume} {10}},\ \bibinfo {pages} {5259} (\bibinfo {year}
  {2019})}\BibitemShut {NoStop}%
\bibitem [{\citenamefont {Puertas~Mart{\'i}nez}\ \emph
  {et~al.}(2019)\citenamefont {Puertas~Mart{\'i}nez}, \citenamefont
  {L{\'e}ger}, \citenamefont {Gheeraert}, \citenamefont {Dassonneville},
  \citenamefont {Planat}, \citenamefont {Foroughi}, \citenamefont {Krupko},
  \citenamefont {Buisson}, \citenamefont {Naud}, \citenamefont
  {Hasch-Guichard}, \citenamefont {Florens}, \citenamefont {Snyman},\ and\
  \citenamefont {Roch}}]{puertas_martinez_tunable_2019}%
  \BibitemOpen
  \bibfield  {author} {\bibinfo {author} {\bibfnamefont {J.}~\bibnamefont
  {Puertas~Mart{\'i}nez}}, \bibinfo {author} {\bibfnamefont {S.}~\bibnamefont
  {L{\'e}ger}}, \bibinfo {author} {\bibfnamefont {N.}~\bibnamefont
  {Gheeraert}}, \bibinfo {author} {\bibfnamefont {R.}~\bibnamefont
  {Dassonneville}}, \bibinfo {author} {\bibfnamefont {L.}~\bibnamefont
  {Planat}}, \bibinfo {author} {\bibfnamefont {F.}~\bibnamefont {Foroughi}},
  \bibinfo {author} {\bibfnamefont {Y.}~\bibnamefont {Krupko}}, \bibinfo
  {author} {\bibfnamefont {O.}~\bibnamefont {Buisson}}, \bibinfo {author}
  {\bibfnamefont {C.}~\bibnamefont {Naud}}, \bibinfo {author} {\bibfnamefont
  {W.}~\bibnamefont {Hasch-Guichard}}, \bibinfo {author} {\bibfnamefont
  {S.}~\bibnamefont {Florens}}, \bibinfo {author} {\bibfnamefont
  {I.}~\bibnamefont {Snyman}},\ and\ \bibinfo {author} {\bibfnamefont
  {N.}~\bibnamefont {Roch}},\ }\bibfield  {title} {\bibinfo {title} {A tunable
  josephson platform to explore many-body quantum optics in circuit-qed},\
  }\href {https://doi.org/10.1038/s41534-018-0104-0} {\bibfield  {journal}
  {\bibinfo  {journal} {npj Quantum Information}\ }\textbf {\bibinfo {volume}
  {5}},\ \bibinfo {pages} {19} (\bibinfo {year} {2019})}\BibitemShut {NoStop}%
\bibitem [{\citenamefont {Goldstein}\ \emph {et~al.}(2013)\citenamefont
  {Goldstein}, \citenamefont {Devoret}, \citenamefont {Houzet},\ and\
  \citenamefont {Glazman}}]{goldstein_inelastic_2013}%
  \BibitemOpen
  \bibfield  {author} {\bibinfo {author} {\bibfnamefont {M.}~\bibnamefont
  {Goldstein}}, \bibinfo {author} {\bibfnamefont {M.~H.}\ \bibnamefont
  {Devoret}}, \bibinfo {author} {\bibfnamefont {M.}~\bibnamefont {Houzet}},\
  and\ \bibinfo {author} {\bibfnamefont {L.~I.}\ \bibnamefont {Glazman}},\
  }\bibfield  {title} {\bibinfo {title} {Inelastic {Microwave} {Photon}
  {Scattering} off a {Quantum} {Impurity} in a {Josephson}-{Junction}
  {Array}},\ }\href {https://doi.org/10.1103/PhysRevLett.110.017002} {\bibfield
   {journal} {\bibinfo  {journal} {Phys. Rev. Lett.}\ }\textbf {\bibinfo
  {volume} {110}},\ \bibinfo {pages} {017002} (\bibinfo {year}
  {2013})}\BibitemShut {NoStop}%
\bibitem [{\citenamefont {Le~Hur}(2012)}]{le_hur_kondo_2012}%
  \BibitemOpen
  \bibfield  {author} {\bibinfo {author} {\bibfnamefont {K.}~\bibnamefont
  {Le~Hur}},\ }\bibfield  {title} {\bibinfo {title} {Kondo resonance of a
  microwave photon},\ }\href {https://doi.org/10.1103/PhysRevB.85.140506}
  {\bibfield  {journal} {\bibinfo  {journal} {Phys. Rev. B}\ }\textbf {\bibinfo
  {volume} {85}},\ \bibinfo {pages} {140506} (\bibinfo {year}
  {2012})}\BibitemShut {NoStop}%
\bibitem [{\citenamefont {Kuzmin}\ \emph {et~al.}(2021)\citenamefont {Kuzmin},
  \citenamefont {Grabon}, \citenamefont {Mehta}, \citenamefont {Burshtein},
  \citenamefont {Goldstein}, \citenamefont {Houzet}, \citenamefont {Glazman},\
  and\ \citenamefont {Manucharyan}}]{kuzmin_inelastic_2021}%
  \BibitemOpen
  \bibfield  {author} {\bibinfo {author} {\bibfnamefont {R.}~\bibnamefont
  {Kuzmin}}, \bibinfo {author} {\bibfnamefont {N.}~\bibnamefont {Grabon}},
  \bibinfo {author} {\bibfnamefont {N.}~\bibnamefont {Mehta}}, \bibinfo
  {author} {\bibfnamefont {A.}~\bibnamefont {Burshtein}}, \bibinfo {author}
  {\bibfnamefont {M.}~\bibnamefont {Goldstein}}, \bibinfo {author}
  {\bibfnamefont {M.}~\bibnamefont {Houzet}}, \bibinfo {author} {\bibfnamefont
  {L.}~\bibnamefont {Glazman}},\ and\ \bibinfo {author} {\bibfnamefont
  {V.}~\bibnamefont {Manucharyan}},\ }\bibfield  {title} {\bibinfo {title}
  {Inelastic {Scattering} of a {Photon} by a {Quantum} {Phase} {Slip}},\ }\href
  {https://doi.org/10.1103/PhysRevLett.126.197701} {\bibfield  {journal}
  {\bibinfo  {journal} {Phys. Rev. Lett.}\ }\textbf {\bibinfo {volume} {126}},\
  \bibinfo {pages} {197701} (\bibinfo {year} {2021})}\BibitemShut {NoStop}%
\bibitem [{\citenamefont {Mehta}\ \emph {et~al.}(2023)\citenamefont {Mehta},
  \citenamefont {Kuzmin}, \citenamefont {Ciuti},\ and\ \citenamefont
  {Manucharyan}}]{mehta_down-conversion_2023}%
  \BibitemOpen
  \bibfield  {author} {\bibinfo {author} {\bibfnamefont {N.}~\bibnamefont
  {Mehta}}, \bibinfo {author} {\bibfnamefont {R.}~\bibnamefont {Kuzmin}},
  \bibinfo {author} {\bibfnamefont {C.}~\bibnamefont {Ciuti}},\ and\ \bibinfo
  {author} {\bibfnamefont {V.~E.}\ \bibnamefont {Manucharyan}},\ }\bibfield
  {title} {{\selectlanguage {English}\bibinfo {title} {Down-conversion of a
  single photon as a probe of many-body localization}},\ }\href
  {https://doi.org/10.1038/s41586-022-05615-y} {\bibfield  {journal} {\bibinfo
  {journal} {Nature}\ }\textbf {\bibinfo {volume} {613}},\ \bibinfo {pages}
  {650} (\bibinfo {year} {2023})}\BibitemShut {NoStop}%
\bibitem [{\citenamefont {Léger}\ \emph {et~al.}(2023)\citenamefont {Léger},
  \citenamefont {Sépulcre}, \citenamefont {Fraudet}, \citenamefont {Buisson},
  \citenamefont {Naud}, \citenamefont {Hasch-Guichard}, \citenamefont
  {Florens}, \citenamefont {Snyman}, \citenamefont {Basko},\ and\ \citenamefont
  {Roch}}]{leger_revealing_2022}%
  \BibitemOpen
  \bibfield  {author} {\bibinfo {author} {\bibfnamefont {S.}~\bibnamefont
  {Léger}}, \bibinfo {author} {\bibfnamefont {T.}~\bibnamefont {Sépulcre}},
  \bibinfo {author} {\bibfnamefont {D.}~\bibnamefont {Fraudet}}, \bibinfo
  {author} {\bibfnamefont {O.}~\bibnamefont {Buisson}}, \bibinfo {author}
  {\bibfnamefont {C.}~\bibnamefont {Naud}}, \bibinfo {author} {\bibfnamefont
  {W.}~\bibnamefont {Hasch-Guichard}}, \bibinfo {author} {\bibfnamefont
  {S.}~\bibnamefont {Florens}}, \bibinfo {author} {\bibfnamefont
  {I.}~\bibnamefont {Snyman}}, \bibinfo {author} {\bibfnamefont {D.~M.}\
  \bibnamefont {Basko}},\ and\ \bibinfo {author} {\bibfnamefont
  {N.}~\bibnamefont {Roch}},\ }\bibfield  {title} {\bibinfo {title} {{Revealing
  the finite-frequency response of a bosonic quantum impurity}},\ }\href
  {https://doi.org/10.21468/SciPostPhys.14.5.130} {\bibfield  {journal}
  {\bibinfo  {journal} {SciPost Phys.}\ }\textbf {\bibinfo {volume} {14}},\
  \bibinfo {pages} {130} (\bibinfo {year} {2023})}\BibitemShut {NoStop}%
\bibitem [{\citenamefont {Kuzmin}\ \emph {et~al.}(2023)\citenamefont {Kuzmin},
  \citenamefont {Mehta}, \citenamefont {Grabon}, \citenamefont {Mencia},
  \citenamefont {Burshtein}, \citenamefont {Goldstein},\ and\ \citenamefont
  {Manucharyan}}]{kuzmin_observation_2023}%
  \BibitemOpen
  \bibfield  {author} {\bibinfo {author} {\bibfnamefont {R.}~\bibnamefont
  {Kuzmin}}, \bibinfo {author} {\bibfnamefont {N.}~\bibnamefont {Mehta}},
  \bibinfo {author} {\bibfnamefont {N.}~\bibnamefont {Grabon}}, \bibinfo
  {author} {\bibfnamefont {R.}~\bibnamefont {Mencia}}, \bibinfo {author}
  {\bibfnamefont {A.}~\bibnamefont {Burshtein}}, \bibinfo {author}
  {\bibfnamefont {M.}~\bibnamefont {Goldstein}},\ and\ \bibinfo {author}
  {\bibfnamefont {V.}~\bibnamefont {Manucharyan}},\ }\href
  {https://arxiv.org/abs/2304.05806} {\bibinfo {title} {Observation of
  {Schmid}-{Bulgadaev} {Dissipative} {Quantum} {Phase} {Transition}}} (\bibinfo
  {year} {2023}),\ \bibinfo {note} {arXiv:2304.05806}\BibitemShut {NoStop}%
\bibitem [{\citenamefont {Ghoshal}\ and\ \citenamefont
  {Zamolodchikov}(1994)}]{ghoshal_boundary_1994}%
  \BibitemOpen
  \bibfield  {author} {\bibinfo {author} {\bibfnamefont {S.}~\bibnamefont
  {Ghoshal}}\ and\ \bibinfo {author} {\bibfnamefont {A.}~\bibnamefont
  {Zamolodchikov}},\ }\bibfield  {title} {\bibinfo {title} {Boundary s matrix
  and boundary state in two-dimensional integrable quantum field theory},\
  }\href {https://doi.org/10.1142/S0217751X94001552} {\bibfield  {journal}
  {\bibinfo  {journal} {International Journal of Modern Physics A}\ }\textbf
  {\bibinfo {volume} {09}},\ \bibinfo {pages} {3841} (\bibinfo {year}
  {1994})}\BibitemShut {NoStop}%
\bibitem [{\citenamefont {Gheeraert}\ \emph {et~al.}(2018)\citenamefont
  {Gheeraert}, \citenamefont {Zhang}, \citenamefont {S\'epulcre}, \citenamefont
  {Bera}, \citenamefont {Roch}, \citenamefont {Baranger},\ and\ \citenamefont
  {Florens}}]{gheeraert_particle_2018}%
  \BibitemOpen
  \bibfield  {author} {\bibinfo {author} {\bibfnamefont {N.}~\bibnamefont
  {Gheeraert}}, \bibinfo {author} {\bibfnamefont {X.~H.~H.}\ \bibnamefont
  {Zhang}}, \bibinfo {author} {\bibfnamefont {T.}~\bibnamefont {S\'epulcre}},
  \bibinfo {author} {\bibfnamefont {S.}~\bibnamefont {Bera}}, \bibinfo {author}
  {\bibfnamefont {N.}~\bibnamefont {Roch}}, \bibinfo {author} {\bibfnamefont
  {H.~U.}\ \bibnamefont {Baranger}},\ and\ \bibinfo {author} {\bibfnamefont
  {S.}~\bibnamefont {Florens}},\ }\bibfield  {title} {\bibinfo {title}
  {Particle production in ultrastrong-coupling waveguide qed},\ }\href
  {https://doi.org/10.1103/PhysRevA.98.043816} {\bibfield  {journal} {\bibinfo
  {journal} {Phys. Rev. A}\ }\textbf {\bibinfo {volume} {98}},\ \bibinfo
  {pages} {043816} (\bibinfo {year} {2018})}\BibitemShut {NoStop}%
\bibitem [{\citenamefont {Houzet}\ and\ \citenamefont
  {Glazman}(2020)}]{houzet_critical_2020}%
  \BibitemOpen
  \bibfield  {author} {\bibinfo {author} {\bibfnamefont {M.}~\bibnamefont
  {Houzet}}\ and\ \bibinfo {author} {\bibfnamefont {L.}~\bibnamefont
  {Glazman}},\ }\bibfield  {title} {\bibinfo {title} {Critical {Fluorescence}
  of a {Transmon} at the {Schmid} {Transition}},\ }\href
  {https://doi.org/10.1103/PhysRevLett.125.267701} {\bibfield  {journal}
  {\bibinfo  {journal} {Phys. Rev. Lett.}\ }\textbf {\bibinfo {volume} {125}},\
  \bibinfo {pages} {267701} (\bibinfo {year} {2020})}\BibitemShut {NoStop}%
\bibitem [{\citenamefont {Burshtein}\ \emph {et~al.}(2021)\citenamefont
  {Burshtein}, \citenamefont {Kuzmin}, \citenamefont {Manucharyan},\ and\
  \citenamefont {Goldstein}}]{burshtein_photon-instanton_2021}%
  \BibitemOpen
  \bibfield  {author} {\bibinfo {author} {\bibfnamefont {A.}~\bibnamefont
  {Burshtein}}, \bibinfo {author} {\bibfnamefont {R.}~\bibnamefont {Kuzmin}},
  \bibinfo {author} {\bibfnamefont {V.~E.}\ \bibnamefont {Manucharyan}},\ and\
  \bibinfo {author} {\bibfnamefont {M.}~\bibnamefont {Goldstein}},\ }\bibfield
  {title} {\bibinfo {title} {Photon-{Instanton} {Collider} {Implemented} by a
  {Superconducting} {Circuit}},\ }\href
  {https://doi.org/10.1103/PhysRevLett.126.137701} {\bibfield  {journal}
  {\bibinfo  {journal} {Phys. Rev. Lett.}\ }\textbf {\bibinfo {volume} {126}},\
  \bibinfo {pages} {137701} (\bibinfo {year} {2021})}\BibitemShut {NoStop}%
\bibitem [{\citenamefont {Mehta}\ \emph {et~al.}(2022)\citenamefont {Mehta},
  \citenamefont {Ciuti}, \citenamefont {Kuzmin},\ and\ \citenamefont
  {Manucharyan}}]{mehta_theory_2022}%
  \BibitemOpen
  \bibfield  {author} {\bibinfo {author} {\bibfnamefont {N.}~\bibnamefont
  {Mehta}}, \bibinfo {author} {\bibfnamefont {C.}~\bibnamefont {Ciuti}},
  \bibinfo {author} {\bibfnamefont {R.}~\bibnamefont {Kuzmin}},\ and\ \bibinfo
  {author} {\bibfnamefont {V.~E.}\ \bibnamefont {Manucharyan}},\ }\href
  {https://doi.org/10.48550/arXiv.2210.14681} {\bibinfo {title} {Theory of
  strong down-conversion in multi-mode cavity and circuit {QED}}} (\bibinfo
  {year} {2022}),\ \bibinfo {note} {arXiv:2210.14681}\BibitemShut {NoStop}%
\bibitem [{\citenamefont {Houzet}\ \emph {et~al.}(2023)\citenamefont {Houzet},
  \citenamefont {Yamamoto},\ and\ \citenamefont
  {Glazman}}]{houzet_microwave_2023}%
  \BibitemOpen
  \bibfield  {author} {\bibinfo {author} {\bibfnamefont {M.}~\bibnamefont
  {Houzet}}, \bibinfo {author} {\bibfnamefont {T.}~\bibnamefont {Yamamoto}},\
  and\ \bibinfo {author} {\bibfnamefont {L.~I.}\ \bibnamefont {Glazman}},\
  }\href {https://arxiv.org/abs/2308.16072} {\bibinfo {title} {Microwave
  spectroscopy of {Schmid} transition}} (\bibinfo {year} {2023}),\ \bibinfo
  {note} {arXiv:2308.16072}\BibitemShut {NoStop}%
\bibitem [{\citenamefont {Mallet}\ \emph {et~al.}(2006)\citenamefont {Mallet},
  \citenamefont {Ericsson}, \citenamefont {Mailly}, \citenamefont
  {\"Unl\"ubayir}, \citenamefont {Reuter}, \citenamefont {Melnikov},
  \citenamefont {Wieck}, \citenamefont {Micklitz}, \citenamefont {Rosch},
  \citenamefont {Costi}, \citenamefont {Saminadayar},\ and\ \citenamefont
  {B\"auerle}}]{mallet_scaling_2006}%
  \BibitemOpen
  \bibfield  {author} {\bibinfo {author} {\bibfnamefont {F.}~\bibnamefont
  {Mallet}}, \bibinfo {author} {\bibfnamefont {J.}~\bibnamefont {Ericsson}},
  \bibinfo {author} {\bibfnamefont {D.}~\bibnamefont {Mailly}}, \bibinfo
  {author} {\bibfnamefont {S.}~\bibnamefont {\"Unl\"ubayir}}, \bibinfo {author}
  {\bibfnamefont {D.}~\bibnamefont {Reuter}}, \bibinfo {author} {\bibfnamefont
  {A.}~\bibnamefont {Melnikov}}, \bibinfo {author} {\bibfnamefont {A.~D.}\
  \bibnamefont {Wieck}}, \bibinfo {author} {\bibfnamefont {T.}~\bibnamefont
  {Micklitz}}, \bibinfo {author} {\bibfnamefont {A.}~\bibnamefont {Rosch}},
  \bibinfo {author} {\bibfnamefont {T.~A.}\ \bibnamefont {Costi}}, \bibinfo
  {author} {\bibfnamefont {L.}~\bibnamefont {Saminadayar}},\ and\ \bibinfo
  {author} {\bibfnamefont {C.}~\bibnamefont {B\"auerle}},\ }\bibfield  {title}
  {\bibinfo {title} {Scaling of the low-temperature dephasing rate in kondo
  systems},\ }\href {https://doi.org/10.1103/PhysRevLett.97.226804} {\bibfield
  {journal} {\bibinfo  {journal} {Phys. Rev. Lett.}\ }\textbf {\bibinfo
  {volume} {97}},\ \bibinfo {pages} {226804} (\bibinfo {year}
  {2006})}\BibitemShut {NoStop}%
\bibitem [{\citenamefont {Borda}\ \emph {et~al.}(2007)\citenamefont {Borda},
  \citenamefont {Fritz}, \citenamefont {Andrei},\ and\ \citenamefont
  {Zar\'and}}]{borda_theory_2007}%
  \BibitemOpen
  \bibfield  {author} {\bibinfo {author} {\bibfnamefont {L.}~\bibnamefont
  {Borda}}, \bibinfo {author} {\bibfnamefont {L.}~\bibnamefont {Fritz}},
  \bibinfo {author} {\bibfnamefont {N.}~\bibnamefont {Andrei}},\ and\ \bibinfo
  {author} {\bibfnamefont {G.}~\bibnamefont {Zar\'and}},\ }\bibfield  {title}
  {\bibinfo {title} {Theory of inelastic scattering from quantum impurities},\
  }\href {https://doi.org/10.1103/PhysRevB.75.235112} {\bibfield  {journal}
  {\bibinfo  {journal} {Phys. Rev. B}\ }\textbf {\bibinfo {volume} {75}},\
  \bibinfo {pages} {235112} (\bibinfo {year} {2007})}\BibitemShut {NoStop}%
\bibitem [{\citenamefont {Freton}\ and\ \citenamefont
  {Boulat}(2014)}]{freton_out_of_eq_2014}%
  \BibitemOpen
  \bibfield  {author} {\bibinfo {author} {\bibfnamefont {L.}~\bibnamefont
  {Freton}}\ and\ \bibinfo {author} {\bibfnamefont {E.}~\bibnamefont
  {Boulat}},\ }\bibfield  {title} {\bibinfo {title} {Out-of-equilibrium
  properties and nonlinear effects for interacting quantum impurity systems in
  the strong-coupling regime},\ }\href
  {https://doi.org/10.1103/PhysRevLett.112.216802} {\bibfield  {journal}
  {\bibinfo  {journal} {Phys. Rev. Lett.}\ }\textbf {\bibinfo {volume} {112}},\
  \bibinfo {pages} {216802} (\bibinfo {year} {2014})}\BibitemShut {NoStop}%
\bibitem [{\citenamefont {Bera}\ \emph {et~al.}(2016)\citenamefont {Bera},
  \citenamefont {Baranger},\ and\ \citenamefont
  {Florens}}]{bera_dynamics_2016}%
  \BibitemOpen
  \bibfield  {author} {\bibinfo {author} {\bibfnamefont {S.}~\bibnamefont
  {Bera}}, \bibinfo {author} {\bibfnamefont {H.~U.}\ \bibnamefont {Baranger}},\
  and\ \bibinfo {author} {\bibfnamefont {S.}~\bibnamefont {Florens}},\
  }\bibfield  {title} {\bibinfo {title} {Dynamics of a qubit in a
  high-impedance transmission line from a bath perspective},\ }\href
  {https://doi.org/10.1103/PhysRevA.93.033847} {\bibfield  {journal} {\bibinfo
  {journal} {Phys. Rev. A}\ }\textbf {\bibinfo {volume} {93}},\ \bibinfo
  {pages} {033847} (\bibinfo {year} {2016})}\BibitemShut {NoStop}%
\bibitem [{\citenamefont {Schmid}(1983)}]{schmid_diffusion_1983}%
  \BibitemOpen
  \bibfield  {author} {\bibinfo {author} {\bibfnamefont {A.}~\bibnamefont
  {Schmid}},\ }\bibfield  {title} {\bibinfo {title} {Diffusion and
  {Localization} in a {Dissipative} {Quantum} {System}},\ }\href
  {https://doi.org/10.1103/PhysRevLett.51.1506} {\bibfield  {journal} {\bibinfo
   {journal} {Phys. Rev. Lett.}\ }\textbf {\bibinfo {volume} {51}},\ \bibinfo
  {pages} {1506} (\bibinfo {year} {1983})}\BibitemShut {NoStop}%
\bibitem [{\citenamefont {Bulgadaev}(1984)}]{bulgadaev_phase_1984}%
  \BibitemOpen
  \bibfield  {author} {\bibinfo {author} {\bibfnamefont {S.~A.}\ \bibnamefont
  {Bulgadaev}},\ }\bibfield  {title} {\bibinfo {title} {Phase {Diagram} of a
  {Dissipative} {Quantum} {System}},\ }\href
  {http://jetpletters.ru/ps/1289/article_19477.shtml} {\bibfield  {journal}
  {\bibinfo  {journal} {{JETP} {Letters}}\ }\textbf {\bibinfo {volume} {39}},\
  \bibinfo {pages} {264} (\bibinfo {year} {1984})}\BibitemShut {NoStop}%
\bibitem [{\citenamefont {Murani}\ \emph {et~al.}(2020)\citenamefont {Murani},
  \citenamefont {Bourlet}, \citenamefont {le~Sueur}, \citenamefont {Portier},
  \citenamefont {Altimiras}, \citenamefont {Esteve}, \citenamefont {Grabert},
  \citenamefont {Stockburger}, \citenamefont {Ankerhold},\ and\ \citenamefont
  {Joyez}}]{murani_absence_2020}%
  \BibitemOpen
  \bibfield  {author} {\bibinfo {author} {\bibfnamefont {A.}~\bibnamefont
  {Murani}}, \bibinfo {author} {\bibfnamefont {N.}~\bibnamefont {Bourlet}},
  \bibinfo {author} {\bibfnamefont {H.}~\bibnamefont {le~Sueur}}, \bibinfo
  {author} {\bibfnamefont {F.}~\bibnamefont {Portier}}, \bibinfo {author}
  {\bibfnamefont {C.}~\bibnamefont {Altimiras}}, \bibinfo {author}
  {\bibfnamefont {D.}~\bibnamefont {Esteve}}, \bibinfo {author} {\bibfnamefont
  {H.}~\bibnamefont {Grabert}}, \bibinfo {author} {\bibfnamefont
  {J.}~\bibnamefont {Stockburger}}, \bibinfo {author} {\bibfnamefont
  {J.}~\bibnamefont {Ankerhold}},\ and\ \bibinfo {author} {\bibfnamefont
  {P.}~\bibnamefont {Joyez}},\ }\bibfield  {title} {\bibinfo {title} {Absence
  of a {Dissipative} {Quantum} {Phase} {Transition} in {Josephson}
  {Junctions}},\ }\href {https://doi.org/10.1103/PhysRevX.10.021003} {\bibfield
   {journal} {\bibinfo  {journal} {Phys. Rev. X}\ }\textbf {\bibinfo {volume}
  {10}},\ \bibinfo {pages} {021003} (\bibinfo {year} {2020})}\BibitemShut
  {NoStop}%
\bibitem [{\citenamefont {Hakonen}\ and\ \citenamefont
  {Sonin}(2021)}]{hakonen_comment_2021}%
  \BibitemOpen
  \bibfield  {author} {\bibinfo {author} {\bibfnamefont {P.~J.}\ \bibnamefont
  {Hakonen}}\ and\ \bibinfo {author} {\bibfnamefont {E.~B.}\ \bibnamefont
  {Sonin}},\ }\bibfield  {title} {\bibinfo {title} {Comment on ``{Absence} of a
  {Dissipative} {Quantum} {Phase} {Transition} in {Josephson} {Junctions}''},\
  }\href {https://doi.org/10.1103/PhysRevX.11.018001} {\bibfield  {journal}
  {\bibinfo  {journal} {Phys. Rev. X}\ }\textbf {\bibinfo {volume} {11}},\
  \bibinfo {pages} {018001} (\bibinfo {year} {2021})}\BibitemShut {NoStop}%
\bibitem [{\citenamefont {Murani}\ \emph {et~al.}(2021)\citenamefont {Murani},
  \citenamefont {Bourlet}, \citenamefont {le~Sueur}, \citenamefont {Portier},
  \citenamefont {Altimiras}, \citenamefont {Esteve}, \citenamefont {Grabert},
  \citenamefont {Stockburger}, \citenamefont {Ankerhold},\ and\ \citenamefont
  {Joyez}}]{murani_reply_2021}%
  \BibitemOpen
  \bibfield  {author} {\bibinfo {author} {\bibfnamefont {A.}~\bibnamefont
  {Murani}}, \bibinfo {author} {\bibfnamefont {N.}~\bibnamefont {Bourlet}},
  \bibinfo {author} {\bibfnamefont {H.}~\bibnamefont {le~Sueur}}, \bibinfo
  {author} {\bibfnamefont {F.}~\bibnamefont {Portier}}, \bibinfo {author}
  {\bibfnamefont {C.}~\bibnamefont {Altimiras}}, \bibinfo {author}
  {\bibfnamefont {D.}~\bibnamefont {Esteve}}, \bibinfo {author} {\bibfnamefont
  {H.}~\bibnamefont {Grabert}}, \bibinfo {author} {\bibfnamefont
  {J.}~\bibnamefont {Stockburger}}, \bibinfo {author} {\bibfnamefont
  {J.}~\bibnamefont {Ankerhold}},\ and\ \bibinfo {author} {\bibfnamefont
  {P.}~\bibnamefont {Joyez}},\ }\bibfield  {title} {\bibinfo {title} {Reply to
  `{Comment} on ``{Absence} of a {Dissipative} {Quantum} {Phase} {Transition}
  in {Josephson} {Junctions}'''},\ }\href
  {https://doi.org/10.1103/PhysRevX.11.018002} {\bibfield  {journal} {\bibinfo
  {journal} {Phys. Rev. X}\ }\textbf {\bibinfo {volume} {11}},\ \bibinfo
  {pages} {018002} (\bibinfo {year} {2021})}\BibitemShut {NoStop}%
\bibitem [{\citenamefont {Masuki}\ \emph {et~al.}(2022)\citenamefont {Masuki},
  \citenamefont {Sudo}, \citenamefont {Oshikawa},\ and\ \citenamefont
  {Ashida}}]{masuki_absence_2022}%
  \BibitemOpen
  \bibfield  {author} {\bibinfo {author} {\bibfnamefont {K.}~\bibnamefont
  {Masuki}}, \bibinfo {author} {\bibfnamefont {H.}~\bibnamefont {Sudo}},
  \bibinfo {author} {\bibfnamefont {M.}~\bibnamefont {Oshikawa}},\ and\
  \bibinfo {author} {\bibfnamefont {Y.}~\bibnamefont {Ashida}},\ }\bibfield
  {title} {\bibinfo {title} {Absence versus {Presence} of {Dissipative}
  {Quantum} {Phase} {Transition} in {Josephson} {Junctions}},\ }\href
  {https://doi.org/10.1103/PhysRevLett.129.087001} {\bibfield  {journal}
  {\bibinfo  {journal} {Phys. Rev. Lett.}\ }\textbf {\bibinfo {volume} {129}},\
  \bibinfo {pages} {087001} (\bibinfo {year} {2022})}\BibitemShut {NoStop}%
\bibitem [{\citenamefont {S\'epulcre}\ \emph {et~al.}(2023)\citenamefont
  {S\'epulcre}, \citenamefont {Florens},\ and\ \citenamefont
  {Snyman}}]{sepulcre_comment_2022}%
  \BibitemOpen
  \bibfield  {author} {\bibinfo {author} {\bibfnamefont {T.}~\bibnamefont
  {S\'epulcre}}, \bibinfo {author} {\bibfnamefont {S.}~\bibnamefont
  {Florens}},\ and\ \bibinfo {author} {\bibfnamefont {I.}~\bibnamefont
  {Snyman}},\ }\bibfield  {title} {\bibinfo {title} {Comment on ``absence
  versus presence of dissipative quantum phase transition in josephson
  junctions''},\ }\href {https://doi.org/10.1103/PhysRevLett.131.199701}
  {\bibfield  {journal} {\bibinfo  {journal} {Phys. Rev. Lett.}\ }\textbf
  {\bibinfo {volume} {131}},\ \bibinfo {pages} {199701} (\bibinfo {year}
  {2023})}\BibitemShut {NoStop}%
\bibitem [{\citenamefont {Masuki}\ \emph {et~al.}(2023)\citenamefont {Masuki},
  \citenamefont {Sudo}, \citenamefont {Oshikawa},\ and\ \citenamefont
  {Ashida}}]{masuki_reply_2022}%
  \BibitemOpen
  \bibfield  {author} {\bibinfo {author} {\bibfnamefont {K.}~\bibnamefont
  {Masuki}}, \bibinfo {author} {\bibfnamefont {H.}~\bibnamefont {Sudo}},
  \bibinfo {author} {\bibfnamefont {M.}~\bibnamefont {Oshikawa}},\ and\
  \bibinfo {author} {\bibfnamefont {Y.}~\bibnamefont {Ashida}},\ }\bibfield
  {title} {\bibinfo {title} {Masuki et al. reply:},\ }\href
  {https://doi.org/10.1103/PhysRevLett.131.199702} {\bibfield  {journal}
  {\bibinfo  {journal} {Phys. Rev. Lett.}\ }\textbf {\bibinfo {volume} {131}},\
  \bibinfo {pages} {199702} (\bibinfo {year} {2023})}\BibitemShut {NoStop}%
\bibitem [{\citenamefont {Subero}\ \emph {et~al.}(2023)\citenamefont {Subero},
  \citenamefont {Maillet}, \citenamefont {Golubev}, \citenamefont {Thomas},
  \citenamefont {Peltonen}, \citenamefont {Karimi}, \citenamefont
  {Mar{\'i}n-Su{\'a}rez}, \citenamefont {Yeyati}, \citenamefont {S{\'a}nchez},
  \citenamefont {Park},\ and\ \citenamefont {Pekola}}]{subero_bolometric_2022}%
  \BibitemOpen
  \bibfield  {author} {\bibinfo {author} {\bibfnamefont {D.}~\bibnamefont
  {Subero}}, \bibinfo {author} {\bibfnamefont {O.}~\bibnamefont {Maillet}},
  \bibinfo {author} {\bibfnamefont {D.~S.}\ \bibnamefont {Golubev}}, \bibinfo
  {author} {\bibfnamefont {G.}~\bibnamefont {Thomas}}, \bibinfo {author}
  {\bibfnamefont {J.~T.}\ \bibnamefont {Peltonen}}, \bibinfo {author}
  {\bibfnamefont {B.}~\bibnamefont {Karimi}}, \bibinfo {author} {\bibfnamefont
  {M.}~\bibnamefont {Mar{\'i}n-Su{\'a}rez}}, \bibinfo {author} {\bibfnamefont
  {A.~L.}\ \bibnamefont {Yeyati}}, \bibinfo {author} {\bibfnamefont
  {R.}~\bibnamefont {S{\'a}nchez}}, \bibinfo {author} {\bibfnamefont
  {S.}~\bibnamefont {Park}},\ and\ \bibinfo {author} {\bibfnamefont {J.~P.}\
  \bibnamefont {Pekola}},\ }\bibfield  {title} {\bibinfo {title} {Bolometric
  detection of josephson inductance in a highly resistive environment},\ }\href
  {https://doi.org/10.1038/s41467-023-43668-3} {\bibfield  {journal} {\bibinfo
  {journal} {Nature Communications}\ }\textbf {\bibinfo {volume} {14}},\
  \bibinfo {pages} {7924} (\bibinfo {year} {2023})}\BibitemShut {NoStop}%
\bibitem [{\citenamefont {Giacomelli}\ and\ \citenamefont
  {Ciuti}(2023)}]{giacomelli_emergent_2023}%
  \BibitemOpen
  \bibfield  {author} {\bibinfo {author} {\bibfnamefont {L.}~\bibnamefont
  {Giacomelli}}\ and\ \bibinfo {author} {\bibfnamefont {C.}~\bibnamefont
  {Ciuti}},\ }\href {https://arxiv.org/abs/2307.06383} {\bibinfo {title}
  {Emergent quantum phase transition of a {Josephson} junction coupled to a
  high-impedance multimode resonator}} (\bibinfo {year} {2023}),\ \bibinfo
  {note} {arXiv:2307.06383}\BibitemShut {NoStop}%
\bibitem [{\citenamefont {Smirnov}(1992)}]{smirnov_form_1992}%
  \BibitemOpen
  \bibfield  {author} {\bibinfo {author} {\bibfnamefont {F.~A.}\ \bibnamefont
  {Smirnov}},\ }\href {https://doi.org/10.1142/1115} {{\selectlanguage
  {English}\emph {\bibinfo {title} {Form {Factors} in {Completely} {Integrable}
  {Models} of {Quantum} {Field} {Theory}}}}},\ \bibinfo {series} {Advanced
  {Series} in {Mathematical} {Physics}}, Vol.~\bibinfo {volume} {14}\ (\bibinfo
   {publisher} {World {Scientific}},\ \bibinfo {year} {1992})\BibitemShut
  {NoStop}%
\bibitem [{\citenamefont {Fendley}\ \emph {et~al.}(1994)\citenamefont
  {Fendley}, \citenamefont {Saleur},\ and\ \citenamefont
  {Warner}}]{fendley_exact_1994}%
  \BibitemOpen
  \bibfield  {author} {\bibinfo {author} {\bibfnamefont {P.}~\bibnamefont
  {Fendley}}, \bibinfo {author} {\bibfnamefont {H.}~\bibnamefont {Saleur}},\
  and\ \bibinfo {author} {\bibfnamefont {N.~P.}\ \bibnamefont {Warner}},\
  }\bibfield  {title} {{\selectlanguage {English}\bibinfo {title} {Exact
  solution of a massless scalar field with a relevant boundary interaction}},\
  }\href {https://doi.org/10.1016/0550-3213(94)90160-0} {\bibfield  {journal}
  {\bibinfo  {journal} {Nuclear Physics B}\ }\textbf {\bibinfo {volume}
  {430}},\ \bibinfo {pages} {577} (\bibinfo {year} {1994})}\BibitemShut
  {NoStop}%
\bibitem [{\citenamefont {Lesage}\ \emph
  {et~al.}(1996{\natexlab{a}})\citenamefont {Lesage}, \citenamefont {Saleur},\
  and\ \citenamefont {Skorik}}]{lesage_form_1996}%
  \BibitemOpen
  \bibfield  {author} {\bibinfo {author} {\bibfnamefont {F.}~\bibnamefont
  {Lesage}}, \bibinfo {author} {\bibfnamefont {H.}~\bibnamefont {Saleur}},\
  and\ \bibinfo {author} {\bibfnamefont {S.}~\bibnamefont {Skorik}},\
  }\bibfield  {title} {{\selectlanguage {English}\bibinfo {title} {Form factors
  approach to current correlations in one-dimensional systems with
  impurities}},\ }\href {https://doi.org/10.1016/0550-3213(96)00234-9}
  {\bibfield  {journal} {\bibinfo  {journal} {Nuclear Physics B}\ }\textbf
  {\bibinfo {volume} {474}},\ \bibinfo {pages} {602} (\bibinfo {year}
  {1996}{\natexlab{a}})}\BibitemShut {NoStop}%
\bibitem [{\citenamefont {Lesage}\ \emph
  {et~al.}(1996{\natexlab{b}})\citenamefont {Lesage}, \citenamefont {Saleur},\
  and\ \citenamefont {Skorik}}]{lesage_time_1996}%
  \BibitemOpen
  \bibfield  {author} {\bibinfo {author} {\bibfnamefont {F.}~\bibnamefont
  {Lesage}}, \bibinfo {author} {\bibfnamefont {H.}~\bibnamefont {Saleur}},\
  and\ \bibinfo {author} {\bibfnamefont {S.}~\bibnamefont {Skorik}},\
  }\bibfield  {title} {\bibinfo {title} {Time {Correlations} in {1D} {Quantum}
  {Impurity} {Problems}},\ }\href {https://doi.org/10.1103/PhysRevLett.76.3388}
  {\bibfield  {journal} {\bibinfo  {journal} {Phys. Rev. Lett.}\ }\textbf
  {\bibinfo {volume} {76}},\ \bibinfo {pages} {3388} (\bibinfo {year}
  {1996}{\natexlab{b}})}\BibitemShut {NoStop}%
\bibitem [{\citenamefont {Babujian}\ \emph {et~al.}(2017)\citenamefont
  {Babujian}, \citenamefont {Karowski},\ and\ \citenamefont
  {Tsvelik}}]{babujian_multipoint_2017}%
  \BibitemOpen
  \bibfield  {author} {\bibinfo {author} {\bibfnamefont {H.~M.}\ \bibnamefont
  {Babujian}}, \bibinfo {author} {\bibfnamefont {M.}~\bibnamefont {Karowski}},\
  and\ \bibinfo {author} {\bibfnamefont {A.~M.}\ \bibnamefont {Tsvelik}},\
  }\bibfield  {title} {{\selectlanguage {English}\bibinfo {title} {Multipoint
  {Green}'s functions in 1+1 dimensional integrable quantum field theories}},\
  }\href {https://doi.org/10.1016/j.nuclphysb.2017.02.002} {\bibfield
  {journal} {\bibinfo  {journal} {Nuclear Physics B}\ }\textbf {\bibinfo
  {volume} {917}},\ \bibinfo {pages} {122} (\bibinfo {year}
  {2017})}\BibitemShut {NoStop}%
\bibitem [{\citenamefont {Squarcini}(2021)}]{squarcini_multipoint_2021}%
  \BibitemOpen
  \bibfield  {author} {\bibinfo {author} {\bibfnamefont {A.}~\bibnamefont
  {Squarcini}},\ }\bibfield  {title} {{\selectlanguage {English}\bibinfo
  {title} {Multipoint correlation functions at phase separation. {Exact}
  results from field theory}},\ }\href
  {https://doi.org/10.1007/JHEP11(2021)096} {\bibfield  {journal} {\bibinfo
  {journal} {Journal of High Energy Physics}\ }\textbf {\bibinfo {volume}
  {2021}},\ \bibinfo {pages} {96} (\bibinfo {year} {2021})}\BibitemShut
  {NoStop}%
\bibitem [{\citenamefont {Fava}\ \emph {et~al.}(2023)\citenamefont {Fava},
  \citenamefont {Gopalakrishnan}, \citenamefont {Vasseur}, \citenamefont
  {Essler},\ and\ \citenamefont {Parameswaran}}]{fava_divergent_2023}%
  \BibitemOpen
  \bibfield  {author} {\bibinfo {author} {\bibfnamefont {M.}~\bibnamefont
  {Fava}}, \bibinfo {author} {\bibfnamefont {S.}~\bibnamefont
  {Gopalakrishnan}}, \bibinfo {author} {\bibfnamefont {R.}~\bibnamefont
  {Vasseur}}, \bibinfo {author} {\bibfnamefont {F.}~\bibnamefont {Essler}},\
  and\ \bibinfo {author} {\bibfnamefont {S.~A.}\ \bibnamefont {Parameswaran}},\
  }\bibfield  {title} {\bibinfo {title} {Divergent nonlinear response from
  quasiparticle interactions},\ }\href
  {https://doi.org/10.1103/PhysRevLett.131.256505} {\bibfield  {journal}
  {\bibinfo  {journal} {Phys. Rev. Lett.}\ }\textbf {\bibinfo {volume} {131}},\
  \bibinfo {pages} {256505} (\bibinfo {year} {2023})}\BibitemShut {NoStop}%
\bibitem [{\citenamefont {Kuzmin}\ \emph
  {et~al.}(2019{\natexlab{b}})\citenamefont {Kuzmin}, \citenamefont {Mehta},
  \citenamefont {Grabon}, \citenamefont {Mencia},\ and\ \citenamefont
  {Manucharyan}}]{kuzmin_superstrong_2019}%
  \BibitemOpen
  \bibfield  {author} {\bibinfo {author} {\bibfnamefont {R.}~\bibnamefont
  {Kuzmin}}, \bibinfo {author} {\bibfnamefont {N.}~\bibnamefont {Mehta}},
  \bibinfo {author} {\bibfnamefont {N.}~\bibnamefont {Grabon}}, \bibinfo
  {author} {\bibfnamefont {R.}~\bibnamefont {Mencia}},\ and\ \bibinfo {author}
  {\bibfnamefont {V.~E.}\ \bibnamefont {Manucharyan}},\ }\bibfield  {title}
  {{\selectlanguage {English}\bibinfo {title} {Superstrong coupling in circuit
  quantum electrodynamics}},\ }\href
  {https://doi.org/10.1038/s41534-019-0134-2} {\bibfield  {journal} {\bibinfo
  {journal} {npj Quantum Information}\ }\textbf {\bibinfo {volume} {5}},\
  \bibinfo {pages} {1} (\bibinfo {year} {2019}{\natexlab{b}})}\BibitemShut
  {NoStop}%
\bibitem [{\citenamefont {Manucharyan}\ \emph {et~al.}(2009)\citenamefont
  {Manucharyan}, \citenamefont {Koch}, \citenamefont {Glazman},\ and\
  \citenamefont {Devoret}}]{manucharyan_fluxonium_2009}%
  \BibitemOpen
  \bibfield  {author} {\bibinfo {author} {\bibfnamefont {V.~E.}\ \bibnamefont
  {Manucharyan}}, \bibinfo {author} {\bibfnamefont {J.}~\bibnamefont {Koch}},
  \bibinfo {author} {\bibfnamefont {L.~I.}\ \bibnamefont {Glazman}},\ and\
  \bibinfo {author} {\bibfnamefont {M.~H.}\ \bibnamefont {Devoret}},\
  }\bibfield  {title} {\bibinfo {title} {Fluxonium: {Single} {Cooper}-{Pair}
  {Circuit} {Free} of {Charge} {Offsets}},\ }\href
  {https://doi.org/10.1126/science.1175552} {\bibfield  {journal} {\bibinfo
  {journal} {Science}\ }\textbf {\bibinfo {volume} {326}},\ \bibinfo {pages}
  {113} (\bibinfo {year} {2009})}\BibitemShut {NoStop}%
\bibitem [{\citenamefont {Leggett}\ \emph {et~al.}(1987)\citenamefont
  {Leggett}, \citenamefont {Chakravarty}, \citenamefont {Dorsey}, \citenamefont
  {Fisher}, \citenamefont {Garg},\ and\ \citenamefont
  {Zwerger}}]{leggett_dynamics_1987}%
  \BibitemOpen
  \bibfield  {author} {\bibinfo {author} {\bibfnamefont {A.~J.}\ \bibnamefont
  {Leggett}}, \bibinfo {author} {\bibfnamefont {S.}~\bibnamefont
  {Chakravarty}}, \bibinfo {author} {\bibfnamefont {A.~T.}\ \bibnamefont
  {Dorsey}}, \bibinfo {author} {\bibfnamefont {M.~P.~A.}\ \bibnamefont
  {Fisher}}, \bibinfo {author} {\bibfnamefont {A.}~\bibnamefont {Garg}},\ and\
  \bibinfo {author} {\bibfnamefont {W.}~\bibnamefont {Zwerger}},\ }\bibfield
  {title} {\bibinfo {title} {Dynamics of the dissipative two-state system},\
  }\href {https://doi.org/10.1103/RevModPhys.59.1} {\bibfield  {journal}
  {\bibinfo  {journal} {Rev. Mod. Phys.}\ }\textbf {\bibinfo {volume} {59}},\
  \bibinfo {pages} {1} (\bibinfo {year} {1987})}\BibitemShut {NoStop}%
\bibitem [{\citenamefont {Forn-D{\'i}az}\ \emph {et~al.}(2017)\citenamefont
  {Forn-D{\'i}az}, \citenamefont {Garc{\'i}a-Ripoll}, \citenamefont
  {Peropadre}, \citenamefont {Orgiazzi}, \citenamefont {Yurtalan},
  \citenamefont {Belyansky}, \citenamefont {Wilson},\ and\ \citenamefont
  {Lupascu}}]{forn_diaz_ultrastrong_2017}%
  \BibitemOpen
  \bibfield  {author} {\bibinfo {author} {\bibfnamefont {P.}~\bibnamefont
  {Forn-D{\'i}az}}, \bibinfo {author} {\bibfnamefont {J.~J.}\ \bibnamefont
  {Garc{\'i}a-Ripoll}}, \bibinfo {author} {\bibfnamefont {B.}~\bibnamefont
  {Peropadre}}, \bibinfo {author} {\bibfnamefont {J.-L.}\ \bibnamefont
  {Orgiazzi}}, \bibinfo {author} {\bibfnamefont {M.~A.}\ \bibnamefont
  {Yurtalan}}, \bibinfo {author} {\bibfnamefont {R.}~\bibnamefont {Belyansky}},
  \bibinfo {author} {\bibfnamefont {C.~M.}\ \bibnamefont {Wilson}},\ and\
  \bibinfo {author} {\bibfnamefont {A.}~\bibnamefont {Lupascu}},\ }\bibfield
  {title} {\bibinfo {title} {Ultrastrong coupling of a single artificial atom
  to an electromagnetic continuum in the nonperturbative regime},\ }\href
  {https://doi.org/10.1038/nphys3905} {\bibfield  {journal} {\bibinfo
  {journal} {Nature Physics}\ }\textbf {\bibinfo {volume} {13}},\ \bibinfo
  {pages} {39} (\bibinfo {year} {2017})}\BibitemShut {NoStop}%
\bibitem [{\citenamefont {Kaur}\ \emph {et~al.}(2021)\citenamefont {Kaur},
  \citenamefont {S\'epulcre}, \citenamefont {Roch}, \citenamefont {Snyman},
  \citenamefont {Florens},\ and\ \citenamefont {Bera}}]{kaur_spin-boson_2021}%
  \BibitemOpen
  \bibfield  {author} {\bibinfo {author} {\bibfnamefont {K.}~\bibnamefont
  {Kaur}}, \bibinfo {author} {\bibfnamefont {T.}~\bibnamefont {S\'epulcre}},
  \bibinfo {author} {\bibfnamefont {N.}~\bibnamefont {Roch}}, \bibinfo {author}
  {\bibfnamefont {I.}~\bibnamefont {Snyman}}, \bibinfo {author} {\bibfnamefont
  {S.}~\bibnamefont {Florens}},\ and\ \bibinfo {author} {\bibfnamefont
  {S.}~\bibnamefont {Bera}},\ }\bibfield  {title} {\bibinfo {title} {Spin-boson
  quantum phase transition in multilevel superconducting qubits},\ }\href
  {https://doi.org/10.1103/PhysRevLett.127.237702} {\bibfield  {journal}
  {\bibinfo  {journal} {Phys. Rev. Lett.}\ }\textbf {\bibinfo {volume} {127}},\
  \bibinfo {pages} {237702} (\bibinfo {year} {2021})}\BibitemShut {NoStop}%
\bibitem [{hou()}]{houzet_priv}%
  \BibitemOpen
  \bibinfo {note} {M. Houzet, private communication}\BibitemShut {NoStop}%
\bibitem [{\citenamefont {Kane}\ and\ \citenamefont
  {Fisher}(1992)}]{kane_transmission_1992}%
  \BibitemOpen
  \bibfield  {author} {\bibinfo {author} {\bibfnamefont {C.~L.}\ \bibnamefont
  {Kane}}\ and\ \bibinfo {author} {\bibfnamefont {M.~P.~A.}\ \bibnamefont
  {Fisher}},\ }\bibfield  {title} {\bibinfo {title} {Transmission through
  barriers and resonant tunneling in an interacting one-dimensional electron
  gas},\ }\href {https://doi.org/10.1103/PhysRevB.46.15233} {\bibfield
  {journal} {\bibinfo  {journal} {Phys. Rev. B}\ }\textbf {\bibinfo {volume}
  {46}},\ \bibinfo {pages} {15233} (\bibinfo {year} {1992})}\BibitemShut
  {NoStop}%
\bibitem [{\citenamefont {Yagi}\ \emph {et~al.}(1997)\citenamefont {Yagi},
  \citenamefont {Kobayashi},\ and\ \citenamefont {Ootuka}}]{yagi_phase_1997}%
  \BibitemOpen
  \bibfield  {author} {\bibinfo {author} {\bibfnamefont {R.}~\bibnamefont
  {Yagi}}, \bibinfo {author} {\bibfnamefont {S.-i.}\ \bibnamefont
  {Kobayashi}},\ and\ \bibinfo {author} {\bibfnamefont {Y.}~\bibnamefont
  {Ootuka}},\ }\bibfield  {title} {\bibinfo {title} {Phase {Diagram} for
  {Superconductor}-{Insulator} {Transition} in {Single} {Small} {Josephson}
  {Junctions} with {Shunt} {Resistor}},\ }\href
  {https://doi.org/10.1143/JPSJ.66.3722} {\bibfield  {journal} {\bibinfo
  {journal} {Journal of the Physical Society of Japan}\ }\textbf {\bibinfo
  {volume} {66}},\ \bibinfo {pages} {3722} (\bibinfo {year}
  {1997})}\BibitemShut {NoStop}%
\bibitem [{\citenamefont {Penttil\"a}\ \emph {et~al.}(1999)\citenamefont
  {Penttil\"a}, \citenamefont {Parts}, \citenamefont {Hakonen}, \citenamefont
  {Paalanen},\ and\ \citenamefont {Sonin}}]{penttila_superconductor_1999}%
  \BibitemOpen
  \bibfield  {author} {\bibinfo {author} {\bibfnamefont {J.~S.}\ \bibnamefont
  {Penttil\"a}}, \bibinfo {author} {\bibfnamefont {U.}~\bibnamefont {Parts}},
  \bibinfo {author} {\bibfnamefont {P.~J.}\ \bibnamefont {Hakonen}}, \bibinfo
  {author} {\bibfnamefont {M.~A.}\ \bibnamefont {Paalanen}},\ and\ \bibinfo
  {author} {\bibfnamefont {E.~B.}\ \bibnamefont {Sonin}},\ }\bibfield  {title}
  {\bibinfo {title} {``superconductor-{Insulator} {Transition}'' in a {Single}
  {Josephson} {Junction}},\ }\href
  {https://doi.org/10.1103/PhysRevLett.82.1004} {\bibfield  {journal} {\bibinfo
   {journal} {Phys. Rev. Lett.}\ }\textbf {\bibinfo {volume} {82}},\ \bibinfo
  {pages} {1004} (\bibinfo {year} {1999})}\BibitemShut {NoStop}%
\bibitem [{\citenamefont {Penttilä}\ \emph {et~al.}(2001)\citenamefont
  {Penttilä}, \citenamefont {Hakonen}, \citenamefont {Sonin},\ and\
  \citenamefont {Paalanen}}]{penttila_experiments_2001}%
  \BibitemOpen
  \bibfield  {author} {\bibinfo {author} {\bibfnamefont {J.~S.}\ \bibnamefont
  {Penttilä}}, \bibinfo {author} {\bibfnamefont {P.~J.}\ \bibnamefont
  {Hakonen}}, \bibinfo {author} {\bibfnamefont {E.~B.}\ \bibnamefont {Sonin}},\
  and\ \bibinfo {author} {\bibfnamefont {M.~A.}\ \bibnamefont {Paalanen}},\
  }\bibfield  {title} {{\selectlanguage {English}\bibinfo {title} {Experiments
  on {Dissipative} {Dynamics} of {Single} {Josephson} {Junctions}}},\ }\href
  {https://doi.org/10.1023/A:1012971500694} {\bibfield  {journal} {\bibinfo
  {journal} {Journal of Low Temperature Physics}\ }\textbf {\bibinfo {volume}
  {125}},\ \bibinfo {pages} {89} (\bibinfo {year} {2001})}\BibitemShut
  {NoStop}%
\bibitem [{\citenamefont {Watanabe}\ and\ \citenamefont
  {Haviland}(2003)}]{watanabe_quantum_2003}%
  \BibitemOpen
  \bibfield  {author} {\bibinfo {author} {\bibfnamefont {M.}~\bibnamefont
  {Watanabe}}\ and\ \bibinfo {author} {\bibfnamefont {D.~B.}\ \bibnamefont
  {Haviland}},\ }\bibfield  {title} {\bibinfo {title} {Quantum {Effects} in
  {Small}-{Capacitance} {Single} {Josephson} {Junctions}},\ }\href
  {https://doi.org/10.1103/PhysRevB.67.094505} {\bibfield  {journal} {\bibinfo
  {journal} {Phys. Rev. B}\ }\textbf {\bibinfo {volume} {67}},\ \bibinfo
  {pages} {094505} (\bibinfo {year} {2003})}\BibitemShut {NoStop}%
\bibitem [{\citenamefont {Dorey}(1997)}]{dorey_exact_1997}%
  \BibitemOpen
  \bibfield  {author} {\bibinfo {author} {\bibfnamefont {P.}~\bibnamefont
  {Dorey}},\ }\bibfield  {title} {{\selectlanguage {English}\bibinfo {title}
  {Exact {S}-matrices}},\ }in\ \href {https://doi.org/10.1007/BFb0105279}
  {{\selectlanguage {English}\emph {\bibinfo {booktitle} {Conformal {Field}
  {Theories} and {Integrable} {Models}}}}},\ \bibinfo {series and number}
  {Lecture {Notes} in {Physics}},\ \bibinfo {editor} {edited by\ \bibinfo
  {editor} {\bibfnamefont {Z.}~\bibnamefont {Horváth}}\ and\ \bibinfo {editor}
  {\bibfnamefont {L.}~\bibnamefont {Palla}}}\ (\bibinfo  {publisher}
  {Springer},\ \bibinfo {address} {Berlin, Heidelberg},\ \bibinfo {year}
  {1997})\ pp.\ \bibinfo {pages} {85--125}\BibitemShut {NoStop}%
\bibitem [{\citenamefont {Bombardelli}(2016)}]{bombardelli_s-matrices_2016}%
  \BibitemOpen
  \bibfield  {author} {\bibinfo {author} {\bibfnamefont {D.}~\bibnamefont
  {Bombardelli}},\ }\bibfield  {title} {{\selectlanguage {English}\bibinfo
  {title} {S-matrices and integrability}},\ }\href
  {https://doi.org/10.1088/1751-8113/49/32/323003} {\bibfield  {journal}
  {\bibinfo  {journal} {Journal of Physics A: Mathematical and Theoretical}\
  }\textbf {\bibinfo {volume} {49}},\ \bibinfo {pages} {323003} (\bibinfo
  {year} {2016})}\BibitemShut {NoStop}%
\bibitem [{\citenamefont {Zamolodchikov}(1995)}]{zamolodchikov_mass_1995}%
  \BibitemOpen
  \bibfield  {author} {\bibinfo {author} {\bibfnamefont {A.~B.}\ \bibnamefont
  {Zamolodchikov}},\ }\bibfield  {title} {\bibinfo {title} {Mass scale in the
  sine–gordon model and its reductions},\ }\href
  {https://doi.org/10.1142/S0217751X9500053X} {\bibfield  {journal} {\bibinfo
  {journal} {International Journal of Modern Physics A}\ }\textbf {\bibinfo
  {volume} {10}},\ \bibinfo {pages} {1125} (\bibinfo {year}
  {1995})}\BibitemShut {NoStop}%
\bibitem [{\citenamefont {Lesage}\ and\ \citenamefont
  {Saleur}(1999)}]{lesage_perturbation_1999}%
  \BibitemOpen
  \bibfield  {author} {\bibinfo {author} {\bibfnamefont {F.}~\bibnamefont
  {Lesage}}\ and\ \bibinfo {author} {\bibfnamefont {H.}~\bibnamefont
  {Saleur}},\ }\bibfield  {title} {\bibinfo {title} {Perturbation of infra-red
  fixed points and duality in quantum impurity problems},\ }\href
  {https://doi.org/https://doi.org/10.1016/S0550-3213(99)00076-0} {\bibfield
  {journal} {\bibinfo  {journal} {Nuclear Physics B}\ }\textbf {\bibinfo
  {volume} {546}},\ \bibinfo {pages} {585} (\bibinfo {year}
  {1999})}\BibitemShut {NoStop}%
\bibitem [{\citenamefont {Abramowitz}\ and\ \citenamefont
  {Stegun}(1965)}]{abramowitz_handbook_1965}%
  \BibitemOpen
  \bibfield  {author} {\bibinfo {author} {\bibfnamefont {M.}~\bibnamefont
  {Abramowitz}}\ and\ \bibinfo {author} {\bibfnamefont {I.~A.}\ \bibnamefont
  {Stegun}},\ }\href@noop {} {\emph {\bibinfo {title} {{Handbook of
  Mathematical Functions}}}}\ (\bibinfo  {publisher} {Dover, New York},\
  \bibinfo {year} {1965})\BibitemShut {NoStop}%
\bibitem [{\citenamefont {Gogolin}\ \emph {et~al.}(2004)\citenamefont
  {Gogolin}, \citenamefont {Nersesyan},\ and\ \citenamefont
  {Tsvelik}}]{gogolin_bosonization_2004}%
  \BibitemOpen
  \bibfield  {author} {\bibinfo {author} {\bibfnamefont {A.~O.}\ \bibnamefont
  {Gogolin}}, \bibinfo {author} {\bibfnamefont {A.~A.}\ \bibnamefont
  {Nersesyan}},\ and\ \bibinfo {author} {\bibfnamefont {A.~M.}\ \bibnamefont
  {Tsvelik}},\ }\href
  {https://www.cambridge.org/il/academic/subjects/physics/condensed-matter-physics-nanoscience-and-mesoscopic-physics/bosonization-and-strongly-correlated-systems?format=PB&isbn=9780521617192}
  {\emph {\bibinfo {title} {Bosonization and {Strongly} {Correlated}
  {Systems}}}},\ Cambridge {Monographs} on {Mathematical} {Physics}\ (\bibinfo
  {publisher} {Cambridge University Press},\ \bibinfo {address} {Cambridge},\
  \bibinfo {year} {2004})\BibitemShut {NoStop}%
\bibitem [{\citenamefont {Kamenev}(2011)}]{kamenev_field_2011}%
  \BibitemOpen
  \bibfield  {author} {\bibinfo {author} {\bibfnamefont {A.}~\bibnamefont
  {Kamenev}},\ }\href {https://doi.org/10.1017/CBO9781139003667} {\emph
  {\bibinfo {title} {Field {Theory} of {Non}-{Equilibrium} {Systems}}}}\
  (\bibinfo  {publisher} {Cambridge University Press},\ \bibinfo {address}
  {Cambridge},\ \bibinfo {year} {2011})\BibitemShut {NoStop}%
\bibitem [{\citenamefont {Chou}\ \emph {et~al.}(1985)\citenamefont {Chou},
  \citenamefont {Su}, \citenamefont {Hao},\ and\ \citenamefont
  {Yu}}]{chou_equilibrium_1985}%
  \BibitemOpen
  \bibfield  {author} {\bibinfo {author} {\bibfnamefont {K.-c.}\ \bibnamefont
  {Chou}}, \bibinfo {author} {\bibfnamefont {Z.-b.}\ \bibnamefont {Su}},
  \bibinfo {author} {\bibfnamefont {B.-l.}\ \bibnamefont {Hao}},\ and\ \bibinfo
  {author} {\bibfnamefont {L.}~\bibnamefont {Yu}},\ }\bibfield  {title}
  {{\selectlanguage {English}\bibinfo {title} {Equilibrium and nonequilibrium
  formalisms made unified}},\ }\href
  {https://doi.org/10.1016/0370-1573(85)90136-X} {\bibfield  {journal}
  {\bibinfo  {journal} {Physics Reports}\ }\textbf {\bibinfo {volume} {118}},\
  \bibinfo {pages} {1} (\bibinfo {year} {1985})}\BibitemShut {NoStop}%
\bibitem [{\citenamefont {Kugler}\ \emph {et~al.}(2021)\citenamefont {Kugler},
  \citenamefont {Lee},\ and\ \citenamefont {von
  Delft}}]{kugler_multipoint_2021}%
  \BibitemOpen
  \bibfield  {author} {\bibinfo {author} {\bibfnamefont {F.~B.}\ \bibnamefont
  {Kugler}}, \bibinfo {author} {\bibfnamefont {S.-S.~B.}\ \bibnamefont {Lee}},\
  and\ \bibinfo {author} {\bibfnamefont {J.}~\bibnamefont {von Delft}},\
  }\bibfield  {title} {\bibinfo {title} {Multipoint {Correlation} {Functions}:
  {Spectral} {Representation} and {Numerical} {Evaluation}},\ }\href
  {https://doi.org/10.1103/PhysRevX.11.041006} {\bibfield  {journal} {\bibinfo
  {journal} {Phys. Rev. X}\ }\textbf {\bibinfo {volume} {11}},\ \bibinfo
  {pages} {041006} (\bibinfo {year} {2021})}\BibitemShut {NoStop}%
\bibitem [{\citenamefont {LeClair}\ and\ \citenamefont
  {Mussardo}(1999)}]{leclair_finite_1999}%
  \BibitemOpen
  \bibfield  {author} {\bibinfo {author} {\bibfnamefont {A.}~\bibnamefont
  {LeClair}}\ and\ \bibinfo {author} {\bibfnamefont {G.}~\bibnamefont
  {Mussardo}},\ }\bibfield  {title} {\bibinfo {title} {Finite {Temperature}
  {Correlation} {Functions} in {Integrable} {QFT}},\ }\href
  {https://doi.org/https://doi.org/10.1016/S0550-3213(99)00280-1} {\bibfield
  {journal} {\bibinfo  {journal} {Nuclear Physics B}\ }\textbf {\bibinfo
  {volume} {552}},\ \bibinfo {pages} {624} (\bibinfo {year}
  {1999})}\BibitemShut {NoStop}%
\bibitem [{\citenamefont {Saleur}(2000)}]{saleur_comment_2000}%
  \BibitemOpen
  \bibfield  {author} {\bibinfo {author} {\bibfnamefont {H.}~\bibnamefont
  {Saleur}},\ }\bibfield  {title} {\bibinfo {title} {A {Comment} on {Finite}
  {Temperature} {Correlations} in {Integrable} {QFT}},\ }\href
  {https://doi.org/https://doi.org/10.1016/S0550-3213(99)00665-3} {\bibfield
  {journal} {\bibinfo  {journal} {Nuclear Physics B}\ }\textbf {\bibinfo
  {volume} {567}},\ \bibinfo {pages} {602} (\bibinfo {year}
  {2000})}\BibitemShut {NoStop}%
\bibitem [{\citenamefont {Doyon}(2007)}]{doyon_2007}%
  \BibitemOpen
  \bibfield  {author} {\bibinfo {author} {\bibfnamefont {B.}~\bibnamefont
  {Doyon}},\ }\bibfield  {title} {\bibinfo {title} {Finite-temperature form
  factors: a review.},\ }\href {http://eudml.org/doc/53921} {\bibfield
  {journal} {\bibinfo  {journal} {SIGMA. Symmetry, Integrability and Geometry:
  Methods and Applications}\ }\textbf {\bibinfo {volume} {3}},\ \bibinfo
  {pages} {Paper 011, 37} (\bibinfo {year} {2007})}\BibitemShut {NoStop}%
\bibitem [{\citenamefont {Essler}\ and\ \citenamefont
  {Konik}(2009)}]{essler_finite-temperature_2009}%
  \BibitemOpen
  \bibfield  {author} {\bibinfo {author} {\bibfnamefont {F.~H.~L.}\
  \bibnamefont {Essler}}\ and\ \bibinfo {author} {\bibfnamefont {R.~M.}\
  \bibnamefont {Konik}},\ }\bibfield  {title} {{\selectlanguage
  {English}\bibinfo {title} {Finite-temperature dynamical correlations in
  massive integrable quantum field theories}},\ }\href
  {https://doi.org/10.1088/1742-5468/2009/09/P09018} {\bibfield  {journal}
  {\bibinfo  {journal} {Journal of Statistical Mechanics: Theory and
  Experiment}\ }\textbf {\bibinfo {volume} {2009}},\ \bibinfo {pages} {P09018}
  (\bibinfo {year} {2009})}\BibitemShut {NoStop}%
\bibitem [{\citenamefont {Pozsgay}\ and\ \citenamefont
  {Takács}(2010)}]{pozsgay_form_factor_2010}%
  \BibitemOpen
  \bibfield  {author} {\bibinfo {author} {\bibfnamefont {B.}~\bibnamefont
  {Pozsgay}}\ and\ \bibinfo {author} {\bibfnamefont {G.}~\bibnamefont
  {Takács}},\ }\bibfield  {title} {\bibinfo {title} {Form {Factor} {Expansion}
  for {Thermal} {Correlators}},\ }\href
  {https://doi.org/10.1088/1742-5468/2010/11/P11012} {\bibfield  {journal}
  {\bibinfo  {journal} {Journal of Statistical Mechanics: Theory and
  Experiment}\ }\textbf {\bibinfo {volume} {2010}},\ \bibinfo {pages} {P11012}
  (\bibinfo {year} {2010})}\BibitemShut {NoStop}%
\bibitem [{\citenamefont {Pozsgay}\ and\ \citenamefont
  {Szécsényi}(2018)}]{pozsgay_leclair-mussardo_2018}%
  \BibitemOpen
  \bibfield  {author} {\bibinfo {author} {\bibfnamefont {B.}~\bibnamefont
  {Pozsgay}}\ and\ \bibinfo {author} {\bibfnamefont {I.}~\bibnamefont
  {Szécsényi}},\ }\bibfield  {title} {{\selectlanguage {English}\bibinfo
  {title} {{LeClair}-{Mussardo} {Series} for {Two}-{Point} {Functions} in
  {Integrable} {QFT}}},\ }\href {https://doi.org/10.1007/JHEP05(2018)170}
  {\bibfield  {journal} {\bibinfo  {journal} {Journal of High Energy Physics}\
  }\textbf {\bibinfo {volume} {2018}},\ \bibinfo {pages} {170} (\bibinfo {year}
  {2018})}\BibitemShut {NoStop}%
\bibitem [{\citenamefont {Cubero}\ and\ \citenamefont
  {Panfil}(2019)}]{cubero_thermodynamic_2019}%
  \BibitemOpen
  \bibfield  {author} {\bibinfo {author} {\bibfnamefont {A.~C.}\ \bibnamefont
  {Cubero}}\ and\ \bibinfo {author} {\bibfnamefont {M.}~\bibnamefont
  {Panfil}},\ }\bibfield  {title} {\bibinfo {title} {Thermodynamic {Bootstrap}
  {Program} for {Integrable} {QFT}'s: {Form} {Factors} and {Correlation}
  {Functions} at {Finite} {Energy} {Density}},\ }\href
  {https://doi.org/10.1007/JHEP01(2019)104} {\bibfield  {journal} {\bibinfo
  {journal} {Journal of High Energy Physics}\ }\textbf {\bibinfo {volume}
  {2019}},\ \bibinfo {pages} {104} (\bibinfo {year} {2019})}\BibitemShut
  {NoStop}%
\bibitem [{\citenamefont {Pozsgay}\ and\ \citenamefont
  {Takács}(2008{\natexlab{a}})}]{pozsgay_form_2008}%
  \BibitemOpen
  \bibfield  {author} {\bibinfo {author} {\bibfnamefont {B.}~\bibnamefont
  {Pozsgay}}\ and\ \bibinfo {author} {\bibfnamefont {G.}~\bibnamefont
  {Takács}},\ }\bibfield  {title} {{\selectlanguage {English}\bibinfo {title}
  {Form factors in finite volume {I}: {Form} factor bootstrap and truncated
  conformal space}},\ }\href {https://doi.org/10.1016/j.nuclphysb.2007.06.027}
  {\bibfield  {journal} {\bibinfo  {journal} {Nuclear Physics B}\ }\textbf
  {\bibinfo {volume} {788}},\ \bibinfo {pages} {167} (\bibinfo {year}
  {2008}{\natexlab{a}})}\BibitemShut {NoStop}%
\bibitem [{\citenamefont {Pozsgay}\ and\ \citenamefont
  {Takács}(2008{\natexlab{b}})}]{pozsgay_form_2008-1}%
  \BibitemOpen
  \bibfield  {author} {\bibinfo {author} {\bibfnamefont {B.}~\bibnamefont
  {Pozsgay}}\ and\ \bibinfo {author} {\bibfnamefont {G.}~\bibnamefont
  {Takács}},\ }\bibfield  {title} {{\selectlanguage {English}\bibinfo {title}
  {Form factors in finite volume {II}: {Disconnected} terms and finite
  temperature correlators}},\ }\href
  {https://doi.org/10.1016/j.nuclphysb.2007.07.008} {\bibfield  {journal}
  {\bibinfo  {journal} {Nuclear Physics B}\ }\textbf {\bibinfo {volume}
  {788}},\ \bibinfo {pages} {209} (\bibinfo {year}
  {2008}{\natexlab{b}})}\BibitemShut {NoStop}%
\bibitem [{\citenamefont {Delfino}\ \emph {et~al.}(1996)\citenamefont
  {Delfino}, \citenamefont {Mussardo},\ and\ \citenamefont
  {Simonetti}}]{delfino_nonintegrable_1996}%
  \BibitemOpen
  \bibfield  {author} {\bibinfo {author} {\bibfnamefont {G.}~\bibnamefont
  {Delfino}}, \bibinfo {author} {\bibfnamefont {G.}~\bibnamefont {Mussardo}},\
  and\ \bibinfo {author} {\bibfnamefont {P.}~\bibnamefont {Simonetti}},\
  }\bibfield  {title} {\bibinfo {title} {Non-integrable quantum field theories
  as perturbations of certain integrable models},\ }\href
  {https://doi.org/https://doi.org/10.1016/0550-3213(96)00265-9} {\bibfield
  {journal} {\bibinfo  {journal} {Nuclear Physics B}\ }\textbf {\bibinfo
  {volume} {473}},\ \bibinfo {pages} {469} (\bibinfo {year}
  {1996})}\BibitemShut {NoStop}%
\bibitem [{\citenamefont {Bard}\ \emph {et~al.}(2018)\citenamefont {Bard},
  \citenamefont {Protopopov},\ and\ \citenamefont {Mirlin}}]{bard_decay_2018}%
  \BibitemOpen
  \bibfield  {author} {\bibinfo {author} {\bibfnamefont {M.}~\bibnamefont
  {Bard}}, \bibinfo {author} {\bibfnamefont {I.~V.}\ \bibnamefont
  {Protopopov}},\ and\ \bibinfo {author} {\bibfnamefont {A.~D.}\ \bibnamefont
  {Mirlin}},\ }\bibfield  {title} {\bibinfo {title} {Decay of plasmonic waves
  in josephson junction chains},\ }\href
  {https://doi.org/10.1103/PhysRevB.98.224513} {\bibfield  {journal} {\bibinfo
  {journal} {Phys. Rev. B}\ }\textbf {\bibinfo {volume} {98}},\ \bibinfo
  {pages} {224513} (\bibinfo {year} {2018})}\BibitemShut {NoStop}%
\bibitem [{\citenamefont {Wu}\ and\ \citenamefont
  {Sau}(2019)}]{huan-kuang_theory_2019}%
  \BibitemOpen
  \bibfield  {author} {\bibinfo {author} {\bibfnamefont {H.-K.}\ \bibnamefont
  {Wu}}\ and\ \bibinfo {author} {\bibfnamefont {J.~D.}\ \bibnamefont {Sau}},\
  }\bibfield  {title} {\bibinfo {title} {Theory of coherent phase modes in
  insulating josephson junction chains},\ }\href
  {https://doi.org/10.1103/PhysRevB.99.214509} {\bibfield  {journal} {\bibinfo
  {journal} {Phys. Rev. B}\ }\textbf {\bibinfo {volume} {99}},\ \bibinfo
  {pages} {214509} (\bibinfo {year} {2019})}\BibitemShut {NoStop}%
\bibitem [{\citenamefont {Houzet}\ and\ \citenamefont
  {Glazman}(2019)}]{houzet_microwave_2019}%
  \BibitemOpen
  \bibfield  {author} {\bibinfo {author} {\bibfnamefont {M.}~\bibnamefont
  {Houzet}}\ and\ \bibinfo {author} {\bibfnamefont {L.~I.}\ \bibnamefont
  {Glazman}},\ }\bibfield  {title} {\bibinfo {title} {Microwave spectroscopy of
  a weakly pinned charge density wave in a superinductor},\ }\href
  {https://doi.org/10.1103/PhysRevLett.122.237701} {\bibfield  {journal}
  {\bibinfo  {journal} {Phys. Rev. Lett.}\ }\textbf {\bibinfo {volume} {122}},\
  \bibinfo {pages} {237701} (\bibinfo {year} {2019})}\BibitemShut {NoStop}%
\bibitem [{\citenamefont {Bloch}\ \emph {et~al.}(2008)\citenamefont {Bloch},
  \citenamefont {Dalibard},\ and\ \citenamefont
  {Zwerger}}]{bloch_many-body_2008}%
  \BibitemOpen
  \bibfield  {author} {\bibinfo {author} {\bibfnamefont {I.}~\bibnamefont
  {Bloch}}, \bibinfo {author} {\bibfnamefont {J.}~\bibnamefont {Dalibard}},\
  and\ \bibinfo {author} {\bibfnamefont {W.}~\bibnamefont {Zwerger}},\
  }\bibfield  {title} {\bibinfo {title} {Many-body physics with ultracold
  gases},\ }\href {https://doi.org/10.1103/RevModPhys.80.885} {\bibfield
  {journal} {\bibinfo  {journal} {Reviews of Modern Physics}\ }\textbf
  {\bibinfo {volume} {80}},\ \bibinfo {pages} {885} (\bibinfo {year}
  {2008})}\BibitemShut {NoStop}%
\bibitem [{\citenamefont {Schweigler}\ \emph {et~al.}(2021)\citenamefont
  {Schweigler}, \citenamefont {Gluza}, \citenamefont {Tajik}, \citenamefont
  {Sotiriadis}, \citenamefont {Cataldini}, \citenamefont {Ji}, \citenamefont
  {M{\o}ller}, \citenamefont {Sabino}, \citenamefont {Rauer}, \citenamefont
  {Eisert},\ and\ \citenamefont {Schmiedmayer}}]{schweigler_decay_2021}%
  \BibitemOpen
  \bibfield  {author} {\bibinfo {author} {\bibfnamefont {T.}~\bibnamefont
  {Schweigler}}, \bibinfo {author} {\bibfnamefont {M.}~\bibnamefont {Gluza}},
  \bibinfo {author} {\bibfnamefont {M.}~\bibnamefont {Tajik}}, \bibinfo
  {author} {\bibfnamefont {S.}~\bibnamefont {Sotiriadis}}, \bibinfo {author}
  {\bibfnamefont {F.}~\bibnamefont {Cataldini}}, \bibinfo {author}
  {\bibfnamefont {S.-C.}\ \bibnamefont {Ji}}, \bibinfo {author} {\bibfnamefont
  {F.~S.}\ \bibnamefont {M{\o}ller}}, \bibinfo {author} {\bibfnamefont
  {J.}~\bibnamefont {Sabino}}, \bibinfo {author} {\bibfnamefont
  {B.}~\bibnamefont {Rauer}}, \bibinfo {author} {\bibfnamefont
  {J.}~\bibnamefont {Eisert}},\ and\ \bibinfo {author} {\bibfnamefont
  {J.}~\bibnamefont {Schmiedmayer}},\ }\bibfield  {title} {\bibinfo {title}
  {Decay and recurrence of non-gaussian correlations in a quantum many-body
  system},\ }\href {https://doi.org/10.1038/s41567-020-01139-2} {\bibfield
  {journal} {\bibinfo  {journal} {Nature Physics}\ }\textbf {\bibinfo {volume}
  {17}},\ \bibinfo {pages} {559} (\bibinfo {year} {2021})}\BibitemShut
  {NoStop}%
\bibitem [{\citenamefont {Kukuljan}\ \emph {et~al.}(2018)\citenamefont
  {Kukuljan}, \citenamefont {Sotiriadis},\ and\ \citenamefont
  {Takacs}}]{kukuljan_correlation_2018}%
  \BibitemOpen
  \bibfield  {author} {\bibinfo {author} {\bibfnamefont {I.}~\bibnamefont
  {Kukuljan}}, \bibinfo {author} {\bibfnamefont {S.}~\bibnamefont
  {Sotiriadis}},\ and\ \bibinfo {author} {\bibfnamefont {G.}~\bibnamefont
  {Takacs}},\ }\bibfield  {title} {\bibinfo {title} {Correlation functions of
  the quantum sine-gordon model in and out of equilibrium},\ }\href
  {https://doi.org/10.1103/PhysRevLett.121.110402} {\bibfield  {journal}
  {\bibinfo  {journal} {Phys. Rev. Lett.}\ }\textbf {\bibinfo {volume} {121}},\
  \bibinfo {pages} {110402} (\bibinfo {year} {2018})}\BibitemShut {NoStop}%
\bibitem [{\citenamefont {\ifmmode \check{Z}\else
  \v{Z}\fi{}nidari\ifmmode~\check{c}\else
  \v{c}\fi{}}(2020)}]{znidaric_weak_2020}%
  \BibitemOpen
  \bibfield  {author} {\bibinfo {author} {\bibfnamefont {M.}~\bibnamefont
  {\ifmmode \check{Z}\else \v{Z}\fi{}nidari\ifmmode~\check{c}\else
  \v{c}\fi{}}},\ }\bibfield  {title} {\bibinfo {title} {Weak integrability
  breaking: Chaos with integrability signature in coherent diffusion},\ }\href
  {https://doi.org/10.1103/PhysRevLett.125.180605} {\bibfield  {journal}
  {\bibinfo  {journal} {Phys. Rev. Lett.}\ }\textbf {\bibinfo {volume} {125}},\
  \bibinfo {pages} {180605} (\bibinfo {year} {2020})}\BibitemShut {NoStop}%
\bibitem [{\citenamefont {Brenes}\ \emph {et~al.}(2020)\citenamefont {Brenes},
  \citenamefont {LeBlond}, \citenamefont {Goold},\ and\ \citenamefont
  {Rigol}}]{brenes_eigenstate_2020}%
  \BibitemOpen
  \bibfield  {author} {\bibinfo {author} {\bibfnamefont {M.}~\bibnamefont
  {Brenes}}, \bibinfo {author} {\bibfnamefont {T.}~\bibnamefont {LeBlond}},
  \bibinfo {author} {\bibfnamefont {J.}~\bibnamefont {Goold}},\ and\ \bibinfo
  {author} {\bibfnamefont {M.}~\bibnamefont {Rigol}},\ }\bibfield  {title}
  {\bibinfo {title} {Eigenstate thermalization in a locally perturbed
  integrable system},\ }\href {https://doi.org/10.1103/PhysRevLett.125.070605}
  {\bibfield  {journal} {\bibinfo  {journal} {Phys. Rev. Lett.}\ }\textbf
  {\bibinfo {volume} {125}},\ \bibinfo {pages} {070605} (\bibinfo {year}
  {2020})}\BibitemShut {NoStop}%
\bibitem [{\citenamefont {Bulchandani}\ \emph {et~al.}(2022)\citenamefont
  {Bulchandani}, \citenamefont {Huse},\ and\ \citenamefont
  {Gopalakrishnan}}]{bulchandani_onset_2022}%
  \BibitemOpen
  \bibfield  {author} {\bibinfo {author} {\bibfnamefont {V.~B.}\ \bibnamefont
  {Bulchandani}}, \bibinfo {author} {\bibfnamefont {D.~A.}\ \bibnamefont
  {Huse}},\ and\ \bibinfo {author} {\bibfnamefont {S.}~\bibnamefont
  {Gopalakrishnan}},\ }\bibfield  {title} {\bibinfo {title} {Onset of many-body
  quantum chaos due to breaking integrability},\ }\href
  {https://doi.org/10.1103/PhysRevB.105.214308} {\bibfield  {journal} {\bibinfo
   {journal} {Phys. Rev. B}\ }\textbf {\bibinfo {volume} {105}},\ \bibinfo
  {pages} {214308} (\bibinfo {year} {2022})}\BibitemShut {NoStop}%
\bibitem [{\citenamefont {D'Hoker}\ and\ \citenamefont
  {Freedman}(2002)}]{dhoker_SYM_2002}%
  \BibitemOpen
  \bibfield  {author} {\bibinfo {author} {\bibfnamefont {E.}~\bibnamefont
  {D'Hoker}}\ and\ \bibinfo {author} {\bibfnamefont {D.~Z.}\ \bibnamefont
  {Freedman}},\ }\bibfield  {title} {\bibinfo {title} {{Supersymmetric gauge
  theories and the AdS / CFT correspondence}},\ }in\ \href@noop {} {\emph
  {\bibinfo {booktitle} {{Theoretical Advanced Study Institute in Elementary
  Particle Physics (TASI 2001): Strings, Branes and EXTRA Dimensions}}}}\
  (\bibinfo {year} {2002})\ pp.\ \bibinfo {pages} {3--158},\ \Eprint
  {https://arxiv.org/abs/hep-th/0201253} {arXiv:hep-th/0201253} \BibitemShut
  {NoStop}%
\bibitem [{\citenamefont {Weiss}(1999)}]{weiss_quantum_1999}%
  \BibitemOpen
  \bibfield  {author} {\bibinfo {author} {\bibfnamefont {U.}~\bibnamefont
  {Weiss}},\ }\href {https://books.google.co.il/books?id=kqZclKUZdq0C} {\emph
  {\bibinfo {title} {Quantum {Dissipative} {Systems}}}},\ Series in modern
  condensed matter physics\ (\bibinfo  {publisher} {World Scientific},\
  \bibinfo {year} {1999})\BibitemShut {NoStop}%
\bibitem [{\citenamefont {Ghoshal}(1994)}]{ghoshal_bound_1994}%
  \BibitemOpen
  \bibfield  {author} {\bibinfo {author} {\bibfnamefont {S.}~\bibnamefont
  {Ghoshal}},\ }\bibfield  {title} {\bibinfo {title} {Bound state boundary s
  matrix of the sine-gordon model},\ }\href
  {https://doi.org/10.1142/S0217751X94001941} {\bibfield  {journal} {\bibinfo
  {journal} {International Journal of Modern Physics A}\ }\textbf {\bibinfo
  {volume} {09}},\ \bibinfo {pages} {4801} (\bibinfo {year}
  {1994})}\BibitemShut {NoStop}%
\bibitem [{\citenamefont {Weisz}(1977)}]{weisz_exact_1977}%
  \BibitemOpen
  \bibfield  {author} {\bibinfo {author} {\bibfnamefont {P.~H.}\ \bibnamefont
  {Weisz}},\ }\bibfield  {title} {{\selectlanguage {English}\bibinfo {title}
  {Exact quantum {Sine}-{Gordon} soliton form factors}},\ }\href
  {https://doi.org/10.1016/0370-2693(77)90097-1} {\bibfield  {journal}
  {\bibinfo  {journal} {Physics Letters B}\ }\textbf {\bibinfo {volume} {67}},\
  \bibinfo {pages} {179} (\bibinfo {year} {1977})}\BibitemShut {NoStop}%
\bibitem [{\citenamefont {Hewson}(1993)}]{hewson_kondo_1993}%
  \BibitemOpen
  \bibfield  {author} {\bibinfo {author} {\bibfnamefont {A.~C.}\ \bibnamefont
  {Hewson}},\ }\href {https://doi.org/10.1017/CBO9780511470752} {\emph
  {\bibinfo {title} {The {Kondo} {Problem} to {Heavy} {Fermions}}}},\ Cambridge
  Studies in Magnetism\ (\bibinfo  {publisher} {Cambridge University Press},\
  \bibinfo {year} {1993})\BibitemShut {NoStop}%
\end{thebibliography}%

\end{document}